\documentstyle[12pt,preprint,floats,epsf,aps,rmp]{revtex}
\newcommand{\et}{{\rm E}_{\scriptscriptstyle\rm T}}
\newcommand{\met}{\mbox{$\protect \raisebox{.3ex}{$\not$}\et$}}
\newcommand{\MET}{\mbox{$\protect \raisebox{.3ex}{$\not$}\et$}}

\def\gepsfcentered#1{
  \def\testit{#1}
  \def\lbracket{[}
  \ifx\testit\lbracket
    \let\dofilecmd=\gepsfwithopt
  \else
    \let\dofilecmd=\gepsfnoopt
  \fi
  \dofilecmd}

\def\gepsfnoopt#1{
  \begin{center}
  \leavevmode
  \epsffile{#1}
  \end{center}}

\def\gepsfwithopt#1 #2 #3 #4]#5{
  \begin{center}
  \leavevmode
  \gepsfmaxx=0.94\textwidth
  \epsffile[#1 #2 #3 #4]{#5}
  \end{center}}
\newdimen\gepsfmaxx
\def\epsfsize#1#2{
  \ifnum \epsfxsize=0
    \ifnum \epsfysize=0
      \ifnum #1 > \gepsfmaxx
        \gepsfmaxx
      \else
        #1
      \fi
    \else
      \epsfxsize
    \fi
  \else
    \epsfxsize
  \fi
}

\tighten
\begin{document}

\input feynman
\bigphotons
\input psfig
\psfull

\title{UCSB-HEP-96-01 \\ HUTP-96/A023 \\ 
{\em To be published in Reviews of Modern Physics} \\ \\
The discovery of the top quark}
\vskip 3cm

\author{Claudio Campagnari}
\address{University of California, Santa Barbara, California 93106}
\author{Melissa Franklin}
\address{Harvard University, Cambridge, Massachussets 02138}

\maketitle

\begin{abstract}
Evidence for pair production of a new particle consistent with 
the Standard Model top quark has been reported recently by groups studying
proton anti-proton collisions at 1.8 TeV center of mass energy at
the Fermi National Accelerator Laboratory. In this paper we review the history
of the search for the top quark in 
electron positron and proton anti-proton collisions.
We report on a number of precise electro-weak measurements 
and the value of the top quark mass which can be extracted from
these measurements within the context of
the Standard Model.
We review the theoretical predictions for top quark production
and the dominant backgrounds.
We describe the collider and the detectors that were used to
measure the pair production process and follow on to describe 
the data from which the existence of the top quark is evinced. 
Finally, we present possible measurements that could be
made in the future with more data, 
measurements of quantities that would confirm the
nature of this particle, the details of its production in hadron collisions,
and its decay properties.
\end{abstract}

\tableofcontents

\section{Introduction}
\label{intro}

The intensive experimental efforts in the search for
the heaviest fundamental fermion culminated in 1995
with the discovery of the top quark in proton-antiproton
annhilations at the Fermilab Tevatron Collider.   Observation
of the top quark is the latest in a long series of triumphs for 
the Standard Model of particles and fields.  
The top quark is the last fundamental
fermion and the next-to-last fundamental particle
predicted by the Standard Model.  Only the Higgs boson remains
unobserved.

The search for the top quark started in the late seventies, soon after
discovery of the companion bottom-quark.  It has been a long and arduous
process because the top quark turned out to be much more massive
than was originally expected.  
The mass of the top quark
is remarkably large, approximately 200 times larger than
the mass of the proton and 40 times higher than the mass of the next lightest
quark.  Whether this property of the top quark is a mere
accident or a manifestation of a deeper physical process is an
unanswered question in particle physics.

The discovery of the top quark has been made possible by the
technological progresses in high energy physics in
the past fifteen years.  In particular, the development of
proton-antiproton colliders, pioneered first at CERN and then
at Fermilab
has been a crucial ingredient in the discovery
of the top.

In this article we review the discovery of the top quark
as well as the developments that led to it.  In Section~\ref{role}
we discuss the top quark within the framework of the Standard Model.
While the top quark mass is a free parameter in the Standard Model, its
value enters in calculations of a number of electroweak
observables.  The top-mass dependence of the theoretical 
predictions is in general rather weak.
However, the accuracy of many of these
measurements is now such that meaningful constraints on the top 
quark mass can be obtained by comparing them with theoretical predictions. 
These constraints constitute a test of the predictive power of
the Standard Model, and are reviewed 
in Section~\ref{ind}.  The top production mechanisms
in proton-antiproton collisions, and
the experimental signatures of top events that 
are crucial to the understanding of
the experimental results are
discussed in Section~\ref{prod} and~\ref{sig}.
Early searches for the top quark are described in 
Section~\ref{pre1a}.
The Tevatron Collider, whose remarkable performance played
a very important role in the discovery of the top quark, will be described
briefly in Section~\ref{accel}.
The data that finally led to the discovery
of the top quark are reviewed in Section~\ref{discintro}.
The value of the top mass is of fundamental importance, and
it is needed to complete precise tests of the Standard Model. 
Furthermore, from an experimental point of view,
the techniques developed to measure the top mass are 
new and particularly interesting.  The top mass 
measurement is described in Section~\ref{mass}.  
Historically, discoveries of new leptons or quarks have opened
up new fields of inquiry which have enhanced our understanding
of elementary particles and their interactions.  Consequently, this
article concludes in Section~\ref{future} with a discussion of the
experimental prospects for top physics.

\section{The role of the top quark in the Standard Model}
\label{role}
\begin{figure}[htb]
\epsfxsize=4.0in
\gepsfcentered[200 200 720 720]{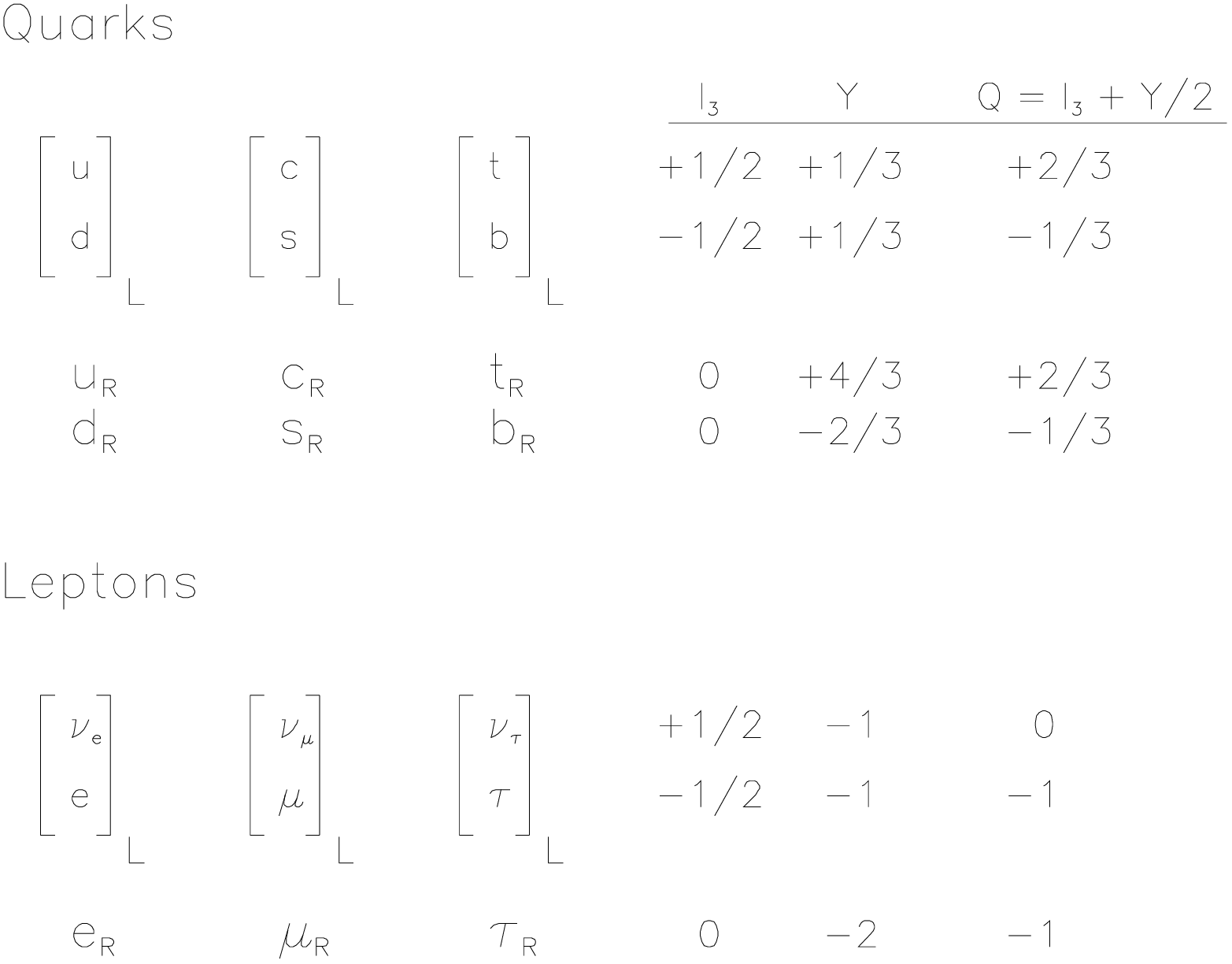}
\caption{Leptons and quarks in SU(2)xU(1) (Standard Model).  Also shown
are the values for the SU(2) weak isospin (I$_{3}$), U(1) weak
hypercharge (Y), and electric charge (Q, in units of the
electron charge).  The subscripts L and R refer to the left
and right-handed components respectively.}
\label{sm}
\end{figure}

Quarks and leptons constitute the basic building blocks of matter
in the Standard Model (SM).  There are three $generations$ of quarks and
leptons in the model, with identical quantum numbers, but different 
masses.  Within each generation, quarks and leptons appear in pairs, 
(see Fig.~\ref{sm}).  The left-handed quarks form weak 
isospin doublets, with the Q$=+2/3$ and Q$=-1/3$ quarks having weak
isospin I$_{3}=+1/2$ and $-1/2$ respectively.

The tau-lepton ($\tau$) was the first particle of the third
generation to be discovered (Perl {\em et al.}, 1975).
A short time later, in 1977,
the $\Upsilon$ was discovered at Fermilab
(Herb {\em et al.}, 1977)
as a resonance in the $\mu^{+}\mu^{-}$
invariant mass spectrum in the reaction p + nucleon $\rightarrow
\mu^{+}\mu^{-} +$ X.  This resonance was interpreted as
a $b\bar{b}$ bound state (the $\Upsilon$),
which subsequently decays into muon pairs.
As will become abundantly clear in the remainder of this paper,
the top signature in hadron collisions is much more complicated.

In the past fifteen years, 
a tremendous amount of experimental data on the properties of
the $b$-quark and of $b$-flavored hadrons have become available, mostly
from experiments at $e^{+}e^{-}$ colliders.  Both the charge
and the weak isospin of the bottom-quark are by now well established
(Q$_{b}=-1/3$, and I$_{3}=-1/2$).

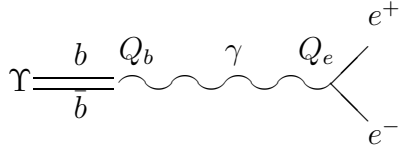
\begin{figure}
\hskip 2cm
\vskip -0.5cm
\begin{picture}(15000,15000)

\put(11000,9500){$\Upsilon$}
\drawline\fermion[\E\REG](12000,10000)[3000]
\put(\pmidx,10500){$b$}
\drawline\fermion[\SE\REG](\particlebackx,\particlebacky)[300]
\put(\pbackx,10700){$Q_{b}$}
\drawline\photon[\E\REG](\particlebackx,\particlebacky)[8]
\put(22000,10700){$Q_{e}$}
\put(\pmidx,10700){$\gamma$}
\drawline\fermion[\SW\REG](\photonfrontx,\photonfronty)[300]
\drawline\fermion[\W\REG](\particlebackx,\particlebacky)[3000]
\put(\pmidx,8500){$\bar{b}$}
\drawline\fermion[\NE\REG](\photonbackx,\photonbacky)[2000]
\put(\pbackx,12000){$e^{+}$}
\drawline\fermion[\SE\REG](\photonbackx,\photonbacky)[2000]
\put(\pbackx,7500){$e^{-}$}

\end{picture}
\label{upswidth}
\vskip -2cm
\caption{Feynman diagram for $\Upsilon \rightarrow$ e$^{+}$e$^{-}$.}
\end{figure}

\begin{table}
\begin{center}
\begin{tabular}{cccc} \hline \hline
Model 1  & Model 2 & Model 3 & Experimental value \\
1.05 keV & 1.07 keV & 1.15 $\pm$ 0.20 keV & 1.22  $\pm$ 0.03 keV \\
\hline \hline
\end{tabular}
\end{center}
\caption{The experimental value of 
$\Gamma(\Upsilon \rightarrow e^{+}e^{-})$ compared with
theoretical expectations from quark-antiquark potential models assuming
Q$_{b}=-$1/3.  Model 1 : Krasemann and Ono, 1979. 
Model 2 : B\"uchmuller, Grunberg and Tye, 1980.
Model 3 : Voloshin and Zakharov, 1980.  The quoted
experimental value is from the compilation of the 
Particle Data Group,
corrected by 7 \% for consistency in comparison with
potential model calculations, which only include the 
lowest order Born term (Montanet {\em et al.}, 1994).}
\label{wdtbl}
\end{table}

The value of the charge was first inferred from measurements
of the $\Upsilon$ leptonic width (Berger {\em et al.}, 1978;
Darden {\em et al.}, 1978; Bienlein {\em et al.}, 1978)
at the DORIS $e^{+}e^{-}$ storage ring.   
This width is proportional to the square of
the charge of the $b$-quark, (see Fig. 2), and can be
quantitatively estimated from heavy quark-antiquark
potential models, see Table~\ref{wdtbl} 
(Quigg and Rosner, 1977; Eichten and Gottfried, 1977;
Rosner, Quigg, and Thacker, 1978; Krasemann and Ono, 1979;
B\"uchmuller, Grunberg and Tye, 1980; Voloshin and
Zakharov, 1980).
The charge assignment
was subsequently confirmed by measurements of 
R $=\sigma(e^{+}e^{-} \rightarrow $ hadrons) $/ \sigma(e^{+}e^{-}
\rightarrow \mu^{+}\mu^{-}$).
At lowest order, and ignoring resonance effects, 
R = $\sum_{quarks} 3Q_{q}^{2}$, where the factor
of three arises from the fact that quarks come in three colors.
The sum is over all quarks that can be produced, i.e. all quarks
with mass below one-half the center of mass energy of the 
e$^{+}$e$^{-}$ system.  Above threshold for $b\bar{b}$ production,
the value of R was found to increase by 
$0.36 \pm 0.09 \pm 0.03$ (Rice {\em et al.}, 1982), in agreement
with the expectations of 3Q$_{b}^{2}$=1/3.
The pole mass of the $b$-quark is estimated to be in the
range 4.5 - 4.9 GeV/c$^{2}$ from knowledge of the $\Upsilon$ and 
$b$-meson masses (Montanet {\em et al.}, 1994).

The weak isospin of the $b$-quark was first extracted 
from the
forward-backward asymmetry  (A$_{FB}$)
in $e^{+}e^{-} \rightarrow b\bar{b}$.
This asymmetry is defined in terms of the $b$-quark production
cross section $\sigma(b)$ as

\begin{eqnarray*}
A_{FB}&=&\frac{\sigma(b,\theta > 90^{o}) - \sigma(b,\theta < 90^{o})}
{\sigma(b,\theta > 90^{o}) + \sigma(b,\theta < 90^{o})}
\end{eqnarray*}
where $\theta$ is the polar angle of the $b$-quark in the
e$^{+}$e$^{-}$ center of mass as measured from the direction
of flight of the e$^{-}$. 
The asymmetry originates from the coupling of the $Z$ to fermions, which
in the Standard Model depends on the weak isospin through a term in
the lagrangian of the form
$\bar{f} \gamma_{\mu}(g_{V} - g_{A}\gamma_{5}) Z^{\mu} f$, 
where the vector and axial couplings $g_{V}$
and $g_{a}$ are given by

\begin{eqnarray*}
g_{V}&=&\frac{I_{3} - 2 Q\sin^{2}\theta_{W}}{2\sin\theta_{W}
\cos\theta_{W}}\cr
\cr
g_{A}&=&\frac{I_{3}}{2\sin\theta_{W}\cos\theta_{W}}
\end{eqnarray*}
and $\theta_{W}$ is the Weinberg angle.  
The first measurement of A$_{FB}$ was performed in the mid-eighties
(A$_{FB} = - 22.8 \pm 6.0 \pm 2.5$\% at $\sqrt{s}$ = 34.6 GeV;
Bartel {\em et al.}, 1984), and was found to be consistent
with the Standard Model prediction (A$_{FB}$ = $-$25\%) assuming
I$_{3} = -1/2$ for the weak isospin of the $b$-quark.
Alternative isospin assignments
(e.g. I=0) for the bottom-quark were also found to be inconsistent with the
observed suppression of flavor-changing neutral current decays of 
$b$-mesons.  If the $b$-quark formed a weak-isospin
singlet, and if there were only five quarks ($u$, $d$, $c$, $s$, $b$),
then it can be shown that 
BR(B$\rightarrow$Xl$^{+}$l$^{-}$) $\ge$ 
0.12 BR(B$\rightarrow$Xl$\nu$) $\approx$ 0.026 (Kane and Peskin, 1982).
This was soon found to be inconsistent with the first upper limits
placed on flavor changing neutral currents in $b$-decays,
BR(B$\rightarrow$Xl$^{+}$l$^{-}$) $< 0.008$ at 90\% C.L. 
(Matteuzzi {\em et al.}, 1983).

The I$_{3}=-$1/2 isospin of the bottom quark 
implies the existence of an additional
quark, the top quark, as the third-generation weak isospin partner
of the bottom quark.    
Furthermore, the existence of such a third generation 
quark doublet, in conjunction with the presence of three lepton generations,
ensures the necessary cancellations in diagrams contributing to 
triangle anomalies.  For the electro-weak theory to be renormalizable,
the sum over fermions for diagrams such as 
the one displayed in Fig. 3 should vanish 
(see for example Leader and Pedrazzi, 1982).
The contributions to this diagram for each fermion in the theory is 
proportional to N$_{c}$g$_{a}^{f}$Q$_{f}^{2}$, where the factor N$_{c}=3$ 
is the number of colors and applies to quarks only.
Hence, the contribution from a lepton isodoublet
exactly cancels that of a quark isodoublet.  With three lepton
generations, the existence of a third
quark isodoublet, whose members are the top and bottom quarks,
results in the desired cancellation of triangle anomalies.

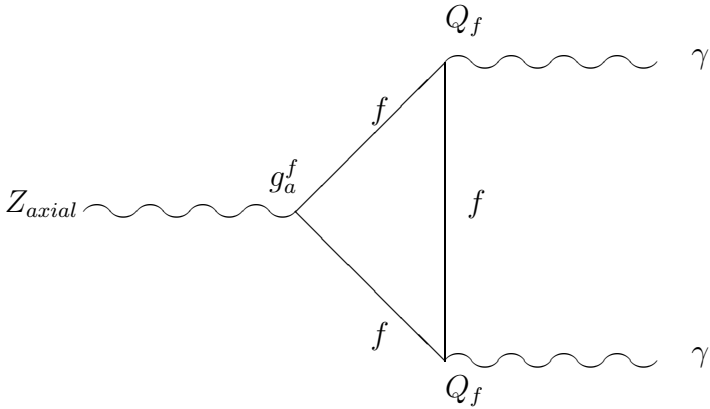
\begin{figure}
\begin{picture}(30000,15000)

\drawline\photon[\E\REG](7000,10000)[8]
\put(4000,10000){$Z_{axial}$}
\put(14000,11000){$g_{a}^{f}$}
\drawline\fermion[\SE\REG](\particlebackx,\particlebacky)[8000]
\put(\pmidx,5000){$f$}
\thicklines
\thinlines
\drawline\fermion[\NE\REG](\particlefrontx,\particlefronty)[8000]
\put(\pmidx,13500){$f$}
\thicklines
\thinlines
\drawline\fermion[\S\REG](\particlebackx,\particlebacky)[11314]
\thicklines
\thinlines
\put(21500,\pmidy){$f$}
\drawline\photon[\E\REG](\fermionbackx,\fermionbacky)[8]
\put(30000,\particlebacky){$\gamma$}
\drawline\photon[\E\REG](\fermionfrontx,\fermionfronty)[8]
\put(30000,\particlebacky){$\gamma$}
\put(\particlefrontx,3000){$Q_{f}$}
\put(\particlefrontx,17000){$Q_{f}$}

\end{picture}
\label{anomf}
\caption{\protect \baselineskip 12pt
An example of a fermion triangle diagram which could cause
an anomaly; g$_{a}^{f}$ is the fermion axial coupling to the $Z$, and
Q$_{f}$ is the fermion charge.}
\end{figure}

Measurements of the $Z$ width at
the LEP and SLC colliders
rule out the existence of a 4th generation neutrino with mass 
M$_{\nu} \lesssim$ M$_{Z}$/2 (Montanet {\em et al.}, 1994).
Unless the 4th generation neutrino is very massive,
no additional generations are allowed
in the context of the Standard Model.  The top quark is therefore
the last fermion expected
in the Standard Model.  Only the Higgs boson is left to be
discovered in order
to complete the particle and field content of the
minimal Standard Model.

While the Standard Model predicts the charge and weak isospin of the
top quark ($Q=2/3$, and I$_{3}=1/2$), its mass remains
a free parameter. As we will discuss in Section~\ref{mass}, the
recent top mass measurements yield
M$_{top} = 175 \pm 8$ GeV/c$^{2}$, a factor of 40 higher than the mass of the
second heaviest fundamental fermion (the $b$-quark).
The reason for such a
high mass for the top quark is a mystery of the Standard Model.
It does however occur quite naturally in local supersymmetric
theories where the electro-weak symmetry is broken through
radiative corrections (Ibanez and Lopez, 1983; J. Ellis
{\em et al.}, 1983; Alvarez-Gaume, Polchinski, and Wise, 1983).

    
The value of the top mass enters in the
calculation of radiative corrections to a large number
of electro-weak observables.   As we will discuss in Section~\ref{ind},
the level of precision achieved in these measurements is good enough
that a comparison between the measured top mass and
the calculation of electro-weak radiative corrections
provides a stringent test of electro-weak theory and is
sensitive to physics beyond the Standard Model.

\section{Top mass and precision electro-weak measurements}
\label{ind}

      Over the past few years, a number of very precise measurements at 
$e^{+}e^{-}$ colliders have been performed using large samples of $Z$ events.
At CERN the four LEP experiments (Aleph, Delphi, L3, and Opal)
have collected of order two million $Z$ events
each, and at the SLC collider at 
SLAC the SLD experiment has collected a data sample of order one 
hundred thousand events. Although the SLAC data sample is considerably smaller
than the LEP data sample, it was possible to polarize the beams at the SLC
and this lead to a competitive measurement of the Weinberg angle. 
The level of accuracy of the LEP and SLC
measurements, as well as of measurements 
of the $W$-boson mass at the Tevatron, is now
such that the data are not well described by tree-level 
theoretical calculations and  
radiative corrections must be included.

If we assume that the Standard Model can be used to 
correctly calculate higher
order electro-weak processes, we can infer the top quark mass by 
comparing these calculations to 
precise measurements
(Altarelli, Kleiss, and Verzegnassi, 1989).
We can then check whether
this value is consistent with the directly measured top quark mass. 
The free parameters in the model are the weak coupling constant G$_{Fermi}$, 
the electromagnetic coupling constant $\alpha$, the Weinberg angle, 
$\theta_{W}$,
the mass of the Higgs boson, M$_{Higgs}$, 
the strong coupling constant $\alpha_{s}$,
the masses of the six quarks, the masses of the six leptons, 
and the four quark mixing
parameters which determine the CKM matrix. 
\par At lowest order the masses of the weak intermediate vector bosons can be 
determined completely from the first three of these parameters, 
G$_{Fermi}$,~$\alpha$,
and $\theta_{W}$. The best measured Standard Model parameters are
$\alpha$, G$_{Fermi}$ and M$_{Z}$. Using these three parameters we can, at 
lowest order predict several measurable quantites. However, 
when higher order corrections are considered,
the fermions and Higgs masses enter into the calculations. The top mass
plays a particularly large role in these radiative corrections 
due to the large mass difference beween the top quark and its weak
isospin partner, the bottom quark.
The dependence of these radiative corrections on the top quark mass 
contains terms quadratic in M$_{top}$ of the form
M$_{top}^{2}$/M$_{Z}^{2}$,
whereas the Higgs mass dependence is logarithmic
($\ln(M_{Higgs}/M_{Z})$). 
Fig. 4 shows two higher order diagrams involving  top quark and  Higgs radiative
corrections which modify observables at the $Z$ resonance.

\begin{figure}[b]
\begin{picture}(20000,14000)(-5000,-4000)
\put(12000,7000){\circle{4000}}
\drawline\photon[\E\REG](14000,7000)[6]
\put(12000,10000){$t$}
\put(12000,3500){$\overline{t}$}
\put(\pmidx,8000){$Z$}
\drawline\photon[\W\REG](10000,7000)[6]
\put(\pmidx,8000){$Z$}
\drawline\photon[\E\REG](22000,7000)[15]
\drawloop\gluon[\N 5](27000,7000)
\put(23000,8000){$Z$}
\put(34000,8000){$Z$}
\put(29000,10500){$H$}
\end{picture}
\label{loop}
\vskip -2cm
\caption{Examples of radiative corrections to the $Z$ mass involving top quarks
or Higgs boson loops.}
\end{figure}
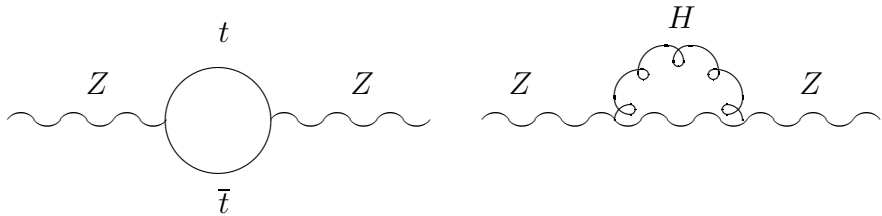
\subsection{Neutral current experiment measurments}
At LEP and SLC, the line shape and asymmetries at the $Z$ have been
precisely measured and can be compared with 
theoretical prediction. The uncertainties on most of these 
quantities are 
smaller from the LEP measurements due to the  size of the LEP data samples. 
These quantities are:

\begin{itemize}
\item  The total width of the $Z$, $\Gamma_{Z}$.
\item  The value of the hadronic cross section at the $Z$ peak, 
$\sigma_{had}^{0}\equiv \frac{12\pi}{m_{Z}^{2}}
\frac{\Gamma_{ee}\Gamma_{had}}{\Gamma_{Z}^{2}}$.
\item The ratio of the hadronic to leptonic widths, 
$R_{l}\equiv \Gamma_{had}/\Gamma_{ll}$.
\item  The forward-backward asymmetry in $Z \rightarrow ll$ decays,
$A_{FB}^{0,l} \equiv \frac{3}{4} A_{e} A_{f}$ where $ A_{f} \equiv
\frac{2g_{Vf}g_{Af}}{g_{Vf}^{2} + g_{Af}^{2}}$, and the leptons ($l$) include 
$e$, $\mu$ and $\tau$.
\item $ A_{\tau}$ as defined above. This is obtained from measurements
of the $\tau$ polarization defined as $ P_{\tau} \equiv 
\frac{\sigma_{R}-\sigma_{L}}{\sigma_{R} + \sigma_{L}}$ where $\sigma_{R}$ and
$\sigma_{L}$ are the $\tau$ pair cross sections for the production of right and
left-handed $\tau$'s respectively.
\item $ A_{e}$, as defined above, i.e. $A_{e} \equiv 
\frac{2g_{Ve}g_{Ae}}{g_{Ve}^{2} + g_{Ae}^{2}}$.
\item The forward-backward asymmetry for decays, $Z\rightarrow b
\overline{b}$ and $Z\rightarrow c\overline{c}$, at the $Z$ pole mass,
 $A_{FB}^{0,b}$ and
$A_{FB}^{0,c}$.
\item The value of 
$\sin^{2}\theta_{eff}^{lept} \equiv 
\frac{1}{4}(1 - \frac{g_{Vl}}{g_{Al}})$
from the hadronic charge asymmetry,
$<Q_{FB}>$ which is the forward-backward asymmetry measured from $Z\rightarrow
q \overline{q}$ decays. The charge of the outgoing quark is determined
using a statistical weighting method.
\item The ratios  $R_{b}\equiv\frac{\Gamma_{Z}^{b\overline{b}}}
{\Gamma_{Z}^{hadrons}}$ and $R_{c}\equiv \frac{\Gamma_{Z}^{c\overline{c}}}
{\Gamma_{Z}^{hadrons}}$.
\item The left-right asymmetry, A$_{LR}\equiv (\sigma_{L}-\sigma_{R})/(\sigma_{L}+\sigma_{R})$,
where $\sigma_{L}$ and $\sigma_{R}$ are the production cross sections
for $Z$ bosons at the $Z$ pole energy with left-handed and right-handed electrons
respectively.  This measurement has been performed only at the SLC, 
since the beams are unpolarized at LEP.
\end{itemize}

\begin{figure}[htb]
\hskip 1cm
\epsfysize=7.0in
\epsffile {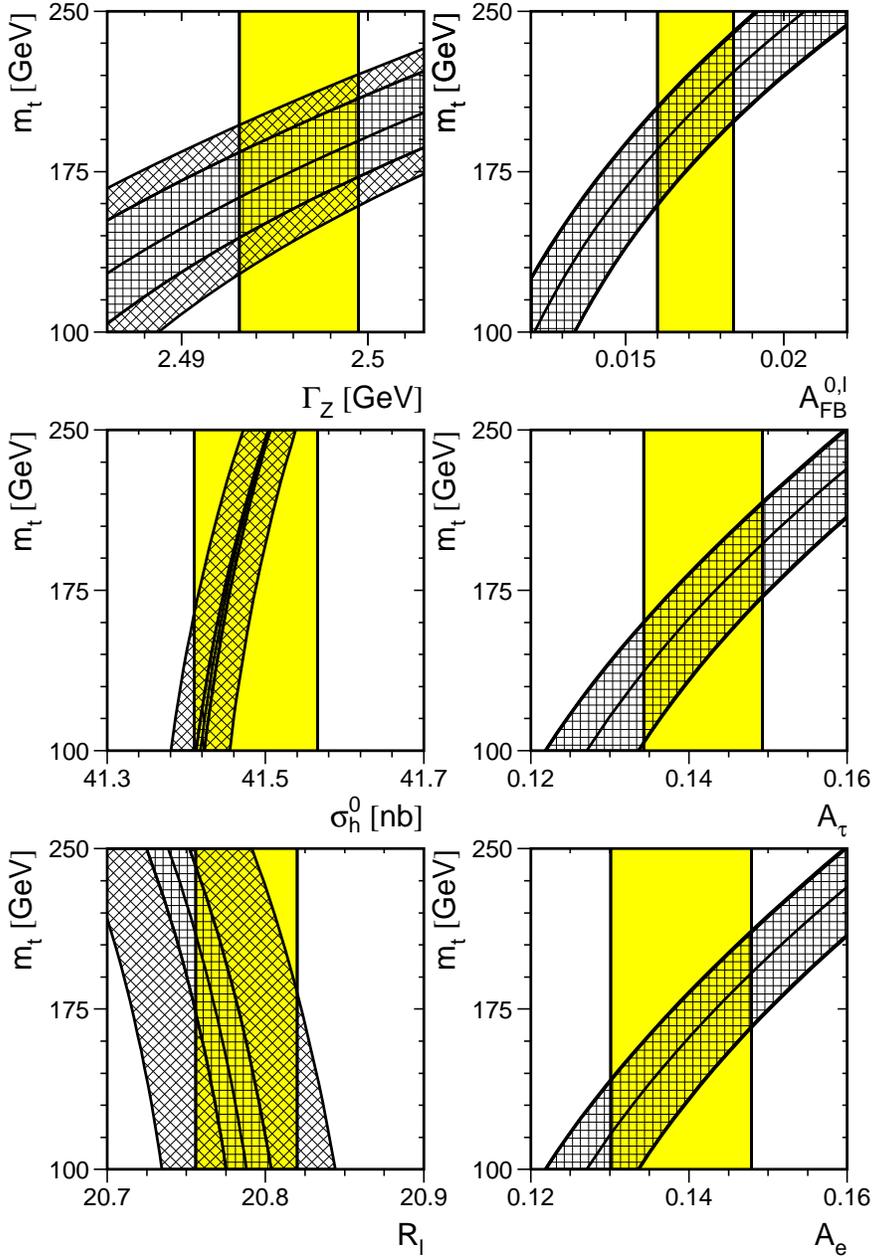}
\vskip 0.5cm
\caption{\protect Comparison of LEP measurements 
with Standard Model predictions as a function of 
M$_{top}$. 
The experimental errors on the parameters are indicated as vertical bands.
The cross-hatch pattern parallel to the axes indicates the variation
of the Standard Model prediction with M$_{Higgs}$ spanning the interval
$60 \leq M_{Higgs} \leq 1000$ GeV/c$^{2}$, and the diagonal cross-hatch pattern 
corresponds to a variation
of $\alpha_{s}(M_{Z}^{2})$ within the interval $\alpha_{s}(M_{Z}^{2}) = 0.123 
\pm 0.006$.
The total width of the band corresponds to the linear sum of both 
uncertainties (The LEP Collaborations, 1995).}
\label{observables}
\end{figure}

\begin{figure}[htb]
\hskip 2cm
\epsfysize=7.0in
\epsffile{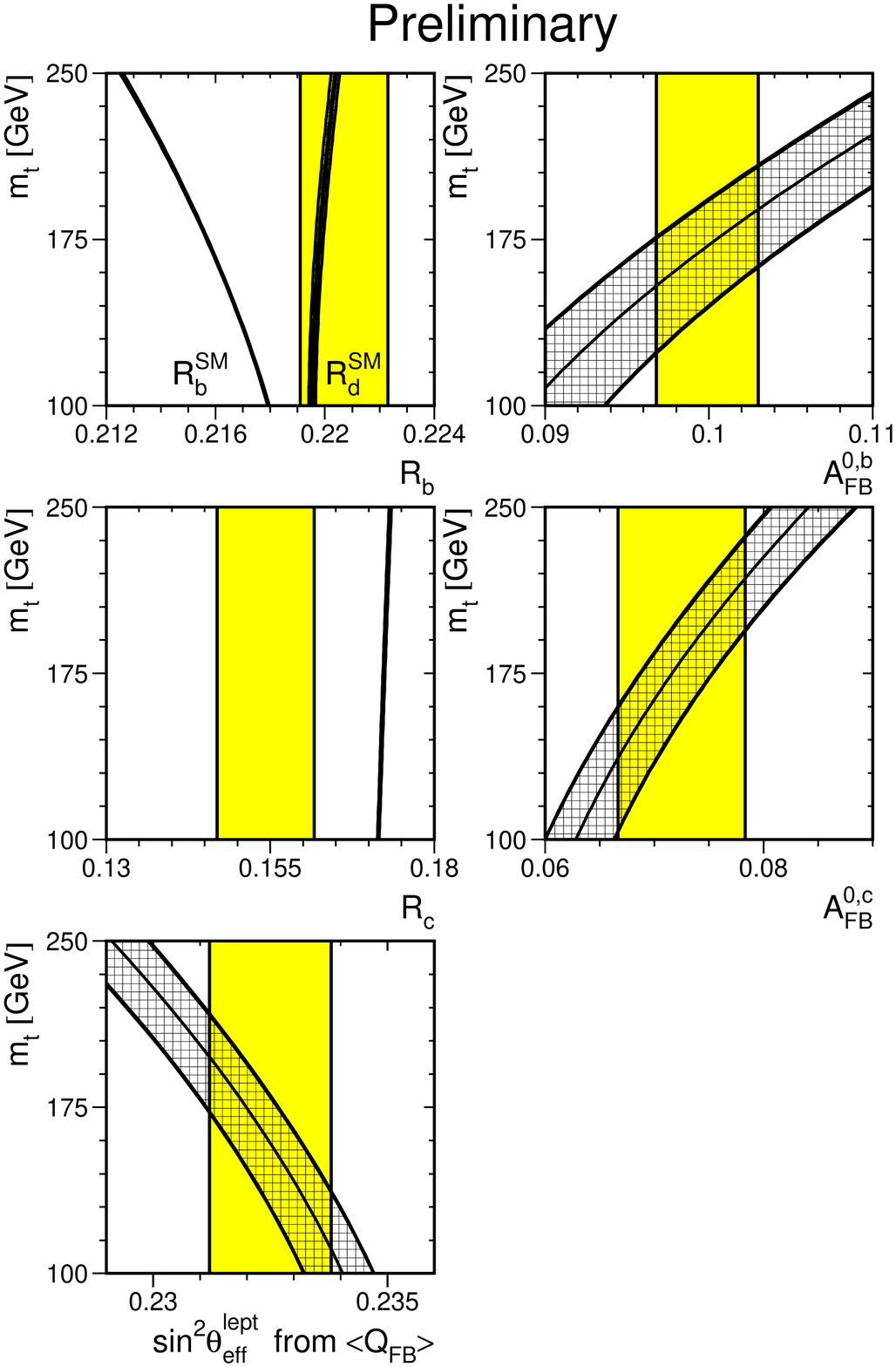}
\vskip 0.5cm
\caption{\protect Comparison of LEP measurements with Standard Model 
predictions as a function 
of M$_{top}$ as in the previous figure. For the ratios of 
the partial widths the variations with 
M$_{Higgs}$ and $\alpha_{s}(M_{Z}^{2})$ nearly cancel. 
For the comparison of R$_{b}$
with the Standard Model the value R$_{c}$ has been fixed to its Standard Model
prediction. To illustrate the impact of special vertex corrections to R$_{b}$,
the Standard Model prediction for R$_{d}$ is also shown
(The LEP Collaborations, 1995).}
\label{observables2}
\end{figure}

In Fig.~\ref{observables} and Fig.~\ref{observables2} we show the measurements
of the variables listed above at LEP (The LEP Collaborations, 1995), 
compared with Standard Model expectations
as a function of the top quark mass.  The vertical bands are the measurements,
where the width of the bands include one standard deviation uncertainties.
The cross-hatched bands show the theoretical predictions 
taking into account the uncertainty
on the Higgs mass (the inner bands) and the 
uncertainty on the value of the strong
coupling constant $\alpha_{s}$ (the outer bands). 
In calculating the theoretical uncertainties,
the Higgs mass is varied from
60 to 1000 GeV/c$^{2}$ and $\alpha_{s}$ is varied within the interval 
$\alpha_{s}(M_{z}^{2})=0.123 \pm 0.006$ (Bethke, 1995). 
The dependence of the $Z$ width measurements
on $\alpha_{s}$ enters through radiative diagrams involving gluons, an example
of which is shown in Fig. 7.
The LEP  measurements
are consistent with theoretical predictions, except for 
the ratios R$_{b}$ and R$_{c}$, which will be discussed in 
Section~\ref{rbrbrb}.


\subsubsection{Asymmetries at the $Z$}
The asymmetries measured at LEP, $A_{FB}^{0,l}$, $A_{\tau}$, $A_{e}$,
$A_{FB}^{0,b}$, $A_{FB}^{0,c}$ and 
$<Q_{FB}>$ are effectively measurements of $\sin^{2}\theta_{eff}^{lept}$. 
Despite the lower statistics SLC $Z$ sample, the A$_{LR}$ measurement
from SLD provides a competitive measurement of 
$\sin^{2}\theta_{eff}^{lept}$.
$\sin^{2}\theta_{eff}^{lept}$ has also been
measured in $\nu N$ experiments at lower center of mass energies
(Abramowicz {\em et al.}, 1986; Blondel {\em et al.}, 1990;
Allaby {\em et al.}, 1986 and 1987; Arroyo {\em et al.}, 1994).
All of these measurements of
$\sin^{2}\theta_{eff}^{lept}$ are found to be 
consistent within their respective uncertainties and 
are combined. The combined result from LEP is 
$\sin^{2}\theta_{eff}^{lept} =
0.23186 \pm 0.00034$, while the corresponding result from the SLD A$_{LR}$ 
measurement is 
$\sin^{2}\theta_{eff}^{lept}= 0.23049 \pm 0.00050$  (Woods, 1996).
Figure~\ref{worldsine} summarizes all the measurements of 
$\sin^{2}\theta_{eff}^{lept}$.

\clearpage

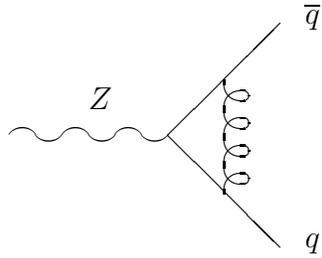
\begin{figure}
\begin{picture}(32000,15000)(0,-3000)
\drawline\photon[\E\REG](10000,6000)[6]
\put(\pmidx,7000){$Z$}
\drawline\fermion[\SE\REG](\pbackx,\pbacky)[6000]
\put(21200,\pbacky){$q$}
\drawline\fermion[\NE\REG](\pfrontx,\pfronty)[6000]
\put(21200,\pbacky){$\overline{q}$}
\drawline\gluon[\S\FLIPPED](\pmidx,\pmidy)[4]
\end{picture}
\label{qcdcor}
\caption{One Feyman diagram illustrating  a QCD correction to the $Z$ width.}
\end{figure}

\begin{figure}
\epsfxsize=4.0in
\hskip 3cm
\epsffile{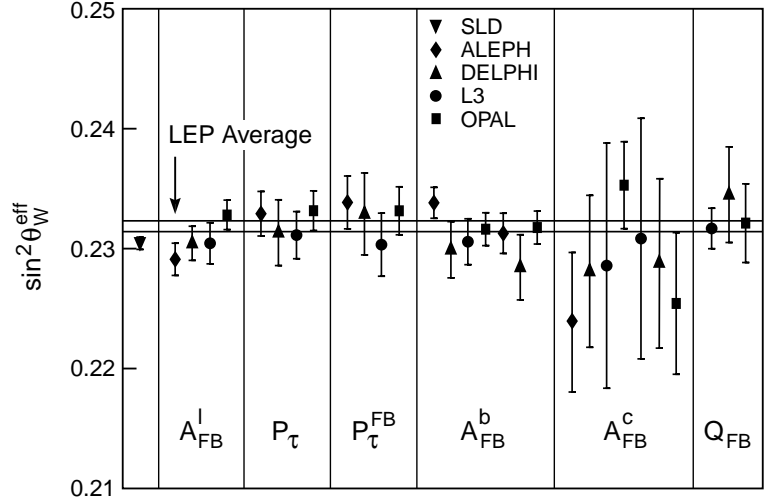}
\vskip 1.cm
\caption{A summary of the measurements of $\sin^{2}\theta_{eff}^{lept}$ 
from LEP and SLC. The leftmost point is the value from A$_{LR}$ at SLD;
$\sin^{2}\theta_{eff}^{lept}$ = 0.2305 $\pm$ 0.0005. The LEP average is
$\sin^{2}\theta_{eff}^{lept}$ = 0.2319 $\pm$ 0.0004.
(Woods, 1996).}
\label{worldsine}
\end{figure}

\subsubsection{R$_{b}$}
\label{rbrbrb}

The decay of the $Z$ into $b \overline{b}$ is of particular
interest, because
it is sensitive through weak vertex corrections
to the top quark mass, (see Fig. 9).
The dependence of all the $Z$ partial widths
on the Higgs mass is due mostly to
corrections to the $Z$ propagator. 
Therefore, in R$_{b}$,
the ratio of $\Gamma_{Z}^{b\overline{b}}$
to $\Gamma_{Z}^{hadrons}$, most of the Higgs and $\alpha_{s}$ 
dependence cancels.
This ratio then gives the only indirect
measurement of the top mass independent of the Higgs mass. 
A comparison between the experimental measurement and the theoretical prediction
is a particularly good test of the Standard Model and possibly the 
first place to look for new (non-Standard Model) physics. 

The value of
R$_{b}$ from the combinations of many measurements at LEP and SLC
is higher than expected, whereas the value of R$_{c}$
is lower than expected.  The measurement of R$_{b}$ depends on
what is assumed for R$_{c}$ because charm quarks are a background to
the bottom quark signal in the data. Because the two are correlated,
R$_{b}$ is 
quoted either
assuming a Standard Model value for R$_{c}$, or the measured value. 
In Fig.~\ref{rbrc}
we show the LEP measurements of R$_{c}$ vs. R$_{b}$, together with the
Standard Model theoretical prediction based on the direct measurement
of the top quark mass performed at the Tevatron (see Section~\ref{mass}).
The disagreement between data and theory may be an indication of new
physics beyond the Standard Model (see for example
Altarelli {\em et al.} 1996;
Wells and Kane, 1995).


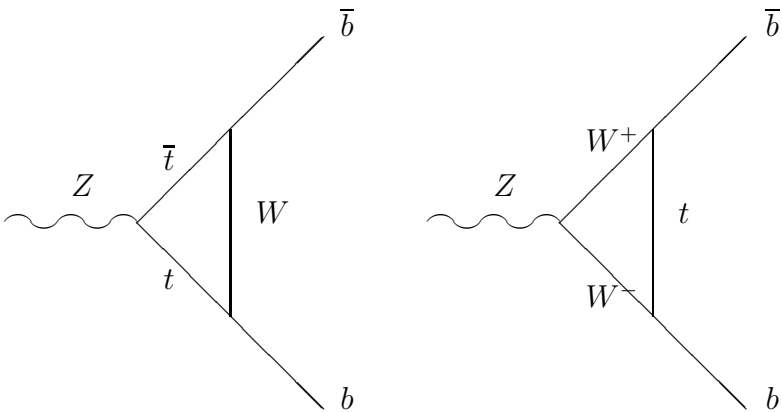
\begin{figure}
\begin{picture}(32000,20000)(0,-3000)
\drawline\photon[\E\REG](4000,6000)[5]
\put(\pmidx,7000){$Z$}
\drawline\fermion[\SE\REG](\pbackx,\pbacky)[10000]
\put(16700,\pbacky){$b$}
\drawline\fermion[\NE\REG](\pfrontx,\pfronty)[10000]
\put(16700,\pbacky){$\overline{b}$}
\drawline\fermion[\S\REG](\pmidx,\pmidy)[7071]
\drawline\photon[\E\REG](20000,6000)[5]
\put(\pmidx,7000){$Z$}
\put(13500,6000){$W$}
\put(10000,3500){$t$}
\put(10000,8000){$\overline{t}$}
\drawline\fermion[\SE\REG](\pbackx,\pbacky)[10000]
\put(32800,\pbacky){$b$}
\drawline\fermion[\NE\REG](\pfrontx,\pfronty)[10000]
\put(26000,8700){$W^{+}$}
\put(26000,2800){$W^{-}$}
\put(29500,6000){$t$}
\put(32800,\pbacky){$\overline{b}$}
\drawline\fermion[\S\REG](\pmidx,\pmidy)[7071]
\end{picture}
\label{gambb}
\caption{Diagrams showing $t$-quark corrections to 
$Z\rightarrow b\overline{b}$.}
\end{figure}

\begin{figure}
\vskip 1.cm
\hskip 3cm
\epsfxsize=4.0in
\epsffile{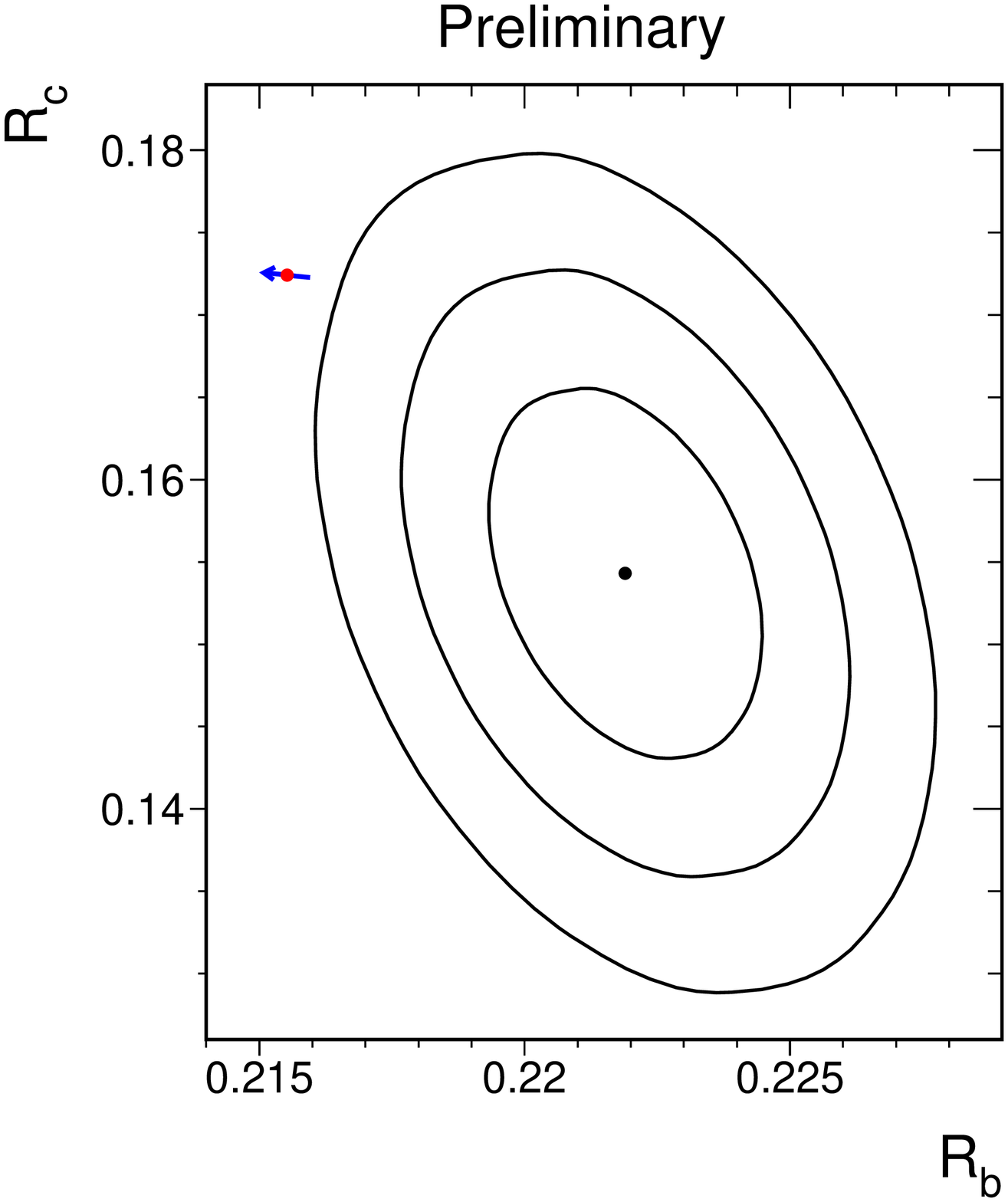}
\caption{\protect 
Contours in the R$_{b}$ - R$_{c}$ plane derived from LEP data,
corresponding to $68\%$, $95\%$ and $99.7\%$ confidence levels assuming Gaussian
systematic errors. The Standard Model prediction for M$_{top} = 180 \pm 12$ 
GeV/c$^{2}$
is also shown as a dot with an arrow through it. 
(This is the average of the 1995 CDF and D0 top mass measurements; the
current average is $175 \pm 8$ GeV/c$^{2}$).
The arrow points in the direction 
of increasing values of M$_{top}$
(The LEP Collaborations, 1995).}
\label{rbrc}
\end{figure}
\subsection{$W$ mass}
The $W$ vector boson mass (M$_{W}$)  also depends on the top quark and 
Higgs masses through loop diagrams
like those shown in Fig. 11, 
in which $W$ $\rightarrow t\overline{b}\rightarrow $W,
or $W\rightarrow WH \rightarrow W$. 
A precise measurement of M$_{W}$ constrains the top mass for a fixed Higgs mass.
When combined with a precise measurement of the top mass, such a measurement
can provide information on the Higgs mass, or, in the case of
disagreement with theory, can signal the presence of new physics.
Even with  precise direct measurements  of M$_{top}$ and M$_{W}$, however,
the constraints on
M$_{Higgs}$ are weak, because the Higgs mass dependence is only logarithmic.
The present status of the $W$ and top mass measurements, and their comparison
with theory, is summarized in Fig.~\ref{topvsw}.


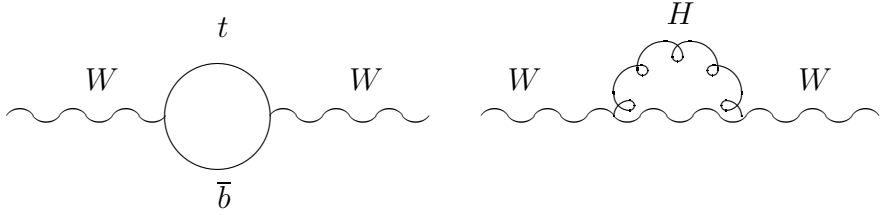
\begin{figure}[htb]
\begin{picture}(20000,14000)(-5000,-4000)
\put(10000,7000){\circle{4000}}
\drawline\photon[\E\REG](12000,7000)[6]
\put(10000,10000){$t$}
\put(10000,3500){$\overline{b}$}
\put(\pmidx,8000){$W$}
\drawline\photon[\W\REG](8000,7000)[6]
\put(\pmidx,8000){$W$}
\drawline\photon[\E\REG](20000,7000)[15]
\drawloop\gluon[\N 5](25000,7000)
\put(21000,8000){$W$}
\put(32000,8000){$W$}
\put(27000,10500){$H$}
\end{picture}
\label{wmass}
\vskip -2cm
\caption{Lowest order radiative corrections to the $W$ mass 
involving top and bottom quarks and the Higgs.}
\end{figure}

\begin{figure}[htb]
\epsfxsize=3.5in
\gepsfcentered[20 200 600 600]{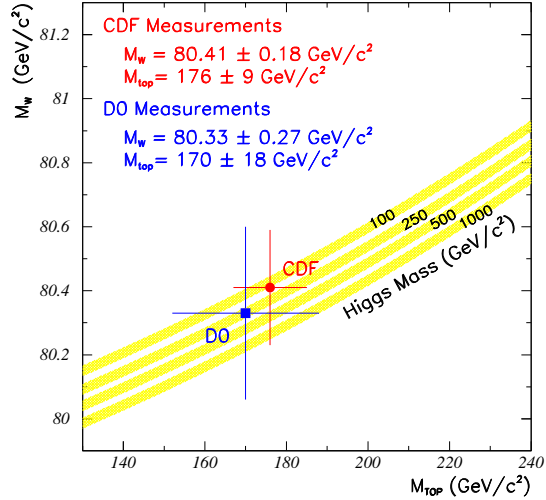}
\caption{$W$ mass and top quark mass measurements from the Fermilab collider
experiments (CDF and D0).  The top mass values are from the full 
Tevatron data sets, with an integrated luminosity of $\approx$ 100 pb$^{-1}$.
The $W$ mass values
are derived from analyses of the first 15-20 pb$^{-1}$ only.
The lines are Standard Model 
predictions for
four different Higgs masses (Flattum,1996).} 
\label{topvsw}
\end{figure}

\subsection{Global fits for M$_{top}$ and M$_{Higgs}$}
\begin{figure}[htb]
\hskip 2cm
\epsfxsize=3.0in
\hspace*{1.0in}
\epsffile{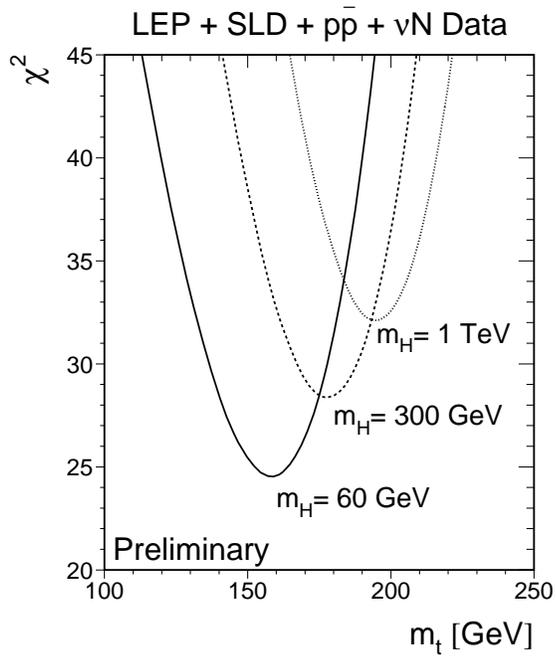}
\caption{\protect The $\chi^{2}$ curves for the Standard Model fit to the 
electro-weak
precision measurements from LEP, SLD, CDF and D0 ($W$ mass only) 
and neutrino scattering
experiments as a function of
M$_{top}$ for three different Higgs mass values spanning the interval 
60 GeV/c$^{2} \leq$ M$_{Higgs} \leq 1000$ GeV/c$^{2}$.  The number of degrees
of freedom is 14 (The LEP Collaborations, 1995).}
\label{chitop}
\end{figure}

	Combining the indirect information from the neutral current experiments
and the $W$ mass measurement, a global fit for M$_{top}$ has been made by 
the LEP
Electro-Weak Working Group (The LEP Collaborations, 1995).
The fits are made with $\alpha_{s}$ and M$_{top}$ as 
free parameters, since $\alpha_{s}$ at the $Z$ mass has a large uncertainty.
The best predicted value for M$_{top}$ using data from LEP,
SLC, the Fermilab collider $W$ mass measurements
and $\nu$ N scattering data, 
is M$_{top} = 178 \pm 8 ^{+17}_{-20}$ GeV/c$^{2}$,
with $\alpha_{s} = 0.123 \pm 0.004 \pm 0.002$ and
with $\chi^{2}/d.o.f. = 28/14$ (where we have chosen M$_{Higgs}=300$ 
GeV/c$^{2}$
to quote the goodness of fit).
The second uncertainty in this fit to the top mass
comes from varying M$_{Higgs}$ from 60 GeV/c$^{2}$ to 1 TeV/c$^{2}$.
The fit results are in good agreement with the directly measured values 
of $\alpha_{s}$ and M$_{top}$, $\alpha_{s}(M_{Z}) = 0.123 \pm
0.006$ (Bethke, 1995) and
M$_{top} = 175 \pm 8$ GeV/c$^{2}$, see Section~\ref{mass}.
The variation of the fit-$\chi^{2}$ as a function of M$_{top}$ for 
three
different choices of M$_{Higgs}$ is displayed in Fig.~\ref{chitop}.

\begin{figure}[h]
\epsffile{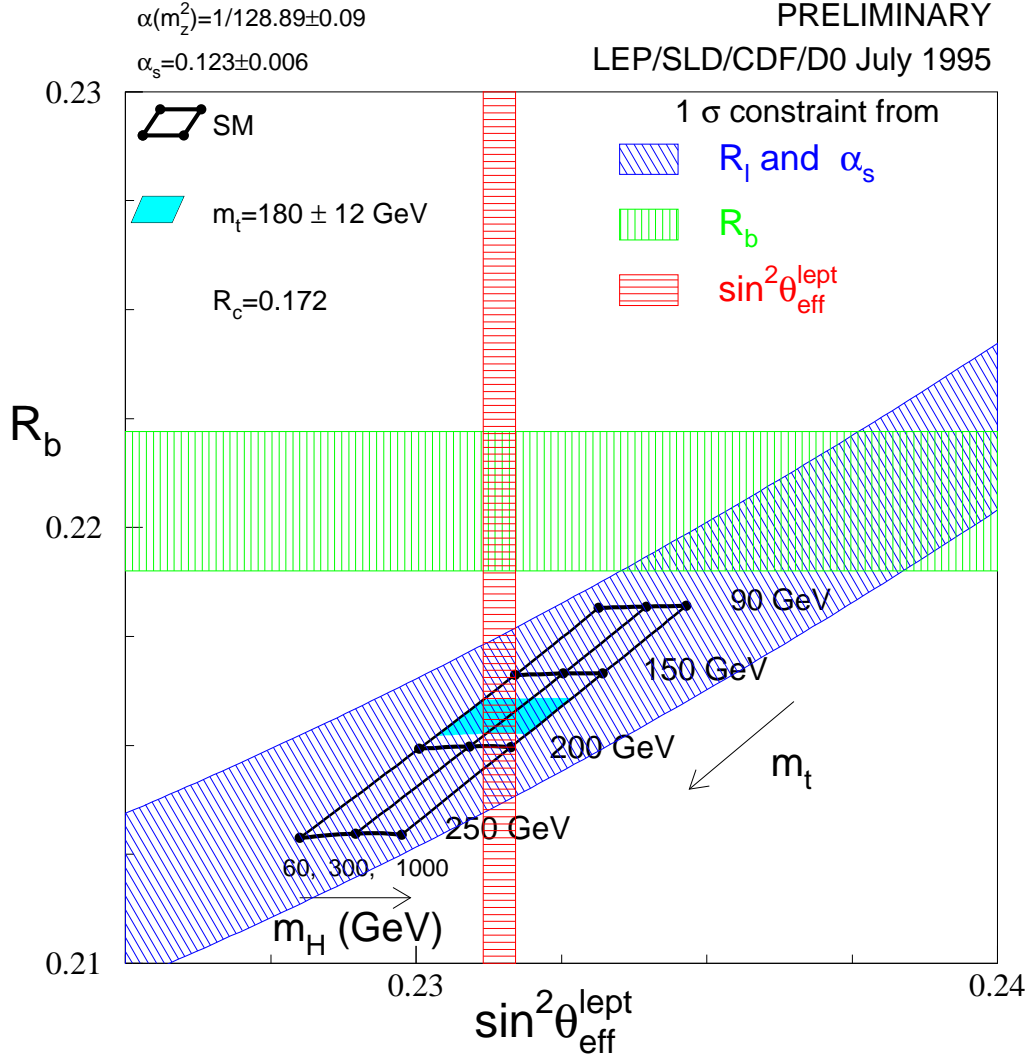}
\vskip 1.cm
\caption{\protect The combined
LEP/SLD measurements of $\sin^{2}\theta_{eff}^{lept}$
and R$_{b}$ assuming the Standard Model value of R$_{c} = 0.172$ and the 
Standard
Model prediction. Also shown is the constraint resulting from the measurement
of R$_{l}$ on these variables, 
assuming $\alpha_{s}(M_{Z}^{2}) = 0.123 \pm 0.006$, 
as well as the Standard Model dependence of light-quark partial widths on 
$\sin^{2}\theta_{eff}^{lept}$ (The LEP Collaborations, 1995).} 
\label{alot}
\end{figure}

In conclusion, all the neutral current data, as well as
the $W$ and top mass measurements
are in agreement with each other, with the exception of the measurement of 
R$_{b}$.  The situation is nicely summarized in Fig.~\ref{alot}. 
In this figure, the correlation between R$_{l}$ and R$_{b}$ is
due to the fact that R$_{l}$ depends on the total hadronic width,
and hence on $\Gamma(Z \rightarrow b\bar{b})$.
Given the measured value of R$_{l}$ and $\alpha_{s}$, and assuming
Standard Model
dependence of the partial widths on $\sin^{2}\theta_{eff}^{lept}$ for all 
but the
$b$-quarks, R$_{l}$ constrains R$_{b}$ and $\sin^{2}\theta_{eff}^{lept}$.
These three measurements are compared with the Standard Model prediction
given the measured top mass. The measured values of
R$_{b}$ and R$_{c}$ are somewhat inconsistent 
with the combination of $\sin^{2}\theta_{eff}^{lept}$, R$_{l}$ and M$_{top}$
within the context of the Standard Model.

\begin{figure}[htb]
\hskip 2cm
\epsfxsize=4.0in
\epsffile{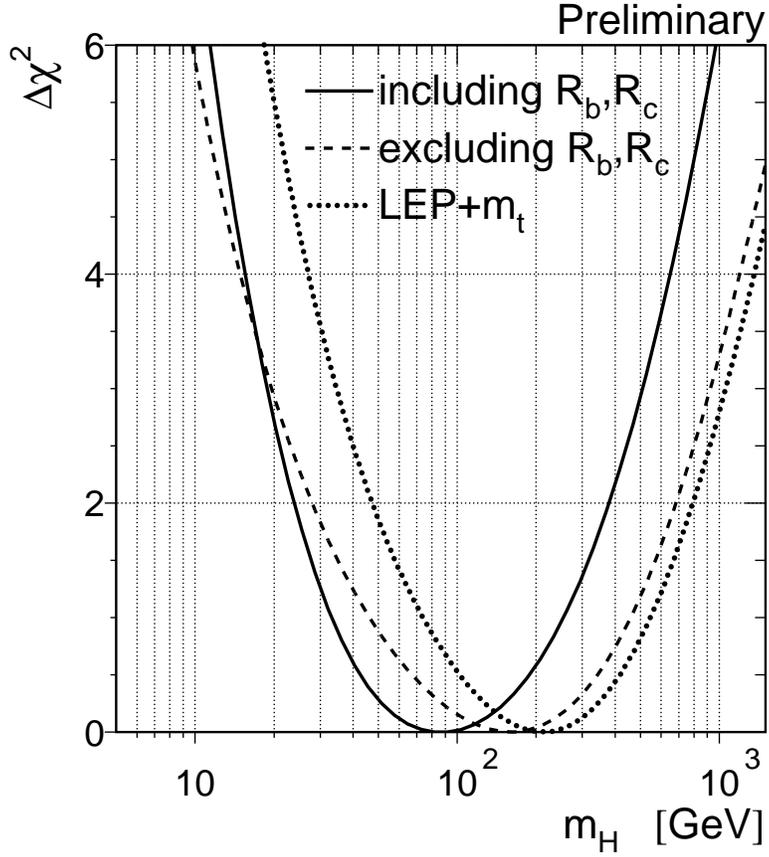}
\vskip 1.cm
\caption{\protect $\Delta\chi^{2} = \chi^{2} - \chi^{2}_{min}$ 
vs M$_{Higgs}$ curves.
The continuous line uses all the data
from neutral currents and 
$p\overline{p}$.
The dashed line excludes the LEP and SLD R$_{b}$ and R$_{c}$ measurements.
The dotted line excludes the SLD data (The LEP Collaborations, 1995).}
\label{higgsest}
\end{figure}

Despite the large number of very precise measurements, there is still
little information on the mass of the Higgs boson, although the data
seems to prefer a low value for M$_{Higgs}$, (see Fig.~\ref{chitop}).
An estimate for the Higgs mass can be made using all the neutral current 
and hadron collider data as shown in Fig.~\ref{higgsest}. The best estimate of
the Higgs mass is shown with and without the R$_{b}$ and R$_{c}$ measurements
included. In all cases a Higgs mass less than 300 GeV/c$^{2}$ is favored with 
large uncertainty.

\clearpage

\section {Top quark production}
\label{prod}
Because of its large mass, the top quark can only be
observed directly in collider experiments, where sufficiently
high center-of-mass energies ($\sqrt{s}$) have been achieved.
In electron-positron
collisions, top quarks are produced in pairs via a photon
or a $Z$. Since at lowest order
this is a purely electro-weak process,
the cross section and production kinematics can be precisely
predicted.  
Today's highest energies $e^{+}e^{-}$ accelerators,
LEP at CERN and SLC at SLAC, operate at 
$\sqrt{s} \approx$ M$_{z} \approx$ 91 GeV, and therefore the mass
region M$_{top} >$ 46 GeV/c$^2$ cannot be explored.
Searches for top in $e^{+}e^{-}$ collisions will be briefly
reviewed in Section~\ref{pre1a}.  The HERA electron-proton collider
at DESY has achieved a center-of-mass energy of $\approx$ 310 GeV.  
However
the top production cross section at HERA
is too small for observation
and study of the top quark.

Significantly higher center-of-mass energies have been
achieved at hadron colliders.  The $p\bar{p}$ collider at CERN
(the S$p\bar{p}S$), which operated between 1981 and 1989,
reached $\sqrt{s}$ = 630 GeV; the $p\bar{p}$ collider at
Fermilab (the Tevatron) came on line in 1987 with 
$\sqrt{s}$ = 1800 GeV = 1.8 TeV; a new $pp$ collider 
(LHC, $\sqrt{s}$ = 14 TeV) is under development at CERN, and 
is expected to begin operation in 2003.
Until a very high energy $e^{+}e^{-}$ machine is built,
top quark physics can be directly pursued only at hadron colliders.
In this Section we will concentrate on the production of top quarks 
in $p\bar{p}$ collisions.  Top quarks are produced by 
colliding partons (quarks, gluons, and antiquarks) from the
proton and anti-proton.  Therefore many aspects of our discussion also
apply to other (e.g. $pp$) hadron-hadron collisions.
Because the partons carry only a fraction of the momentum of 
the hadron, the center-of-mass energies of parton-parton
collisions span a wide range of energies, see the discussion of 
parton luminosities in Section~\ref{mech}.
 
\subsection{Production mechanisms and cross sections}
\label{mech}
There are three mechanisms for top production in $p\bar{p}$
collisions :
\begin{itemize}
\item Pair production of top quarks,  $p\bar{p} \rightarrow t\bar{t}~+$ X.
The leading order Feynman diagrams for this process are shown in 
Fig. 16. At higher order, 
gluon-quark scattering also contributes.
$t\bar{t}$ pairs can also be produced through
a $Z$ or a photon, however the cross section is much smaller 
and we will
not consider this possibility further.

\item Drell-Yan production of a $W$ boson, with subsequent decay into 
$t\bar{b}$, 
i.e. $p\bar{p} \rightarrow  $W + X; $W \rightarrow
t\bar{b}$, (see Fig. 17).
Except for small contributions from off mass-shell $W$
boson production, this mechanism only contributes for 
top masses smaller than M$_{W} - $M$_{b}$.

\item Single top-quark production via $W$-gluon fusion, see 
Fig 18.  Photon-gluon and
$Z$-gluon fusion are also allowed, with a much lower cross section.

\end{itemize}

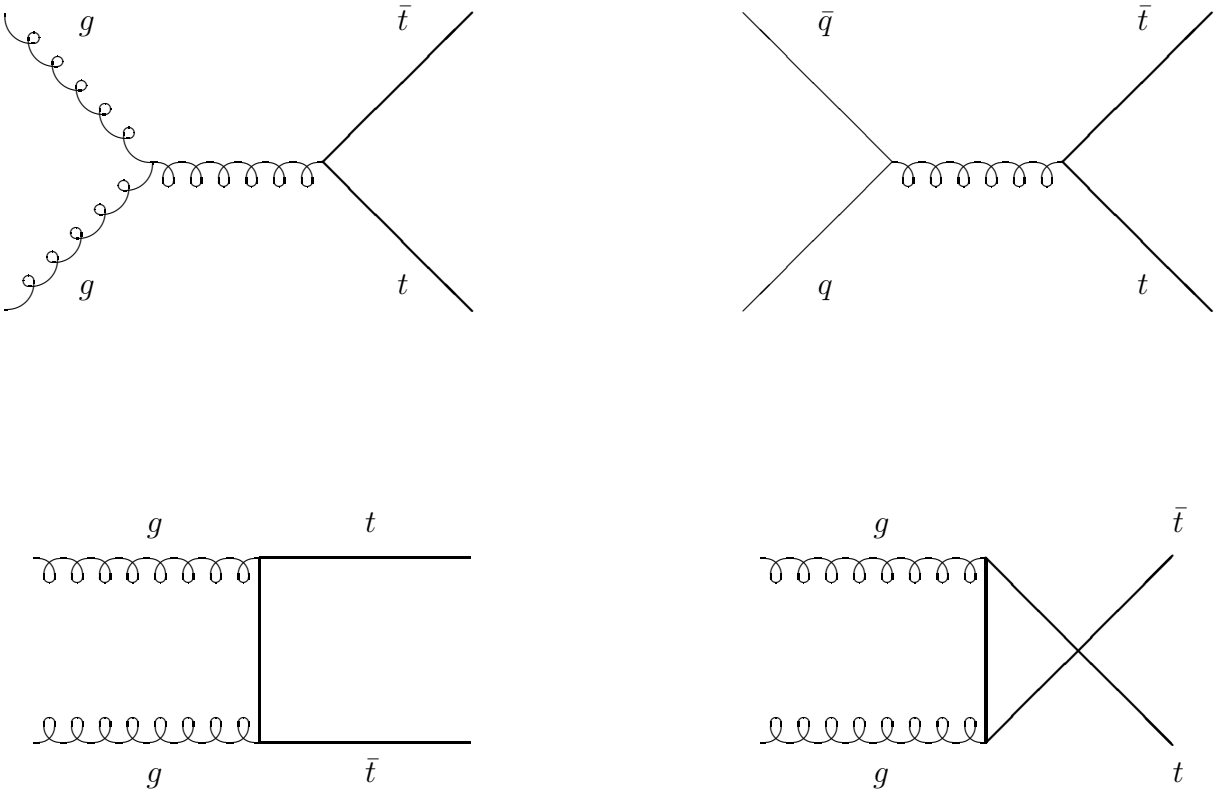
\begin{figure}
\hskip 1.5cm
\begin{picture}(30000,35000)(0,-10000)

\drawline\gluon[\E\REG](0,15000)[6]
\THICKLINES
\drawline\fermion[\SE\REG](\particlebackx,\particlebacky)[8000]
\THINLINES
\put(\pmidx,20000){$\bar{t}$}
\THICKLINES
\drawline\fermion[\NE\REG](\gluonbackx,\gluonbacky)[8000]
\THINLINES
\put(\pmidx,10000){$t$}
\drawline\gluon[\NW\REG](\gluonfrontx,\gluonfronty)[5]
\drawline\gluon[\SW\REG](\gluonfrontx,\gluonfronty)[5]
\put(\pmidx,20000){$g$}
\put(\pmidx,10000){$g$}

\drawline\gluon[\E\REG](28000,15000)[6]
\THICKLINES
\drawline\fermion[\SE\REG](\particlebackx,\particlebacky)[8000]
\THINLINES
\put(\pmidx,20000){$\bar{t}$}
\THICKLINES
\drawline\fermion[\NE\REG](\gluonbackx,\gluonbacky)[8000]
\THINLINES
\put(\pmidx,10000){$t$}
\drawline\fermion[\NW\REG](\gluonfrontx,\gluonfronty)[8000]
\drawline\fermion[\SW\REG](\gluonfrontx,\gluonfronty)[8000]
\put(\pmidx,20000){$\bar{q}$}
\put(\pmidx,10000){$q$}

\drawline\gluon[\E\REG](-4500,0)[8]
\put(\pmidx,1000){$g$}
\THICKLINES
\drawline\fermion[\E\REG](\gluonbackx,\gluonbacky)[8000]
\THINLINES
\put(\pmidx,1000){$t$}
\THICKLINES
\drawline\fermion[\S\REG](\particlefrontx,\particlefronty)[7000]
\drawline\fermion[\E\REG](\particlebackx,\particlebacky)[8000]
\put(\pmidx,-8500){$\bar{t}$}
\THINLINES
\drawline\gluon[\W\REG](\particlefrontx,\particlefronty)[8]
\put(\pmidx,-8500){$g$}

\drawline\gluon[\E\REG](23000,0)[8]
\put(\pmidx,1000){$g$}
\THICKLINES
\drawline\fermion[\SE\REG](\gluonbackx,\gluonbacky)[10000]
\THINLINES
\put(\pbackx,1000){$\bar{t}$}
\THICKLINES
\drawline\fermion[\S\REG](\particlefrontx,\particlefronty)[7000]
\drawline\fermion[\NE\REG](\particlebackx,\particlebacky)[10000]
\THINLINES
\put(\pbackx,-8500){${t}$}
\drawline\gluon[\W\REG](\particlefrontx,\particlefronty)[8]
\put(\pmidx,-8500){$g$}

\end{picture}
\label{toppair}
\caption{\protect\baselineskip 12pt
Lowest order Feynman diagrams for production of $t\bar{t}$
pairs in $p\bar{p}$ collisions.}
\end{figure}

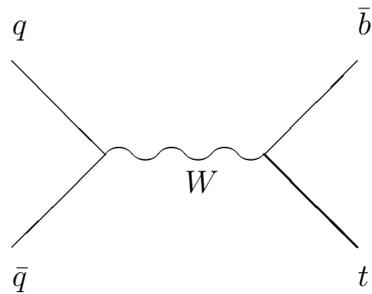
\begin{figure}
\begin{picture}(32000,10000)(0,-3000)

\drawline\fermion[\SE\REG](10000,6000)[5000]
\put(\particlefrontx,7000){$q$}
\drawline\fermion[\SW\REG](\particlebackx,\particlebacky)[5000]
\put(\particlebackx,-2500){$\bar{q}$}
\drawline\photon[\E\REG](\particlefrontx,\particlefronty)[6]
\put(\particlemidx,1000){$W$}
\THICKLINES
\drawline\fermion[\SE\REG](\particlebackx,\particlebacky)[5000]
\THINLINES
\put(\particlebackx,7000){$\bar{b}$}
\drawline\fermion[\NE\REG](\particlefrontx,\particlefronty)[5000]
\put(\particlebackx,-2500){$t$}

\end{picture}
\label{dyt}
\caption{\protect \baselineskip 12pt
Lowest order Feynman diagram for
Drell-Yan production
of $t\bar{b}$ pairs, $p\bar{p} \rightarrow W \rightarrow t\bar{b}$.}
\end{figure}

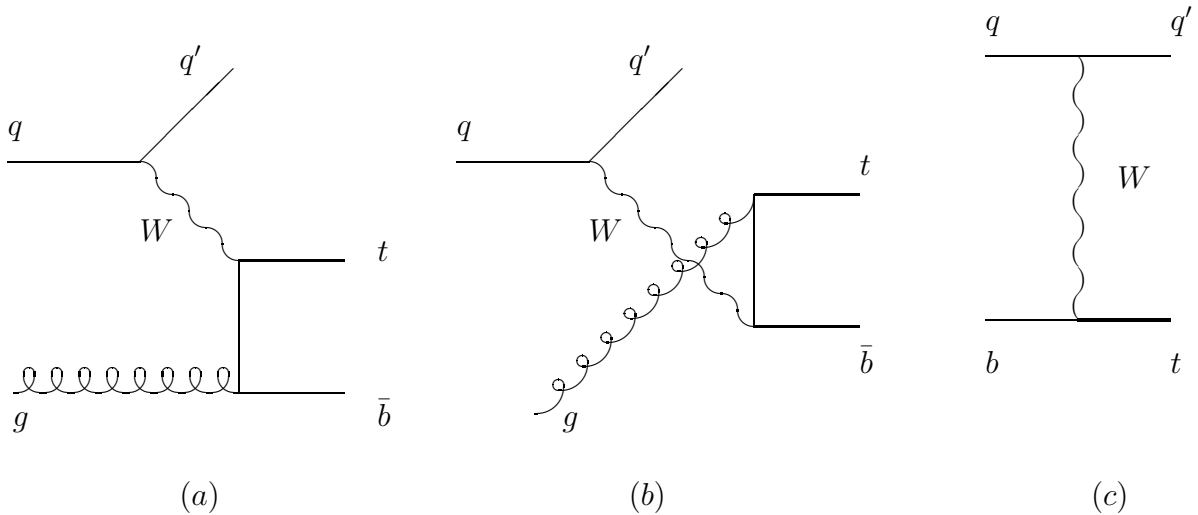
\begin{figure}
\hskip 3cm
\begin{picture}(32000,20000)(0,-5000)



\drawline\fermion[\E\REG](-7000,8000)[5000]
\put(\particlefrontx,9000){$q$}
\drawline\fermion[\NE\REG](\particlebackx,\particlebacky)[5000]
\put(-500,\particlebacky){$q'$}
\drawline\photon[\SE\REG](\particlefrontx,\particlefronty)[6]
\put(\fermionfrontx,5000){$W$}
\THICKLINES
\drawline\fermion[\E\REG](\particlebackx,\particlebacky)[4000]
\THINLINES
\put(7000,\particlebacky){$t$}
\drawline\fermion[\S\REG](\photonbackx,\photonbacky)[5000]
\drawline\fermion[\E\REG](\particlebackx,\particlebacky)[4000]
\put(7000,-2000){$\bar{b}$}
\drawline\gluon[\W\REG](\particlefrontx,\particlefronty)[8]
\put(\particlebackx,-2000){$g$}
\put(-500,-5000){$(a)$}

\drawline\fermion[\E\REG](10000,8000)[5000]
\put(\particlefrontx,9000){$q$}
\drawline\fermion[\NE\REG](\particlebackx,\particlebacky)[5000]
\put(16500,\particlebacky){$q'$}
\drawline\photon[\SE\REG](\particlefrontx,\particlefronty)[10]
\put(\fermionfrontx,5000){$W$}
\drawline\fermion[\E\REG](\particlebackx,\particlebacky)[4000]
\put(\particlebackx,7500){$t$}
\drawline\fermion[\N\REG](\photonbackx,\photonbacky)[5000]
\THICKLINES
\drawline\fermion[\E\REG](\particlebackx,\particlebacky)[4000]
\THINLINES
\put(\particlebackx,0){$\bar{b}$}
\drawline\gluon[\SW\REG](\particlefrontx,\particlefronty)[8]
\put(14000,-2000){$g$}
\put(16500,-5000){$(b)$}

\drawline\fermion[\E\REG](30000,12000)[7000]
\put(\particlefrontx,13000){$q$}
\put(\particlebackx,13000){$q'$}
\drawline\photon[\S\REG](\particlemidx,\particlemidy)[10]
\put(35000,\particlemidy){$W$}
\drawline\fermion[\W\REG](\particlebackx,\particlebacky)[3500]
\put(\particlebackx,0){$b$}
\THICKLINES
\drawline\fermion[\E\REG](\particlefrontx,\particlefronty)[3500]
\THINLINES
\put(\particlebackx,0){$t$}
\put(34000,-5000){$(c)$}

\end{picture}
\label{wgfus}
\vskip 1cm
\caption{\protect \baselineskip 12pt
Lowest order Feynman diagrams for production of a single top quark
via $W$-gluon fusion in $p\bar{p}$ collisions.  When combining
diagrams (a) and (c), care must be exercised to avoid 
double counting
and to define the $b$-quark distribution inside the proton in
a consistent fashion (Heinson, Belyaev, and Boos, 1995;
Carlson and Yuan, 1995).}
\end{figure}

As will be illustrated in this Section, at Fermilab's Tevatron the 
strong-interaction pair production process 
($p\bar{p} \rightarrow t\bar{t}$)
is dominant over a wide range of top masses. 
For
M$_{top} \approx$ 60 GeV/c$^2$, the top production
rate from $W \rightarrow t\bar{b}$ is comparable to that
of $t\bar{t}$.
For very high top mass, above M$_{top} \approx$ 220 GeV/c$^2$,
the expected cross section for single top production through
$W$-gluon fusion 
becomes larger than the pair production cross section due to 
the very high parton center-of-mass energy required to
produce a $t\bar{t}$ pair (see discussion below).  

The pair production
cross section for heavy quarks such as the top
can be calculated in perturbative QCD.  It factorizes
as a product of the parton distribution functions inside the 
protons and the parton-parton point cross section,
and is written as a sum over contributions from partons inside the
proton and anti-proton
(Collins, Soper, and Sterman, 1986) :

$$\sigma(p\bar{p}\rightarrow t\bar{t}) = \sum_{i,j} \int
dx_{i} F_{i}(x_{i},\mu^{2}) \int dx_{j} F_{j}(x_{j},\mu^{2}) 
\hat{\sigma}_{ij}(\hat{s},\mu^{2},M_{top}).$$

\begin{figure}[htb]
\vskip 1cm
\epsfxsize=6.0in
\gepsfcentered[20 200 600 600]{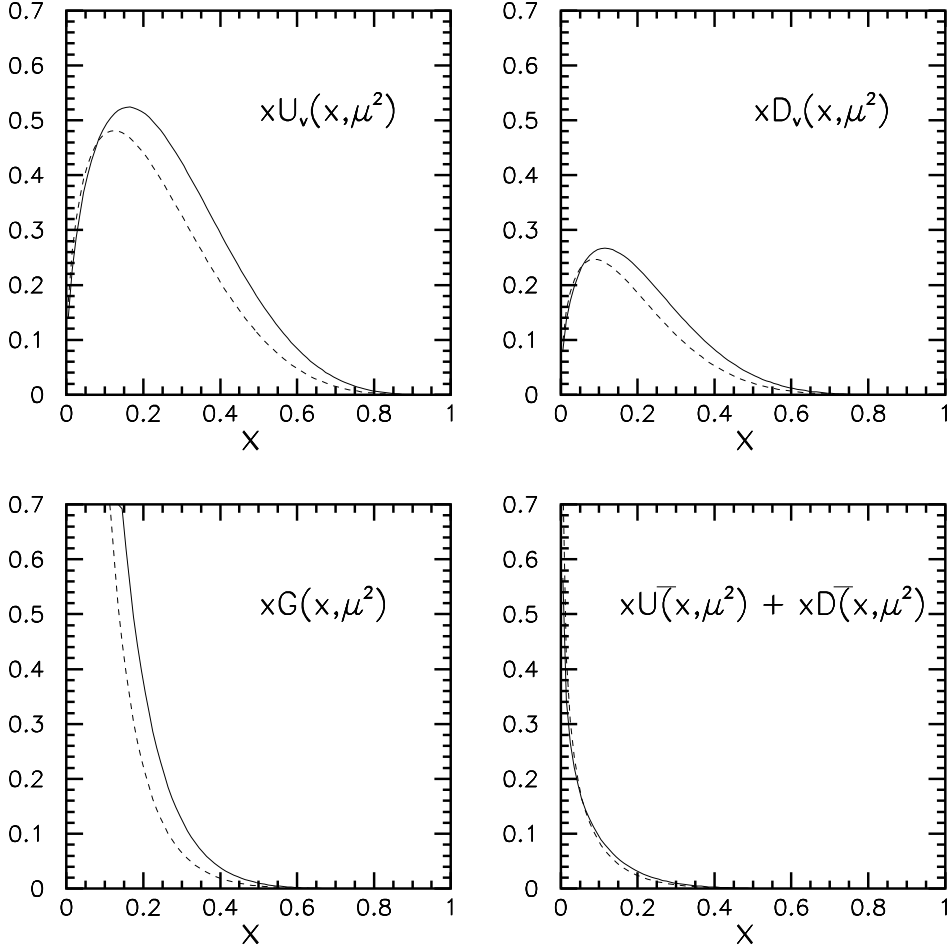}
\vskip 1cm
\caption{\protect \baselineskip 12pt
MT-B2 parametrization (Morfin and Tung, 1991) of $xF_{i}$ for
valence $u$-quarks, $xU_{v}(x,\mu^{2})$, valence $d$-quarks,
$xD_{v}(x,\mu^{2})$, gluons, $xG(x,\mu^2)$, and sea $u$- and
$d$-quarks, $x\bar{U}(x,\mu^{2}) + x\bar{D}(x,\mu^{2})$.
These are calculated at $\mu^{2} =$ (20 GeV)$^{2}$ 
(solid line) and
$\mu^{2} =$ (400 GeV)$^{2}$ (dashed line). The $t\bar{t}$ 
cross section is usually calculated with $\mu^{2}$~= M$_{top}^{2}$.}
\label{pdf}                                  
\end{figure}

The functions $F_{i}$ and $F_{j}$ are the number densities
of light partons (quarks, antiquarks, and gluons) evaluated
at a scale $\mu$ in the proton and anti-proton;  $x_{i}$ and 
$x_{j}$ are the momentum fractions of the incoming partons, i.e. 
parton i (j) has momentum $x_{i}P$ ($-x_{j}P$), where $P$ is the
magnitude of the proton momentum in the center of mass frame
(which in colliding beam experiments coincides with the lab frame);
$\hat{\sigma}_{ij}$ is the point cross section for $i + j \rightarrow t \bar{t}$
and $\hat{s} = 4 x_{i} x_{j} P^{2} = x_{i} x_{j} s$ 
is the square of the
center of mass energy of the parton-parton collision.
The factorization and renormalization
scale $\mu$ is an arbitrary parameter with dimensions of energy, which
is introduced in
the renormalization procedure.   The exact result for the cross section
should be independent of the value of $\mu$.  However, since 
calculations are performed to finite order in perturbative QCD,
cross-section predictions are in general dependent on the
choice of scale, which is usually taken to be of the order of M$_{top}$.
The sensitivity of perturbative calculations to reasonable variations in
$\mu$ is used to estimate the accuracy of the prediction.
Parametrizations of the parton number densities ($F_{i}$ and $F_{j}$) are 
extracted from fits to a large number of experimental results,
mostly from deep inelastic scattering, see for example Fig.~\ref{pdf}.

The cross section for $p\bar{p} \rightarrow t\bar{t}$
can also be written as (Eichten {\em et al.}, 1984) :

$$\frac{d\sigma}{d\tau} = \sum_{ij} \frac{dL_{ij}}{d\tau}
\hat{\sigma}_{ij}(\hat{s},\mu^{2},M_{top})$$
where $\tau = \hat{s}/s$ and $dL_{ij}/d\tau$ are the differential parton
luminosities defined as :

$$\frac{dL_{ij}}{d\tau} = \frac{1}{1+\delta_{ij}} \int\limits^{1}_{\tau}
\frac{dx}{x}~ [F_{i}(x,\mu^{2}) F_{j}(\tau/x,\mu^{2}) +
F_{j}(x,\mu^{2}) F_{i}(\tau/x,\mu^{2})].$$
          
\begin{figure}[htb]
\vskip 1cm
\epsfxsize=4.0in
\gepsfcentered[20 200 600 600]{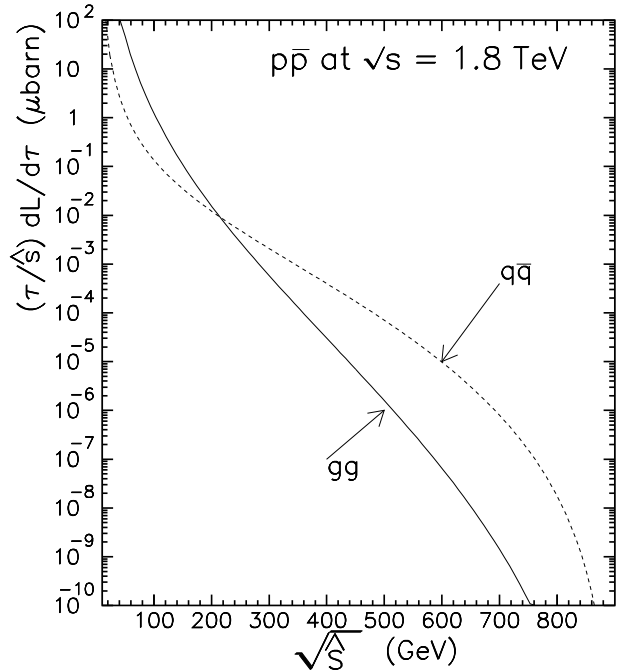}
\vskip 1cm
\caption{ \protect \baselineskip 12pt
Gluon-gluon ($gg$) and quark-antiquark \protect ($q\bar{q} = u\bar{u} +
d\bar{d} + s\bar{s}$) 
parton luminosities in \protect $p\bar{p}$ collisions at  \protect 
$\surd{s} = 1.8$ TeV.
These are calculated using the MT-B2 parametrization
of the parton distribution functions (Morfin and Tung, 1991)
evaluated at a scale $\mu^{2} = \hat{s}$.}
\label{lum}                                  
\end{figure}

The parton luminosities for quark-antiquark and gluon-gluon processes
at Tevatron energies are displayed in Fig.~\ref{lum}.
The sharp fall-off of these luminosities with increasing
$\hat{s}$, as well as the asymptotic
$1/\hat{s}$ dependence of $\hat{\sigma}$, 
result in predictions for the $t\bar{t}$ cross-sections
which fall steeply as a function of M$_{top}$.  At
$\sqrt{s} = 1.8$ TeV 
the $gg$ luminosity is larger than the $q\bar{q}$ luminosity up to 
$\hat{s} \approx$ (220 GeV)$^{2}$.  
As a result, 
top pair production is dominated by
the $gg \rightarrow t\bar{t}$ process up to 
M$_{top} \approx$ 90 GeV/c$^{2}$.  For
higher top quark mass, $q\bar{q}$ initial states are the most important
source of $t\bar{t}$ pairs.
                                                           
The leading order (LO) cross section for producing a pair of heavy quarks in
parton-parton collisions was calculated in the late 1970's 
(Gluck, Owens, and Reya, 1978; Combridge, 1979; Babcock, Sivers, and 
Wolfram, 1978; Hagiwara and Yoshino, 1979; Jones and Wyld, 1978;
Georgi {\em et al.}, 1978).
The full next-to-leading order (NLO) calculation was performed by  Nason, 
Dawson, and Ellis in 1988, and shortly afterwards by 
Beenaker {\em et al.}, 1991.
On the basis of their result,
cross-section predictions were then made (Altarelli {\em et al.},
1988; Ellis, 1991) 
by convoluting
the partonic cross-section with parametrizations of the
parton distribution functions.

There are two sources of uncertainty in such a
calculation of the $p\bar{p} \rightarrow t\bar{t}$ cross-section 
as a function of M$_{top}$.
As mentioned above, the first uncertainty is due to the nature of
the perturbative QCD calculation for the partonic 
cross-section ($\hat{\sigma}$).
The size of the uncertainty is customarily quantified by varying the
arbitrary value of the scale $\mu$ by a factor of two
around the top mass.  Note that this is not a rigorous procedure, and
it merely results in a {\em reasonable} estimate of the systematic
uncertainty due to the missing higher order terms in the calculation.
An additional uncertainty arises
from the limited knowledge of the input parton distribution
functions, and the assumed value of the QCD parameter $\Lambda_{QCD}$.
The $\Lambda_{QCD}$-dependence arises from the fact that the
assumed value of
$\Lambda_{QCD}$ 
affects the $\mu^{2}$ evolution of both $\alpha_{s}$ and
the quark and gluon distributions.  In particular, the extraction of 
the gluon distribution from deep inelastic data also depends on 
$\Lambda_{QCD}$.
The uncertainty on the cross-section calculation
due to the parton distribution uncertainties are very hard to quantify.
This uncertainty is usually estimated by
studying the variations of the calculated cross-section
using different parametrizations for the parton distribution functions
and different values of $\Lambda_{QCD}$.  As a result
of these studies, the total theoretical uncertainty on
the $t\bar{t}$ production cross-section at
$\sqrt{s}$ = 1.8 TeV is estimated
to be of order $\pm$~20\%.  The uncertainties due to the choice
of scale and to the parton distribution assumptions are
found to contribute approximately the same amount to the
total uncertainty.
       
The next-to-leading-order, O($\alpha_{s}^{3}$), predictions for the pair
production cross-section have been subsequently refined by Laenen,
Smith, and van Neerven (1992 and 1994).  In their calculation
the corrections due to initial state gluon brehmstrahlung,
which are large near $t\bar{t}$ threshold, have been resummed to all
orders in perturbative QCD, and have been included in
the computation.   This procedure introduces a new scale
$\mu_{0} >> \Lambda_{QCD}$, where the resummation is terminated
since the calculation diverges as 
$\mu_{0} \rightarrow 0$, where
nonperturbative effects are expected to dominate.
Given that the corrections due to soft gluons have been
shown to be positive at all orders in perturbative QCD for
$\mu$ = M$_{top}$,
they estimate the lower limit on the $t\bar{t}$ cross-section
as the sum of the full O($\alpha_{s}^{3}$) prediction
and the O($\alpha_{s}^{4}$) soft gluon correction, using
the conservative value of $\Lambda_{QCD}$ = 105 MeV.
Their best estimate of the cross section includes the
full gluon resummation contributions, and the uncertainty arises
mostly from the choice of $\mu_{0}$, which is allowed to become
as small as 0.05M$_{top}$ and 0.2M$_{top}$ for the 
$q\bar{q} \rightarrow t\bar{t}$
and $gg \rightarrow t\bar{t}$ channels respectively.  

A separate calculation of
the $t\bar{t}$ cross-section including the
perturbative resummation of gluon radiative corrections
has become available in the past year
(Berger and Contopanagos, 1995).  This
calculation is based on Principal Value Resummation (PVR)
techniques (Contopanagos and Sterman, 1993 and 1994), and is
independent of the arbitrary infrared cutoff $\mu_{0}$.
Theoretical uncertainties are estimated
by varying the renormalization and factorization scale 
$\mu$ by a factor of two around the
top mass.  A more recent evaluation of the effects
of gluon resummation suggests that its contribution is much smaller
than previously thought (Catani {\em et al.}, 1996).  
In Table~\ref{xsectheo} we summarize the 
results of the various calculations of the $p\bar{p} \rightarrow
t\bar{t}$ cross-section at Tevatron energies for a top mass of 175 
GeV/c$^{2}$ which, as we will discuss in Section~\ref{mass},
corresponds to the directly measured value of the top mass.


\begin{table}
\begin{center}
\begin{tabular}{ccc} \hline \hline
Calculation & Order & $\sigma(t\bar{t})$ \\
\hline
(1) Ellis, 1991 & NLO & $4.20^{+0.28}_{-0.54}$ pb \\ 
(2) Laenen, Smith and van Neerven, 1994 & NLO $+$ gluon resummation &  
$4.94^{+0.71}_{-0.45}$ pb \\
(3) Berger and Contopanagos, 1995 & NLO $+$ gluon resummation & 
$5.52^{+0.07}_{-0.45}$ pb \\
(4) Catani {\em et al.}, 1996 & NLO $+$ gluon resummation &
$4.75^{+0.63}_{-0.68}$ pb \\
\hline \hline
\end{tabular}
\end{center}
\caption{Calculations of the $p\bar{p} \rightarrow t\bar{t}$ cross-sections
at Tevatron energies for M$_{top}$ = 175 GeV/c$^{2}$.
Note that these cross-section calculations use different sets of parton
distribution functions (PDFs).  The systematic uncertainties in
(2) and (3) do not include the effects of varying the input PDFs.} 
\label{xsectheo}
\end{table}

\begin{figure}[htb]
\vskip 1cm
\epsfxsize=4.0in
\gepsfcentered[20 200 600 600]{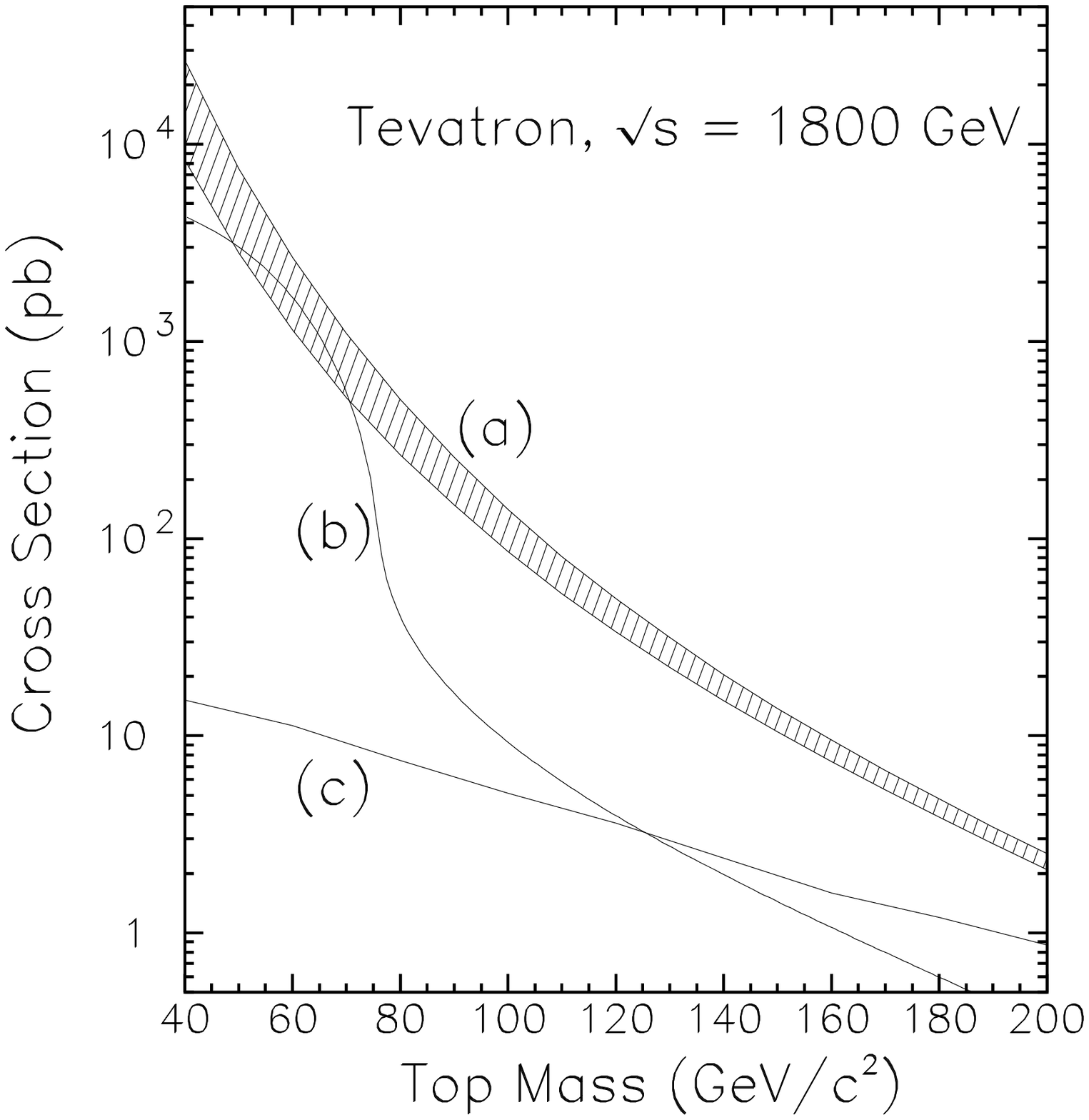}
\vskip 1cm
\caption{\protect \baselineskip 12pt
Top production cross-sections in 
$p\bar{p}$ collisions at $\surd{s}$ = 1.8 TeV.
(a) $ p\bar{p} \rightarrow t\bar{t}$ from 
Laenen, Smith and van Neerveen, 1994.  The band represents the
estimated theoretical uncertainty; 
(b) sum of $t\bar{b}$ and $\bar{t}b$ from $W$ decay (Drell-Yan);
(c) sum of $t\bar{b}$ and $\bar{t}b$ from $W$-gluon fusion.  See 
text for details.}
\label{xsec}
\end{figure}

\begin{figure}[htb]
\vskip 1cm
\epsfxsize=4.0in
\gepsfcentered[20 200 600 600]{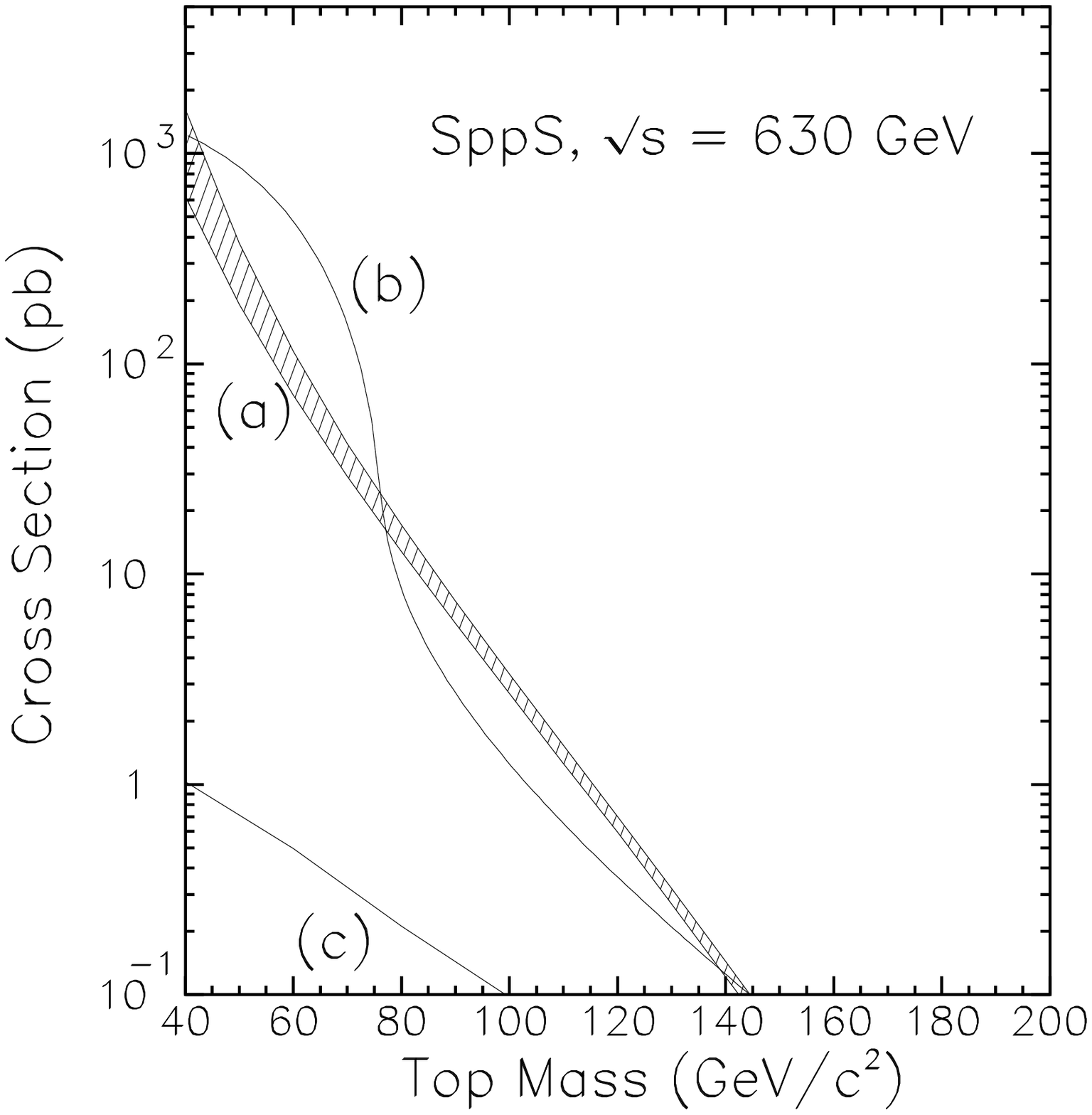}
\vskip 1cm
\caption{\protect \baselineskip 12pt
Top production cross-sections in 
$p\bar{p}$ collisions at $\surd{s}$ = 0.63 TeV.
(a) $ p\bar{p} \rightarrow t\bar{t}$ from 
Laenen, Smith and van Neerveen, 1994.  The band represents the
estimated theoretical uncertainty; 
(b) sum of $t\bar{b}$ and $\bar{t}b$ from $W$ decay (Drell-Yan);
(c) sum of $t\bar{b}$ and $\bar{t}b$ from $W$-gluon fusion.  See
text for details.}
\label{xsec2}
\end{figure}
           
\begin{figure}
\begin{picture}(32000,15000)(0,-3000)
\drawline\fermion[\E\REG](10000,6000)[6000]
\put(\particlefrontx,7000){$q$}
\drawline\photon[\E\REG](\particlebackx,\particlebacky)[6]
\put(\particlemidx,7000){$W$}
\THICKLINES
\drawline\fermion[\NE\REG](\particlebackx,\particlebacky)[5000]
\THINLINES
\put(26500,\particlebacky){$t$}
\drawline\fermion[\SE\REG](\particlefrontx,\particlefronty)[5000]
\put(26500,\particlebacky){$\bar{b}$}
\drawline\fermion[\S\REG](16000,6000)[8000]
\drawline\gluon[\W\REG](\particlebackx,\particlebacky)[6]
\put(\particlebackx,0){$g$}
\drawline\fermion[\E\REG](\particlefrontx,\particlefronty)[10000]
\put(\particlebackx,-1000){${q'}$}

\end{picture}
\caption{\protect \baselineskip 12pt
Diagram contributing to the O($\alpha_{s}$) corrections to
the Drell-Yan process $p\bar{p} \rightarrow W \rightarrow t\bar{b}$.}
\label{hior}
\end{figure}
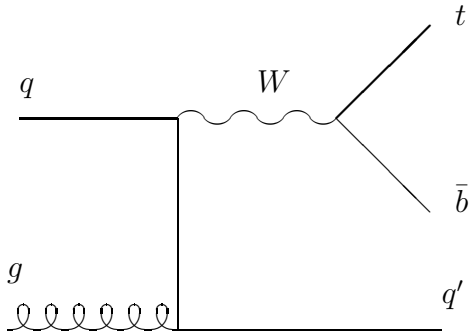

The expected top production cross-sections at Tevatron 
and S$p\bar{p}$S energies,
for the three production mechanisms,
are displayed in Fig.~\ref{xsec} and Fig.~\ref{xsec2}.
The Drell-Yan
cross section,
$\sigma(p\bar{p}\rightarrow W \rightarrow t\bar{b}$),
is calculated from the diagram shown in Fig. 17.
The value of this cross section is
normalized to the rate of $W$ 
production through measurements of
$p\bar{p} \rightarrow W \rightarrow e\nu$
(F. Abe {\em et al.}, 1991b; Alitti, {\em et al.}, 1990a; 
Albajar {\em et al.}, 1991a),
including corrections 
for the phase-space suppression of a $t\bar{b}$ pair and
the finite $W$ width.
The tree-level $W$-gluon fusion cross-section has been 
calculated by several authors (Willenbrock and Dicus, 1986; Yuan, 1990;
Anselmo, van Eijk, and Bordes, 1992;
Ellis and Parke, 1992; Bordes and van Eijk, 1993); 
the cross sections shown in Figs.~\ref{xsec}
and~\ref{xsec2}
are obtained using the PYTHIA 
Monte Carlo event
generator
(Sj\"ostrand and Bengtsson, 1987).
Since the $W$-gluon fusion
matrix element is calculated
at tree level, the systematic uncertainties
on the absolute rate prediction can be large, see for example
Ellis and Parke, 1992.  Here we have used the default PYTHIA scale
$\mu^{2}$ = 0.5 (M$_{t1}^{2}+$ M$_{t2}^{2}$), where
M$_{t1}$ and M$_{t2}$ are the transverse masses of the outgoing
partons.  Recently, a calculation of the next-to-leading order QCD
corrections for the $W$-gluon fusion process
has been performed (Bordes and van Eijk, 1995), and the
enhancement of the cross section over the Born-level result
has been found to be of order 30\%. 
It is worth mentioning that, despite their apparent similarities,
the Drell Yan and $W$-gluon mechanisms are quite distinct.
The higher order Drell-Yan diagram, $qg \rightarrow qW^{*}$,
$W^{*} \rightarrow t\bar{b}$ (see Fig.~\ref{hior}) 
and the $W$-gluon fusion diagrams (see Fig. 18) have
the same initial and final state partons.  
However, in one case the $W$ is space-like, and in the other case
it is time-like.  Furthermore,
the $t\bar{b}$ pairs in the two processes are in
different color states.

As anticipated, the pair production cross-section at the Tevatron 
(see Fig.~\ref{xsec}) is
dominant up to very high mass, except for the top mass
region around 60 GeV/c$^{2}$, where the cross sections for
the $t\bar{t}$ and 
$W \rightarrow t\bar{b}$ processes are approximately equal.
On the other hand, at the lower energy of the 
S$\bar{p}p$S collider, top production through $W$ decay
dominates in the mass region 40-80 GeV/c$^{2}$, (see Fig.~\ref{xsec2}).
Once the experimental evidence
started pointing towards higher top masses (see Section~\ref{pre1a}),
it became clear that top searches at the S$\bar{p}p$S were not
competitive with those at the Tevatron, due to the lower
$t\bar{t}$ cross-section at 
$\sqrt{s}$ = 630 GeV.  

It should be emphasized that top
production is a very rare process in $p\bar{p}$ collisions.
The total inelastic cross-section at the Tevatron is
approximately 60 mb (F. Abe {\em et al.}, 1994b), 
ten orders of magnitude higher than 
$\sigma(t\bar{t})$ for M$_{top}$ = 175 GeV/c$^2$.  Therefore,
in trying to isolate a top signal, both excellent background rejection
and high luminosities are critical.

In the remainder of this Section and in most of this review
we will concentrate on the $ p\bar{p} \rightarrow t\bar{t}$
reaction.  We will however revisit the $W$-gluon fusion 
process in Section~\ref{future}, since it is interesting in its
own right and its study will become accessible in the not too
distant future.

\subsection{Top quark hadronization}
\label{hadronization}
\begin{figure}[htb]
\vskip 1cm
\epsfxsize=4.0in
\gepsfcentered[20 200 600 600]{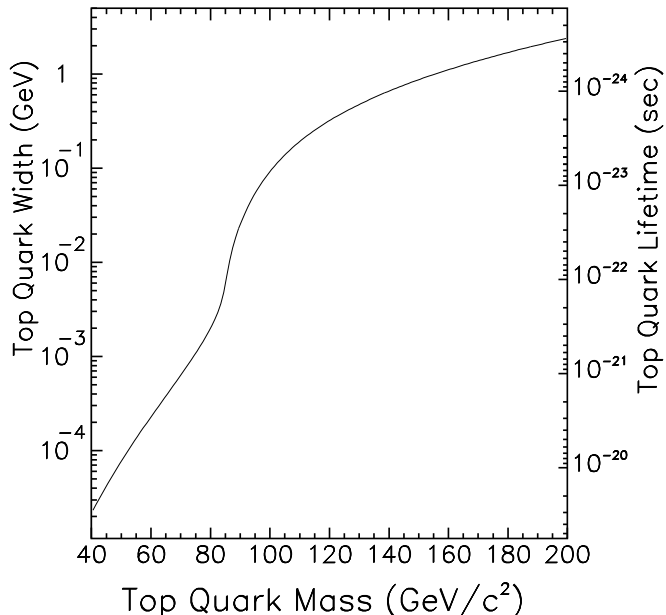}
\caption{\protect \baselineskip 12pt
The Standard Model width of the top quark
as a function of its mass (Bigi, 1986).  Note the transition
between the region of virtual and real $W$ decays which occurs
at M$_{top} \approx$ M$_{W}$ $+$ M$_{b}$.}
\label{topw}
\end{figure}

Quarks are not observed as free particles but
are confined to form hadronic bound states.   
The top
quark however is unique in that its mass is high
enough that it can decay before hadronization.  According to
the Standard Model, top quarks undergo the
weak decay $t \rightarrow W b$, where 
the $W$ boson is real if M$_{top} >$~M$_{W} +$ M$_{b}$,
and virtual otherwise.  Decay modes such as $t \rightarrow W s$
and $t \rightarrow W d$ are also allowed. They
are suppressed by factors of $|V_{ts}|^{2} / |V_{tb}|^{2}
\approx 10^{-3}$ and
$|V_{td}|^{2} / |V_{tb}|^{2} \approx$ 5 x 10$^{-4}$ 
respectively, where $V_{ij}$ is a Cabibbo-Kobayashi-Maskawa 
(CKM) mixing-matrix element (Montanet {\em et al.}, 1994).
The expected
width of the top quark, and hence the lifetime, as a function
of its mass, is shown in Fig.~\ref{topw}.

The hadronization process, which is non-perturbative in nature,
is not well understood.  However, the formation of hadrons
is estimated to take place in a time of order
$\Lambda_{QCD}^{-1} \approx$ O(100 MeV)$^{-1} \approx$ O($10^{-23}$)
seconds (Bigi, 1986).   As can be seen in Fig.~\ref{topw},
the top lifetime becomes shorter than this characteristic time 
if the top mass is higher than approximately 100 GeV/c$^{2}$.
A more quantitative treatment is given by Orr (1991),
and is briefly summarized here.  In this model, the $t$ and 
the $\bar{t}$ emerging from the hard scatter are linked by color
strings to the remnants of the proton and anti-proton.
When the separation between the outgoing quarks and
the color-connected remnants exceeds a distance of
order 1 fm, the stretched color string is expected to break,
resulting in
the creation of fragmentation particles from the vacuum, and
possibly the formation 
of a bound state top hadron.
For M$_{top} >$ 165 GeV/c$^{2}$,
the top production kinematics at the Tevatron are such
that essentially all top quarks are expected to
decay before having travelled that minimum distance.
Conversely, for top masses
below 120 GeV/c$^{2}$, the overwhelming majority
of top quarks will survive to a distance of 1 fm, and hadronization
effects are expected to occur.  Varying the assumption on the
hadronization distance by a factor of two results in a mass-shift of order 
20 GeV/c$^{2}$ for the transition region between hadronization and
free-quark decay.

The exceedingly short top-quark
lifetime is due not only to the very high mass, but also to the fact
that M$_{top} >$~M$_{b}$, so that the the top can decay into $Wb$, and
this decay mode is not CKM suppressed.
A heavy I$_{3}=-1/2$ fourth generation quark ($b'$)
would decay into
$Wu'$, where $u'$ here stands for an I$_{3}=+1/2$ up-type
quark.  The decay rate would be proportional to the
square of the CKM mixing-matrix element that connects
$b'$ and $u'$.  
If the $b'$ were lighter than the fourth-generation
up-type quark, then only generation-changing, CKM
suppressed, decays would be allowed.  As a result, the
lifetime of such a fourth generation quark could
be considerably longer than that of a top-quark of the same mass.
Fourth generation heavy hadrons, as well as new quarkonia states,
could then still be allowed to form.

Even if hadronization effects do occur in $t\bar{t}$ production,
their effects, although potentially interesting, are not expected 
to be experimentally observable, at least in the forseeable
future.  The reason is that the fragmentation of heavy quarks is
{\em hard},  i.e.
the fractional energy loss of the top quark as it hadronizes
is small
(Peterson {\em et al.}, 1983).  
Distortions
to the kinematics of the top quark from the
perturbative partonic calculation are
minimal.  Additional particles produced in the hadronization process
have little effect on the overall event 
topology.  If a top hadron is indeed produced, the 
kinematics of the top decay will not be very different from that of
a free quark decay since the companion quark is so much lighter.
The fragmentation of the $b$-quark produced in top decay
could potentially be more seriously affected. This is because the
color string would link the $b$-quark to the 
light quark produced in the top fragmentation
rather than the proton or anti-proton
remnant (Orr, 1991).  However, all top-quark experimental studies to date
have not been precise enough to be sensitive to fragmentation
assumptions.

\begin{figure}[htb]
\vskip 1cm
\epsfxsize=5.0in
\gepsfcentered[20 200 600 600]{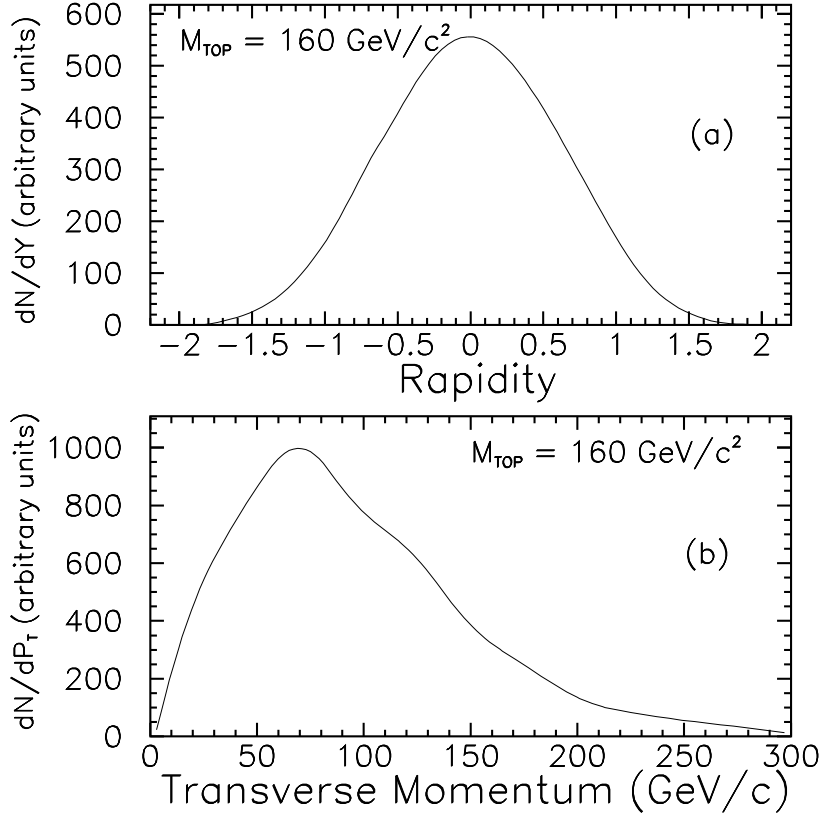}
\vskip 1cm
\caption{\protect \baselineskip 12pt
The expected rapidity (y) and transverse momentum (P$_{T}$)
distributions for top quarks at the Tevatron.  The predictions
are form the ISAJET Monte Carlo program.}
\label{toppt}
\end{figure}

\subsection{Underlying event}
After the hard collision, the remnants of the proton and anti-proton 
also hadronize.   This process cannot be described within
the framework of perturbative QCD, and is therefore poorly
understood.  The particles from the remnant hadronization 
form what is usually referred to as the {\em underlying event}.
The structure of the underlying
event is similar to that of the bulk of soft $p\bar{p}$
collisions (the so-called minimum bias events).

Minimum bias 
events are events collected
with a simple interaction trigger.
This trigger
usually consists of a coincidence
between large banks of scintillator counters in the very forward and 
backward regions, and is highly efficient for all types of inelastic 
$p\bar{p}$ collisions, except for singly diffractive events.
In minimum bias events, the average
transverse momentum of the hadrons is P$_{T} \approx$ 500 MeV/c
(Para, 1988).  (Transverse momentum is the component of momentum 
perpendicular to the direction of the beams).
Most of the energy is carried away
by particles which remain inside the beam pipe and are not seen in the
detector.  The charged particle multiplicity
in the central region per unit pseudo-rapidity (dN$^{ch}$/d$\eta$) grows
approximately
logarithmically with the center-of-mass energy, and
at $\sqrt{s}$ = 1.8 TeV is dN$^{ch}$/d$\eta$ $\approx$ 4 
(F. Abe {\em et al.}, 1990e). (NB: the rapidity $y$ of a particle
is defined in terms of the longitudinal Lorentz boost,
with $\beta$~= tanh$y$, to the frame in which
the particle's momentum
is purely transverse.  Rapidity can be written as

$$y~~ = \frac{1}{2}~\ln\frac{E + P_{z}}{E - P_{z}}~~ =~~
\frac{1}{2}~\ln\frac{(E + P_{z})^{2}}{M^{2} + P_{T}^{2}}$$
where $E$ is the energy of the particle, $M$ is its mass,
and 
$P_{z}$ and $P_{T}$ are the components of momenta parallel and
transverse to the beam direction.
Pseudo-rapidity ($\eta$) is the
rapidity calculated neglecting the
particle's mass, $\eta = -$ ln tan($\theta /2$), where
$\theta$ is the polar angle with respect to the proton direction.
Zero rapidity or pseudorapidity corresponds to particles moving
at 90$^{o}$ from the beamline; high values of $|y|$ or $|\eta|$
implies very forward or backward going particles).

\subsection{Modelling of top quark production}
\label{modelling}
The reliability of the modeling of $t\bar{t}$ production in 
is an important issue.
Top production is usually modelled using a QCD shower Monte
Carlo program, such as ISAJET (Paige and Protopopescu, 1986), 
HERWIG (Marchesini and Webber, 1984 and 1988)
or PYTHIA (Sj\"ostrand and Bengtsson, 1987).  
These Monte Carlos are used by the experimental groups
to calculate the $t\bar{t}$ acceptance and kinematics, and to model
the resolution of the top quark mass measurement, see 
Section~\ref{mass}.

In all these Monte Carlo programs, the
initial hard scatter is generated from tree-level matrix
elements convoluted with parametrizations of the parton
distribution functions.  Initial and final state partons
are then developed into a gluon and $q\bar{q}$ 
radiation cascade, with angular and energy spectra
based on the QCD Altarelli-Parisi splitting function.  
The QCD shower is terminated when the virtual invariant
mass of the parton in the cascade becomes smaller than
a minimum value which is 
of order 1 GeV for HERWIG and PYTHIA
and 6 GeV for ISAJET, 
at which point
perturbative QCD is expected to break down.  
Phenomenological
models are then employed to combine the remaining partons
into hadrons.  The underlying event is also modelled in a
phenomenological way, with a number of parameters tuned
to reproduce the hadron multiplicities and 
transverse momentum spectra measured in soft $p\bar{p}$ collisions
(minimum bias events).
Short lived particles are made to decay, 
with branching ratios
and decay models based on the compilation from the Particle
Data Group (Montanet {\em et al.}, 1994).

The main differences
between these Monte Carlo event generators reside in the
modeling of the radiation processes.  
ISAJET employs
an independent fragmentation model, i.e. radiation from each
parton occurs independently from the structure of the 
rest of the event, whereas in both HERWIG and PYTHIA radiation
is more realistically emitted 
taking into account color-correlations between all
partons in the initial and final state.
The output of these Monte Carlo event generators consists
of a list of stable particles, which can then be fed
to a detector simulation for detailed studies of the 
expected signature of a top event.

In the pair production process, the $t$ and $\bar{t}$ quarks
are produced in the central rapidity region, with P$_{T}$ of
order M$_{top}$/2, (see Fig.~\ref{toppt}).   These features
of $t\bar{t}$ production 
can be simply understood from the properties of the 
$i+j \rightarrow t\bar{t}$ process.
The cross-section for the $q\bar{q} \rightarrow t\bar{t}$ 
subprocess, which dominates at high top mass, is given
at lowest order by (see for example Nason, Dawson, and Ellis, 1988) :

$$\hat{\sigma}~~ = \frac{8\pi\alpha_{s}^{2}}{27\hat{s}}
\sqrt{1-\frac{4M_{top}^{2}}{\hat{s}}}~~ 
(1 + \frac{2M_{top}^{2}}{\hat{s}})$$
The parton-parton cross-section as a function of $\hat{s}$
rises from zero at threshold ($\hat{s} = 4$M$_{top}^{2}$),
reaches a maximum at $\hat{s} = 5.6$M$_{top}^{2}$, and then
falls off asymptotically as 1$/\hat{s}$.  When convoluted with
the falling $q\bar{q}$ luminosity, 
(see Fig.~\ref{lum}),
the maximum of the 
$q\bar{q} \rightarrow t\bar{t}$ cross-section is shifted
down to $\hat{s} \approx 4.5$M$_{top}^{2}$.  Therefore, the most
probable energy for a top quark is E $\approx \sqrt{4.5}$M$_{top}$/2
$\approx 1.1$M$_{top}$, and the most probable momentum
is P $\approx 0.4$M$_{top}$.
%

From the definition of rapidity, it is clear that 
the maximum of $|y|$ occurs as $P_{T} \rightarrow$~0 and at maximum $E$,
which for pair-produced objects is
$E = 0.5\sqrt{s}$ = 900 GeV at the Tevatron.  
For M$_{top}$ = 160 GeV/c$^{2}$,
the kinematic limit is then $|y| < 2.4$; however, as can be seen
from Fig.~\ref{toppt}, most top quarks have $|y| < 1.5$.
High values of $|y|$ are suppressed because they 
require $P_{T} \rightarrow 0$, where the phase space factor
also $\rightarrow 0$, and they require
high values
of $E$, i.e. high values of $\hat{s}$.  Both the parton-parton
luminosities and parton-parton
cross-section fall off with increasing $\hat{s}$.

The lowest
order diagrams (Fig. 16) lead to a back-to-back
topology for the $t$ and the $\bar{t}$
in the transverse plane, which is slightly modified
by higher order corrections.   Because the top quark momentum
is not large compared to M$_{top}$, the decay products are
not significantly boosted along the original top quark flight path,
leading to nearly spherical events.  
In the next Section we will
turn to the discussion of top decays and signatures.

\begin{figure}[htb]
\vskip 1cm
\epsfxsize=5.0in
\gepsfcentered[20 200 600 600]{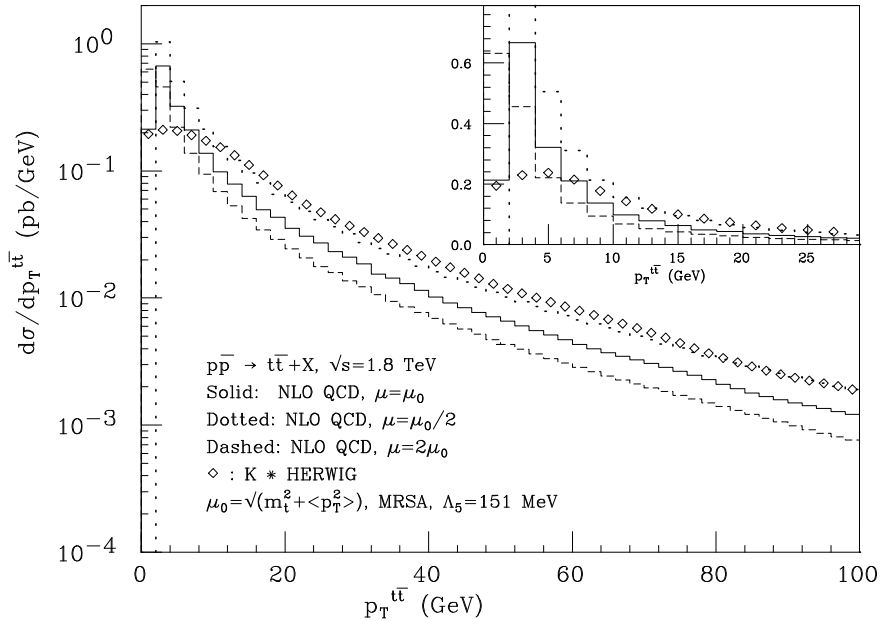}
\vskip 1cm
\caption{\protect \baselineskip 12pt
Comparison of the transverse momentum distributions of the $t\bar{t}$ pair
as predicted from the Herwig Monte Carlo and next-to-leading order (NLO)
QCD.  At Tevatron energies, and for M$_{top}$ = 176 GeV/c$^{2}$.
From Frixione {\em et al.}, 1995.}
\label{frixione}
\end{figure}

As was mentioned above, shower Monte Carlos are based on the LO
matrix element for $t\bar{t}$ production and models of initial and
final state radiation.  It is interesting to compare these
models with higher order QCD calculations.  The earlier NLO order
calculations of $t\bar{t}$ production (Nason, Dawson, and
Ellis, 1988; Beenaker {\em et al.}, 1991) are not sufficient, since
these are 
calculations of single quark kinematic distributions
such as P$_{T}$ and rapidity, integrated over the whole phase-space
for the other quark.  
More recently, a NLO calculation of the doubly-inclusive cross-section
for heavy quark production has become available (Mangano, 
Nason and Ridolfi, 1992).    This
calculation allows for comparisons not only of single $t$ or
$\bar{t}$ distributions, but also of correlated distributions, e.g. 
the $t\bar{t}$ invariant mass (M($t\bar{t}$)), the 
$t\bar{t}$ transverse momentum (P$_{T}$($t\bar{t}$)), and
the azimuthal separation between t and $\bar{t}$ ($\Delta\phi$).
Detailed comparisons between the Herwig model and NLO-QCD have
been performed  (Frixione {\em et al.}, 1995). 
Excellent agreement is found  
in the shapes of distributions of quark rapidity, P$_{T}$, and 
M($t\bar{t}$), except for very large values of the latter two quantities.
In this kinematic regime, multiple gluon emission
from the final state top quarks becomes important, and this process is
not modelled by the NLO QCD calculation.  Disagreement between
NLO QCD and HERWIG is also observed in 
distributions of P$_{T}$($t\bar{t}$) and $\Delta\phi$, 
(see Fig.~\ref{frixione}).
Note that at LO these distributions are delta-functions, with
P$_{T}$($t\bar{t}$) = 0 and $\Delta \phi$ = 180$^{o}$, and
deviations from the delta-function behavior are due entirely
to higher order corrections.
For small P$_{T}$($t\bar{t}$), multiple gluon emission is expected to dominate,
and the HERWIG model is expected to be more realistic.

Studies of the expected single top quark P$_{T}$ and rapidity
distributions, calculated in NLO QCD and including the 
resummation of the leading soft gluon corrections, have also
been performed (Kidonakis and Smith, 1995).   The shapes of
these distributions are found to be essentially
identical to those calculated at NLO, which were shown
to agree with the Herwig model.
Comparisons of
gluon emission
in $t\bar{t}$ events 
from HERWIG and from a O($\alpha_{s}^{3}$) matrix element calculation,
including initial and final state gluon radiation have also been made
(Orr, Stelzer, and Stirling, 1995).
These authors find larger contributions of gluon radiation 
in HERWIG than in the matrix element calculation.  
This effect may also be due to the absence of multi-gluon emission
in the calculation.

As we will discuss in Section~\ref{mass}, understanding gluon
radiation in $t\bar{t}$ events is crucial for a precise determination
of the top mass.  We expect that this subject will attract more
and more attention in the next few years.


\section{Top quark signatures}
\label{sig}
Since the top quark decays with a very short lifetime, only its
decay products can be detected.  Therefore, to understand 
the experimental signature for a top event, we first
discuss the decay modes of the top quark.  In this Section we 
review the top quark decay properties and we 
present a general discussion of how the top quark can be
observed, paying particular attention to the background sources.
We will concentrate on the signature for $p\bar{p} \rightarrow
t\bar{t}$, since this is the most important
production mechanism at Tevatron Collider energies.

\subsection {Standard Model top quark decay modes}
\label{SMdecay}
As mentioned in 
the previous Section, according to 
the Standard Model the top quark decays as $t \rightarrow Wb$, where
the $W$ boson is real or virtual depending on the top mass. (Non-
Standard Model decay modes of the top quark will be reviewed in
Section~\ref{NSMdecay}).  The $W$ will subsequently decay into fermion pairs,
either $W \rightarrow l\nu$ or $W \rightarrow q\bar{q}$, where $l$
denotes a charged lepton, and
$q\bar{q}$ denotes a light-quark pair, 
$u\bar{d}$ or $c\bar{s}$.
At tree level, the $W$ couples with equal strength to 
leptons and quarks, so each 
$W$ decay mode occurs with equal probability.  There are
three leptonic channels ($e\nu$, $\mu\nu$, and $\tau\nu$), and
six hadronic channels ($u\bar{d}$ and $c\bar{s}$, with three possible
color assignments), hence each decay mode has a branching ratio
of $1/9$.  QCD corrections enhance the branching ratios of
the hadronic modes by a factor of ($1 + \alpha_{s}/\pi$) 
$\approx 1.05$.  Given the $W$ branching fractions, it is
a simple matter to list the $t\bar{t}$ decay modes,
see Table~\ref{tdec}.

\begin{table}
\begin{center}
\begin{tabular}{cc} \hline \hline
Decay mode & Branching ratio \\ \hline
$t\bar{t} \rightarrow q\bar{q}~ q\bar{q}~ b \bar{b} $ & $36/81$ \\
$t\bar{t} \rightarrow q\bar{q}~ e\nu~ b \bar{b} $     & $12/81$ \\
$t\bar{t} \rightarrow q\bar{q}~ \mu\nu~ b \bar{b} $   & $12/81$ \\
$t\bar{t} \rightarrow q\bar{q}~ \tau\nu~ b \bar{b} $  & $12/81$ \\
$t\bar{t} \rightarrow e\nu~ \mu\nu~ b \bar{b} $       & $2/81$ \\
$t\bar{t} \rightarrow e\nu~ \tau\nu~ b \bar{b} $      & $2/81$ \\
$t\bar{t} \rightarrow \mu\nu~ \tau\nu~ b \bar{b} $    & $2/81$ \\
$t\bar{t} \rightarrow e\nu~ e\nu~ b \bar{b} $         & $1/81$ \\
$t\bar{t} \rightarrow \mu\nu~ \mu\nu~ b \bar{b} $     & $1/81$ \\
$t\bar{t} \rightarrow \tau\nu~ \tau\nu~ b \bar{b} $   & $1/81$ \\
\hline\hline
\end{tabular}
\end{center}
\caption{\protect \baselineskip 12pt
Decay modes for a $t\bar{t}$ pair and their lowest order 
branching ratios assuming Standard Model decays.}
\label{tdec}
\end{table}

\subsection{Detection of the top decay products}
\label{detect}
The possible final states contain combinations of
electrons, muons, taus, neutrinos, and quarks.  
Here we briefly illustrate techniques for detection of the
top-quark decay products.

\begin{figure}[htb]
\hspace*{1.0in}
\epsffile{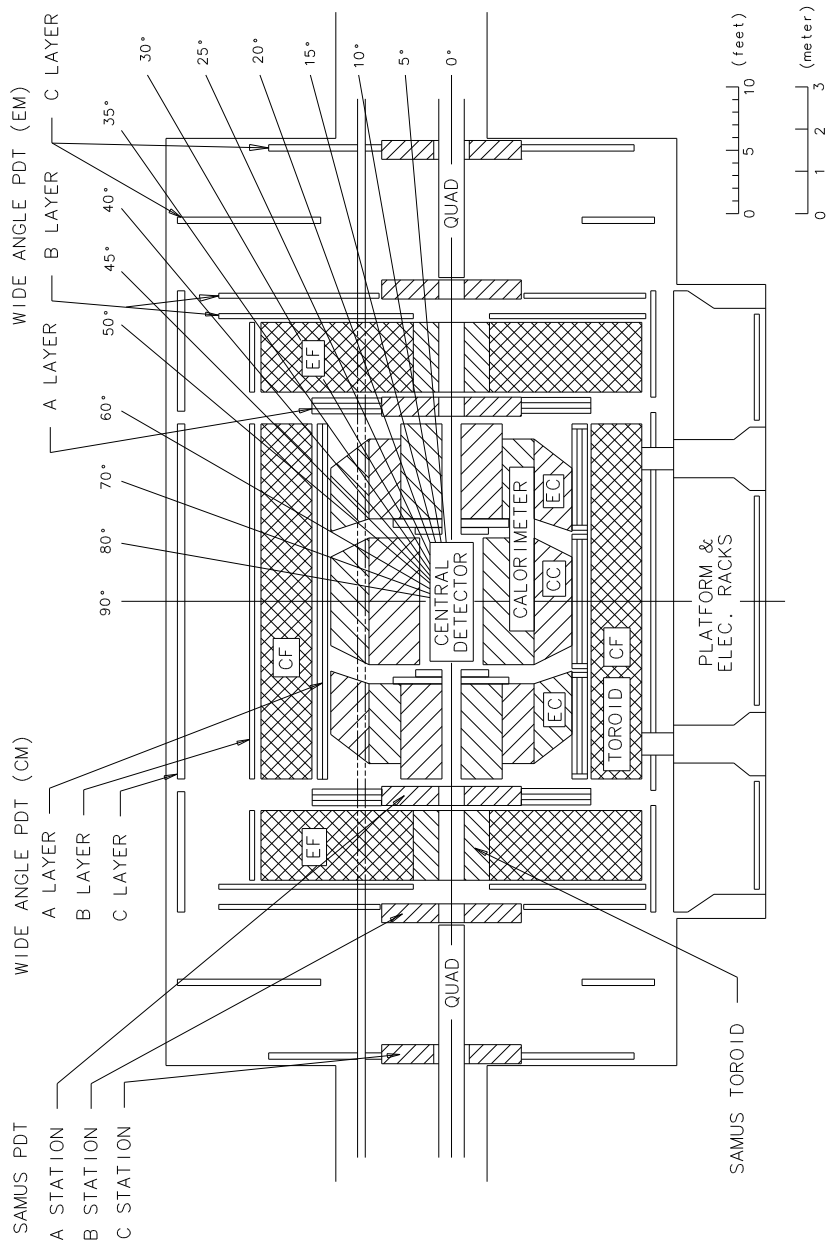}
\vskip 1cm
\caption{\protect \baselineskip 12pt
The D0 detector at the Tevatron.  EC and CC are liquid-argon-uranium 
calorimeters.  The 
central detector provides tracking information.  Muons are detected
using the five toroids (CF, EF, SAMUS) 
and the proportional drift tube (PDT) systems, 
CM and EM. From Snyder, 1995a.}
\label{d0det}
\end{figure}

\begin{figure}[hbt]
\epsfxsize=4.0in
\hspace*{1.5in}
\epsffile{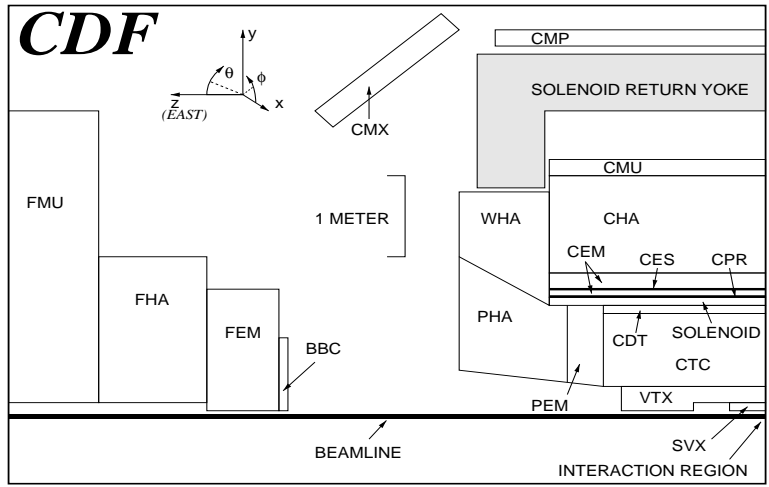}
\vskip 3cm
\vspace*{-1.0in}
\caption{\protect \baselineskip 12pt    
A side-view cross-section of one quadrant of 
the CDF detector at the Tevatron. The detector is forward-backward symmetric 
about the interaction region, which is at the lower-right corner of
the figure.
SVX, VTX, CTC, and CDT are tracking detectors.
CEM, CHA, WHA, PEM, PHA, FEM, and FHA are calorimeters.  CMU, CMP, CMX, and
FMU are muon detectors.  BBC is a bank of scintillators.  CPR and CES are 
multiwire proportional chambers placed in front and in the middle of the 
central electromagnetic calorimeter (CEM). From F. Abe {\em et al.}, 1994a.}
\label{cdfdet}
\end{figure}

General-purpose $p\bar{p}$ collider detectors are needed for top quark
physics.  These detectors are designed to cover as much as possible
of the solid angle around the interaction point, and are
composed of a number of sub-detectors optimized for study of different
aspects of the event.  A number of such detectors have been used
(UA1 and UA2 at the CERN S$p\bar{p}$S),
are still in operation (CDF and D0 at Fermilab's Tevatron),
or are now being designed and constructed (CMS and ATLAS, at the
proposed LHC $pp$ collider).
While the details of the design of these detectors are
different, their overall structure is in general
quite similar.  The region immediately surrounding the
interaction region is instrumented with detectors designed
to measure the trajectories of charged particles.  Except
for UA2 and D0, the tracking volume is immersed in 
a magnetic field for momentum measurement.
The tracking volume is surrounded by calorimeters, where 
measurements of the
energy of electromagnetic and hadronic showers are performed.
Calorimeters are segmented both longitudinally and transversely
to the direction of flight of particles originating from
the interaction point.  Transverse segmentation is necessary
to measure the position of the showers, while longitudinal
information is used to distinguish between electromagnetic (EM) 
and hadronic (HAD) showers.
%
%
%
%
Calorimeters cover
most of the 4$\pi$ solid angle around the interaction region,
however the very forward and backward regions 
must be left uninstrumented to allow for the passage of the beam pipe.
Muon detectors consisting of additional tracking
devices, hadron absorbers, and possibly 
magnets
for momentum measurement, are placed outside the calorimeter.
Drawings of the two collider detectors at Fermilab's Tevatron 
are shown in Fig.~\ref{d0det} and Fig.~\ref{cdfdet}.


\clearpage

\subsubsection{Detection of electrons and muons}
\label{lepdetect}

Electrons are identified as highly electromagnetic showers
in the calorimeters.  If momentum information from the
tracking system is available, consistency between the
measured momentum of the electron candidate and the energy
of the corresponding EM-shower provides a powerful handle
for rejection of backgrounds from e.g. hadronic shower fluctuations
and overlaps between hadron tracks and photons from 
$\pi^{0}$ decay.   Information from the transverse 
and longitudinal shapes of
the shower, from ionization measurements in the tracking
chamber (dE/dX), and from the response of transition radiation
and preshower detectors, are also used for electron identification.

\begin{figure}[htb]
\epsfxsize=4.0in
\vskip 1cm
\gepsfcentered[20 200 600 600]{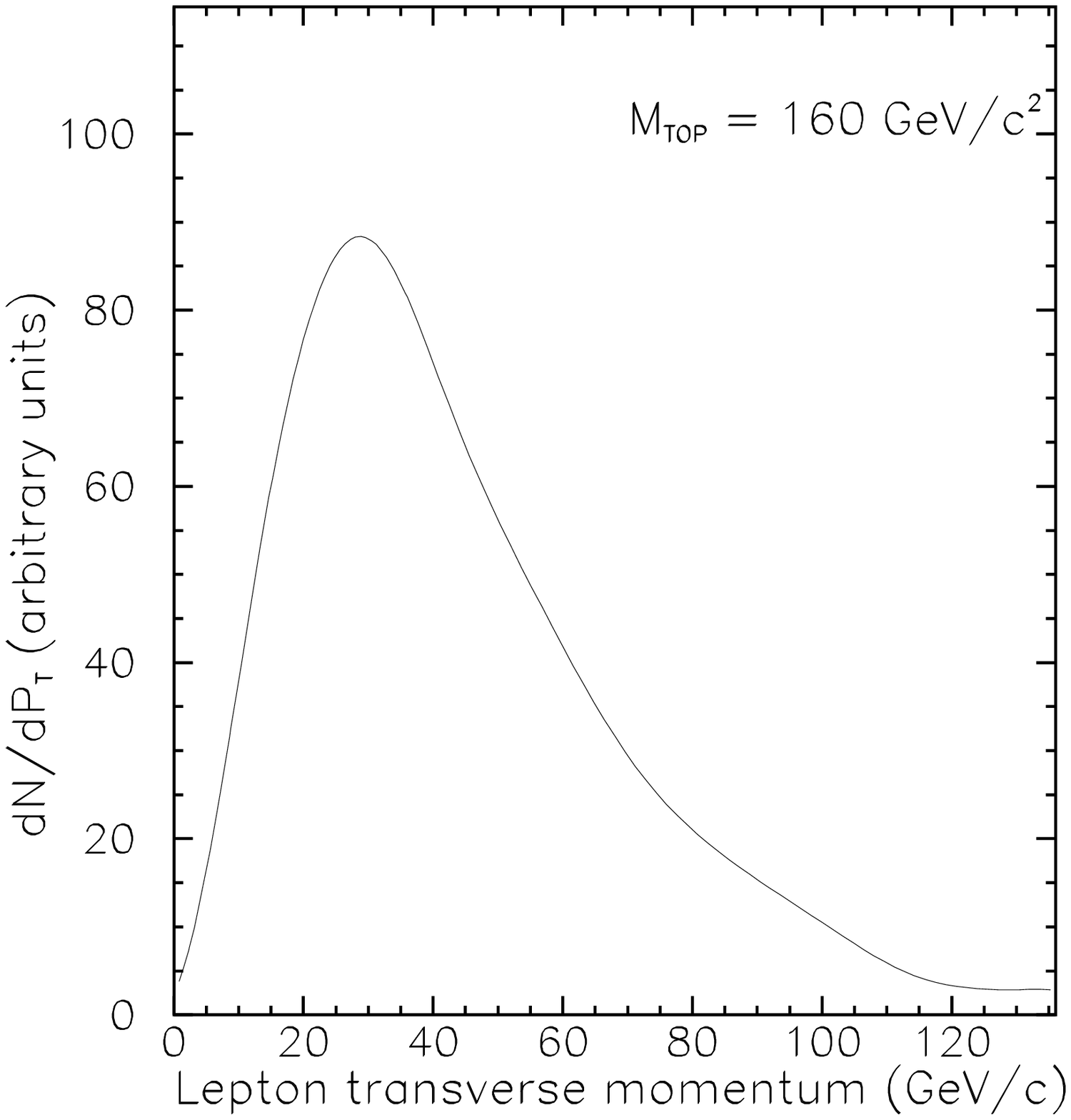}
\vskip 1cm
\caption{\protect \baselineskip 12pt
The expected lepton transverse momentum from 
$t\rightarrow Wb \rightarrow l\nu$ from the ISAJET Monte Carlo generator.
This is for $p\bar{p} \rightarrow t\bar{t}$ at $\surd{s}$ = 1.8 TeV.
Detector effects are not included.}
\label{wlep}
\end{figure}

Electrons from $W$ decays in top events have high transverse
momentum, (P$_{T}$, see Fig.~\ref{wlep}), and are expected to be isolated,
i.e. well separated from the other decay products of the two
top quarks in a $t\bar{t}$ event. These electrons can be
identified with high efficiency,
and their energy can be measured very precisely in the calorimeter
(see Table~\ref{lepres}).
On the other hand, identification of electrons from 
$b \rightarrow ce\nu$ in a top event
is much more problematic.  These electrons have lower transverse momentum
than electrons from $W$ decays and,
since the $b$ is highly boosted, the nearby
hadrons from the $b$ fragmentation and $b$ or $c$ decay 
may deposit their energy in the same calorimeter cells as these electrons.

\begin{table}
\begin{center}
\begin{tabular}{ccc}
\hline \hline
Detector & Electron energy resolution & Muon momentum resolution \\
\hline
UA1  & $\sigma$(E)/E = (0.15-0.21)/$\sqrt{E}$~~$\oplus$ .03
& $\sigma$(P)/P = 0.005P \\
UA2  & $\sigma$(E)/E = 0.17/$\sqrt{E}$~~$\oplus$ .02
& - \\
CDF  & $\sigma$(E$_{T}$)/E$_{T} = 0.14/\sqrt{E_{T}}$~~$\oplus$ .02
& $\sigma$(P$_{T}$)/P$_{T}$ = 0.0009P$_{T}$~~$\oplus$ 0.0066 \\
D0   & $\sigma$(E)/E = 0.15/$\sqrt{E}$~~$\oplus$ .01
& $\sigma$(P)/P = 0.01P~~$\oplus$ .2 \\
\hline \hline
\end{tabular}
\end{center}
\caption{\protect \baselineskip 12pt
Electron energy (GeV) and muon momentum (GeV/c) resolutions 
in the central region for the UA1
(Albajar {\em et al.}, 1989), UA2 (Alitti {\em et al.}, 1992a),
CDF (F. Abe {\em et al.}, 1994a), and D0 (Abachi {\em et al.}, 1994)
detectors.
The UA1 muon momentum resolution is for measurements
in the central-detector, for muons at 90$^{o}$ from the direction of 
the dipole field.  The UA1 electron energy resolution changed
between 1983 and 1985 due to radiation damage of the scintillator. The symbol
$\oplus$ indicates that the two terms are added in quadrature.
The subscript $T$ refers to components transverse to the beam direction.}
\label{lepres}
\end{table}

Muons can also be reliably identified 
as charged particles that penetrate the
calorimeter and reach the outside muon detectors.
Backgrounds to the muon signal arise from decays in
flight of pions and kaons, and from hadrons that 
traverse the calorimeter and hadron absorbers 
without interacting ({\em punchthrough}).  
If there is a magnetic field in the inner
tracking system (before the calorimeter), then
the muon momentum is precisely measured;
otherwise, (e.g. in the D0 experiment) it is measured, with
worse resolution, in the outer muon detector
(see Table~\ref{lepres}).

\begin{figure}[htb]
\hspace*{1.0in}
\epsfxsize=4.0in
\epsffile{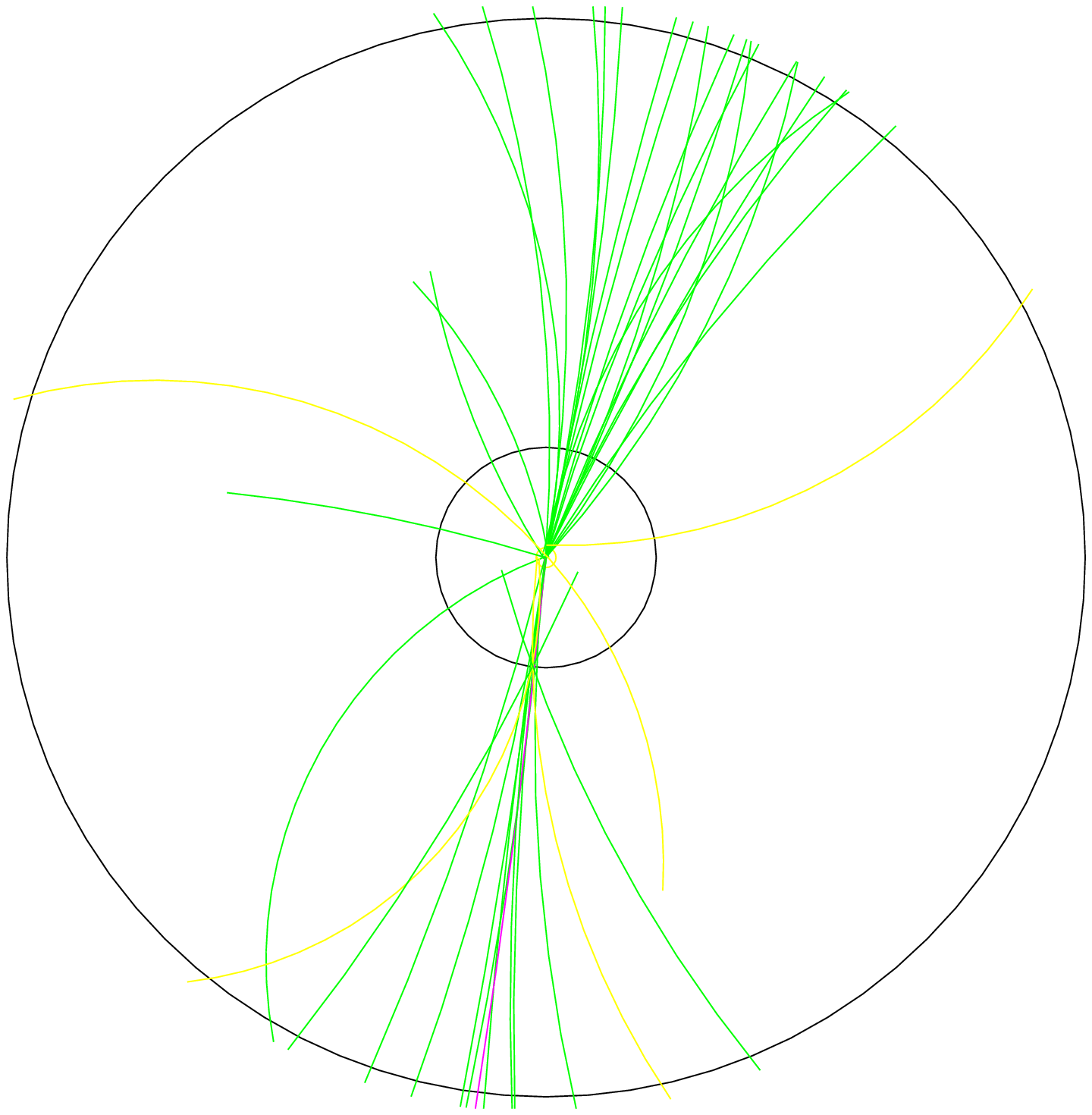}
\vskip 1cm
\caption{\protect \baselineskip 12pt
A $p\bar{p} \rightarrow $ jet-jet event in CDF.  Here we show the reconstructed
tracks in the transverse plane.  The {\em two-jet} structure is apparent.}
\label{jetjet}
\end{figure}

\begin{figure}[hbt]
\epsfxsize=4.0in
\vskip 1cm
\hskip 4cm
\epsffile{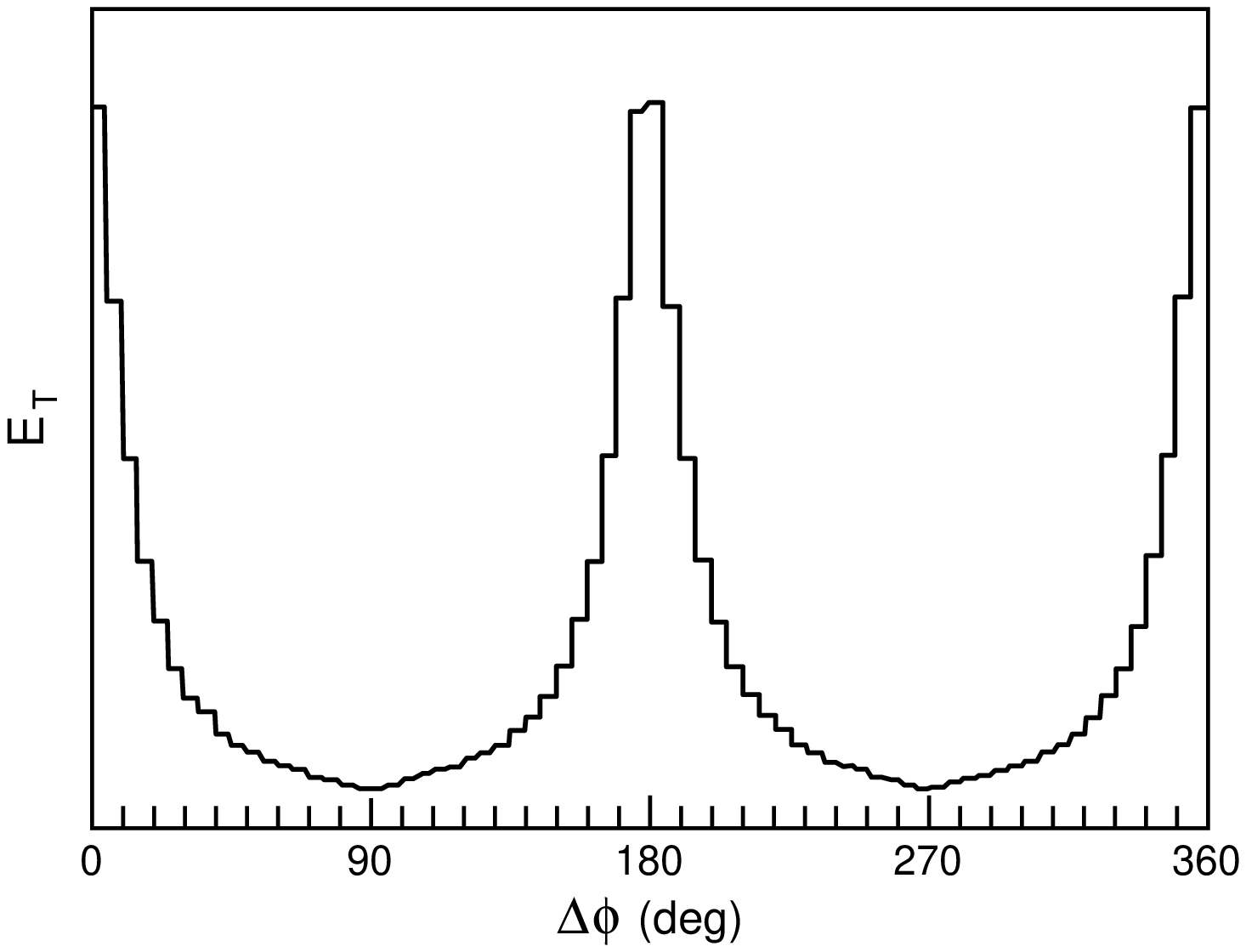}
\vskip 1cm
\caption{\protect \baselineskip 12pt
The relative E$_{T}$ distribution in calorimeter cells at an angle $\phi$ 
with respect to the transverse jet thrust axis
for CDF di-jet data with a 30 GeV E$_{T}$ jet trigger threshold.
Note that in these di-jet events one jet is at $\phi$ = 0$^{o}$,
and the other jet is at $\phi$ = 180$^{o}$.
From F. Abe {\em et. al}, 1991c.}
\label{cdf_jet_shape}
\end{figure}

\subsubsection{Detection of quarks}
\label{quarkdetect}

Quarks hadronize and 
are detected as collimated {\em jets} of particles, (see Fig.~\ref{jetjet}).
Jets in $p\bar{p}$ collisions are reconstructed
by summing up the energy deposited in the calorimeter cells
within a fixed cone in $\eta-\phi$ space, where $\eta$ is the
pseudorapidity, and $\phi$ is the azimuthal angle around the beamline.
The fixed cone algorithm is used because jets are 
are approximately {\em circular} in $\eta-\phi$ space,
because the $\eta-\phi$ size of a jet of a given P$_{T}$ is independent
of the rapidity of a jet, and because this size is
only weakly dependent on the transverse momentum of the jet, as we
will briefly discuss below.

If the typical longitudinal and transverse momentum components
of a fragmentation particle
with respect to the jet axis are $q_{T}$ and $q_{l}$, then
the typical spread of the jet will be $\Delta\theta \approx q_{T}/q_{l}$
and $\Delta\phi \approx q_{T}/$($q_{l}$ sin$\theta$), for $q_{T} << q_{l}$.
Then,
$\Delta\eta = $ (d$\eta$/d$\theta)~\Delta\theta \approx - q_{T}/$($q_{l}$ sin$\theta$)
= $- \Delta\phi$, i.e. jets are approximately circular in $\eta-\phi$ space.
Since the rapidity of a massless
particle under a longitudinal boost changes as y $\rightarrow$ y $+$ $\Delta$y,
where $\Delta$y only depends on the boost, in the limit that
the mass of the fragmentation hadrons is small, the $\eta$ size
of a jet of a given P$_{T}$ is invariant under longitudinal boosts,
i.e. independent of the $\eta$ of the jet itself.  The size of
a jet does vary slightly with its transverse momentum; for example
a simple model of jet fragmentation uniform in rapidity
along the jet axis predicts that the angular size of the cone
containing half of the particles in the jet varies as 1/$\sqrt{E}$,
where $E$ is the energy of the jet.  



The size of the cone used in jet reconstruction
must be matched to the size of a jet.  On average,
of order 70\% of the jet energy is contained within a 
cone of radius $\Delta$R = 
$\sqrt{\Delta\eta^{2} + \Delta\phi^{2}}$ = 0.4
(F. Abe {\em et al.}, 1991c and 1993a;
Linneman, 1995.). See also Fig.~\ref{cdf_jet_shape}; 
NB: $\Delta$R$ < 0.4$ translates into $\Delta\phi < 23^{o}$ 
in Fig.~\ref{cdf_jet_shape}.

\begin{figure}[htb]
\centerline{\psfig{figure=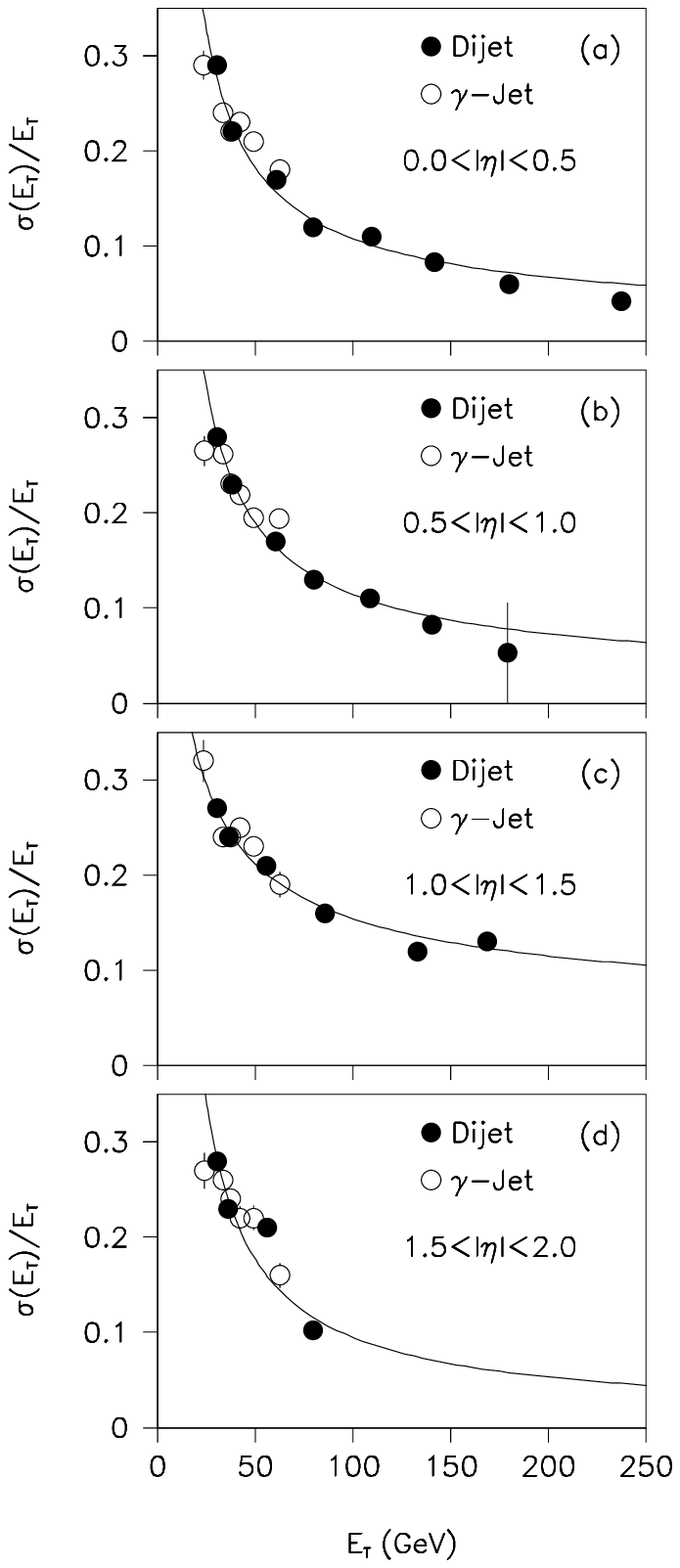,height=6.0in,width=5.0in}}
\caption{\protect \baselineskip 12pt
From the D0 experiment, Abachi 1995d; jet energy resolution
as a function of jet transverse energy (E$_{T}$) as computed
from di-jet and photon-jet events in four pseudo-rapidity
regions.  The fits are of the form
$(\sigma_{E}/E)^{2} = (N/E)^{2}
+ (S/\surd{E})^{2}
+ C^{2}$.
The fitted values of $N$, $S$, and $C$ are :
(a) $N = 7.07$, $S = 0.81$, $C = 0.0$;
(b) $N = 6.92$, $S = 0.91$, $C = 0.0$;
(c) $N = 0.0$, $S = 1.45$, $C = 0.052$;
(d) $N = 8.15$, $S = 0.48$, $C = 0.0$.  Jets are reconstructed using a 
cone-size of 0.5.}
\label{jetres}
\end{figure}

The energy of a jet is defined as the energy of the
corresponding calorimeter cluster.  The resolution 
in the measurement is typically only of order 
$\sigma$(E$_{T}$)/E$_{T} \approx$ 1.0/$\sqrt{E_{T}}$ (E$_{T}$ in GeV),
(see Fig.~\ref{jetres}).
This poor resolution is due to (i) the intrinsic
large fluctuations in the response of calorimeters to
hadronic showers, 
(ii) differences in the calorimeter response between charged
hadrons and electrons or photons,
(iii) energy loss in uninstrumented
calorimeter regions, e.g. in the vicinity of boundaries
between calorimeter modules, (iv) energy loss due to
the use of a finite cone-size in jet reconstruction,
and (v) overlaps between the jet
and hadrons from the underlying event.  The direction of
a jet is measured by linking
the position of the energy cluster in
the calorimeter with the position of the interaction point. 
The resolution on the angular measurement is a few
degrees.

The number of detected jets for a given
decay mode in a $t\bar{t}$ event is not expected to 
correspond to the number of quarks in the final states listed in
Table~\ref{tdec}.  There are a number of reasons for this.
First, as the
P$_{T}$ of the parton becomes small, identification of the
corresponding
jet becomes more and more problematic, as it tends to blend with
the underlying event.  In practice, 
one imposes a minimum cutoff
on the jet transverse momentum which is set 
to a value at least of order
10 GeV/c.  Furthermore, as will be discussed further in this Section,
backgrounds to the top signal
consist mainly of events with low P$_{T}$
jets.  Therefore, to achieve the needed background rejection,
the minimum jet P$_{T}$ requirement is often chosen to
be higher than 10 GeV/c.
A second source of jet reconstruction inefficiency is jet merging.
Nearby jets can be resolved
only if their separation in $\eta-\phi$ space is larger than
a minimum distance of the order of the clustering radius 
used in jet reconstruction.
Top events have a large number of partons in the final state,
and the probability
that at least two of them will be too close to be separately
identified is substantial.  In those cases, the two nearby jets are 
merged and are reconstructed as a single jet.  

\begin{figure}[htb]
\vskip 1cm
\epsfxsize=4.0in
\gepsfcentered[20 200 600 600]{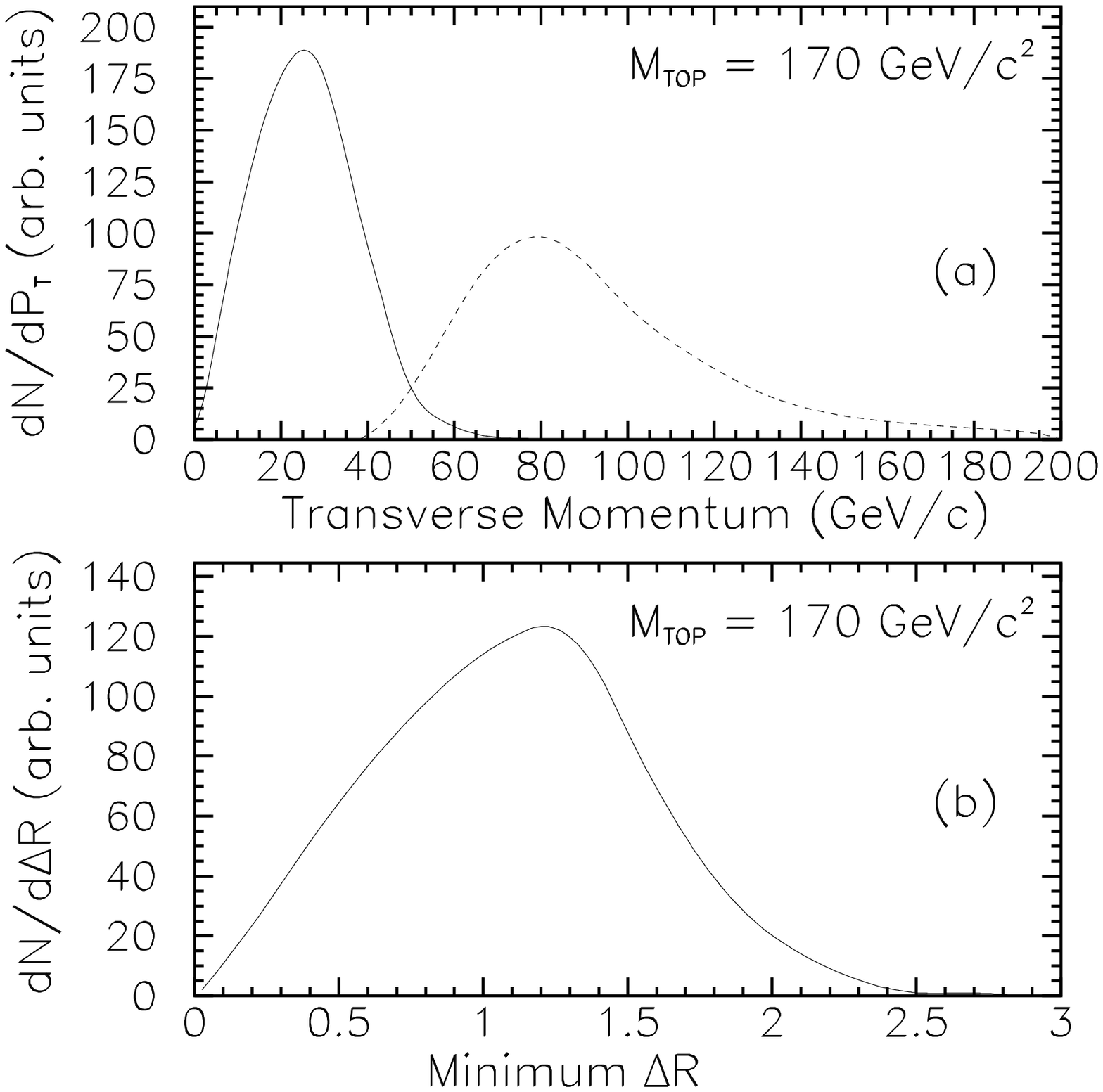}
\vskip 1cm
\caption{\protect \baselineskip 12pt
(a) The expected transverse momentum of the lowest (solid)
and highest (dashed) P$_{T}$ quark in lepton $+$ jets 
events; (b) $\Delta$R between the closest two quarks in
$p\bar{p} \rightarrow t\bar{t} \rightarrow q\bar{q}~ l\nu~ b \bar{b}$.
Results are from the ISAJET Monte 
Carlo event generator at $\surd{s}$ = 1.8 TeV.
Detector effects are not included.}
\label{tj1}
\end{figure}

\begin{figure}[htb]
\vskip 1cm
\epsfxsize=4.0in
\gepsfcentered[20 200 600 600]{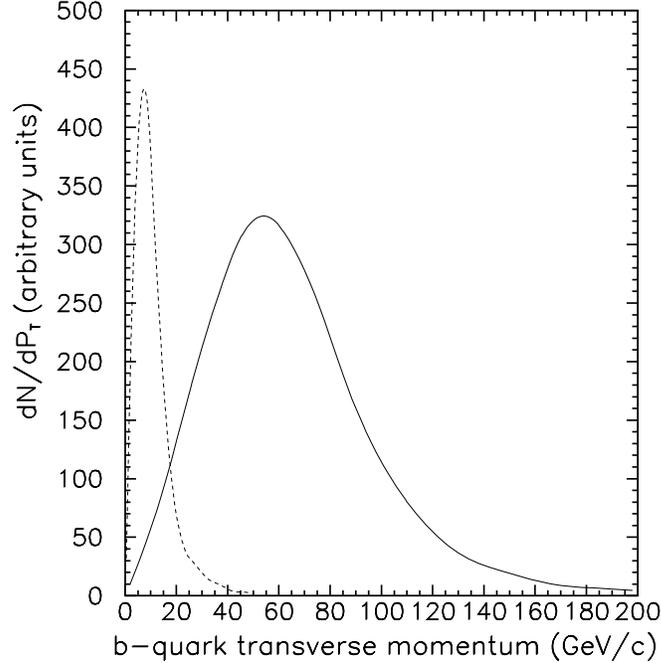}
\vskip 1cm
\caption{\protect \baselineskip 12pt
The expected transverse momentum distribution of
$b$-quarks from top decay for M$_{top}$ = 170 GeV/c$^{2}$ (solid)
and M$_{top}$ = 90 GeV/c$^{2}$ (dashed).
Results are from the ISAJET Monte Carlo event generator
for the process $p\bar{p} \rightarrow t\bar{t}$ at $\surd{s}$ = 1.8 TeV.
Detector effects are not included.}
\label{ptb}
\end{figure}

To illustrate some of these effects, we show
in Fig.~\ref{tj1} the expected transverse momenta
and separation ($\Delta$R) in $\eta-\phi$ space for quarks in 
$t\bar{t} \rightarrow q\bar{q}~ l\nu~ b \bar{b}$
events for 
M$_{top} = 170$ GeV/c$^{2}$.  
Different experiments and different analyses use different
cone clustering radii, typically between 0.3 and 1.0.
In a significant fraction of events, the minimum $\Delta$R between 
quarks is small enough that at least two of the quark-jets 
are expected to be merged.
Even for a high top mass, the
fraction of events with at least one relatively soft jet
is substantial (again, see Fig.~\ref{tj1}).  
For lower top masses of course the
transverse momenta will be even lower.  The situation for
M$_{top}$ close to M$_{W}$ is particularly difficult.
In that case the kinetic energy liberated in the 
$t \rightarrow Wb$ decay is low and the $b$ momentum in the
top rest frame is small.  Even after boosting to the
laboratory frame, the $b$ momentum remains soft, (see Fig.~\ref{ptb}).

\clearpage

The situation gets even more complicated when initial and final state
radiation are taken into account.  
Initial state gluon radiations gives a net P$_{T}$
to the $t\bar{t}$ system, and therefore alters the jet P$_{T}$ spectrum
calculated at tree-level; the radiated gluons can be detected
as additional jets in the final state;
large angle radiation from the final state quarks
softens the spectrum of reconstructed jets, and 
can also result in additional jets.

\begin{figure}[htb]
\vskip 1cm
\epsfxsize=4.0in
\gepsfcentered[20 200 600 600]{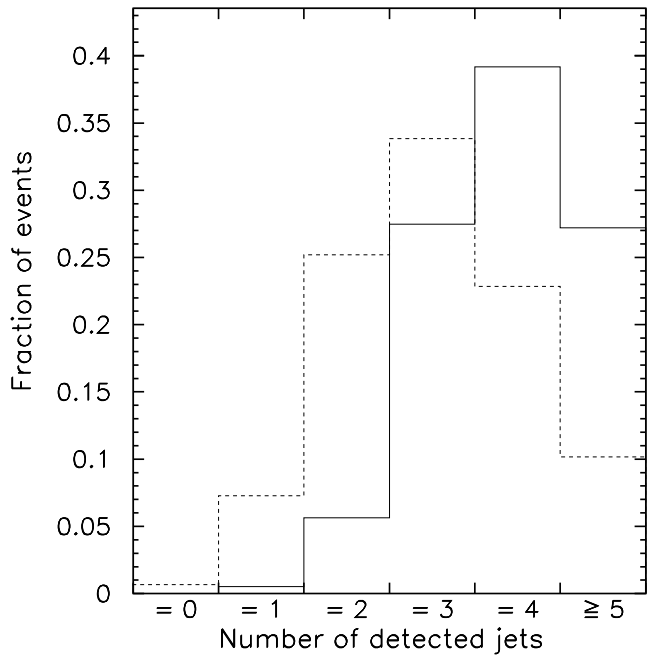}
\vskip 1cm
\caption{\protect \baselineskip 12pt
Jet multiplicity in $t\bar{t} \rightarrow l \nu b q\bar{q} b$ 
(lepton + jets) events 
from the ISAJET Monte Carlo and the CDF detector simulation.  
Solid line : M$_{top}$ = 200 GeV/c$^{2}$;
dashed line : M$_{top}$ = 120 GeV/c$^{2}$. Jets are reconstructed
using a cone-size $\Delta$R=0.4, and must have $|\eta| <$ 2.
The jet transverse energy threshold is 15 GeV, without
application of jet energy corrections. (The
jet energy corrections will be described in Section IX.A.3; 
a 15 GeV jet in CDF is corrected on average to $\approx$ 23 GeV).
Note that in the absence of gluon radiation these events should
have N$_{jets} \leq 4$; the significant fraction of events with
additional jets in the final state is an indication of the importance
of gluon radiation.}
\label{njets_mc}
\end{figure}

All of these jet reconstruction effects are important,
and must be studied using 
QCD-inspired Monte Carlo event generators, in conjunction with
a simulation of the detector response, see for example Fig.~\ref{njets_mc}.
They
constitute one of the major systematic uncertainties in the 
calculation of the $t\bar{t}$ acceptance and, more importantly, 
in the determination of the top mass, see Section~\ref{mass}.


\subsubsection{Detection of neutrinos}
\label{neutdetect}

Neutrinos are detected by missing momentum
techniques: since the initial center-of-mass momentum is zero, the 
vector sum of the momenta of all of the neutrinos
in the event is inferred as the negative of the vector
sum of the momenta of all the detected particles.  
However, because the very forward and backward detector
regions are uninstrumented,
longitudinal information is lost, and only
the transverse components of the momenta of neutrinos can be measured.   
In practice, what is measured is not the momentum of all of the particles, 
but rather the energy deposited in the calorimeter. 
The missing transverse energy vector ($\vec{\MET}$) is defined
%
%
as $\vec{\MET} \equiv - \Sigma_{i} \vec{E^{i}_{T}}$, where the sum is 
over all calorimeter cellss, and $\vec{E^{i}_{T}}$ is a vector 
whose direction points to the $ith$ cell and whose
magnitude is equal to the transverse energy deposited in the $ith$ tower.
The missing transverse energy vector must be corrected for 
detected muons which only loose a minimal amount of energy in the
calorimeter, and is then
associated with the neutrino transverse momentum.  

The resolution on the neutrino energy measurement is very much
dependent on the event topology, since it depends directly on
the resolution in the measurements of all the leptons
and jets in the event.  Since leptons are in general well
measured, the uncertainty in the measurement of \MET~ arises mostly
from errors in the measurements of jet energies.  It is customary
to parametrize the resolution in \MET, $\sigma$(\MET),
as a function                                
of the total transverse energy in the event, $\Sigma E_{T}$, see
Table~\ref{metres}.            

\begin{table}
\begin{center}
\begin{tabular}{ccc}                               
\hline \hline
Detector & $\sigma$(\MET) & Reference \\
\hline
UA1  & 0.7 $\sqrt{\Sigma E_{T}}$~~~ (in GeV)  & Albajar {\em et al.}, 1989 \\
UA2  & 0.8 $(\Sigma E_{T})^{0.4}$~~~ (in GeV) & Alitti {\em et al.},  1990b \\
CDF  & 0.7 $\sqrt{\Sigma E_{T}}$~~~ (in GeV)  & F. Abe {\em et al.}, 1994a \\
D0   & 1.08 GeV $+$ 0.019 $\Sigma E_{T}$   & Abachi {\em et al.}, 1995d \\
\hline \hline
\end{tabular}
\end{center}
\caption{\protect \baselineskip 12pt
Missing transverse energy resolution for minimum bias events.}
\label{metres}
\end{table}

\subsubsection{Detection of tau leptons}
\label{taudetect}

Taus are very hard to identify.  Approximately
36\% of the time, a tau lepton will decay into a muon or an electron.
The signature for an event with a 
$t \rightarrow W \rightarrow \tau \nu \rightarrow l \nu \nu$,
where $l = e$ or $\mu$, is very similar to that of an event with 
a $t \rightarrow W \rightarrow l \nu$ decay, except that the final state 
lepton will
in general have lower momentum.  Taus that decay hadronically are
detected as jets.  Separation between jets from
hadronic decays of taus and quarks
or gluon jets in $p\bar{p}$ collisions has been achieved in
e.g. measurements of the 
$p\bar{p} \rightarrow W \rightarrow \tau\nu$ 
cross-section (Albajar {\em et al.}, 1989;
Alitti {\em et al.}, 1991b; 
F. Abe {\em et al.}, 1992b), 
and in searches for non
Standard-Model top quark decays, see Section~\ref{NSMdecay}.  
The separation is based on the distinctive narrowness of 
a jet from the hadronic decay of a high P$_{T}$ tau-lepton,
and/or the characteristic one- and three-prong track
multiplicities.
However, the efficiency for detecting hadronic taus is so low,
and the backgrounds from jet fluctuations are so high, 
that
these techniques are only now just beginning to be applied successfully in the
context of a Standard Model top search.  

For the remainder of this article, we will refer
to $t\bar{t}$
final states with zero, one, or two leptons (e or $\mu$) from $W$ decay
as {\em all hadronic, lepton + jets}, and {\em dilepton} respectively.
Signatures that include the explicit identification of hadronic tau-decays
will not be considered in this review.
We now turn to a discussion of the $t\bar{t}$ signatures in these
three channels. 

\subsection{All hadronic mode}
\label{fullyhad}
The all hadronic final state,
($t\bar{t} \rightarrow q\bar{q}~ q\bar{q}~ b \bar{b}$, see 
Table~\ref{tdec})
is the most common, but it competes
with very high backgrounds from $p\bar{p} \rightarrow $ 6 
jets (see for example Benlloch, Wainer, and Giele, 1993). 
The cross-sections for this QCD process
at the Tevatron is higher than the top cross-section
by approximately three orders of magnitude.
Despite the extremely high background levels, it may be
possible, with sufficient statistics, to isolate
a top signal in this mode by applying further
kinematic cuts, and by identifying the $b$-quark(s) in the 
final state.  These issues are being carefully
studied by the experimenters (Castro, 1994; Tartarelli, 1996; Narain 1996).
In this review we will concentrate on the dilepton
and lepton $+$ jets modes.

%
%
%

\subsection{Dilepton mode}
\label{dilepton}
\begin{figure}[htb]
\vskip 1cm
\epsfxsize=4.0in
\gepsfcentered[20 200 600 600]{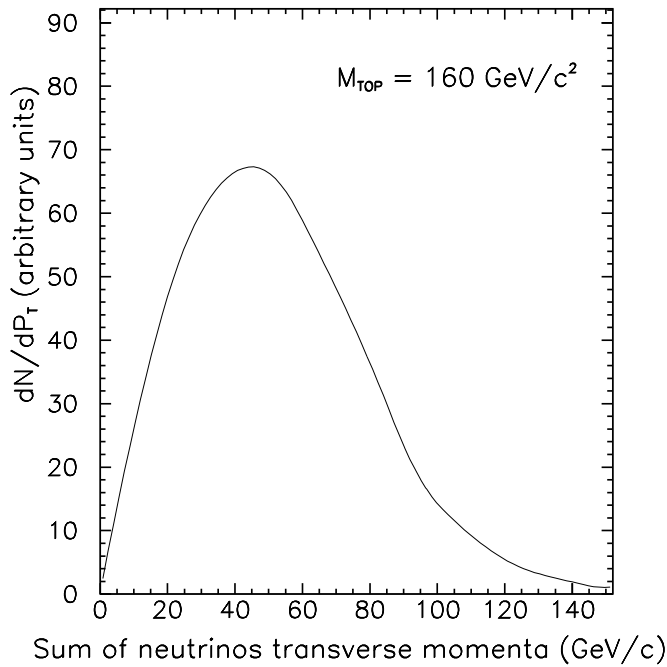}
\vskip 1cm
\caption{\protect \baselineskip 12pt
Expected sum of neutrinos' transverse momenta in the dilepton
channel 
($p\bar{p} \rightarrow t\bar{t} \rightarrow l\nu b l\nu \bar{b}$)
at $\surd{s}$ = 1.8 TeV.
From the ISAJET Monte Carlo event generator, for M$_{top} =$ 160 GeV/c$^{2}$.}
\label{dilepmet}
\end{figure}

The signature for the dilepton final state 
($t\bar{t} \rightarrow l\nu~ l\nu~ b \bar{b}$, see Table~\ref{tdec})
consists of two 
leptons, two $b$-jets, and \MET~from the two neutrinos.  
Since the
leptons originate from $W$ decay,
they tend to be isolated and
to have high transverse momenta.  The \MET~ is also
expected to be high, (see Fig.~\ref{dilepmet}).
%
%
%
%
Typical minimum
lepton transverse momentum or \MET~ requirements are set around
20 GeV/c.
Despite the low branching ratio, this mode turns out to be 
very important because background levels are very low.

\begin{figure}
\begin{picture}(32000,10000)(0,-3000)

\drawline\fermion[\SE\REG](10000,6000)[5000]
\put(\particlefrontx,7000){$q$}
\drawline\fermion[\SW\REG](\particlebackx,\particlebacky)[5000]
\put(\particlebackx,-2500){$\bar{q}$}
\drawline\photon[\E\REG](\particlefrontx,\particlefronty)[6]
\put(\particlemidx,1000){$Z,\gamma$}
\drawline\fermion[\SE\REG](\particlebackx,\particlebacky)[5000]
\put(\particlebackx,7000){$l$}
\drawline\fermion[\NE\REG](\particlefrontx,\particlefronty)[5000]
\put(\particlebackx,-2500){$\bar{l}$}

\end{picture}
\label{DY}
\caption{\protect \baselineskip 12pt
Lowest order Feynman diagram for Drell-Yan production
of lepton pairs.}
\end{figure}
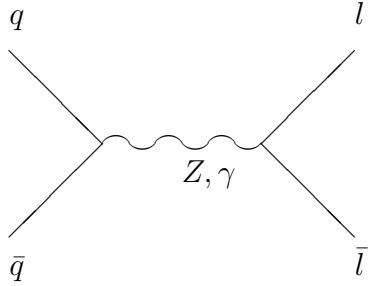

The most probable way to obtain two isolated leptons in
$p\bar{p}$ collisions is through the Drell-Yan process, see 
Fig. 37.
This mechanism yields $e^{+}e^{-}$ and $\mu^{+}\mu^{-}$
pairs, but not $e\mu$ pairs, except through $\tau\tau$
production, with both taus decaying leptonically.  
Here we begin by
addressing backgrounds from direct Drell-Yan production of 
$ee$ and $\mu\mu$ pairs.

The dominant Drell-Yan $p\bar{p} \rightarrow Z \rightarrow l^{+}l^{-}$
resonance can be easily eliminated
by a $l^{+}l^{-}$ invariant mass cut, with a modest ($\approx$ 25 \%)
loss in top acceptance.  After the $Z$ removal, the rate of 
high transverse momentum Drell-Yan
pairs is still approximately two orders of magnitude higher than
the $t\bar{t}$ dilepton rate 
(for M$_{top} \approx $ 150 GeV/c$^{2}$).
Additional background rejection
can be obtained because (i) in Drell-Yan events there are no additional 
emitted jets at lowest order, and (ii) 
there are no neutrinos, and hence 
zero \MET~ except for resolution effects.  Higher order QCD corrections
to the diagram shown in Fig. 37
give rise to final state
jets, for example from gluons radiated off the $q$ and $\bar{q}$ lines.
For each additional jet, the rate is reduced by a factor of 
O($\alpha_{s}$) $\approx 0.15$.  Since there are two $b$-quark jets in 
$t\bar{t}$ dilepton events, one can achieve a significant background 
rejection factor, while 
maintaining efficiency for top, 
by demanding that at least two jets be detected in addition to the
$l^{+}l^{-}$ pair.  In conjunction with a \MET~ requirement,
the Drell-Yan background can then be reduced to a tenth or less of
the expected top signal for top masses as high as 200 GeV. 

\begin{figure}[htb]
\epsfxsize=4.0in
\vskip 1cm
\gepsfcentered[20 200 600 600]{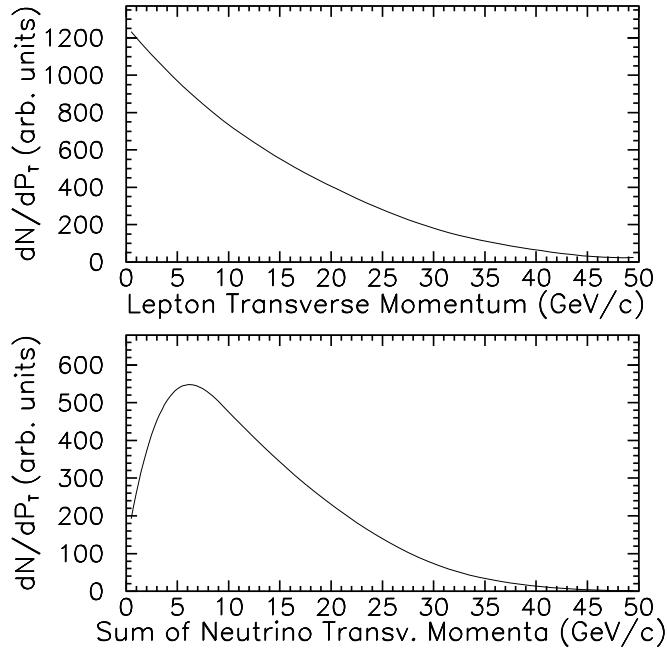}
\vskip 1cm
\caption{\protect \baselineskip 12pt
The expected lepton and neutrino transverse momenta in
$p\bar{p} \rightarrow Z \rightarrow \tau^{+}\tau^{-}$,
$\tau^{+} \rightarrow l^{+}\nu\nu$,
$\tau^{-} \rightarrow l^{-}\nu\nu$ at $\surd{s}$ = 1.8 TeV.
From the ISAJET Monte Carlo generator.}
\label{tau}
\end{figure}

As was mentioned above, Drell-Yan $\tau\tau$ production,
followed by leptonic decays of both taus 
($\tau \rightarrow l \nu \nu$),
constitutes an additional source of lepton pairs.
Events from the $Z \rightarrow \tau \tau$ resonance cannot
be easily removed because the invariant mass information is
lost due to the presence of four neutrinos in the final state.
However, the transverse momenta of the leptons and the \MET~ 
for these events are significantly lower than in $t\bar{t}$
dilepton events, (see Figs.~\ref{tau},~\ref{wlep}, and~\ref{dilepmet}).
By requiring high transverse
momentum leptons, high \MET~, and two jets, this background can
be reduced to approximately the same level as the 
$ee$ and $\mu\mu$ Drell-Yan background.
We note here that,
unlike in the case of direct Drell-Yan production of $ee$
and $\mu\mu$ pairs, this
process can result in $e\mu$ final states.

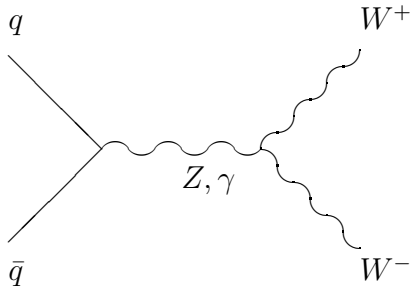
\begin{figure}
\hskip 2cm
\begin{picture}(32000,10000)(0,-3000)

\drawline\fermion[\SE\REG](10000,6000)[5000]
\put(\particlefrontx,7000){$q$}
\drawline\fermion[\SW\REG](\particlebackx,\particlebacky)[5000]
\put(\particlebackx,-2500){$\bar{q}$}
\drawline\photon[\E\REG](\particlefrontx,\particlefronty)[6]
\put(\particlemidx,1000){$Z,\gamma$}
\drawline\photon[\SE\REG](\particlebackx,\particlebacky)[6]
\put(\particlebackx,7000){$W^{+}$}
\drawline\photon[\NE\REG](\particlefrontx,\particlefronty)[6]
\put(\particlebackx,-2500){$W^{-}$}

\end{picture}
\label{dib}
\caption{\protect \baselineskip 12pt
Lowest order Feynman diagram for production of $WW$ pairs in 
$p\bar{p}$ collisions;
similar diagrams lead to $WZ$ and $ZZ$ production.}
\end{figure}

Diboson production, 
(see Fig. 39), constitutes
an additional source of high P$_{T}$ dileptons and \MET~.
These are exceedingly rare processes, which are 
extremely interesting in their own right. 
The cross-section for $WW$ production has been calculated to next-to-leading
order, and at $\sqrt{s}$ = 1.8 TeV it is estimated to be
$\sigma(WW) \approx$ 10 pb (Ohnemus, 1991a).  
This is the same as $\sigma(t\bar{t})$,
for M$_{top} \approx $ 160 GeV/c$^{2}$.  When both $W$-bosons decay 
leptonically, the kinematics for the leptons and the neutrinos are very 
similar as those expected from $t\bar{t}$, which also results in
a $WW$ pair in the final state. 
The most efficient method that can be used
suppress this background is to require that there be jets in the
event.  Just as in the Drell-Yan process, there are no jets 
at leading order in $WW$ events.  By demanding that there be
two jets, the background is reduced by a factor of order 
$\alpha_{s}^{2} \approx 0.02$.  Backgrounds from $WZ$ and $ZZ$ production are
smaller by over one order of magnitude because (i) the $WZ$ and $ZZ$
production cross-sections are significantly
lower than that of $WW$ (Ohnemus, 1991b; Ohnemus and Owens, 1991), 
(ii) the leptonic branching ratios
of the $Z$ are a factor of three smaller than those of the $W$, and
(iii) $l^{+}l^{-}$ pairs from $Z$ decays can be eliminated with
an invariant mass cut.

Additional backgrounds to the dilepton signal from fake leptons
as well as doubly-semileptonic decays of $b\bar{b}$ pairs also have
to be considered, but are generally found to be small. 
In a given analysis the signal-to-background
level can be tuned by the choice of requirements.  In general,
raising the minimum P$_{T}$ cut on the jets eliminates more
background events than signal events.  The reason for this is
that for sufficiently high M$_{top}$ the P$_{T}$ spectrum
of $b$-jets in 
dilepton top events is harder than the brehmstrahlung-like
spectrum of jets in all the processes listed above.  Higher
\MET~ or lepton P$_{T}$ requirements would considerably lower
all backgrounds except the diboson background; the $Z \rightarrow
\tau \tau$ background could be entirely eliminated by
requiring the invariant mass of the $l^{+}l^{-}$ pair to
be higher than the $Z$ mass; requiring that jets be $b$-tagged
would reduce all backgrounds by about two orders of magnitude.
With sufficient luminosity, it should be possible
to obtain very pure $t\bar{t}$ samples in the dilepton mode.

\subsection {Lepton $+$ jets mode}
\label{ljets}
The branching ratio for this mode 
($t\bar{t} \rightarrow q\bar{q} l\nu b\bar{b}$)
is quite large, $24/81$, see 
Table~\ref{tdec}.  The signature consists of one isolated,
high P$_{T}$ lepton (e or $\mu$), \MET~ from the neutrino, two light quark
jets ($u, d, c,$ or $s$), and two $b$-quark jets.  

As was discussed in Section~\ref{quarkdetect}, in practice 
the number of detected jets is not expected to be always four.
In order to maintain high efficiency, in most analyses
the number of jets requirement is usually relaxed to 
$\geq$ 2 or $\geq$ 3, except for very high top mass, or where 
detection of a fourth
jet is essential, e.g. for the determination of the top
mass, see Section~\ref{mass}.

\begin{figure}
\begin{picture}(32000,15000)(0,-7000)

\drawline\fermion[\E\REG](10000,6000)[6000]
\put(\particlefrontx,7000){$q$}
\drawline\photon[\E\REG](\particlebackx,\particlebacky)[8]
\put(\particlemidx,7000){$W$}
\drawline\fermion[\S\REG](16000,6000)[8000]
\drawline\gluon[\W\REG](\particlebackx,\particlebacky)[6]
\put(\particlebackx,0){$g$}
\drawline\fermion[\E\REG](\particlefrontx,\particlefronty)[10000]
\drawline\gluon[\SE\REG](\particlemidx,\particlemidy)[4]
\drawline\gluon[\E\REG](16000,2000)[5]
\drawline\fermion[\NE\REG](\particlebackx,\particlebacky)[3000]
\drawline\fermion[\SE\REG](\particlefrontx,\particlefronty)[3000]

\end{picture}
\caption{\protect \baselineskip 12pt
Feynman diagram for $W +$ 4 jet production in $p\bar{p}$ collisions.  
Many other diagrams
also contribute.}
\label{w4j}
\end{figure}
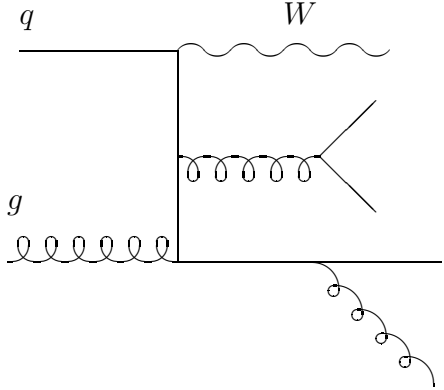
       
\subsubsection{$W +$ jets background}
\label{bg}
For sufficiently high top mass, the dominant background to $t\bar{t}$ in the
lepton $+$ jets channels is due to $W$ $+$ multijets production,
(see Fig.~\ref{w4j}).  The inclusive $W$ production cross-section
at the Tevatron is 
$\sigma(p\bar{p} \rightarrow W) \approx$ 20 nb, over three orders of magnitude
higher than the $t\bar{t}$ cross-section for M$_{top} > 150$ GeV/c$^{2}$.
The cross-section for $W +$~N jets is suppressed by factors of order
$\alpha_{s}^{N}$.

Other backgrounds, such as $p\bar{p} \rightarrow b\bar{b}$~$+$~jets
followed by $b \rightarrow c l \nu$,
and backgrounds from fake leptons also need to be carefully
evaluated.  For top masses above $\approx$ 40 GeV/c$^{2}$,
where an efficient lepton $+$ jets top selection can be
devised based on {\bf both} a high P$_{T}$ lepton and
high missing transverse energy, these backgrounds
are in general found to be much smaller.

The importance of the $W$~+ multijets
process as a background to top production, as well
as to more exotic phenomena (e.g. supersymmetry),
was noted soon after the first results from the
S$p\bar{p}$S collider became available.  As a result,
a large theoretical effort was directed towards the 
calculation
of the cross-sections and kinematic properties for
$p\bar{p} \rightarrow W$ (or $Z$) $+$ N jets.
The first calculations, for N=1 or 2, were performed in
the mid-eighties (Ellis and Gonsalves, 1985; Ellis, Kleiss,
and Stirling, 1985; Kleiss and Stirling, 1985; Gunion and
Kunstz, 1985).  These were then extended to final states
with 3 (Berends, Giele, and Kuijf, 1989; Hagiwara and Zeppenfeld, 1989)
and 4 jets (Berends {\em et al.}, 1991).

\begin{figure}[htb]
\epsfxsize=4.0in
\vskip 1cm
\gepsfcentered[20 200 600 600]{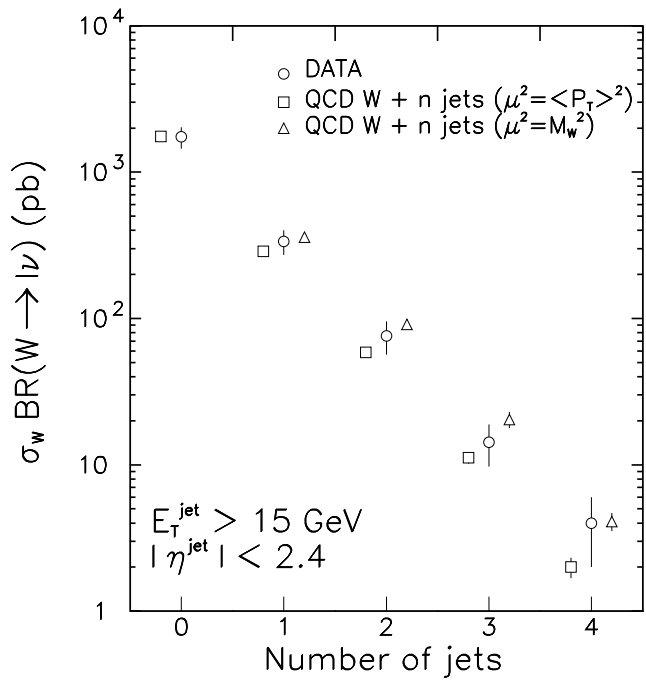}
\vskip 1cm
\caption{\protect \baselineskip 12pt
Product of $W$ cross-section ($\sigma_{W}$)
times leptonic branching ratio
as a function of jet multiplicity in $p\bar{p}$ collisions
at $\surd{s}$ = 1.8 TeV.
The LO QCD predictions are
shown for two different choices of the renormalization and
factorization scale $\mu$. The jet E$_{T}$ threshold was set at 15 GeV.
From F. Abe {\em et al.}, 1993b.}
\label{ale}
\end{figure}

These calculations are performed at tree-level, and therefore 
they diverge as the angular separation between outgoing partons
becomes very small 
or as their transverse momenta tend to zero
({\em collinear} and {\em infrared} divergences).
However, far enough away from the regions of divergence, 
calculations are expected to be quite reliable, within the
estimated theoretical uncertainties due to the missing
higher order terms.
Agreement
is found between the measured $W$ $+$
jets cross-section and the theoretical prediction, (see Fig.~\ref{ale}).
The degree of confidence in these LO calculations
is such that the theoretical predictions have been used
to extract the value of $\alpha_{s}$ from the $W +$ jet data
(Lindgren, 1992; Alitti {\em et al.}, 
1991c).  More recent measurements of $\alpha_{s}$
(Abachi {\em et al.}, 1995c) have been based on the full NLO (order 
$\alpha_{s}$) calculation for $p\bar{p} \rightarrow W$
(Giele, Glover, and Kosower, 1993).

The LO theoretical calculation for 
$p\bar{p} \rightarrow W$ (or $Z$) $+$ N jets,
for N up to 4, is implemented in the
VECBOS Monte Carlo event generator (Berends {\em et al.}, 1991), which is 
extensively used
by the experimenters to model the $W$ $+$ jets background.  
VECBOS is a parton-level
Monte Carlo generator, i.e. its output consists of parton four-vectors
only.  In order to properly simulate the response of the detector,
hadronization effects are included by interfacing the
VECBOS event generator with hadronization models based e.g.,
on independent parton-fragmentation (Field
and Feynman, 1978),
or the HERWIG model (Marchesini and Webber, 1984 and 1988).  A 
model of the underlying event also needs to be included.

In what follows we will show several comparisons of theoretical 
expectations for $t\bar{t}$ and $W +$ jets at Tevatron energies.  
These comparisons
are performed at the parton-level only, i.e. 
hadronization, as well as detector effects,
such as resolution smearing, efficiencies etc.,
are not included.  
Therefore, the discussion 
presented here is intended only as a general
illustration of the issues involved.

The two processes are modeled with the ISAJET
(for $t\bar{t}$) and VECBOS (for $W +$ jets) event generators.
The ISAJET model
employed here does not include initial and final state gluon 
radiation effects; the $t\bar{t}$ rates are normalized
to a recent $p\bar{p} \rightarrow t\bar{t}$ cross-section
calculation (Laenen, Smith, and van Neerven, 1994).
In order to mimic actual
experimental conditions, and to avoid the 
infrared and collinear divergencies in the $W +$ jets calculation,
we impose the following requirements :
\begin{itemize}
\item P$_{T}$ of partons (quarks or gluons) $>$ 15 GeV/c
\item $|\eta|$  of jets $<$ 2
\item P$_{T}$ of leptons (electrons, muons and neutrinos) $>$ 20 GeV/c
\item $|\eta|$ of electron or muon $<$ 1
\item $\Delta$R between jets $>$ 0.5
\end{itemize}

\begin{figure}[htb]
\epsfxsize=4.0in
\vskip 1cm
\gepsfcentered[20 200 600 600]{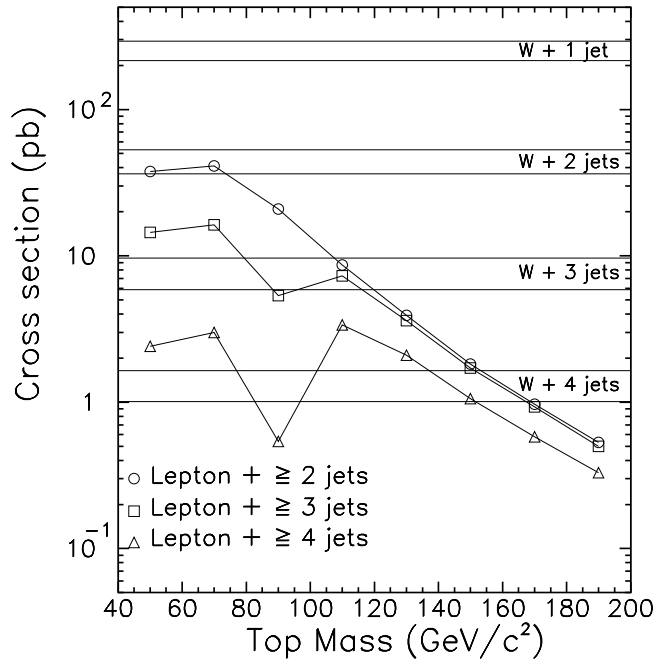}
\vskip 1cm
\caption{\protect \baselineskip 12pt
Comparison of expected lepton $+$ jets rates at the Tevatron
for $t\bar{t}$ and $W +$ jets as a function of jet multiplicity
and top mass. Detector and $t\bar{t}$ gluon radiation 
effects are not included.  The $W +$ jets
theoretical predictions are given as bands which reflect the
effects of reasonable variations in
the $\mu^{2}$ scale.
See text for details.}
\label{jmult}
\end{figure}

In Fig.~\ref{jmult} we show the expected rates of lepton $+$ jets
events from $t\bar{t}$ and $W +$ jets.  
(To get a feeling for the effects of gluon radiation, 
which are not included here, see for example
Fig.~\ref{njets_mc}). 
Note the drop in efficiency for detecting a fourth jet in $t\bar{t}$
events, even at top masses of 200 GeV.  
As discussed in Section~\ref{quarkdetect}, this is due to (i)
final state quarks 
having P$_{T}$ below the threshold (15 GeV/c in this case), and
(ii) to the effect of jet merging.
The expected signal-to-background, for a lepton $+$ $\geq$ 3 jets selection,
with a 15 GeV jet P$_{T}$ threshold, varies
between about 1/1 at
M$_{top}$ = 100 GeV/c$^{2}$ and 1/10 at
M$_{top}$ = 200 GeV/c$^{2}$.  By requiring $\geq$ 4 jets,
the signal-to-background for high top quark mass is significantly
improved
(by approximately a factor of 3 for M$_{top}$  = 200 GeV/c$^{2}$).
As we will show,
further
improvements in signal-to-background can be achieved by raising the
P$_{T}$ threshold on the jets.
For top masses in the neighborhood
of the $W$-mass, $b$-quarks are expected to have 
a soft transverse momentum spectrum, (see Fig.~\ref{ptb}).  As a result,
the probability of detecting more than two jets in this
mass region is particularly low.  The required number of detected
jets in a lepton $+$ jets
top search is therefore in general dependent
on the top mass region that is being explored.

The cross-section for $W +$ N jets is 
proportional to $\alpha_{s}^{N}$, therefore for each additional
jet the $W$ cross-section drops by a factor of order $\alpha_{s}$.
The $W +$ jets predictions in Fig.~\ref{jmult}
are derived from VECBOS, which is based on 
a tree-level calculation
with significant uncertainties.
These uncertainties can be partially quantified by
the stability of the calculation under changes in
the factorization and renormalization 
scale $\mu$.  Here we
present these predictions as bands that reflect the variation
between the choices $\mu^{2} =$ M$_{W}^{2}$ and 
$\mu^{2} = <P_{T}>^{2}$, where $<P_{T}>$ is the average
transverse momentum of the partons in the event.

The relative uncertainty on the $W +$ N jets cross-section due to the
choice of $\mu$ grows with the number of jets. This is because
$\sigma$($W +$ N jets) is proportional to $\alpha_{s}^{N}$($\mu^{2}$),
so that when choosing a different $\mu^{2}$,
$\delta\sigma / \sigma$ becomes proportional to
N $\delta\alpha_{s} / \alpha_{s}$
(where we have ignored the $\mu^{2}$ dependence of the parton
distribution functions that need to be convoluted with the
partonic cross-section).  

Because of these theoretical uncertainties,
the existence of the top
quark cannot be firmly established based on just the
observation of an excess of $W +$ jets events.
Additional information need to be employed to isolate
a potential top signal.  
A number of possibilities will be discussed next.

\begin{figure}[htb]
\epsfxsize=4.0in
\vskip 1cm
\gepsfcentered[20 200 600 600]{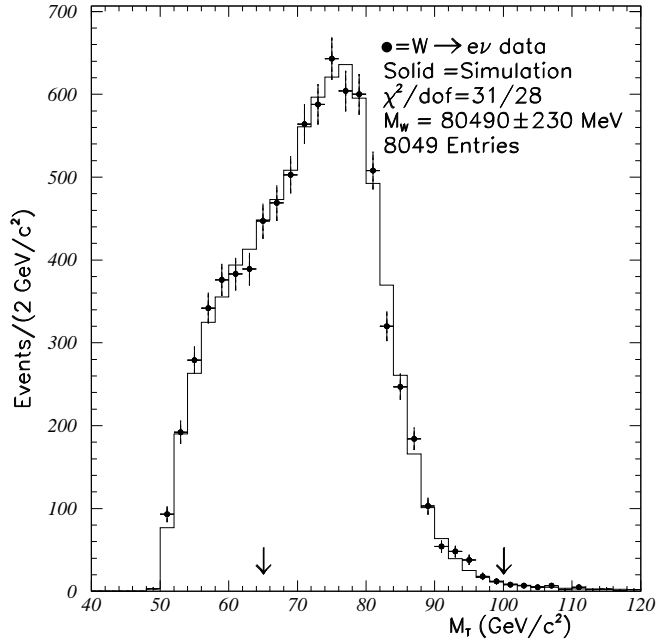}
\vskip 1cm
\caption{\protect \baselineskip 12pt
Transverse mass distribution in $W \rightarrow e\nu$ decays.  From
the CDF collaboration, Frisch, 1995.  This distribution is used to measure
the $W$ mass by fitting to Monte Carlo expectations; the arrows delimit
the range of the fit.}
\label{hfwmass}
\end{figure}

\subsubsection{Separation of $W +$ jets and $t\bar{t}$ for 
M$_{top} <$ M$_{W} +$ M$_{b}$}
\label{lowmass}
As discussed previously, if M$_{top} <$~M$_{W}$~+~M$_{b}$, the 
$W$ from the decay $t \rightarrow Wb$
will be virtual ($W^{*}$).  In this case, the 
invariant mass of the $l\nu$ pair in $t\bar{t}$ events from 
$W^{*} \rightarrow l\nu$ will be smaller than both M$_{W}$ and M$_{top}$.
In contrast, the $l\nu$ pair in $W +$ jets events,
which originates from real $W$ decays, has invariant mass
equal to the $W$ mass.
However, as was mentioned in Section~\ref{neutdetect}, 
the longitudinal component of the neutrino momentum cannot
be measured, so the $l\nu$ invariant mass cannot be calculated.
Fortunately, the {\em transverse mass} (M$_{T}$)
of the $l\nu$ pair still provides significant 
discrimination between real and virtual $W$-decays (Rosner, 1989).
The transverse mass is defined as the
pseudo-invariant mass of the lepton and the neutrino constructed
from the transverse components only :                         

$${\rm M}_T^2 \equiv ({\rm P}_{T\nu} + {\rm P}_{Tl})^2 - 
(\vec{P}_{T\nu} + \vec{P}_{Tl})^2 = 
2 {\rm P}_{T\nu} {\rm P}_{Tl} (1 - {\rm cos}\Delta\phi)$$
where P$_{T\nu}$ and P$_{Tl}$ are the neutrino and lepton transverse
momenta respectively, $\Delta\phi$ is the angle between the two
transverse momentum vectors,
and where we have ignored the mass of the lepton
compared to its momentum.                          
Transverse mass distributions, (see Fig.~\ref{hfwmass}), have the following
properties : (i) M$_{T}$ is less than or equal the invariant mass
of the lepton-neutrino pair;
(ii) M$_{T}$ distributions  
are invariant under longitudinal boosts, hence
they are independent of the longitudinal momentum of the pair; (iii) they
result in Jacobian peaks at the original $l\nu$ invariant mass;
(iv) they are fairly insensitive to the total transverse momentum of the 
pair. (The transverse mass distribution of a pair of invariant mass M
and transverse momentum P$_{T}$ differs from that of a P$_{T} = 0$
pair by corrections of order $\approx$ (P$_{T}$/M)$^{2}$, see
for example Barger and Phillips, 1987).

\begin{figure}[htb]
\epsfxsize=4.0in
\vskip 1cm
\gepsfcentered[20 200 600 600]{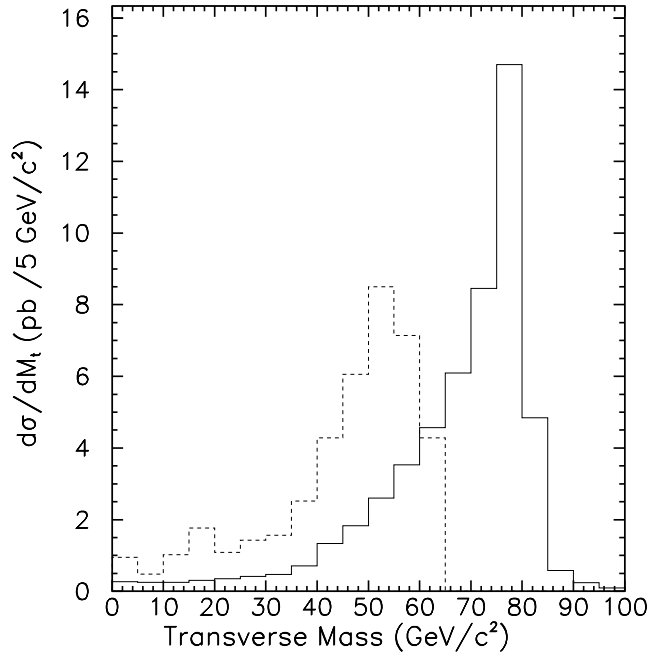}
\vskip 1cm
\caption{\protect \baselineskip 12pt
Expected transverse mass distributions for $W +$ 2 jets (solid)
and $p\bar{p} \rightarrow t\bar{t} \rightarrow$~lepton $+$ 2 jets, 
for M$_{top} =$ 70 GeV/c$^{2}$ (dashed).  This is at Tevatron energies 
(1.8 TeV).
The $W +$ 2 jets calculation has been
performed with a scale $\mu^{2} = <P_{T}>^{2}$.  Detector 
and $t\bar{t}$ gluon radiation effects
are not included.  Relative normalizations are from Monte Carlo.
See text for details.}
\label{trmass}
\end{figure}

In Fig.~\ref{trmass} we show expected transverse mass distributions
for 
$W +$ jets and $t\bar{t}$, with M$_{top} =$ 70 GeV/c$^{2}$.  A top
signal would result in a significant distortion of the transverse
mass spectrum of $W +$ 2 jets events, even after accounting for 
smearing due to resolution effects (see Section~\ref{pre1a}).
By concentrating on just the shape of the transverse mass
distribution, uncertainties due to the theoretical expectation of
the $W +$ jets rate do not enter in the analysis.  
Furthermore, the shape
of the transverse mass distribution for the $W +$ jets
background is dominated by the kinematics of the $W \rightarrow l\nu$
decay, and depends only weakly on the modeling of the $W$ production
properties.
We note that the transverse mass method can be used to separate
a top signal from the $W +$ jets background for both
the $t\bar{t}$ and $W \rightarrow t\bar{b}$ 
production mechanisms.

For higher top masses, the $l\nu$ transverse mass distributions
in $t\bar{t}$ and $W +$ jets events become indistinguishable,
since top quarks will decay into real $W$
bosons.  In order to separate signal from background in the
lepton $+$ jets mode, one then has to rely on the
kinematic differences between the two processes, and/or the fact
that $t\bar{t}$ events always contain two $b$-quarks in the
final state.  

\subsubsection{Kinematic differences between $t\bar{t}$ and $W +$
jets}
\label{kinematics}
Several possible ways of extracting a top signal from the 
$W +$ jets background using kinematic signatures 
have been suggested in the  literature (Berends {\em et al.}, 1989;
Baer, Barger and Phillips, 1989; Agrawal and Ellis, 1989;
Giele and Stirling, 1990; Berends, Tausk, and Giele, 1993;
Barger, Ohnemus, and Phillips, 1993; Benlloch, Sumorok, and
Giele, 1994; Cobal, Grassmann, and Leone, 1994; Barger {\em et al.}, 1995).  
Briefly, the differences between the two processes are the
following :

\begin{itemize}
\item In $t\bar{t}$ events, the invariant mass of
three of the jets should reconstruct 
to the top mass, and two out of these three jets should 
have invariant mass equal 
to the $W$ mass.  This is the consequence of the decay chain
$t \rightarrow Wb$ followed by
$W \rightarrow q\bar{q}$.  Of course no such invariant mass 
enhancements occurs for the $W +$ jets background.
\item Jets in $W +$ jets events tend to have lower transverse momentum
than jets in $t\bar{t}$ events.  This is due to the fact that
jets in $W$ events arise from a brehmstrahlung-like process.  For the
same reason, these jets also tend to be emitted more in the forward
direction than jets from the decay
of centrally produced 
top quarks. 
\item Top events tend to be more spherical and aplanar than 
$W +$ jets events.  The reason for this is that
the QCD $W +$ jets matrix element introduces
significant spatial correlations between
the jets, e.g. gluon brehmstrahlung ($q \rightarrow qg$)
and gluon splitting ($g \rightarrow q\bar{q}$)
favor small opening angles between partons in the final state. 
\end{itemize}

At first glance two-jet and three-jet invariant masses appear
to be the most attractive discriminators.  Unfortunately, because
of the poor jet energy resolution and the number of possible
jet combinations that may be present in a given event, 
this method turns out to be useful only if very large
statistics data sets are available.
We will not further discuss jet invariant 
masses in this Section.  We will however revisit the issue in 
Section~\ref{mass},
where the measurements of the top mass will be reviewed. 

\begin{figure}[htb]
\epsfxsize=6.0in
\vskip 2cm
\gepsfcentered[20 200 600 600]{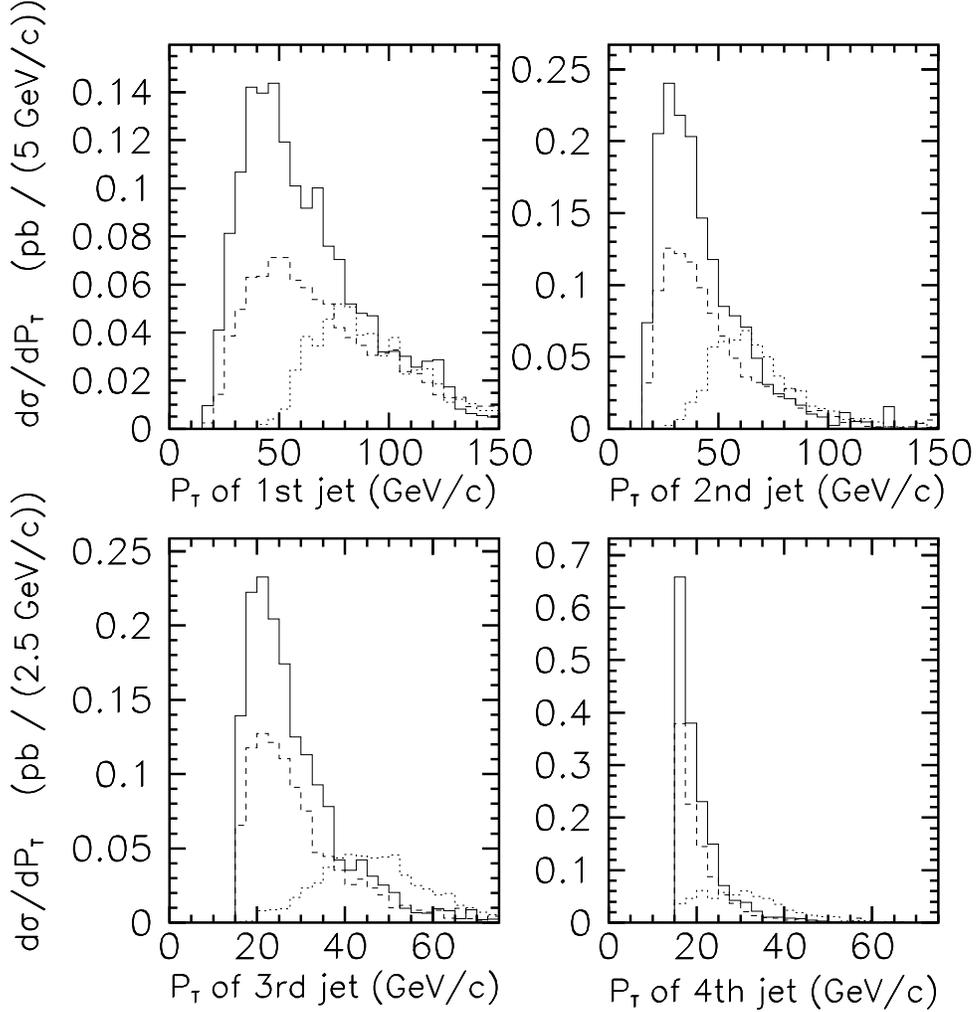}
\vskip 1cm
\caption{\protect \baselineskip 12pt
Expected transverse momentum (P$_{T}$) distributions of the
first, second, third, and fourth highest P$_{T}$ jet in
$W +$ 4 jets and $t\bar{t}$ events.  For $p\bar{p}$ collisions
at $\surd{s}$ = 1.8 TeV.
Solid line : $W +$ 4 jets,
scale $\mu^{2} = <P_{T}>^{2}$ ; dashed line : $W +$ 4 jets,
scale $\mu^{2}$ = M$_{W}^{2}$ ; dots : $t\bar{t}$,  M$_{top} =$
170 GeV/c$^{2}$. Note the expanded horizontal scale for the
third and fourth jets and that all jets have been required to have 
E$_{T} > 15$ GeV. Detector 
and $t\bar{t}$ gluon radiation  effects
are not included.  See text for details.}
\label{j1234}
\end{figure}

\begin{figure}[htb]
\epsfxsize=4.0in
\vskip 1cm
\gepsfcentered[20 200 600 600]{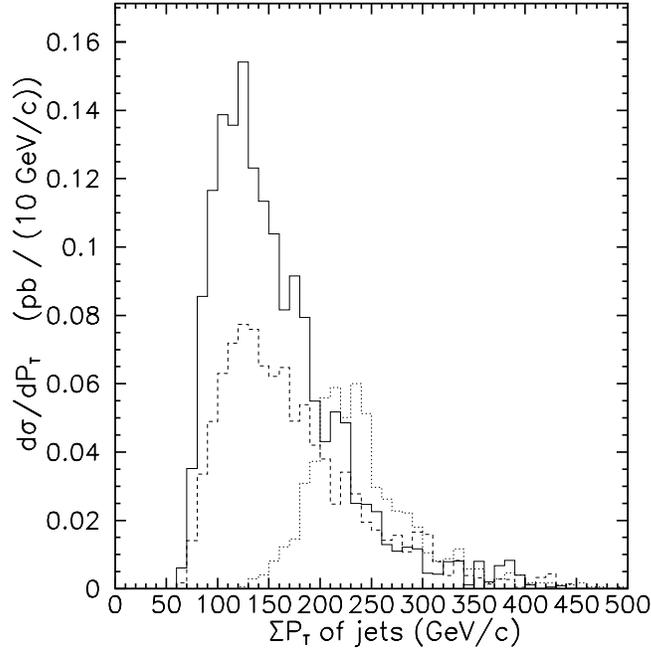}
\vskip 1cm
\caption{\protect \baselineskip 12pt
Expected scalar sum of transverse momenta of 4 jets
in $W +$ 4 jets and $t\bar{t}$ events.  
For $p\bar{p}$ collisions
at $\surd{s}$ = 1.8 TeV.
Solid line : $W +$ 4 jets,
scale $\mu^{2} = <P_{T}>^{2}$ ; dashed line : $W +$ 4 jets,
scale $\mu^{2}$ = M$_{W}^{2}$ ; dots : $t\bar{t}$,  M$_{top} =$
170 GeV/c$^{2}$. Detector and $t\bar{t}$ gluon radiation effects
are not included.  See text for details.}
\label{htpart}
\end{figure}

\begin{figure}[htb]
\epsfxsize=4.0in
\vskip 1cm
\gepsfcentered[20 200 600 600]{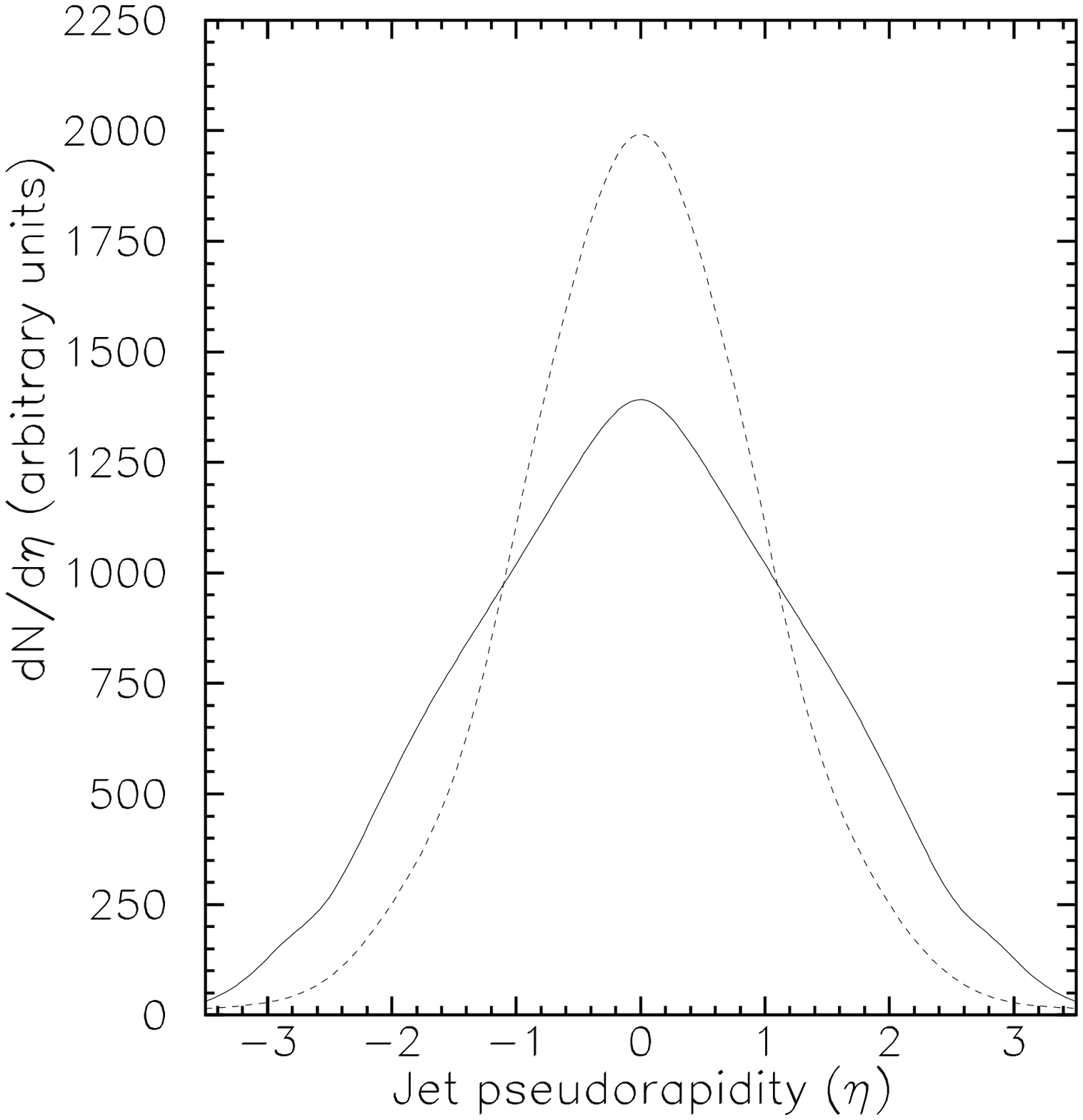}
\vskip 1cm
\caption{\protect \baselineskip 12pt
Pseudorapidity distribution of jets in 
$W +$ 4 jets and $t\bar{t}$ events. 
For $p\bar{p}$ collisions
at $\surd{s}$ = 1.8 TeV.
Solid line : $W +$ 4 jets,
scale $\mu^{2} = <P_{T}>^{2}$ ; dashed line :  $t\bar{t}$,  M$_{top} =$
170 GeV/c$^{2}$.}
\label{eta}
\end{figure}

As an illustration of the kinematic differences between
$W$ and top events at the Tevatron, 
we show the Monte Carlo transverse
momentum and pseudorapidity
distributions of jets in lepton $+$ 4 jets events for $W +$ 4 jets and 
$t\bar{t}$ (see Figs.~\ref{j1234},~\ref{htpart}, and~\ref{eta}).
As anticipated, jets in top events tend to be more central and to have
higher transverse momenta.  The scalar sum of the transverse momenta
of all of the jets in the event
is a simple global variable that is expected to provide good
discrimination between signal and background, (see Fig.~\ref{htpart}).

Note that the choice of scale in
the $W +$ jets calculation affects not only the expected rate,
but also the shape of kinematic distributions, particularly
the jet transverse momentum spectrum.   Comparing the
distributions for the
two equally arbitrary, and a-priori equally reasonable choices, 
$\mu^{2} = <P_{T}>^{2}$ and 
$\mu^{2}$ = M$_{W}^{2}$, it is found that the $\mu^{2}$ = M$_{W}^{2}$
choice results in a harder jet P$_{T}$ spectrum, see
Figs.~\ref{j1234} and~\ref{htpart}.
This can simply be understood as follows.  
The cross-section for a 
$W +$ N jets event is just given by the convolution of the
relevant matrix element with the parton distribution functions.
Neglecting
the $\mu^{2}$-dependence of the parton distribution functions,
the only scale dependence is due to the factor
$\alpha_{s}^{N}(\mu^{2})$ which appears in the 
$W +$ N jets matrix element.  Because of the running of the
strong coupling constant, the choice of an
event-by-event scale such as $\mu^{2} =~ <P_{T}>^{2}$
results in a higher probability for events with
low P$_{T}$ jets as compared to what one would obtain
by choosing a global scale like $\mu^{2}$ = M$_{W}^{2}$.


\clearpage

\begin{figure}[h]
\epsfxsize=4.0in
\vskip 1cm
\gepsfcentered[20 200 600 600]{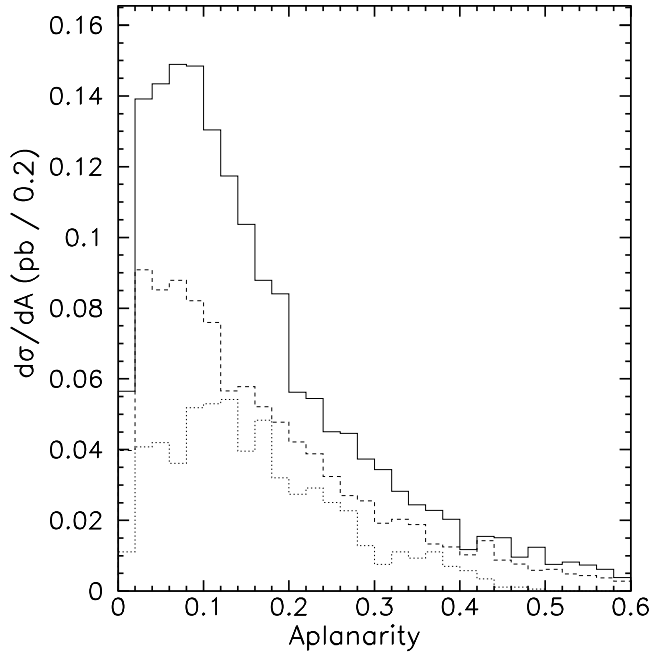}
\vskip 1cm
\caption{\protect \baselineskip 12pt
Expected aplanarity
for $W +$ 4 jets and $t\bar{t}$ events.  
For $p\bar{p}$ collisions
at $\surd{s}$ = 1.8 TeV.
Solid line : $W +$ 4 jets,
scale $\mu^{2} = <P_{T}>^{2}$ ; dashed line : $W +$ 4 jets,
scale $\mu^{2}$ = M$_{W}^{2}$ ; dots : $t\bar{t}$,  M$_{top} =$
170 GeV/c$^{2}$. Aplanarity here is calculated in the
laboratory frame from the four 
jets, the lepton, and the neutrino.
Detector and $t\bar{t}$ gluon radiation effects
are not included.  See text for details.}
\label{aplan}
\end{figure}

Another discriminant that can be employed is {\em aplanarity}.
Aplanarity is defined as $\cal A$ $\equiv 3/2 \lambda_{1}$,
where $\lambda_{1}$ is the smallest eigenvalue of
the matrix $M_{ab} = \sum P_{a}P_{b} / \sum P^{2}$, where
$P_{i}$ are the cartesian components of momentum of the parton, 
$P$ is the magnitude of the three-momentum, and the sum is over
all final state objects: jets, electrons, muons, and neutrinos. 
Zero aplanarity corresponds to planar events.
$W +$ jets events are expected to be more planar than $t\bar{t}$
events, (see Fig.~\ref{aplan}).

It is clear from this discussion that in order to extract 
a top signal from kinematic distributions, special
attention must be payed to the systematic uncertainties 
associated with the theoretical modeling of the $W +$ jets
backgrounds.   
The theoretical uncertainties can be bounded by 
comparing the data with theoretical predictions in 
kinematic regions where the top signal is small, e.g.
in samples of events with 
low jet multiplicity and/or low transverse momentum
jets.   Deviations from the $W +$ jets expectations 
in kinematic regions where the top quark is expected
to contribute would then signal the presence of $t\bar{t}$
events in the sample (or 
possibly of some other source of $W +$ jets events beyond
standard QCD production).  

Samples of $Z +$ jets events
in principle provide the ideal testing ground for the
$W +$ jets calculation.  Unfortunately, at $\sqrt{s}$ = 1.8 TeV
the cross-section
for $p \bar{p} \rightarrow Z$ is a factor of 3.3 smaller
than that for $p \bar{p} \rightarrow W$.  Furthermore, 
the leptonic branching ratio of the $Z$ is also a factor
of three smaller than that of the $W$.  As a result,
the number of reconstructed $Z$ events is approximately
one order of magnitude smaller than that of $W$ events.
Because of the limited statistics, $Z$ events, although 
useful as a first order check, do not provide 
stringent bounds on the modeling of vector boson $+ \geq 3$
jets production.  We will discuss these issues in more detail
in Section~\ref{discintro}, where the experimental results will be reviewed.

\subsubsection{$b$-quark tagging}
\label{tagging}
An alternative method that can be employed
to separate the $t\bar{t}$ signal from the
$W +$ jets background is to {\em tag} the $b$-quarks in top events.
Each $t\bar{t}$ event contains 
one $b$- and one $\bar{b}$-quark from the decays
$t \rightarrow Wb$ and $\bar{t} \rightarrow W\bar{b}$, whereas
the jets in $W$ events arise mostly from the fragmentation
of gluons and light quarks.

\begin{figure}[htb]
\epsfxsize=4.0in
\vskip 1cm
\gepsfcentered[20 200 600 600]{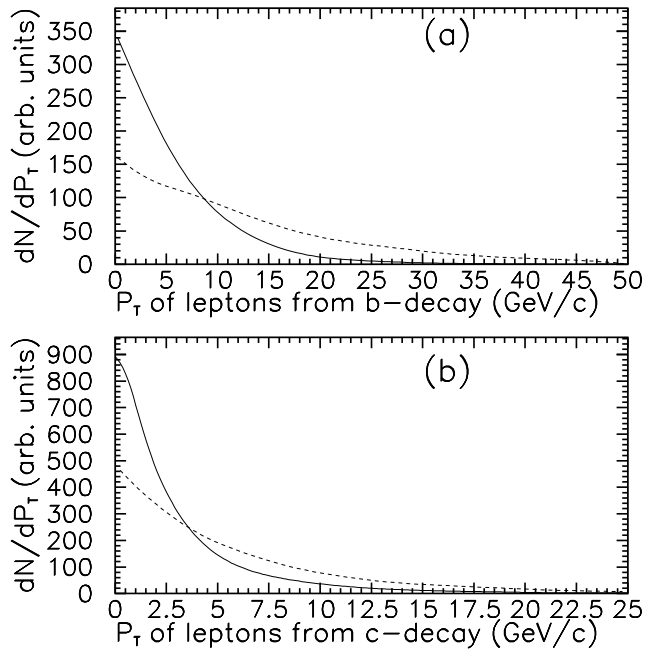}
\vskip 1cm
\caption{\protect \baselineskip 12pt
(a) Expected transverse momentum of leptons from 
$t \rightarrow b \rightarrow cl\nu$ 
Solid line :
M$_{top} =$ 110 GeV/c$^{2}$; dashed line : M$_{top} =$ 170 GeV/c$^{2}$.
(b) Same as (a), but for 
$t \rightarrow b \rightarrow c \rightarrow sl\nu$.
For $p\bar{p} \rightarrow t\bar{t}$ at $\surd{s}$ = 1.8 TeV, from
the ISAJET Monte Carlo generator.
Note the different 
horizontal scales in (a) and (b).}
\label{sltpt}
\end{figure}

There are two ways to detect the presence of $b$-quarks.
The first method is based on the detection of additional 
leptons from the semileptonic decays
$b \rightarrow cl\nu$ or $b \rightarrow c \rightarrow sl\nu$.
The semileptonic branching ratios of bottom and charm quarks 
are approximately 10\% per lepton species (Montanet {\em et al.}, 1994).
There is on average about one lepton (electron $+$ muon)
from $b$ or $c$ decay in each top event, where we have also included
the contributions
from $c$-quarks from $W \rightarrow c\bar{s}$, which occurs in
one half of all hadronic $W$ decays.
From an experimental point
of view, detection of these leptons is more difficult than
detection of leptons from $W$ decays, because these leptons
tend to have a much lower transverse momentum (compare
Figs.~\ref{sltpt} and~\ref{wlep}).  Furthermore, these
leptons are not isolated but are accompanied by nearby hadrons
from the $b$-quark fragmentation and the $b$-hadron decay.
This makes efficient detection of electrons particularly
challenging.

\begin{figure}[htb]
\epsfxsize=4.0in
\vskip 1cm
\gepsfcentered[20 200 600 600]{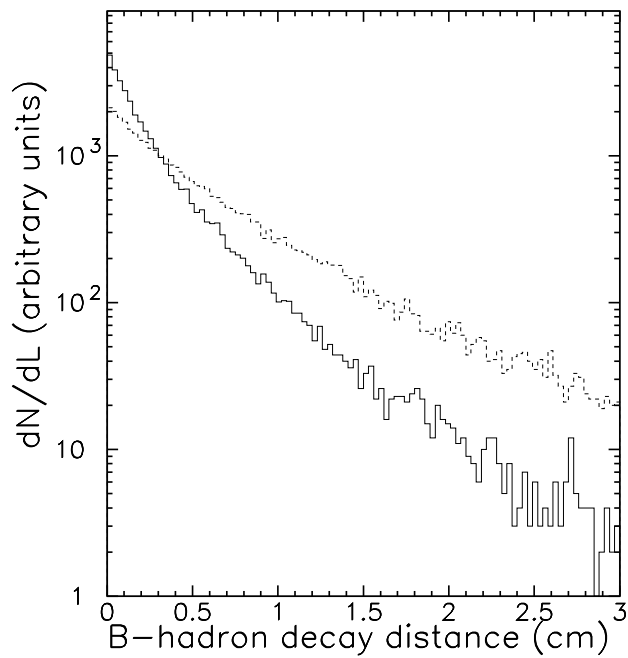}
\vskip 1cm
\caption{\protect \baselineskip 12pt
Distance travelled by $b$-hadrons in top events before decaying.
Solid line : M$_{top} =$ 110 GeV/c$^{2}$; 
dashed line : M$_{top} =$ 170 GeV/c$^{2}$.  This distance is calculated
by convoluting the momentum spectrum of $b$-hadrons from the
ISAJET $p\bar{p} \rightarrow t\bar{t}$ ($\surd{s}$ = 1.8 TeV)
Monte Carlo with their lifetime.}
\label{dl}
\end{figure}

Bottom and charm
quarks can also be tagged by exploiting the long lifetime 
of $b$- and $c$-hadrons.  The recent compilation
from the Particle Data Group (Montanet {\em et al.}) reports
a lifetime of 1.537 $\pm$ 0.021 ps for $b$-hadrons, and
0.415 $\pm$ 0.004 ps and 1.057 $\pm$ 0.015 ps for 
neutral and charged D mesons respectively.
As a consequence of the long lifetime,
$b$-hadrons in a top event
are expected to travel several mm before decaying, (see Fig.~\ref{dl}).
With the advent of silicon microstrip-vertex detectors, 
the position of decay vertices
can be measured with resolutions of order 100-150 $\mu$m.  It then
becomes possible to separate with good efficiency the secondary vertex
where the $b$-decay occurs from the primary $p\bar{p}$ interaction point.

There are two main sources of background to the 
lepton- or vertex-tagged $t\bar{t}$
signal.  The first one is instrumental.  Hadrons originating
from the fragmentation of gluons and light quarks in $W +$ jets
events can be misidentified as muons or electrons.  This happens for 
example when hadronic
showers in the calorimeter fluctuate to mimic the electron signature,
or when kaons and pions decay to muons.
Also, track 
mismeasurements or decays of other long-lived particles such as 
$\Lambda$ and K$_{s}$
can result in the reconstruction of spurious 
detached vertices.  
Needless to say, these effects have to be considered very carefully.
As we will discuss in Section~\ref{discintro}, sufficiently good instrumental 
background rejection has been achieved, and methods to precisely 
estimate the remaining background have been developed.

\begin{figure}[htb]
\begin{picture}(32000,10000)(0,-3000)

\drawline\fermion[\E\REG](10000,6000)[6000]
\put(\particlefrontx,7000){$\bar{d},\bar{s}$}
\drawline\photon[\E\REG](\particlebackx,\particlebacky)[10]
\put(\particlebackx,7000){$W$}
\drawline\fermion[\S\REG](16000,6000)[8000]
\drawline\gluon[\W\REG](\particlebackx,\particlebacky)[6]
\put(\particlebackx,0){$g$}
\drawline\fermion[\E\REG](\particlefrontx,\particlefronty)[10000]
\put(\particlebackx,-1000){$c$}

\end{picture}
\caption{\protect \baselineskip 12pt
Lowest order Feynman diagram for production of $W +$~charm in
$p\bar{p}$ collisions.
At higher order, $gg \rightarrow Wc\bar{s}$ diagrams also
contribute.}
\label{wc}
\end{figure}
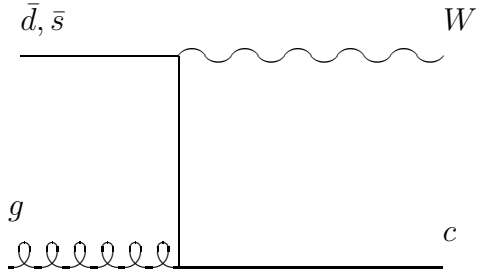

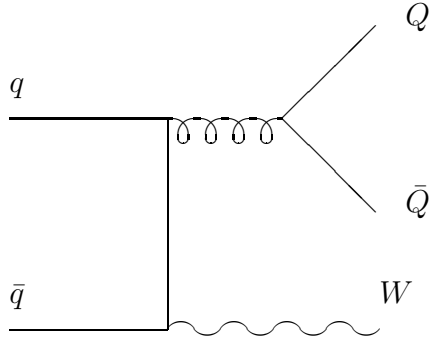
\begin{figure}[hbt]
\begin{picture}(32000,12000)(0,-3000)

\drawline\fermion[\E\REG](10000,6000)[6000]
\put(\particlefrontx,7000){$q$}
\drawline\gluon[\E\REG](\particlebackx,\particlebacky)[4]
\drawline\fermion[\NE\REG](\particlebackx,\particlebacky)[5000]
\put(25000,\particlebacky){$Q$}
\drawline\fermion[\SE\REG](\particlefrontx,\particlefronty)[5000]
\put(25000,\particlebacky){$\bar{Q}$}
\drawline\fermion[\S\REG](16000,6000)[8000]
\drawline\photon[\E\REG](\particlebackx,\particlebacky)[8]
\put(\particlebackx,-1000){$W$}
\drawline\fermion[\W\REG](\particlefrontx,\particlefronty)[6000]
\put(\particlebackx,-1000){$\bar{q}$}

\end{picture}
\caption{\protect \baselineskip 12pt
Lowest order Feynman diagram for production of $WQ\bar{Q}$
in $p\bar{p}$ collisions. $Q = c$ or $b$.}
\label{gspfig}
\end{figure}

The second background source stems
from the fact that a small fraction of
jets in $W +$ jets events will contain heavy-quarks ($b$ or $c$).
In the absence of additional kinematic information,
these events constitute an irreducible physical background to the
lepton $+$ jets $+$ $b$-tag $t\bar{t}$ signature which needs to
be carefully evaluated.  A comprehensive discussion of these backgrounds,
in the context of the CDF experiment, is given by
F. Abe {\em et al.}, 1994a, and will be summarized below.  The discussion
here applies equally well to lepton or vertex $b$-tagging methods.

Heavy-quarks in $W +$ jets event can be produced singly, in the 
process $\bar{s}g \rightarrow Wc$ or $\bar{d}g \rightarrow Wc$,
(see Fig.~\ref{wc}), or in pairs, when a gluon in
a $W +$ jets event {\em splits}
into a $c\bar{c}$ or $b\bar{b}$ pair, (see Fig.~\ref{gspfig}).
The gluon splitting probability, $g \rightarrow Q\bar{Q}$,
$Q = c$ or $b$,
is estimated to be of order a few percent,
with significant theoretical uncertainties
(Mangano and Nason, 1992; M\"uller and Nason, 1985 and 1986). 
Experimental extractions of the $g \rightarrow c\bar{c}$ probability
in both $p\bar{p}$ (Ikeda, 1990, F. Abe {\em et al.}, 1990d)
and $e^{+}e^{-}$ collisions (Akers, 1995) are found
to be in agreement 
with the results of the theoretical calculations, within
the large experimental and theoretical uncertainties.
In what follows we will refer to these two 
processes as $Wc$ and $WQ\bar{Q}$ respectively.

The $Wc$ background estimation begins 
by computing
the fraction of $W +$ jets events that contain
a single $c$-quark, from diagrams like the one shown in Fig.~\ref{wc},
using the VECBOS and HERWIG event 
generators.  This fraction is found to be of order 8\%,
with small variations
depending on the jet multiplicity and the choice of 
the input parton distribution function for strange quarks in
the proton.  To obtain an absolute background prediction,
this fraction is then multiplied by the number of observed
$W +$ jets event and the tagging efficiency for $c$-quarks.
The tagging efficiency for these events clearly depends on the
details of the tagging algorithm.  However, it is in general
much smaller (typically by a factor of order 5)
than the tagging efficiency for $t\bar{t}$ since
(i) the tagging efficiency for $c$-quarks is lower
than that of $b$-quarks because e.g. there are fewer tracks in a $c$
decay than in a $b$ decay, 
and (ii) there are multiple $b$- and
$c$-quarks in a top event that can potentially be tagged.
Therefore, as a result of the small probability for a $Wc$
event and its small tagging efficiency, 
the $Wc$ background
is much smaller than the expected tagged $t\bar{t}$ rate.
The highly uncertain
overall normalization of the $W +$ jets 
theoretical calculation does not enter in the background
estimate, since only the {\bf fraction} of $Wc$ events
is taken from theory.

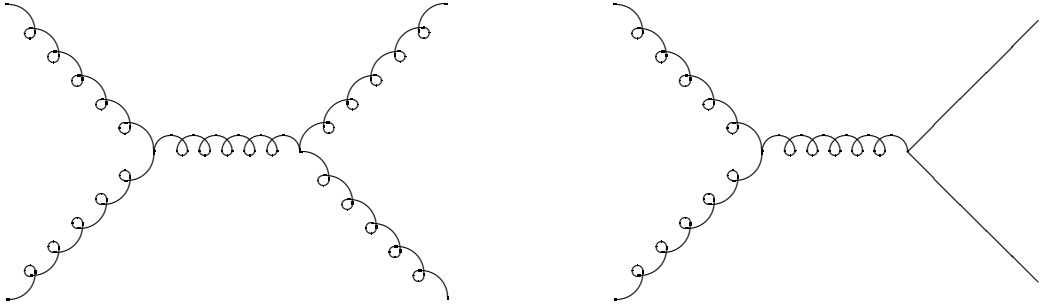
\begin{figure}
\hskip 2cm
\begin{picture}(32000,15000)(0,-3000)
\drawvertex\gluon[\W 3](8000,5000)[5]
\drawline\gluon[\NE\REG](8000,5000)[5]
\drawline\gluon[\SE\REG](8000,5000)[5]

\drawvertex\gluon[\W 3](31000,5000)[5]
\drawline\fermion[\NE\REG](31000,5000)[7000]
\drawline\fermion[\SE\REG](31000,5000)[7000]

\end{picture}
\caption{\protect \baselineskip 12pt
Two of the possible LO 
Feynman diagrams for jet production in
$p\bar{p}$ collisions: $gg \rightarrow gg$ and $gg \rightarrow q\bar{q}$.}
\label{dijet}
\end{figure}

The $WQ\bar{Q}$ background is more important because the
tagging efficiency for $Wb\bar{b}$ is comparable to that of
$t\bar{t}$.  This background can be estimated in two ways.
The first method ({\em Method I}) requires a minimum 
of theoretical input, and is expected to yield an overestimate
of the $WQ\bar{Q}$ background.  The alternative method,
({\em Method II}), is based on the state-of-the-art
theoretical understanding of heavy quark production,

At the heart of the Method I background calculation is 
the assumption that the heavy
flavor content ($b$ or $c$) of jets in $p\bar{p} \rightarrow$~jets
({\em generic-jets}, see Fig.~\ref{dijet})
is the same or larger than
the heavy flavor content of jets in $W +$ jets
events.  Accepting this assumption for now, we proceed to
describe the Method I background calculation.  

The generic-jet sample includes gluon and
light-quark jets, as well as a small fraction of heavy-quark
jets.  A generic-jet can be tagged due to instrumental effects
resulting in a false tag of a light-quark or gluon jet,
or due to the $b$- or $c$-quark contribution.
Operationally, a {\em tag-rate} is measured for generic-jets
as the probability of tagging a generic-jet
as a function of several relevant variables (e.g. jet P$_{T}$,
track multiplicity).
The generic-jet tag-rate is then applied to the sample
of jets in $W +$ jets events, to predict an upper limit for
the sum of the instrumental and $WQ\bar{Q}$ backgrounds to 
the $t\bar{t} \rightarrow W +$ jets $+$ $b$-tag signature.
From an experimental
point of view this has the advantage that both the instrumental
and $WQ\bar{Q}$ backgrounds are estimated simultaneously
directly from the data.
No a-priori knowledge of the tagging efficiency, or the $WQ\bar{Q}$
content of the sample, is needed.


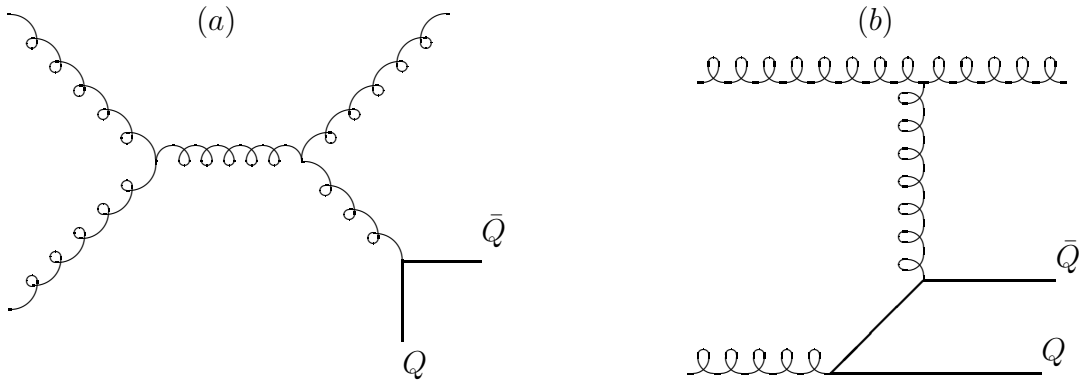
\begin{figure}
\hskip 2cm
\begin{picture}(32000,15000)(0,-3000)
\drawvertex\gluon[\W 3](8000,5000)[5]
\drawline\gluon[\NE\REG](8000,5000)[5]
\drawline\gluon[\SE\REG](8000,5000)[3]
\THICKLINES
\drawline\fermion[\E\REG](\particlebackx,\particlebacky)[3000]
\put(\pbackx,2000){$\bar{Q}$}
\drawline\fermion[\S\REG](\particlefrontx,\particlefronty)[3000]
\put(\pbackx,-3000){$Q$}
\put(4000,10000){$(a)$}
\THINLINES

\drawline\gluon[\E\FLIPPED](23000,8000)[8]
\drawline\gluon[\E\FLIPPED](\pbackx,\pbacky)[5]
\drawline\gluon[\S\REG](\pfrontx,\pfronty)[7]
\THICKLINES
\drawline\fermion[\E\REG](\pbackx,\pbacky)[5000]
\put(\pbackx,1000){$\bar{Q}$}
\drawline\fermion[\SW\REG](\pfrontx,\pfronty)[5000]
\drawline\fermion[\E\REG](\pbackx,\pbacky)[8000]
\put(\pbackx,-2500){$Q$}
\THINLINES
\drawline\gluon[\W\REG](\pfrontx,\pfronty)[5]
\put(29000,10000){$(b)$}

\end{picture}
\vskip 1cm
\caption{\protect \baselineskip 12pt
Higher order Feynman diagrams for $Q\bar{Q}$ production
in $p\bar{p}$ collisions: (a) gluon
splitting; (b) flavor excitation.}
\label{gspl}
\end{figure}

We now turn to a discussion of the theoretical assumption
on which the Method I $WQ\bar{Q}$ background calculation is based.
At Tevatron energies, generic-jets in the relevant
P$_{T}$ range (20-150 GeV/c) consist predominantly
of gluon-jets.  There will also of course be
a contribution from light-quark jets as well as
$p\bar{p} \rightarrow Q\bar{Q}$.
The lowest order Feynman diagrams for direct production
of $b\bar{b}$ 
and $c\bar{c}$ pairs in $p\bar{p}$ collisions
are identical to the ones 
for $t\bar{t}$ production shown in 
Fig. 16.
For the $b$- and $c$-quarks however, it is found that 
higher order diagrams such as the ones displayed
in Fig.~\ref{gspl} contribute
a very significant amount to the 
$p\bar{p} \rightarrow Q\bar{Q} + X$
cross-section.  For instance, the 
gluon splitting diagram is believed to account for of order 70\%
of the $b\bar{b}$ and $c\bar{c}$ production rate (Mangano,
Nason, and Ridolfi, 1992).

The heavy-flavor content of generic-jets is expected to
be higher than that of jets in $W +$ jets events for two reasons :
(i) in $W +$ jets events
there are no contributions from lowest order, direct,
$Q\bar{Q}$ production (Fig. 16)
and next-to-leading order flavor excitation (Fig.~\ref{gspl}b),
and (ii) generic-jets consist mostly of gluon jets
which can result in $Q\bar{Q}$ pairs via gluon splitting, whereas
the jets in $W +$ jets events consist of an approximately
equal mixture of gluon and light-quark jets.  These very
simple qualitative arguments are borne out by a more 
quantitative calculation, as we will illustrate below.

The second way of estimating the $WQ\bar{Q}$ background
(Method II)
is based on an explicit calculation of the $WQ\bar{Q}$
process.  Just as in the $Wc$ case, uncertainties on the
overall normalization of the $W +$ jets calculation are minimized
by using as the theoretical 
input the fraction of $W +$ jets events that contain 
a $Q\bar{Q}$ pair, rather than the absolute rate prediction.
The absolute background level is then
estimated by multiplying this fraction by the number of observed 
$W +$ jets events and the tagging efficiencies for $Wc\bar{c}$
and $Wb\bar{b}$.

The $WQ\bar{Q}$ fraction is estimated using a combination
of the HERWIG Monte Carlo event generator and 
the lowest order $WQ\bar{Q}$ matrix element (Fig.~\ref{gspfig})
calculation (Mangano, 1993). 
This calculation differs from the $W +$ jets calculation 
in that it includes all
mass effects and is therefore free from collinear and infrared 
divergencies.
Since the calculation does not include higher order terms, it
cannot be used directly to estimate the $WQ\bar{Q}$ rate
in events with more than two jets, which are the
most relevant for the top search.
Instead, the results of the exact calculation are compared to those
of HERWIG,
where the $Q\bar{Q}$ pairs are produced by gluon splitting
from initial- and final-state parton evolution.  The HERWIG
results are found to be in good agreement with those of the $WQ\bar{Q}$
matrix element calculation.
The gluon splitting process
in HERWIG also provides a good description 
of the measured tagging rate in generic-jet events, giving further
evidence for the validity of the HERWIG model.
Thus, this model is used to predict the fraction of $WQ\bar{Q}$
events as a function of jet multiplicity.  For the sample
of $W + \geq 3$ jets, it is estimated that
the fractions of $W$ events containing a $b\bar{b}$ or $c\bar{c}$
pair are approximately 3\% and 5\% respectively.
The systematic uncertainties associated with these predictions are
clearly significant and are estimated to be at the level 
of 80\%. 
The $WQ\bar{Q}$ background estimated
with this method turns out to be approximately a factor of three
lower than the conservative estimate made under the assumption
that the heavy flavor content of jets in $W +$ jets events is the
same as that of generic-jets.

To assess the impact of the $WQ\bar{Q}$ background on the top
search, we re-examine the predicted rates of lepton $+$ jets
from $t\bar{t}$ and $W +$ jets.   For example, 
from Fig.~\ref{jmult}, the signal-to-background for a $W +$ 4 jets
selection, with jet P$_{T}$ threshold set at 15 GeV/c,
is expected to vary from approximately $3/1$ at M$_{top}$ = 120 GeV/c$^{2}$
to approximately $1/3$ at M$_{top}$ = 200 GeV/c$^{2}$.
With the fraction of $W +$ jets events containing a 
$Q\bar{Q}$ pair given above, the signal-to-background,
($t\bar{t}$ vs. $WQ\bar{Q}$) will be at least of the order of $7/1$.
Thus, provided the
instrumental backgrounds can be adequately controlled,
and enough luminosity is available to the experimenters, 
the top signal can be separated from the $W +$ jets background
with $b$-tagging methods.


\section{Early searches for the top quark}
\label{pre1a}
Experimental searches for the top quark began
immediately following the discovery of the $b$-quark.
In this Section we will review these searches up
to approximately 1990.

\subsection{Searches in $e^{+}e^{-}$ collisions}
\label{epluseminus}
In $e^{+}e^{-}$ collisions, top quarks would be produced in pairs 
through $e^{+}e^{-}$ annihilation into a photon or a $Z$.
Since at leading order this is a
purely electroweak process,                       
the production
cross-section can be accurately calculated.  
At center-of-mass energies well below the
$Z$ mass, where annihilation of the $e^{+}e^{-}$ pair into 
a photon dominates, 
$t\bar{t}$ production would manifest
itself as an increase by an amount 
$\delta$R $\approx$ 3Q$^{2}_{top}$ = $4/3$ in the
ratio R $=\sigma(e^{+}e^{-} \rightarrow $ hadrons) $/ \sigma(e^{+}e^{-}
\rightarrow \mu^{+}\mu^{-})$ well above
the energy threshold for the production
of a $t\bar{t}$ pair.  
This expected increase in
R is largely independent of the top quark
decay mode, as long as the top-quark decays 
into final states containing hadrons.

Between 1979 and 1984, measurements of R 
were performed at the PETRA $e^{+}e^{-}$ collider
in the center-of-mass energy range between
12 and 46.8 GeV (Barber {\em et al.}, 1979 and 1980;
Berger {\em et al.}, 1979 and 1980;
Bartel {\em et al.}, 1979a, 1979b, and 1981;
Brandelik {\em et al.}, 1982;
Adeva {\em et al.}, 1983a, 1983b, 1985, and 1986;
Behrend {\em et al.}, 1984; Althoff {\em et al.}, 1984a and 1984b).
The value of R was found to be consistent
with Standard Model expectations without a top-quark
contribution.  Event topology studies
gave no evidence for excesses of
spherical, aplanar, or low thrust events that could
be attributed to $t\bar{t}$ production.  The measured rate
of prompt muons was also found to be in agreement with
expectations from models of $c$- and $b$-quark production
and decay, and could not accommodate a
contribution from semileptonic decays of 
top quarks.  Existence of the top quark with mass
below 23.3 GeV/c$^{2}$ was ruled out at the 95\% confidence level (C.L.).

Similar searches were later performed at 
the TRISTAN collider, which reached a center-of-mass
energy of 61.4 GeV
(Yoshida {\em et al.}, 1987; Sagawa {\em et al.}, 1988;
Adachi {\em et al.}, 1988; Igarashi {\em et al.}, 1988;
K. Abe {\em et al.}, 1990).  No evidence 
for top quark production was reported, resulting
in a lower limit on the top mass of 30.2 GeV/c$^{2}$.

In 1989-90, the SLC and LEP $e^{+}e^{-}$ colliders,
with $\sqrt{s} \approx $ M$_{z}$ became operational.
Studies of event topologies and measurements of
the $Z$ width were found to be inconsistent with
a $Z \rightarrow t\bar{t}$ contribution, and resulted
in lower limits on the top quark mass as high as 
45.8 GeV/c$^{2}$ (Abrams {\em et al.}, 1989;
Akrawi {\em et al.}, 1990a; Decamp {\em et al.}, 1990;
Abreu {\em et al.}, 1990b and 1991; Adriani {\em et al.},
1993).

\subsection{Early searches in $p\bar{p}$ collisions assuming Standard Model
top-quark decay}
\label{early}
With the emergence in the 1980's of $p\bar{p}$ colliders,
first at CERN and then at Fermilab, and with
evidence from $e^{+}e^{-}$ experiments pointing towards a 
very high mass for the top quark, focus in the search for
top rapidly shifted to hadron colliders.
The obvious advantage of hadron colliders 
for top physics is the high
center of mass energy, which enables the
exploration of higher mass regions.  However, 
unlike in $e^{+}e^{-}$ collisions, the large
backgrounds make it impossible to directly search for the
top quark in a model-independent way.  
It is necessary to
concentrate on particular signatures, based for example
on the Standard Model decay modes of the top quark
discussed in the previous Section.  

In this Section we will discuss searches for the top
quark in $p\bar{p}$ collisions, assuming Standard
Model top-quark decay,
that were carried out 
in the mid to late 1980's at CERN and Fermilab.
No evidence for top
quark production was uncovered, leading to lower
limits on the top quark mass as high as 91 GeV/c$^{2}$, at the
95\% C.L..  These searches do not result directly
in limits on the top mass, but rather in upper limits on
the product
of top production cross-section and 
branching ratio for top quark decay.  To turn
these limits into mass limits for the top quark, it is necessary to
(i) assume that the top quark decays 
as $t \rightarrow Wb$, as prescribed by the Standard Model,
and (ii) use theoretical expectations
for the production cross-section.  As discussed in Section~\ref{prod},
the main production mechanisms are $W \rightarrow t\bar{b}$ decays 
at S$p\bar{p}$S energies, 
and $t\bar{t}$ pair production at Tevatron energies.
The expected cross-section for 
$p\bar{p} \rightarrow W \rightarrow t\bar{b}$ can be 
reliably predicted from direct measurements of 
$p\bar{p} \rightarrow W \rightarrow l\nu$; on the other hand,
there are significant theoretical uncertainties on the
predicted top-pair production cross-section as a function
of the top mass, see Section~\ref{prod}, Figs.~\ref{xsec}
and~\ref{xsec2}.
Since these theoretical uncertainties
are difficult to quantify, limits on M$_{top}$
in the absence of a top signal in the data are placed based on
the lower range of the calculation of the top production 
cross-section.  This
results in conservative 95\% C.L. lower limits
on M$_{top}$.

Initial results reported by the UA1 collaboration (Arnison
{\em et al.}, 1984; Revol, 1985)
at the CERN
S$p\bar{p}$S collider ($\sqrt{s} = 630$ GeV/c), 
seemed to be consistent with production
of a top quark of mass $40 \pm 10$ GeV/c$^{2}$.
These results were based on the observation of 12 
isolated lepton $+$
2 jets events, with an expected background of approximately 3.5
events in an exposure with an integrated luminosity of
200 nb$^{-1}$ (Revol, 1985).
In these events, the invariant mass of the lepton,
neutrino and one of the jets was found to cluster 
around a common value of
approximately 40 GeV/c$^{2}$, while the invariant mass
of the two jets, the lepton, and the neutrino was
found to be consistent with the $W$ mass.
This is the expected signature for the process
$p\bar{p} \rightarrow W \rightarrow t\bar{b}$,
followed by $t \rightarrow bl\nu$.  
The excess of events over
the background prediction was also consistent with the
expected top production cross-section, providing 
further evidence for a top quark with
M$_{top} \approx$ 40 GeV/c$^{2}$.

These first results were however not supported by a
subsequent UA1 analysis (Albajar {\em et al.}, 1988), with
a higher statistics data sample, as well as a more
complete evaluation of the backgrounds.
This analysis was based on samples of events with 
one isolated muon $+ \geq 2$ jets or
one isolated electron $+ \geq 1$ jet.
The integrated luminosity was 700 nb$^{-1}$.
Given this integrated luminosity, and the expected top
production rate, this search was sensitive 
to a top quark with M$_{top} <$ 55 GeV/c$^{2}$.
A maximum lepton-neutrino transverse mass 
requirement of 40 (45) GeV/c$^{2}$ was imposed in the muon
(electron) sample to substantially reduce the
$W \rightarrow l\nu$ background, while maintaining good
efficiency for top quarks in the relevant mass range.
No missing transverse energy (\MET) requirement was imposed,
since for low top mass the probability for {\bf both}
the lepton and the neutrino in $t \rightarrow bl\nu$
to have high transverse momentum is low.

With no \MET~requirement, the backgrounds from 
fake leptons, Drell-Yan $l^{+}l^{-}$ pairs,
as well as
$b\bar{b}$ and $c\bar{c}$ production are
significant.
The number of observed events was found to be fully consistent
with the 
non-top background contribution only.  
Furthermore,
it was shown that the event selection requirements resulted in
background invariant mass distributions that were
similar to those expected from top-quark 
production and decay (for M$_{top} \approx$ 40 GeV/c$^{2}$).
Thus, this property of the background accounted for the features
of the invariant mass distributions observed in the previous analysis.
Dilepton events were also studied, and their 
properties were found to be
in agreement with expectations from semileptonic decays in
$b\bar{b}$ and $c\bar{c}$ events.  

An upper limit on the combined
$W \rightarrow t\bar{b}$ and $t\bar{t}$ cross-sections 
was extracted from the background-subtracted 
lepton $+$ jets event rates, as well
as kinematic distributions such as those of jet E$_{T}$, \MET,
and lepton isolation.  
At the time that these results were obtained,
only a LO calculation for $p\bar{p} \rightarrow t\bar{t}$ was
available.  Using a conservative value for this cross-section
the limit was inferred to be M$_{top} > 44$ GeV/c$^{2}$ at
the 95\% confidence level.  A subsequent 
re-evaluation (Altarelli {\em et al.}, 1988), based on the NLO 
calculation of heavy-quark pair-production (Nason, Dawson, and
Ellis, 1988)
resulted in a slightly modified lower limit 
M$_{top} > 41$ GeV/c$^{2}$.  This limit is obtained
from the $t\bar{t}$ cross-section corresponding to 
the lower range of the theoretical prediction.  

The lesson to be learned from the UA1 experience is that the 
reliability of background estimates is of paramount importance.
Their first analysis did not include the J/$\Psi$, $\Upsilon$,
and Drell-Yan backgrounds, and the $b\bar{b} + c\bar{c}$ backgrounds
were underestimated by a factor of four.
The top signature at hadron colliders is complicated, and involves
comparing the number of observed events and/or kinematic distributions
with background expectations.  If at all possible, dependences
on uncertain theoretical models of background processes should
be minimized.

More sensitive searches for the top quark were
performed in the period 1988-89.  
At CERN, the new Antiproton Accumulator
Complex (AAC) was commissioned, resulting 
in luminosities as high as 3 x 10$^{30}$ cm$^{-2}$ s$^{-1}$
for the S$p\bar{p}$S. 
The UA1 electromagnetic
calorimeter was removed to allow for a replacement
based on Uranium-TMP (tetramethylpentane) technology.
This new calorimeter was unfortunately not ready 
to be installed, resulting in the loss of
electron identification in UA1.
The UA2 detector
at the CERN S$p\bar{p}$S was significantly
upgraded, with improved calorimeter coverage, and enhanced
electron detection capabilities.
In the US, the first high-luminosity 
(up to 2 x 10$^{30}$ cm$^{-2}$ s$^{-1}$)
run of the Tevatron Collider
also started in 1988, with the CDF detector ready for
data-taking.

The 1988-89 UA1 top search (Albajar {\em et al.}, 1990)
was based on an integrated luminosity of 4.7 pb$^{-1}$, a
factor of seven higher than had been 
previously available.
Because of the missing EM calorimeter,
only $\mu$ $+ \geq$ 2 jets and 
$\mu\mu$ final states were considered.  
The $\mu$ $+$ jets analysis was essentially
an extension of the previous UA1 search (Albajar {\em et al.}, 1988), 
with the maximum 
transverse mass requirement raised to 60 GeV/c$^{2}$.
For each event, a likelihood-like variable (${\cal L}_{1}$)
was defined to discriminate between top and background on
an average basis.
This likelihood was based
on the \MET,  the isolation of the muon, and the opening angle between the
muon and the highest E$_{T}$ jet in the event.  

%
%
%
%
%
%
%
%
\begin{figure}[htb]
\vskip 1cm
\epsfxsize=4.0in
\hskip 3cm
\epsffile{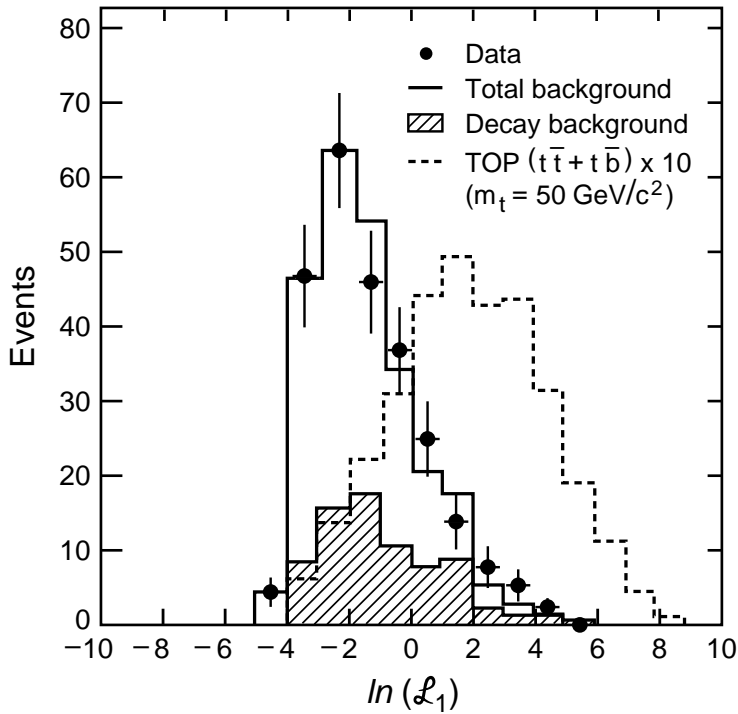}
\vskip 1mm
\caption{\protect \baselineskip 12pt
The ln(${\cal L}_{1}$) distribution compared with 
the expected background and top contributions.
The shaded histogram is for the K and $\pi \rightarrow \mu\nu$
background.  The other major background contribution is
from $b\bar{b}$ and $c\bar{c}$ events.  The expected top
contribution, scaled up by a factor of 10, is shown in the dashed histogram.
From the UA1 collaboration, Albajar {\em et al.}, 1990.}
\label{ln_l1}
\end{figure}

The ${\cal L}_{1}$ distributions of data, expected background, and
expected signal, are shown in Fig.~\ref{ln_l1}.   
The data are well-described by the background contribution only.
Based on the 
number of observed events with ln(${\cal L}_{1}$) $>$ 4,
and the expected top production
cross-section, the $\mu +$ jets data from UA1 results in a lower limit
on the mass of the top quark of 52 GeV/c$^{2}$ at the 95\% 
confidence level.

A similar likelihood variable was defined for $\mu\mu$
events, based on the transverse momentum and isolation of 
the highest P$_{T}$ muon, and the azimuthal opening angle
between the two muons.
Recall that
at $\sqrt{s}$ = 0.63 TeV for M$_{top} > 40$ GeV/c$^{2}$
the major source of top events is $W$ decays, $W \rightarrow t\bar{b}$,
see Section~\ref{prod}.
Hence, the dilepton final state arises from semileptonic decays of both the 
$t$- and $b$-quarks, and the muon from $b \rightarrow c\mu\nu$ is
not expected to be isolated.
%
%
%
%
Studies of
the likelihood distribution for $\mu\mu$ events also gave a null result.
Results from the $\mu +$ jets and the $\mu\mu$ data were combined with the
lower statistics earlier UA1 results, yielding
a lower limit on the top mass of 60 GeV/c$^{2}$ (95\% C.L.).

A better limit on the top-quark mass was also obtained
at the same time
by the UA2 collaboration, based on an integrated
luminosity of 7.5 pb$^{-1}$ at the S$p\bar{p}$S (Akesson {\em et al.}, 1990).
Without muon detection capabilities, the UA2 results
were based entirely on the electron $+ \geq 1$ jet channel.
The UA2 search strategy differed considerably from that of UA1,
with a \MET~ $>$ 15 GeV and an electron isolation
requirement imposed in the event 
selection.  As a result, backgrounds from $b\bar{b}$, $c\bar{c}$,
and fake electrons were highly suppressed, and contributed only 
of order 10\% to the data sample.  The bulk of the background
was due to $W \rightarrow e\nu +$ jets events.  As was discussed
in Section~\ref{lowmass}, for M$_{top} < $ M$_{W} +$ M$_{b}$, the
W in the decay $t \rightarrow Wb$ is virtual.
The differences between the lepton-neutrino
transverse mass distributions 
of real and virtual $W \rightarrow l\nu$ can then be exploited to
separate a top signal from the background.

\begin{figure}[htb]
\vskip 1cm
\hskip 3cm
\epsfxsize=3.0in
\epsffile{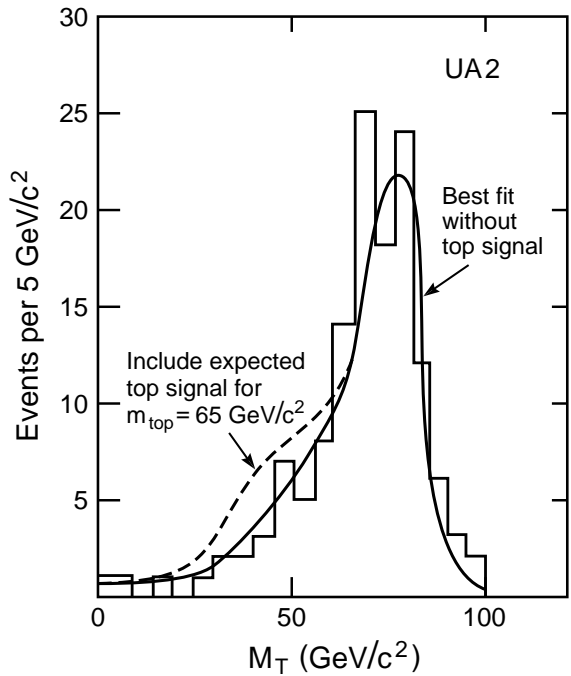}
\vskip 1mm
\caption{\protect \baselineskip 12pt
The electron-neutrino transverse mass distribution in electron $+$ jet $+$
\MET~events.  The data are inconsistent with a top
contribution.  From the UA2 collaboration, Akesson {\em et al.}, 1990.}
\label{ua2tmass}
\end{figure}

In Fig.~\ref{ua2tmass} we show the transverse mass distribution
for electron $+$ jet $+$ \MET~ events in UA2. 
No evidence for an excess of low transverse mass events was 
found, resulting in a lower limit of M$_{top} >$ 69 GeV/c$^{2}$.

The first high statistics run of the Tevatron Collider
also took place in 1988-89, with an integrated luminosity
of 4 pb$^{-1}$ recorded by the CDF collaboration.
As was discussed in Section~\ref{prod}, at the center-of-mass
energy of the Tevatron ($\sqrt{s}$ = 1800 GeV), top production
in the relevant M$_{top}$ range is dominated by the 
$p\bar{p} \rightarrow t\bar{t}$ process.  The CDF collaboration
searched for top in both the lepton $+$ jets (see Section~\ref{ljets})
and dilepton (see Section~\ref{dilepton}) modes.

\begin{figure}[htb]
\vskip 1cm
\hskip 4cm
\epsfxsize=3.0in
\epsffile{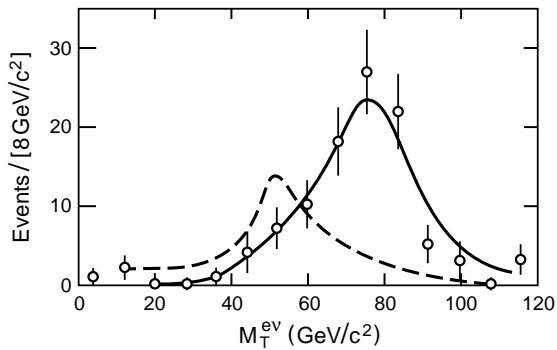}
\vskip 1mm
\caption{\protect \baselineskip 12pt
The transverse mass distribution of electron $+$ two or more jets $+$
\MET~events.  From the CDF collaboration, (F. Abe {\em et al.}, 1990a).
The solid and dashed lines represent expectations from the $W + 2$ jets 
and $t\bar{t}$ (M$_{top} = 70$ GeV/c$^{2}$) Monte Carlo calculations.}
\label{CDFmtrans}
\end{figure}

The CDF top search in the isolated electron $+ \geq$ 2 jets
channel (F. Abe {\em et al.}, 1990a and 1991a)
was qualitatively similar to the
UA2 search.  An explicit \MET~ requirement was imposed, yielding
a data sample containing $W \rightarrow e\nu$ + 2 jets events, 
with a small contamination from semileptonic $b$- and $c$-quark
decays, as well as fake electrons.  The 
resulting transverse mass distribution was found to be 
consistent with no top contribution, (see Fig.~\ref{CDFmtrans}).
The existence
of a Standard Model 
top quark with 40 GeV/c$^{2} < $ M$_{top} < 77$ GeV/c$^{2}$
was ruled out.  (The limit was not extended below 40 GeV/c$^{2}$
because of poor acceptance at low top quark mass).
Consistent results were also found in
a subsequent study of $\mu +$ jets events (Demortier, 1991).

Both the UA2 and CDF
results were based on studies of the shape of the transverse
mass distribution in lepton $+$ jets events and were
therefore independent of uncertainties in the
theoretical predictions for the $p\bar{p} \rightarrow W +$ jets
cross-section.  Furthermore, the shape of the transverse mass
distribution 
depends only weakly on 
the details
of $W +$ jets production.  It is sensitive mostly to
the kinematics of the
$W$ decay, and the \MET~ resolution in the detector, and it
is fairly insensitive to the transverse momentum of the 
$W$, see the discussion in Section~\ref{lowmass}.
Therefore, the influence of the theoretical modelling of the background
was minimized in the CDF and UA2 lepton $+$ jets top searches.

The CDF collaboration also searched for $t\bar{t}$ production in
the dilepton channel.  Initially, only the $e\mu$ channel
was considered (F. Abe {\em et al.}, 1990b).  
As discussed in Section~\ref{dilepton},
the $t\bar{t}$ dilepton signature consists in principle of two 
isolated high P$_{T}$
leptons, \MET~and two jets.  Since the backgrounds in this channel
are small, in order to maximize the top acceptance, no \MET, isolation,
or jet requirements were imposed.   One event was observed,
with an expected background of 1.4 events, mostly from
$Z \rightarrow \tau\tau$ followed by leptonic decays of both taus.
An upper limit on the $t\bar{t}$ production cross-sections was
then obtained based on the observation of one event, under the
conservative assumption that this one event was due to
$t\bar{t}$ production and decay.
The mass region 28 $<$ M$_{top} < 72$ GeV/c$^{2}$
was excluded at the 95\% C.L.

A subsequent CDF search (F. Abe {\em et al.}, 1992a) 
in the dilepton channel, based on the same 
integrated luminosity of 4 pb$^{-1}$, also
included the $ee$ and $\mu\mu$ channels, resulting in an
increase in the top acceptance by a factor of two.
As discussed in Section~\ref{dilepton}, backgrounds in the
$ee$ and $\mu\mu$ channels are in general higher than in
the $e\mu$ channel.  $Z \rightarrow ee$ and $Z \rightarrow \mu\mu$
decays were eliminated by an invariant mass cut, leaving a large number
of events from off-shell Drell-Yan production of
$ee$ and $\mu\mu$ pairs.  In order to control this background,
additional requirements were imposed on the \MET~ and the
azimuthal opening angle between the two leptons.

%
%
%
%

No events consistent with $t\bar{t}$
were found in the $ee$ and $\mu\mu$ channels.
The resulting limit from the dilepton ($ee$ + $\mu\mu$ + $e\mu$)
channel was M$_{top} > 85$ GeV/c$^{2}$,
at the 95\% confidence level.  

\begin{figure}[htb]
\vskip 1cm
\epsfxsize=4.0in
\gepsfcentered[20 200 600 600]{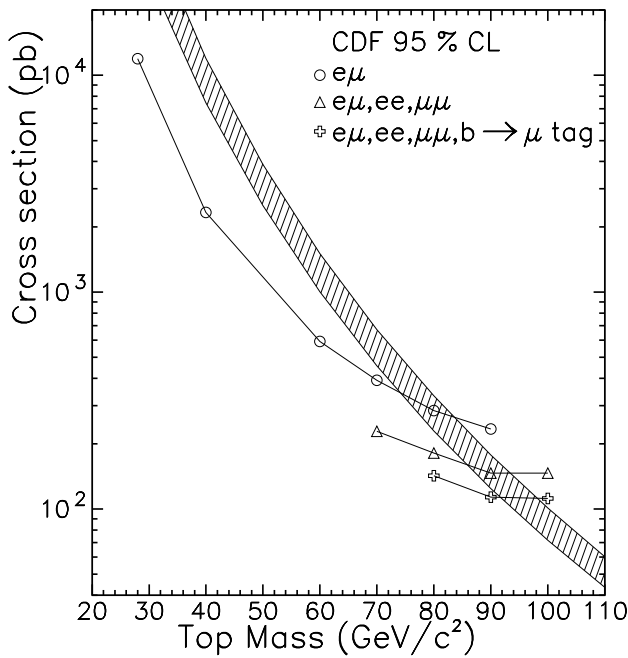}
\vskip 1cm
\caption{\protect \baselineskip 12pt
The 95\% C.L. upper limits on the $t\bar{t}$
cross-section from F. Abe {\em et al.}, 1992a.
Circles : $e\mu$ channel only. Triangles : dilepton
channel ($e\mu$, $ee$, and $\mu\mu$).
Crosses : dilepton channel $+$ lepton $+$ jets with
a $b \rightarrow \mu$ tag.  The band represents the 
NLO theoretical prediction for the $t\bar{t}$ cross-section
from Ellis, 1991.  The integrated luminosity was 4 pb$^{-1}$.
The cross-section
limits are a function of the top mass because the 
$t\bar{t}$ acceptance depends
on the top mass.}
\label{cdf91lim}
\end{figure}

The CDF experiment
also searched for top-quarks in the electron or muon $+ \geq 2$
jets channel, where the dominant $W +$ jets
background was reduced by attempting to tag $b$-quarks
through their semileptonic decay into muons, $b \rightarrow \mu$
or $b \rightarrow c \rightarrow \mu$ (F. Abe {\em et al.}, 1992a).
No events consistent with the top hypothesis were found.
Despite the high branching ratio for the lepton $+$ jets
mode, the acceptance in this search was approximately 
a factor of 3 smaller than that of the search in the
dilepton channel for M$_{top} \sim$ 90 GeV/c$^{2}$.
The low acceptance was the consequence of
the low ($\approx$ 4.5\%) muon tagging efficiency 
for lepton $+$ jets top events.  This low tagging
efficiency was due to (i) the semileptonic $b$-quark
branching ratio, (ii) the limited ($|\eta|$ $< 0.6$)
muon coverage of the 
CDF detector as configured for the 1988-89 run,
and (iii) the low 
transverse momentum of $b$-quarks in top decays for 
M$_{top} \sim 90$ GeV/c$^{2}$.
Combining results from this search and the dilepton
search, resulted in a 95\% C.L. lower limit on the top quark
of 91 GeV/c$^{2}$, (see Fig.~\ref{cdf91lim}).
This limit was extracted using the NLO calculation
of the $t\bar{t}$ cross-section (Ellis, 1991).
The 91 GeV/c$^{2}$ limit corresponds to the point where
the cross-section limit curve crosses the lower (i.e.
more pessimistic) bound of
the theoretical prediction.
Using a more up-to-date calculation of
$\sigma(t\bar{t})$
(Laenen, Smith, and van Neerven, 1994),
would result in a limit of 95 GeV/c$^{2}$.

\begin{table}
\begin{center}
\begin{tabular}{ccccc} \hline \hline
Experiment & Integ. luminosity & Mode & Mass limit (95\% C.L.)& Ref. \\ \hline
UA1 & 5.4 pb$^{-1}$ & $\mu$~$+$ jets,          & $>$ 52 GeV/c$^{2}$ &
Albajar {\em et al.}, 1988\\
UA1 & 5.4 pb$^{-1}$ & $\mu$~$+$ jets, $\mu\mu$ & $>$ 60 GeV/c$^{2}$ &  
Albajar {\em et al.}, 1988\\
UA2 & 7.5 pb$^{-1}$ & $e$~$+$ jets    & $>$ 69 GeV/c$^{2}$ & 
Akesson {\em et al.}, 1990\\
CDF & 4 pb$^{-1}$   & $e\mu$        & $>$ 72 GeV/c$^{2}$ & 
F. Abe {\em et al.}, 1990b\\
CDF & 4 pb$^{-1}$   & $e +$ jets    & $>$ 77 GeV/c$^{2}$ & 
F. Abe {\em et al.}, 1990a\\
CDF & 4 pb$^{-1}$   & dileptons ($ee, \mu\mu, e\mu$) & $>$ 85 GeV/c$^{2}$&
F. Abe {\em et al.}, 1992a \\
CDF & 4 pb$^{-1}$   & dileptons and l $+$ jets $+ b-$tag & $>$ 91 GeV/c$^{2}$
&F. Abe {\em et al.}, 1992a \\
\hline \hline
\end{tabular}
\end{center}
\caption{\protect \baselineskip 12pt
Summary of lower limits on the top quark mass from $p\bar{p}$ 
collisions, circa 1992. See text for details.}
\label{limtable}
\end{table}

The lower limits on the top mass from the UA1, UA2, and CDF experiments
are
summarized in Table~\ref{limtable}. 
With the top quark being so massive,
hopes to observe the top quark at the CERN S$p\bar{p}$S 
were abandoned, because of the small top production
cross-section at $\sqrt{s} = 630$,
(see Fig.~\ref{xsec2}).
In Section~\ref{discintro} we will
discuss the most recent, higher statistics, 
searches for the top quark
at the Tevatron Collider, which finally resulted in the
discovery of the top quark.

\subsection{Searches for non-Standard Model top quark decay modes}
\label{NSMdecay}
The lower limits on M$_{top}$ from $p\bar{p}$ collisions 
discussed in Section~\ref{early} are
not valid if the top quark does not decay as $t \rightarrow Wb$.
If M$_{top} +$ M$_{b} <$ M$_{W}$, the decay  
$W \rightarrow t\bar{b}$ is possible and contributes to
the width of the $W$ boson, $\Gamma(W)$. 

The width of the $W$ is difficult to measure
directly since the mass signature of the $W$ is a Jacobian and not 
a Breit-Wigner (see Fig.~\ref{hfwmass}),
and because the \met~ resolution is much larger than the natural width
of the W.
However,
constraints on the top quark            
mass independent of decay mode can be obtained from measurements
of the leptonic $W$ branching ratio
$Br(W \rightarrow l\nu) = \Gamma(W \rightarrow l\nu) / \Gamma(W)$,
based on theoretical expectations for $\Gamma(W \rightarrow l\nu)$.
At lowest order the branching ratio varies between 
$Br(W \rightarrow l\nu) = 1/9$ for 
M$_{W} <$ M$_{top} +$ M$_{b}$  and 
$Br(W \rightarrow l\nu) = 1/12$ for 
for very light
M$_{top}$.  Experimentally, a value for this branching ratio
can be extracted from measurements of the ratio 
                                                                      
\begin{eqnarray*}
R&=&\frac{\sigma(p\bar{p}\rightarrow W)~~ Br(W \rightarrow l\nu)}
{\sigma(p\bar{p}\rightarrow Z)~~ Br(Z \rightarrow l^{+}l^{-})}
\end{eqnarray*}
(Cabibbo, 1983; Halzen and Marsula, 1983; Hikasa, 1984;
Deshpande {\em et al.}, 1985; Martin, Roberts, and Stirling, 1987;
Berger {\em et al.}, 1989).  The value of the 
$Z$ leptonic branching ratio 
is taken from the very precise measurements
in $e^{+}e^{-}$ annhilations at LEP and SLC
(Montanet {\em et al.}, 1994). The values of 
$\sigma(p\bar{p}\rightarrow W)$ and
$\sigma(p\bar{p}\rightarrow Z)$ can be calculated in QCD; note that
theoretical uncertainties in these
calculations largely cancel in
the ratio, and the uncertainty in
$\sigma(p\bar{p}\rightarrow W)/\sigma(p\bar{p}\rightarrow Z)$ is
only of order 1\%.  (Martin, Roberts, and Stirling, 1989;
Hamberg, van Neerven and Matsuura, 1991; van Neerven and Zijlstra, 1992).

\begin{figure}[htb]
\vskip 1cm
\epsfxsize=4.0in
\gepsfcentered[20 200 600 600]{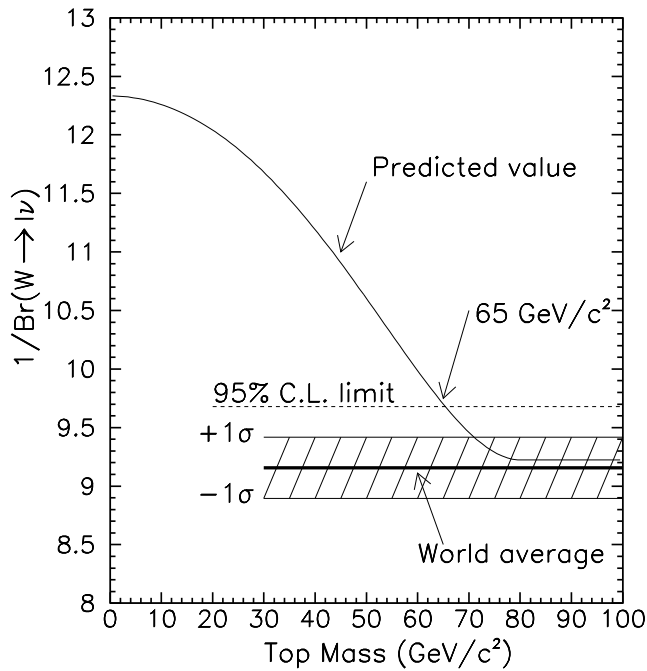}
\vskip 1cm
\caption{\protect \baselineskip 12pt
World average value for 
the inverse $W \rightarrow l\nu$ branching
ratio (Abachi {\em et al.}, 1995e), compared with Standard Model
expectations as a function of the top quark mass. Also shown is the
95\% lower limit on M$_{top}$,
independent of top quark decay mode, which can be extracted from 
the measurement.}
\label{gammaw}
\end{figure}

Early measurements of
this ratio from the $Sp\bar{p}S$ were found to be
more consistent with a small top quark mass (Albajar {\em et al.}, 1987;
Ansari {\em et al.}, 1987). 
These results were used by several authors  
to set upper limits on the top quark mass, e.g.
M$_{top} < 63$ GeV/c$^{2}$ at 90\% C.L.
(Martin, Roberts, and Stirling, 1987);
M$_{top} < 60$ GeV/c$^{2}$ at 95\% C.L.
(Halzen, Kim and Willenbrock, 1988);
M$_{top} < 70 \pm 5$ GeV/c$^{2}$ at 90\% C.L.
(Colas, Denegri and Stubenrauch, 1988).
However, subsequent higher statistics studies at both the 
$Sp\bar{p}S$  (Albajar {\em et al.}, 1991a; Alitti {\em et al.}, 1992b) 
and the Tevatron (F. Abe {\em et al.}, 1990c, 1992c, 1994d; 
Abachi {\em et al.}, 1995e), found results consistent 
with expectations for a high top quark mass.  A lower limit
M$_{top} > 65$ GeV/c$^{2}$ at 95\% C.L.
can be extracted from
the world average value of $Br(W \rightarrow l\nu$), (see Fig.~\ref{gammaw}).

Direct searches for top quarks decaying 
into final states different from $Wb$
have also been carried out at $p\bar{p}$ colliders.
The decay
$t \rightarrow H^{+}b$ is allowed
in several models in which the Higgs sector is
extended to include charged Higgs scalars ($H^{+}$).
This includes non-minimal 
Standard Models, 
such as supersymmetry (Gunion {\em et al.}, 1990).
In the simplest version of these models,
there are two Higgs doublets,
and the decay $t \rightarrow H^{+}b$ dominates for
M$_{top} <$ M$_{W} +$ M$_{b}$.  At higher top masses, both 
$t \rightarrow H^{+}b$ and $t \rightarrow Wb$ are allowed.
The branching
ratios for the two modes are functions of
M$_{top}$, M$_{H^{+}}$,
and a theoretically unconstrained                              
parameter tan$\beta$, which is the ratio of vacuum
expectation values for the two Higgs doublets.  The charged
Higgs scalar from top decay
would then decay into the heaviest lepton or quark
pair ($\tau\nu$ or $c\bar{s}$), with branching fractions which also
depend on the value of tan$\beta$ (Glashow and Jenkins, 1987;
Barger, Hewett, and Phillips, 1990; Dress and Roy, 1991).

\begin{figure}[htb]
\vskip 1cm
\epsfxsize=4.0in
\gepsfcentered[20 200 600 600]{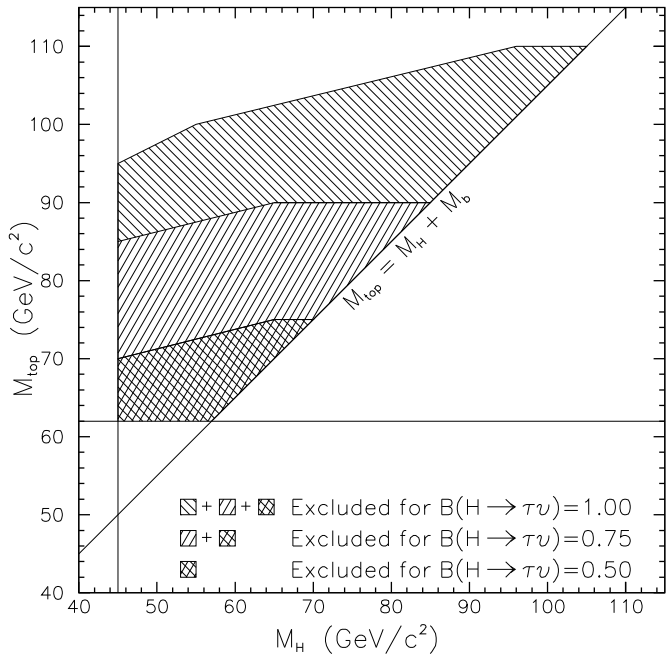}
\vskip 1cm
\caption{\protect \baselineskip 12pt
Regions of the (M$_{top}~-$M$_{H^{+}}$) plane excluded at 95\% C.L.
for different values of the branching ratio for 
H$^{+} \rightarrow \tau\nu$.  From the CDF collaboration,   
F. Abe {\em et al.}, 1994e.  The vertical line reflects the lower
limit M$_{H^{+}} >$ 45 GeV/c$^{2}$ from the LEP experiments,
(Buskulic {\em et al.}, 1992; Abreu {\em et al.}, 1990a;
Adriani {\em et al.}, 1992; Akrawi {\em et al.}, 1990b).
The horizontal line reflects the lower limit
M$_{top} >$ 62 GeV/c$^{2}$ 
from F. Abe {\em et al.}, 1994d, based on $Br(W \rightarrow l\nu$).
The integrated luminosity is 19 pb$^{-1}$.}
\label{higgs}
\end{figure}

Searches for $t \rightarrow H^{+}b$, $H^{+} \rightarrow \tau\nu$
based on the identification of hadronic tau decays
have been carried out by the
UA1 (Albajar {\em et al.}, 1991b), UA2 (Alitti {\em et al.}, 1992c),
and CDF (F. Abe {\em et al.}, 1994c) collaborations.  
No evidence for this process was found, and limits
were placed as a function of M$_{top}$, M$_{H^{+}}$, and the 
$H^{+} \rightarrow \tau\nu$ branching ratio.  The most stringent
limits are the result of a higher statistics analysis from the CDF
collaboration (F. Abe {\em et al.}, 1994e) based on the
$\tau \rightarrow e$ or $\mu$ signature, (see Fig.~\ref{higgs}).

\clearpage

\section{The Fermilab proton anti-proton collider}
\label{accel}

\begin{figure}[hbt]
\epsfxsize=4.0in
\vskip 2cm
\gepsfcentered[20 200 600 600]{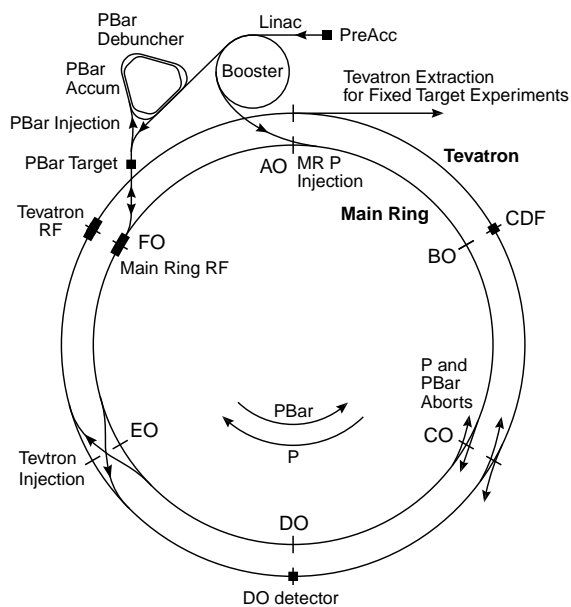}
\caption{\protect \baselineskip 12pt
The Fermilab accelerator complex. From Thompson, 1994.}
\label{tev}
\end{figure}

The Fermilab $p\bar{p}$ Collider in Batavia, Illinois
consists of seven accelerating structures, see
Fig.~\ref{tev}.  It is the highest energy particle accelerator in the
world, with a center-of-mass energy of 1800 GeV (Fermilab, 1984).

\subsection{Linear accelerators and synchrotrons}	
Negatively charged hydrogen ions
(H$^{-}$) are initially accelerated by a 
Cockroft-Walton pre-accelerator to 
750~keV. These ions are then accelerated in a 145 meter linear accelerator
to 400~MeV. They are stripped of their two electrons 
as they are injected into
the first proton synchrotron called the {\em Booster}. The Booster 
is a ring of
150 meter diameter, where 18 accelerating cavities are used to
accelerate the
protons to 8 GeV. The protons are then
extracted and injected into the {\em Main Ring},
which is the original 2 kilometer diameter 
Fermilab proton synchrotron comissioned in 1972.

The Main Ring was built with conventional electromagnets with a maximum
dipole bending field of .65 Tesla, 
with 18 accelerating cavities operating
at about 53 MHz, giving a boost of 2.5 MeV per turn. The Main Ring has 
accelerated protons up to 500 GeV.  It is now routinely used 
to accelerate protons to 120-150 GeV
which are then injected into the final synchrotron, the {\em Tevatron}, 
or into the anti-proton cooling ring.
The Tevatron built in 1983 uses superconducting magnets to bend and
focus the beams, and is situated a meter below the Main Ring.
The dipoles have  a maximum dipole bending field of 4.4 Tesla
and the focussing quadrupoles have a maximum field gradient of 67 Tesla/m.
With 8 RF accelerating cavities the Tevatron 
gives the protons a boost of 1 MeV per turn 
and can accelerate the protons up to 900 GeV with an energy resolution of 
$\pm$ 0.9 GeV. 
\subsection{Anti-proton source}
As was mentioned above, the Main Ring also 
acts as an injector to the $\bar{p}$ source (Church and Marriner, 1993). 
The protons are extracted 
from the Main Ring
at 120 Gev and delivered to a target where anti-protons are produced 
with peak momentum of 8 GeV. The  $\bar{p}$ 's are then focussed in a lithium
lens and
fed into the {\em Debuncher}, which is a ring with a circumference of 500 meters. 
The $\bar{p}$ bunches are debunched, by turning 
the narrow time spread of 
the bunch and 
large momentum spread into a large time spread and a narrow momentum spread.
After this bunch rotation, the $\bar{p}$'s 
are stochastically cooled (van der Meer, 1972), further 
reducing the phase space of the beam. The beam is then transferred to the
{\em Accumulator}, added to the anti-protons already stored there using a process
called stochastic stacking. When enough $\bar{p}$'s have been accumulated to 
make an intense beam, the beam is extracted and injected back into the 
Main Ring, where it is then accelerated and injected into the 
Tevatron. 

During the period when the top data was collected, the accelerator operated
with 6 bunches of protons and 6 bunches of anti-protons
circulating in the machine.
It takes approximately 2.5 hours of shining protons on the $\bar{p}$ source
target every 2.4 seconds in order to accumulate enough 
$\bar{p}$'s to ensure a reasonable luminosity for $p\bar{p}$ collision.
To optimize collider livetime, this process of creating and
stacking the anti-protons 
took place during collider operation.  This meant
that there was often beam in the Main Ring while the collider experiments
(CDF and D0) were taking data. 
The halo from this beam interacted with the walls of the accelerator and
sprayed particles into the two collider 
detectors, depositing
large amounts of energy in the calorimeters.
In the case of CDF, the Main Ring was diverted via a dogleg up above the 
detector and shielded by a steel structure approximately one meter thick. 
Nevertheless, occasionally extra
energy was deposited in the calorimeters.
These events were rejected offline.
The D0 experiment had no such dogleg or
beam diversion
and consequently the Main Ring ran straight through the D0 detector, 
approximately 2.5 meters
above the Tevatron beampipe.  The D0 collaboration 
was forced to turn off data acquisition
while the 
Main Ring beam was passing through their detector, resulting in a 
15\% livetime loss.

\subsection{Collider}

The $p$ and $\bar{p}$ 
bunches are approximately 50 cm in length due to the 
RF frequency of the accelerator. This bunch length determines the long luminous region at
the interaction points, which is roughly
Gaussian and has a sigma of approximately 30 cm.
The length of the luminous region has many limiting features in 
terms of triggering
and solid angle coverage by the detectors. For instance, as will be
discussed later in this review, the silicon vertex detector in the CDF
experiment
was 50 cm long and thus, due to the size of 
the luminous region, had a limited acceptance of roughly 60\% for bottom quarks 
from top quark decays.  Furthermore, the measurements of transverse energies
at the trigger level was smeared by the lack of knowledge of the
event-by-event interaction position.

The Tevatron Collider complex was first commissioned with a short run
in 1985.  The first high-luminosity run took place in 1988-89, at which time 
only the CDF detector was ready for data-taking.  The peak luminosity
for that run was approximately
2 x 10$^{30}$ cm$^{-2}$sec$^{-1}$, a factor of two
higher than the initial design.  The second set of high luminosity
collider runs, Run Ia and Run Ib, begun in 1992 with data taken by
both the D0 and CDF detectors.  During these runs,  
the peak luminosity 
was  2 x 10$^{31}$ cm$^{-2}$sec$^{-1}$,  
while the average luminosity was about 1.4 x 10$^{31}$ cm$^{-2}$sec$^{-1}$.
The lifetime of the beams in the Collider and the $\bar{p}$ stacking rate
was such that new protons and anti-protons were injected into the Collider
once a day. Table VII lists the relevant parameters of the
Collider.

\begin{table}
\begin{center}
\begin{tabular}{c|c}
\hline
Accelerator radius  &1000 m \\
\hline
Maximum beam energy  &900 GeV\\
\hline
Injection energy & 150 GeV\\
\hline
Peak luminosity &  2 x 10$^{31}$ cm$^{-2}$ s$^{-1}$\\
\hline
Number of bunches & $ 6 p,6 \bar{p}$ \\
\hline
Intensity per bunch  & $\approx 10^{11} p$, 5 x 10$^{10}\bar{p}$\\
\hline
Crossing angle  & 0$^{o}$\\
\hline
Bunch length (1 $\sigma$) &  50 cm\\
\hline
Transverse beam radius (1 $\sigma$) & $\approx$ 25 $\mu$m\\
\hline
Energy spread  & 0.15 x $10^{-3}$ GeV\\
\hline
RF frequency &  53 MHz\\
\hline
$\bar{p}$ stacking rate  & $\approx$3.5 x $10^{10}$/hour \\
\hline
Beam crossing frequency    & 290 kHz \\
\hline
Period between crossings   & 3.5 $\mu$s \\
\hline
\end{tabular}
\end{center}
\label{snyder}
\caption{Parameters of the Fermilab Tevatron Collider.}
\end{table}

\section{Discovery of the top quark}
\label{discintro}
In this Section we review the recent results that 
finally led to the establishment of the existence of the top quark.
The discovery of the top quark was made possible by the remarkable
success of the Tevatron Collider project, see Section~\ref{accel}.
These results come from the two collider experiments (CDF and D0)
at Fermilab's Tevatron.  They are based on data from the
1992-93 (Run Ia) and 1994-1995 (Run Ib) runs of the collider.
Analysis of data from the remainder of Run Ib was
in progress at the time that this review was being written.
Preliminary results from both CDF and D0
are fully consistent with those from the earlier data sets.

The data sets that were used to discover the top quark
were collected during Run Ia and the first half of Run Ib.
The total integrated luminosities were 67 pb$^{-1}$
for CDF, and between 44 and 56 pb$^{-1}$, depending on top
decay mode, for D0.  The difference in integrated
luminosities between the two experiments is due mostly to the fact that
the Main Ring accelerator at Fermilab
which operates asynchronously
from the Tevatron,
runs through the D0 calorimeter, see Section~\ref{accel}.  
Data taking in the
D0 experiment must be disabled whenever a Main Ring proton bunch crosses
the detector.

Given the $t\bar{t}$ production cross-section at the Tevatron
(see Section~\ref{prod}), the number of $t\bar{t}$ pairs produced 
in an exposure of 67 pb$^{-1}$ is expected to be
between approximately
6800 (for M$_{top}$ = 100 GeV/c$^{2}$) and 
150 (for M$_{top}$ = 200 GeV/c$^{2}$).
This is a small number of events, especially in light of the fact
that not all the the possible $t\bar{t}$ decay channels
can be exploited, see the discussion in Section~\ref{sig}.
Thus, the high luminosity delivered by the Tevatron was crucial in
enabling the experimenters to isolate the very rare top signal.

Both the CDF and D0 top searches assumed Standard Model decay 
($t \rightarrow Wb$) of the top quark, and were based on the 
$t\bar{t}$ dilepton
and lepton $+$ jets signatures discussed in Section~\ref{sig}.  
In all cases,
the number of observed events, after a selection procedure
designed to maximize the acceptance to top quarks, was compared
with the number of expected events from non-$t\bar{t}$ sources.
An excess of events over the background prediction then
constitutes evidence for the top quark.

At the beginning of Run Ia, the best lower limit on the mass of the top
quark was 91 GeV/c$^{2}$, from the 1989 CDF top search, see 
Section~\ref{pre1a}.  The initial top selection criteria
for both CDF and D0 were optimized
for detection of a low mass 
(M$_{top} \approx$ 120 GeV/c$^{2}$) top quark.
A lower limit of M$_{top} >$ 131 GeV/c$^{2}$ at the 95 \% C.L. 
was established by D0 from the Run Ia data with an integrated
luminosity of 13.5 pb$^{-1}$ (Abachi {\em et al.}, 1994).

Initial evidence for the top quark was reported by the CDF collaboration
(F. Abe {\em et al.}, 1994a) 
based on analysis of the Run Ia data set only, which had
an integrated luminosity of 19.3 pb$^{-1}$.
The excess
in the number of top candidate
events over the background prediction was 2.8 standard deviations.  Additional
kinematic features of the data, such as a reconstructed mass peak,
also supported
the $t\bar{t}$ hypothesis.  However, these results
were not deemed sufficient to unambiguously establish the existence 
of the top quark.  Following the publication of a lower limit
on M$_{top}$, the D0 requirements were reoptimized, i.e. tightened,
for higher top masses.  With the optimized requirements,
a statistically not very significant excess of events 
(1.9 standard deviations) was also found in the Run Ia D0 data
(Abachi {\em et al.}, 1995a and 1995d).   
With the addition
of the data from the first half of Run Ib, a statistically convincing
excess of events emerged from the analyses of the data sets from both
collaborations (F. Abe {\em et al.}, 1995a; Abachi {\em et al.}, 1995b).

The CDF and D0 searches for $t\bar{t}$ in the dilepton channel were
similar and are discussed in Section~\ref{dildisc}.  On the other hand,
the search strategies in the lepton $+$ jets channel were quite
different.  At the beginning of Run Ia, a silicon vertex detector
was installed at CDF (Amidei {\em et al.}, 1994).  Given the excellent
$b$-tagging capabilities of this device (see Section V.E.4), $b$-tagging was used
to separate the top signal in the lepton $+$ jets channel
from the $W +$ jets background.
Both vertex tagging and lepton tagging ($b \rightarrow e$
as well as $b \rightarrow \mu$) were used in CDF.   In contrast,
a large fraction of the top sensitivity in this channel
for the D0 experiment came from kinematic separation of
$t\bar{t}$ and $W +$ jets, although lepton tagging 
($b \rightarrow \mu$ only) was also employed.  
Analyses of the kinematic properties of lepton $+$ jets
events were also performed on the CDF data (F. Abe {\em et al.}, 
1994a, 1995b, and 1995c).
Excesses of top-like events were also seen in these CDF
studies, and were used to confirm the results of the
$b$-tagging observations.

In the remainder of this Section, 
we will describe in some detail the results from 
the CDF and D0 top
searches.  Dilepton searches will be reviewed in Section~\ref{dildisc}, 
followed by a summary of the searches in
the lepton $+$ jets $+$ $b$-tag channel in Section~\ref{ljtag}, 
and of kinematic separation of $t\bar{t}$ and $W +$ 
jets in Section~\ref{ljkin}.
The results from the two experiments will then be summarized
in Section VIII.D, and their
measurements of
the $p\bar{p} \rightarrow t\bar{t}$ cross-section will be presented 
in Section~\ref{xsecmeas}.

Before beginning the discussion of the CDF and D0 results, 
we wish to point out a difference in when the measured jet energies are corrected
back to the parent parton momenta in CDF and D0. 
In all D0 analyses,
correction to the measured jet energies are made prior to the selection of
the sample.
(The jet energy correction procedure
will be discussed extensively in Section~\ref{systematics}).
For most CDF analyses, the
measured jet energies are not 
corrected at the event sample selection stage, or in the
following analysis. Corrections are only applied when
the measurement of the quark or gluon energy is needed, 
for example in
the measurement of the top mass, see Section~\ref{mass}.
Unfortunately, 
this makes it
somewhat difficult to compare results from the two experiments.
For the jet energies relevant to the top search (15 GeV $\lesssim $ E$_{T}
\lesssim $ 150 GeV)
the multiplicative correction
factor for a CDF jet energy cluster is of order 1.5.

\subsection{Dilepton analysis results from CDF and D0}
\label{dildisc}

The searches for top quarks  performed in the dilepton channel by D0 and CDF
are similar. We have discussed the dilepton analysis approach at a general
level in Section~\ref{dilepton}.
What follows are the results of these
two analyses. The main difference in analysis strategy between the two 
experiments is due to the fact that CDF
has momentum determination from charged particle tracking in the central 
region whereas
D0 does not, and that D0 imposes more stringent requirements on the 
transverse energies of the jets.

\begin{table}[htb]
\begin{tabular}{c|c|c|c|c|c|c}
& \multicolumn{2}{c|}{Leptons} & \multicolumn{2}{c|}{Jets} & & \\
\cline{2-5}
Channel & E$_{T}$(e) &  P$_{T}$($\mu$) & $N_{\rm jet}$ & $E_T $
& \met & H$_{T}$  \\
\hline
\hline
$e\mu$ + jets & $\geq$ 15 GeV & $\geq$ 12 GeV/c & $\geq$ 2
& $\geq$ 15 GeV & $\geq$ 20 GeV & $\geq$ 120 GeV \\
$ee$ + jets   & $\geq$ 20 GeV & - & $\geq$ 2
& $\geq$ 15 GeV & $\geq$ 25 GeV & $\geq$ 120  GeV \\
$\mu\mu$ + jets & - & $\geq$ 15 GeV/c & $\geq$ 2
& $\geq$ 15 GeV & - & $\geq$ 100 GeV \\
\hline
\end{tabular}
\label{d0cuts}
\caption{A partial summary of selection criteria for $t\bar{t} \rightarrow$
dileptons in D0.  Jets must have $|\eta| < 2.5$ and their energies are
corrected.
The lepton pseudorapidity coverages are $|\eta| < 2.5$ for electrons,
$|\eta| < 1.7$ for muons in Run Ia, and $|\eta| < 1.0$ for muons in Run Ib.
The loss of muon coverage is due to ageing of the forward muon
chambers. The H$_{T}$ requirement is discussed in the text.}
\end{table}

\begin{table}[hbt]
\begin{tabular}{c|c|c|c|c|c|c}
& \multicolumn{2}{c|}{Leptons} & \multicolumn{2}{c|}{Jets} & & \\
\cline{2-5}
Channel & E$_{T}$(e) & P$_{T}$($\mu$) & N$_{\rm jet}$ & E$_{T} $
& \met & $\Delta\phi$(\met,lepton or jet)$> 20^{o}$  \\
\hline
\hline
$e\mu$ + jets & $\geq$ 20 GeV & $\geq$ 20 GeV/c & $\geq 2$ &
$\geq$ 10 GeV & $\geq$ 25 GeV & $\rightarrow$ \MET $\geq$ 50 \\
$ee$ + jets   & $\geq$ 20 GeV & - & $\geq 2$ &
$\geq$ 10 GeV & $\geq$ 25 GeV & $\rightarrow$ \MET $\geq$ 50  \\
$\mu\mu$ + jets &- & $\geq$ 20 GeV/c  & $\geq 2$ &  $\geq$ 10 GeV &
$\geq$ 25 GeV & $\rightarrow$ \MET $\geq$ 50 \\
\hline
\end{tabular}
\caption{A partial summary of selection criteria for $t\bar{t} \rightarrow$
dileptons in CDF.  Jets must have $|\eta| < 2.5$ and their energies are
not corrected.
The lepton pseudorapidity coverage is $|\eta| < 1.0$. The \MET~
requirement is tightened when the missing transverse energy vector is
collinear with either a lepton or a jet.  See the discussion in the text.}
\label{cdfcuts}
\end{table}

Both analyses require
two high P$_{T}$ leptons, at least two jets and in most cases, 
high \met. The exceptional case is the D0 search for decays in the $\mu\mu$
channel, where the \met~ requirement is removed.
The resolution of the D0 muon momentum 
measurement is limited to of order 20\% by multiple scattering in the 
toroids (see Table~\ref{lepres}),
and therefore in the $\mu\mu$ channel the measurement of the \met~ is 
poor and is not used in the D0 analysis.
The major backgrounds to the top quark signature in the dilepton channel 
are due to events in which jets fake leptons, $Z\rightarrow ee/\mu\mu/\tau\tau$, 
Drell-Yan lepton pair production, $WW$ and
$b\bar{b}$ production.  These backgrounds are reduced dramatically
by the requirements listed in 
Table VIII
and Table~\ref{cdfcuts}.
In particular, as discussed in Section~\ref{dilepton},
asking for the presence of two jets in the event is a very powerful
discriminator between signal and all sources of background. 

\begin{figure}
\vspace*{0.5in}
\mbox{\hspace*{1.0in}\epsfysize=3.6in\epsffile{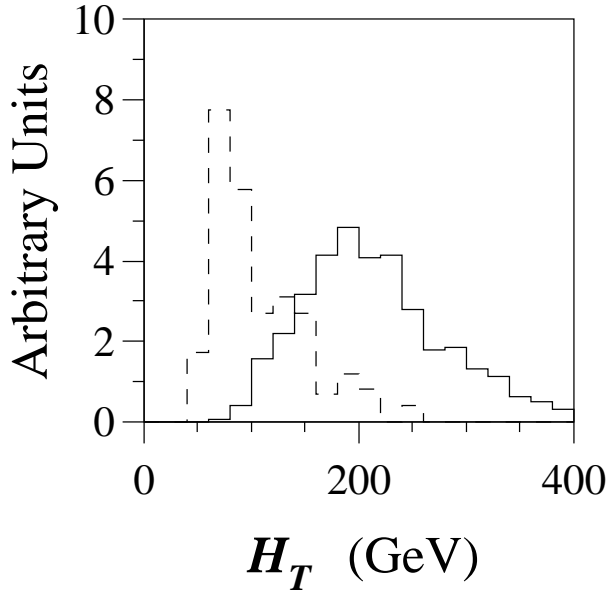}}
\vspace*{-0.6in}
\caption{\protect D0 Monte Carlo predictions of the variable H$_{T}$ for 
dilepton decays of a
200~GeV/c$^{2}$ top quark (solid) and principal dilepton backgrounds (dashed).
From Abachi {\em et al.}, 1995b.}
\label{ht10}
\end{figure}

There are two other differences in the kinematic requirements applied
by CDF and D0.  In the D0 analysis, there is a requirement
on H$_{T}$, defined as H$_{T} \equiv
\Sigma_{jets}(E_{T}) + E_{T}^{e}$ which is the scalar sum of 
all jet transverse
energies plus the highest transverse
energy electron in the event, if there is one.  
This is a kinematic variable 
that attempts to discriminate between the top signal and background, 
by exploiting the difference between the total transverse energy of the jets
in top and background events.
In Fig.~\ref{ht10} we show 
Monte Carlo predictions of H$_{T}$ for the principal backgrounds
and for $t\bar{t}$ events with  M$_{top}$ = 200~GeV/c$^{2}$. 

For the CDF analysis, in events where the 
\MET~ is nearly collinear with the P$_{T}$
of the leptons or jets, the \MET~ requirement is tightened.
As we will show
below, this procedure is somewhat
effective at suppressing a number
of background sources.  We will now discuss the dilepton CDF and D0 results,
starting with the
background sources, and the specific CDF and D0 selection requirements
designed to discriminate against them.

\subsubsection{$Z \rightarrow ee/\mu\mu$ background}
$Z$ decays constitute an important
background to the $t\bar{t}$ signal in the dilepton channel.  
Following the E$_{T}(e)$, P$_{T}(\mu)$ and N$_{jet}$ 
requirements listed in 
Table VIII and Table~\ref{cdfcuts} 
there is still a substantial 
background from $Z$ + $\geq$ 2 jets, in which the $Z$ 
decays to $e^{+}e^{-}$ or 
$\mu^{+}\mu^{-}$. Since $Z$ decays to charged leptons do not produce any
neutrinos, no \met~ is expected in these events. Therefore, requiring
large \met~ eliminates these backgrounds completely, except
for the effects of
mismeasurements of the missing transverse energy in the detector.

To eliminate the $Z \rightarrow ll$ background, $e^{+}e^{-}$ and
$\mu^{+}\mu^{-}$ events with dilepton invariant mass between
75 GeV/c$^{2}$ and 105 GeV/c$^{2}$ are rejected in the
CDF top search.  This requirement is expected to be
approximately 75\% efficient, roughly independent of top mass,
for dileptons ($e^{+}e^{-}$ and 
$\mu^{+}\mu^{-}$) from $t\bar{t}$ decays.
In the D0 analysis, the requirements used to discriminate against
the $Z$ background differ in the $e^{+}e^{-}$ and $\mu^{+}\mu^{-}$
channels because the resolution on the lepton energy/momentum
measurement is so much better for electrons than it is for muons.
In the $e^{+}e^{-}$ channel, if the invariant mass (M$_{ee}$) of the pair
is consistent with the $Z$ mass 
(79 GeV/c$^{2} \leq$ M$_{ee} \leq 103$ GeV/c$^{2}$), the
missing transverse energy requirement is tightened to 
\met~ $\geq$ 40 GeV.   Note that this is in contrast to the CDF
approach of rejecting any event that may be consistent with $Z$ decays,
regardless of the size of the missing transverse energy in the event.
As was discussed earlier, no \MET~ information is used 
by D0 in the $\mu^{+}\mu^{-}$ channel.  $Z$ decays into muons
are then rejected on the basis of an overall kinematic 
likelihood fit to the $Z \rightarrow \mu^{+}\mu^{-}$ hypothesis.

\begin{figure}[htb]
\epsfxsize= 4.in
\hspace*{2.0in}
\epsffile{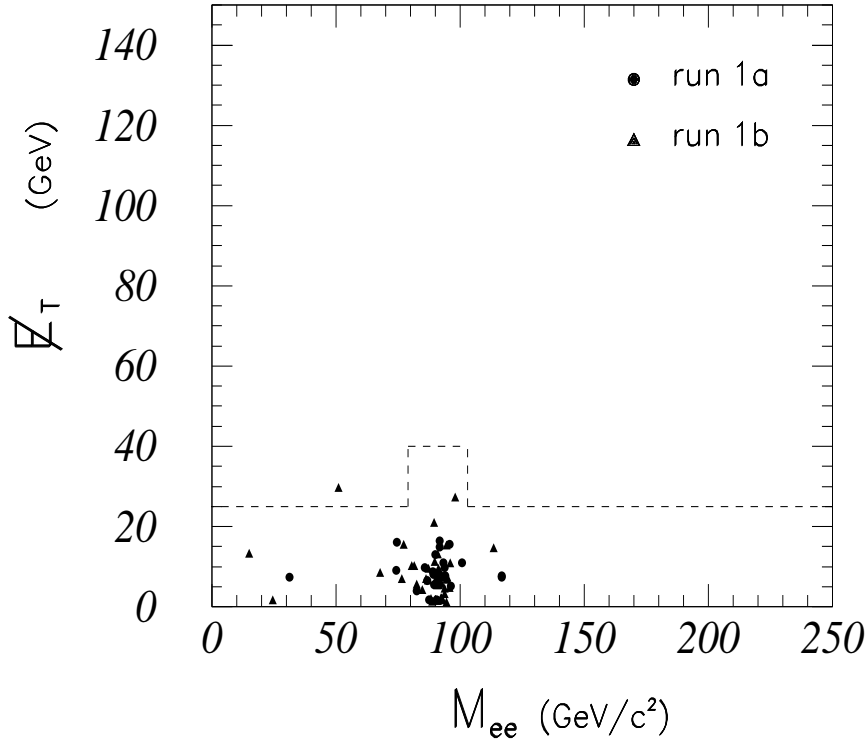}
\vskip -2cm
\caption{\protect\baselineskip 12pt Dielectron mass versus missing transverse 
energy in $ee + $~2 jet events from D0 (Grannis, 1995).  
The integrated luminosity is 55.7 pb$^{-1}$.
Also shown
is the D0 \MET~ requirement in the dielectron channel.}
\label{cuts_14}
\end{figure}

\begin{figure}[htb]
\epsfxsize= 4.in
\gepsfcentered[25 130 440 440]{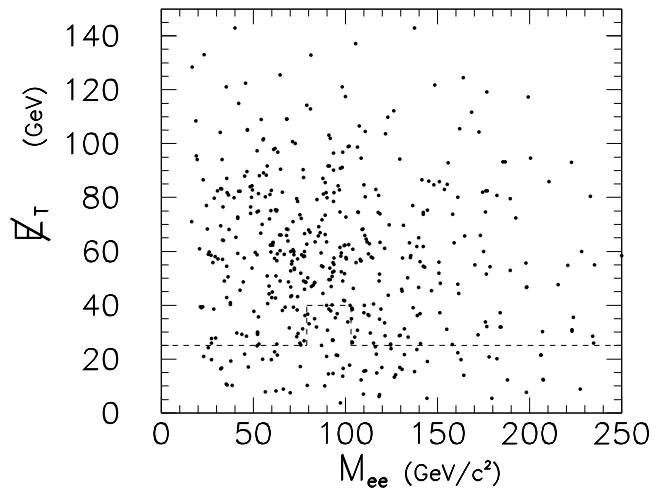}                                   
\caption{\protect\baselineskip 12pt Dielectron mass versus missing transverse 
energy in $t\bar{t} \rightarrow ee + $~2 jets events from the
D0 Monte Carlo (Grannis, 1995).  
Also shown
is the D0 \MET~ requirement in the dielectron channel.}
\label{eetop}
\end{figure}

Since the Monte Carlo simulation
is unlikely to correctly model the tails of the 
\met~ resolution, the
D0 collaboration relies on multijet data to estimate the remaining 
$Z \rightarrow ee +$ jets background after the \met~ requirement. 
As a result of \MET~ studies in the multijet sample, 
the fraction of $Z \rightarrow ee$
events with \MET $>$ 40 GeV is estimated to be less than 2 x 10$^{-4}$
(Abachi {\em et al.}, 1995d).         
The CDF requirement is more conservative
and loses acceptance in return for certainty that there will be
no $Z$ decays in the top 
candidate sample.
In Fig.~\ref{cuts_14}  we show the distribution
of the invariant mass of the electron-positron pair 
versus the \met~ for D0 events.  Expectations for $t\bar{t}$ are
displayed in Fig.~\ref{eetop}.

It is difficult to compare directly the \met~resolutions of CDF and D0. The
parametrizations of the \met~ resolutions in the two experiments
do not have the same functional form. The CDF resolution grows linearly
with $\sqrt{\Sigma E_{T}}$  
whereas the
D0 resolution appears to grow linearly with 
$\Sigma E_{T}$, see Table~\ref{metres}.
The tails of the distribution
in CDF are dominated by cracks in the calorimeter while the D0 detector 
design minimized cracks.  The radioactive noise from the uranium plates
in the D0 calorimeter
adds of order 25 GeV per event to the total energy making difficult a 
direct comparison
of the resolution at a particular $\Sigma E_{T}$. 

\subsubsection{$Z \rightarrow \tau\tau$ background}
\label{tautaubg}
\begin{figure}[htb]
\gepsfcentered[20 400 420 720]{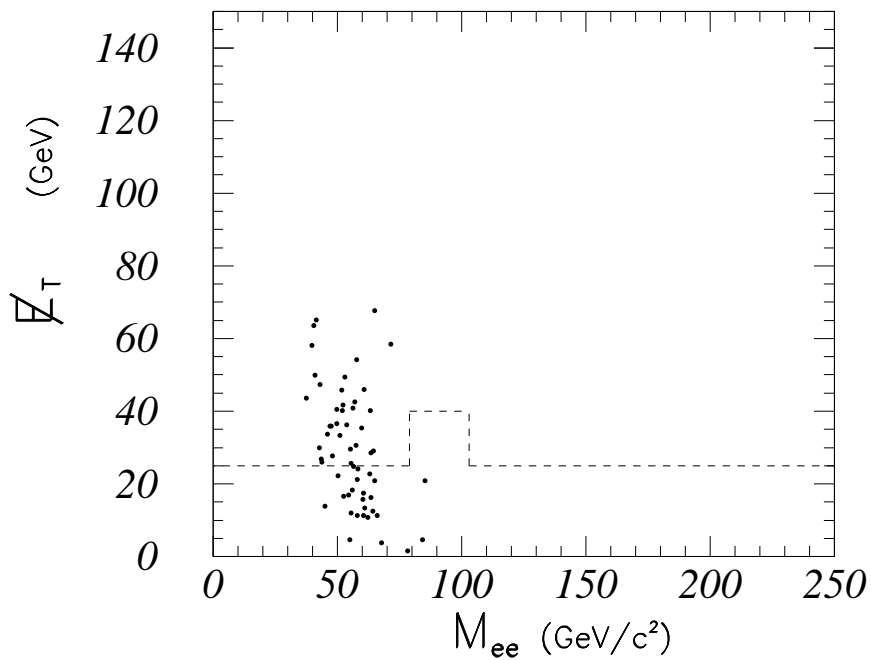}
\caption{\protect\baselineskip 12pt Missing transverse energy versus dilepton
mass for $Z\rightarrow \tau\tau + 2$ jets Monte Carlo events.  From 
the D0 collaboration, Grannis, 1995.
Also shown
is the D0 \MET~ requirement in the dielectron channel.}
\label{tau2}
\end{figure}

$Z \rightarrow \tau\tau$, followed by leptonic decays of both taus,
is a very important background source.  These events cannot be eliminated
by an invariant mass cut, due to the presence of unmeasured neutrinos
from $\tau \rightarrow l\nu\nu$.  Furthermore, these neutrinos can give
rise to \met~in the event.
The requirements on jet multiplicity,
high E$_{T}$ or P$_{T}$ leptons, and high \MET~are all effective at reducing
the background, see the discussion in Section~\ref{dilepton}, and
also Fig.~\ref{tau}.
In Fig.~\ref{tau2} we show the \met~ versus dilepton mass
for leptons from $Z\rightarrow \tau\tau$ decays from the ISAJET Monte Carlo. 
This clearly indicates that the background is not 
eradicated by the \met~ requirement alone.

\begin{figure}[htb]
\epsfxsize=6.0in
\gepsfcentered[10 140 300 420]{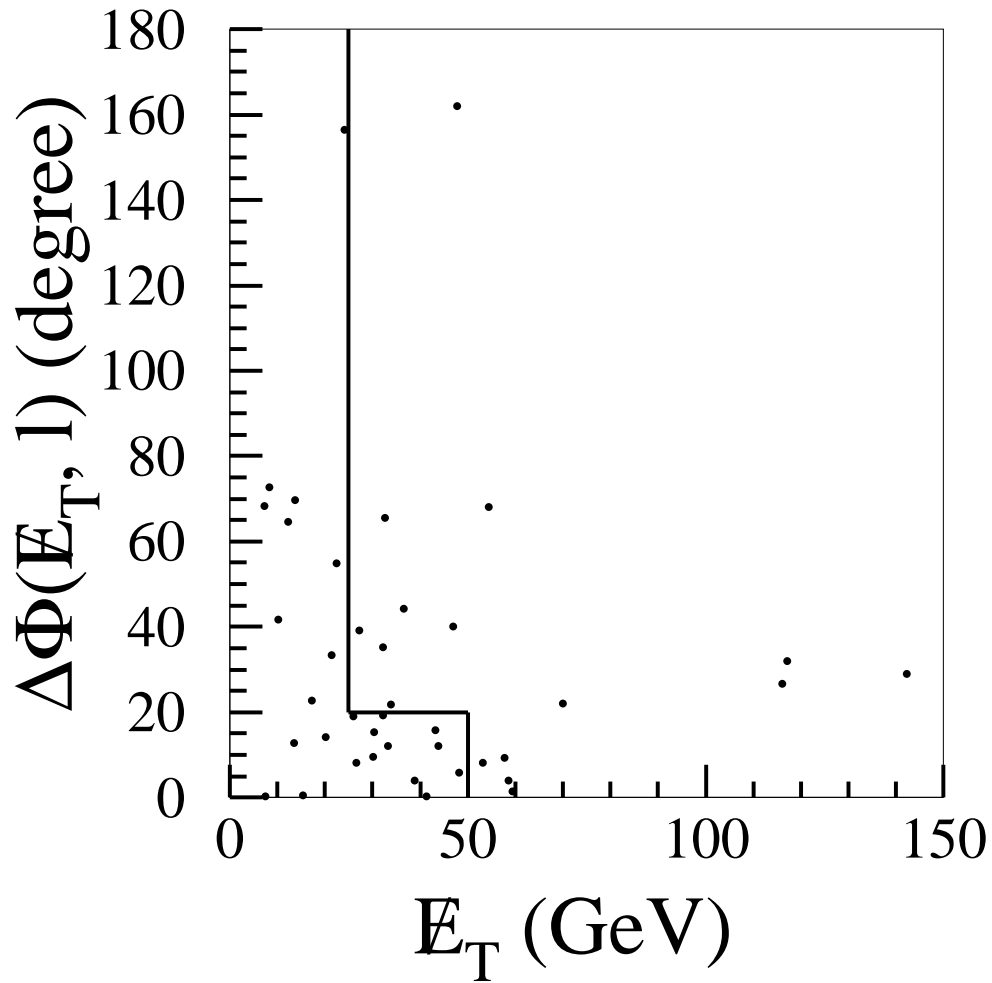}
\caption{\protect CDF Monte Carlo 
distribution of the azimuthal angle between the \MET~ and
the closest lepton as a function of \MET~~
in $Z \rightarrow \tau \tau$ events with two
jets.  Also shown is the boundary of the CDF \MET~ requirement.
From the CDF collaboration, F. Abe {\em et al.}, 1994a.}
\label{cdftaudphi}
\end{figure}

In the decay of a high momentum $\tau$, the decay products are
highly collimated,
due to the small mass of the $\tau$ lepton.
As a result, in these events the \MET~ will often point in the 
direction of one of the two leptons in the transverse plane.
Therefore, in the CDF analysis the \MET~ requirement is
tightened to \MET~ $\geq$ 50 GeV 
for events in which the azimuthal angle ($\phi_{\met}$) of the
\MET~ vector is within $20^{\circ}$ of the azimuthal angle ($\phi_{lepton}$)
of either of the two leptons, (see Fig.~\ref{cdftaudphi}).

There is no such 
$\Delta\phi$ requirement for D0, see 
Table VIII.
However the requirement made on H$_{T}$, see 
Table VIII,
has a rejection power of approximately 2.5 for 
$Z\rightarrow \tau\tau +$ two or more jets.

\clearpage

\subsubsection{Drell-Yan Background}
\begin{figure}[htb]
\epsfxsize=4.0in
\hskip -3cm
\hspace*{1.0in}                  
\epsffile{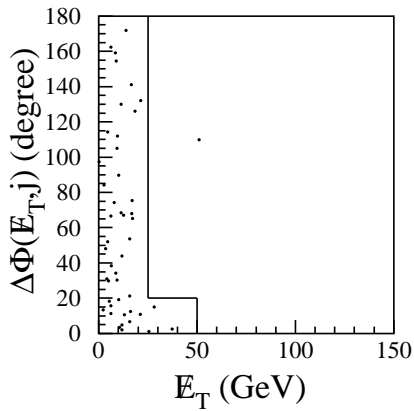}
\vskip -10cm                                                            
\vspace*{1.2in}                  
\caption{\protect Distribution of the 
azimuthal angle between the \MET~ and
the closest jet as a function \MET~~ for 
$Z \rightarrow ll$ events with two
jets.   The \MET~ characteristics of these events should closely resemble
those of Drell-Yan events.  This is CDF data with an integrated luminosity
of 19 pb$^{-1}$.
Also shown is the boundary of the CDF \MET~ requirement.
From the CDF collaboration, F. Abe {\em et al.}, 1994a.}
\label{CDFdphiZ}
\end{figure}

Continuum
Drell-Yan production of lepton pairs accompanied by two jets is a background 
not addressed by the
$Z$ mass requirement.   
In the CDF analysis, the
$\Delta\phi$ requirement between the \MET~ and the closest jet
(see Table~\ref{cdfcuts})
is designed to reduce this background. 
There is no intrinsic \met~ in these events, however jet energy can be
lost in cracks, and the measurement of the jet energy can fluctuate, faking
a \met~ signal.   In these cases, the \met~ tends to point towards
one of the two jets in the event.  In order to reduce this background,
the \met~ requirement in CDF is raised to 50 GeV when
the azimuthal angle between the 
\met~ direction and a jet is less than $20^{\circ}$, see 
Fig.~\ref{CDFdphiZ}.
 
D0's approach is different.  First of all, the H$_{T}$ requirement
is more effective than a minimum number of jets requirement because 
the jets in Drell-Yan $+$ jets events are in general softer
than the jets in $t\bar{t}$ events, for sufficiently high top mass.
Furthermore, to reduce the Drell-Yan background,
the \met~ requirement is raised to 25 GeV for
dielectron events.   Recall that in
the $\mu\mu$ channel, no cut is made on \met~ because the muon energy
measurement is poor. 

\subsubsection{$WW$ background}
The expected background from diboson production, $WW$,
$WZ$, and $ZZ$,
is small but not at all negligible.  
The $WW$ process
is the dominant diboson background to the $t\bar{t}$ dilepton signature.
The $WZ$ and $ZZ$ production cross-sections are expected to be lower 
and are further suppressed by the small
$Z \rightarrow ll$ branching ratio, see the discussion 
in Section~\ref{dilepton}.
Neither collaboration has published evidence for the process 
$p\overline{p}\rightarrow W^{+}W^{-}$, however the cross-section expected from
theory is slightly higher, of order 10 pb, than the equivalent 
$t\bar{t}$  cross-section, which is expected to be 
$\sigma(t\bar{t}) \approx$ 5 pb for 
M$_{top}$ = 175 GeV/c$^{2}$, (see Fig.~\ref{xsec}).  

\begin{figure}[htb]
\mbox{\hspace*{0.1in}\epsfysize=3.6in\epsfxsize=3.in\epsffile{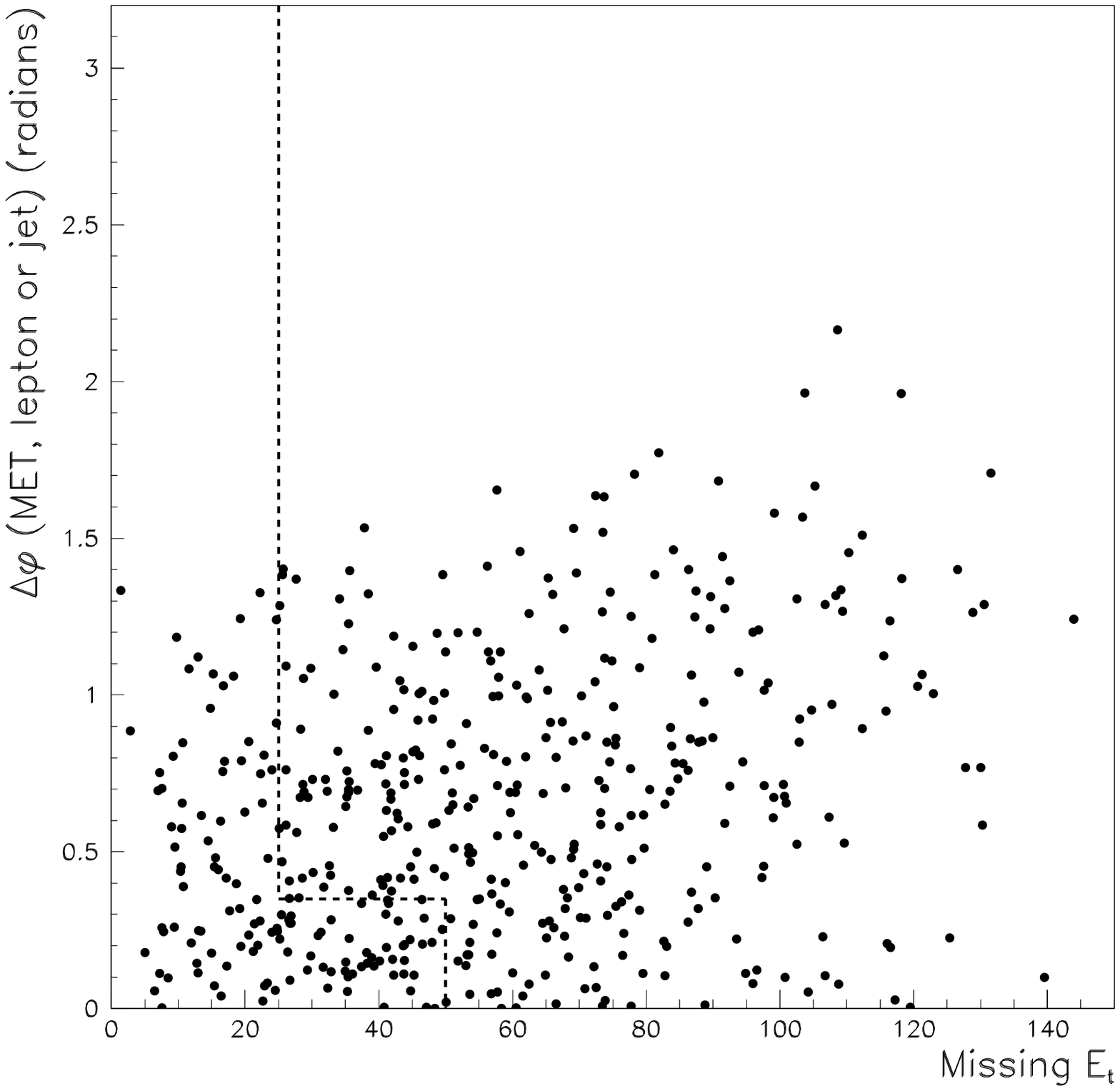}
\hspace*{0.3in}\epsfysize=3.6in\epsfxsize=3.in\epsffile{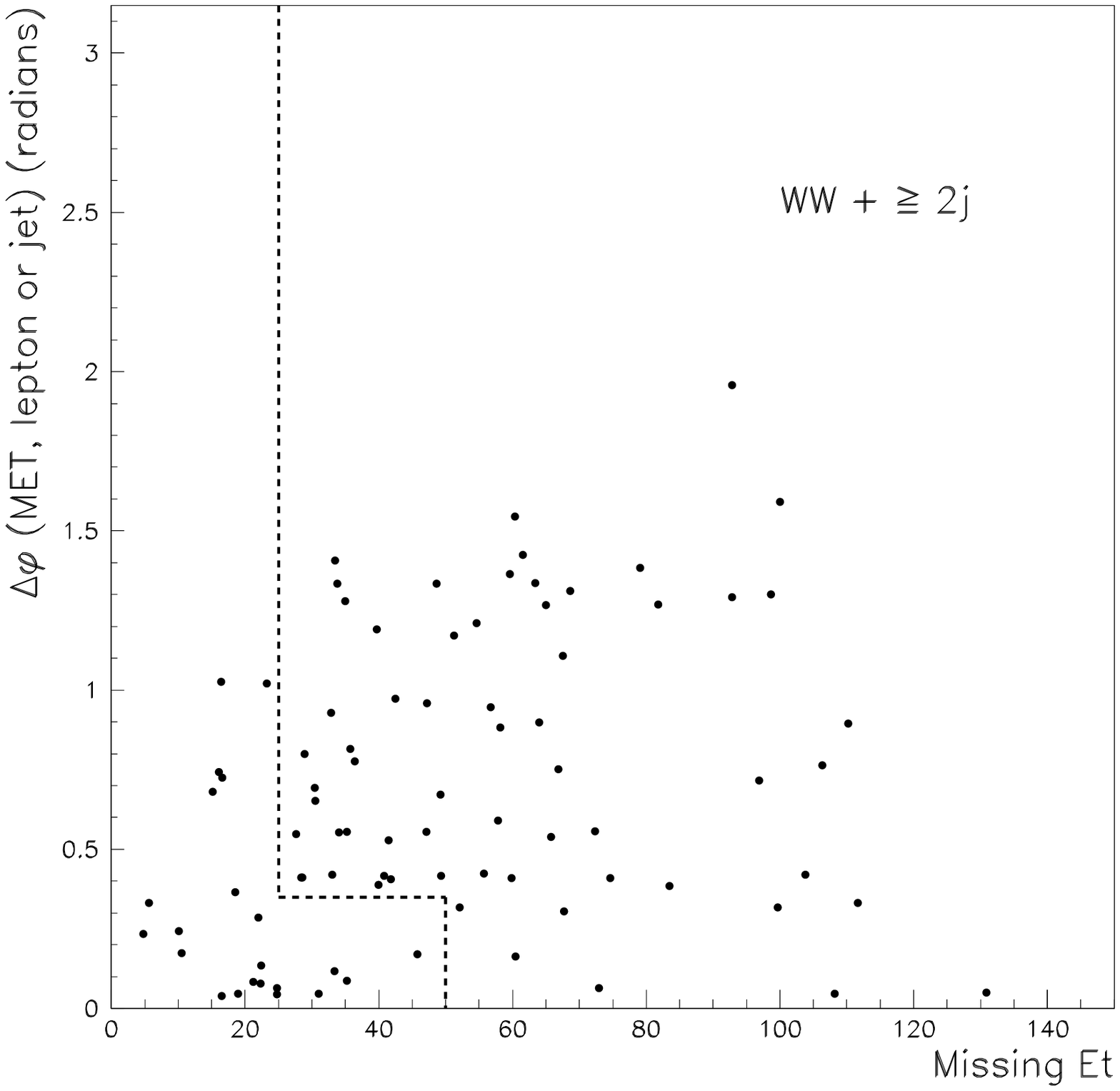}}
\caption{\protect\baselineskip 12pt  CDF Monte Carlo expectations
for $\Delta\phi$ versus \met~for 
(a) $t\bar{t}$ with M$_{top}$ = 180 Gev/c$^{2}$,	
and (b) WW + $\geq$ + 2 jet production.
The vertical axis is the smallest of the azimuthal angles between the
\MET~ vector and the leptons or jets. The symbol MET in the 
vertical-axis label
refers to the \MET.}
\label{cuts_dphi}
\end{figure}

$WW$ events include large transverse
momentum leptons and 
\met.~~In Fig.~\ref{cuts_dphi} we show the expected
distributions of \MET~ and $\Delta\phi$
for $t\bar{t}$ and $WW$ events.  The kinematic properties
of the two processes are very similar.
The most effective way to reduce these backgrounds is to require the
presence of high transverse momentum jets.  The H$_{T}$ requirement
imposed by D0 is more effective in discriminating against $WW$ 
than the simple minimum-number-of-jets requirement employed by CDF.
The remaining background is estimated entirely from Monte Carlo,
using the theoretical expectations for the $WW$ cross-section.

\subsubsection{$b\bar{b}$ and fake lepton backgrounds}
Doubly semileptonic decays in $b\bar{b}$ events and events with
fake leptons constitute another important background
to the dilepton signature.  These backgrounds are considerably reduced
by requiring two isolated high transverse 
momentum leptons, and high \MET.  The $b\bar{b}$ background is calculated
from Monte Carlo, normalized to the rate of lower momentum 
$b\bar{b} \rightarrow$~dileptons events.
The probability of finding two high momentum isolated fake leptons in
an event is exceedingly small.  The main fake dilepton background is due to 
$W +$ jets events, where one of the leptons is from the $W \rightarrow l\nu$
decay, and one of the jets is misidentified as a lepton.  The background
is then estimated by multiplying the number of observed $W +$ jets
events by the probability for a jet to fake the isolated electron or
muon signature as determined from samples of jet events.  This probability
is typically of order 10$^{-4}$.

\begin{figure}[h]
\epsffile{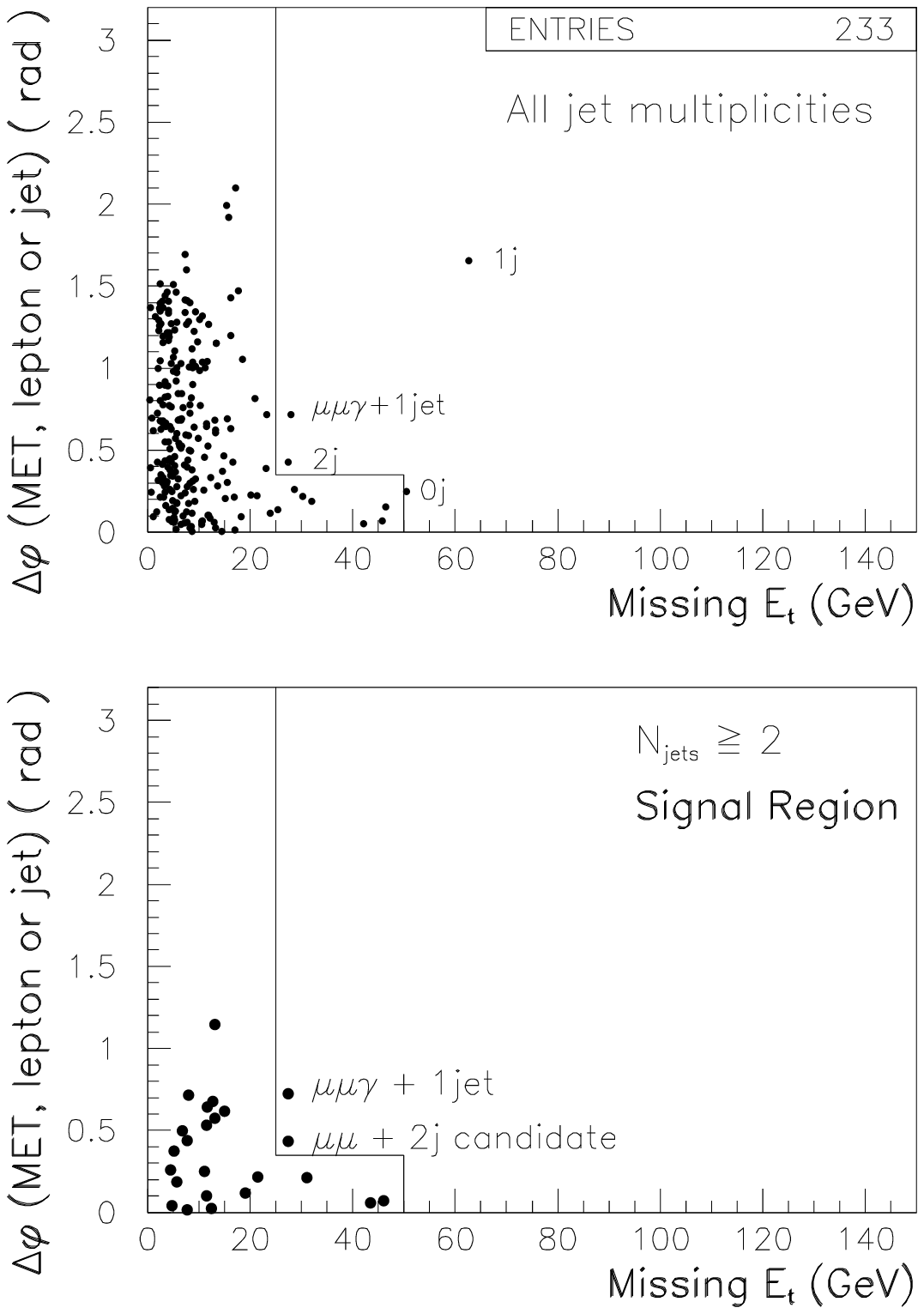}
\vspace{-5cm}
\caption{\protect\baselineskip 12pt  CDF dilepton results in the $\mu\mu$
channel. The integrated luminosity is 67 pb$^{-1}$. From Roser, 1995.
The vertical axis is the smallest of the azimuthal angles between the
\MET~ vector and the leptons or jets.  The symbol MET in the 
vertical-axis label
refers to the \MET.}
\label{mm}
\end{figure}

\begin{figure}[h]
\epsffile{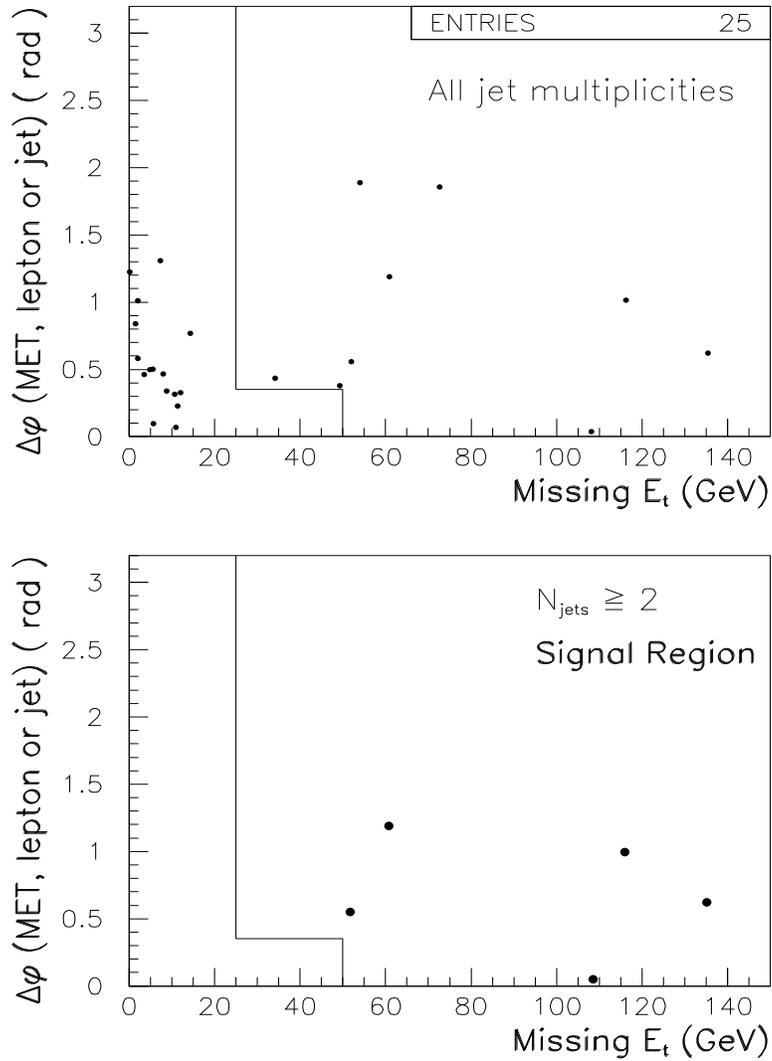}
\vspace{-5cm}
\caption{\protect\baselineskip 12pt  CDF dilepton results in the $e\mu$
channel.  The integrated luminosity is 67 pb$^{-1}$. From Roser, 1995.
The vertical axis is the smallest of the azimuthal angles between the
\MET~ vector and the leptons or jets.  The symbol MET in the 
vertical-axis label
refers to the \MET.}
\label{em}
\end{figure}

\clearpage

\subsubsection{Results}
The acceptance for the CDF analysis for M$_{top} = 180$ GeV 
is $\epsilon = (0.87 \pm 0.10)\%$, where this acceptance 
includes the branching ratio for the dilepton mode (4/81, see 
Table~\ref{tdec}).
The equivalent acceptance for D0 is $(0.55 \pm 0.04)\%$.
With these acceptances, and given the integrated luminosities and
the theoretical expectations for $\sigma(t\bar{t})$, both experiments
are sensitive to $t\bar{t}$ production for top masses up to
M$_{top} \approx$ 200 GeV/c$^{2}$.  The results of the CDF and
D0 $t\bar{t}$ searches in the dilepton channel are summarized in 
Table~\ref{dilsummary}.

\begin{table}
\begin{center}
\begin{tabular}{c|c|c}
\hline
\hline
   &  CDF  &  D0 \\
\hline
Background Source &  &  \\
$Z \rightarrow \tau\tau$       &  0.38 $\pm$ 0.07  & 0.24 $\pm$ 0.07 \\
$WW$                           &  0.21 $\pm$ 0.07  & 0.06 $\pm$ 0.03 \\
Fakes, $b\bar{b}$              &  0.26 $\pm$ 0.16  & 0.12 $\pm$ 0.04 \\
$Z \rightarrow ee$ or $\mu\mu$ &        -        & 0.24 $\pm$ 0.03 \\
Drell Yan                      & 0.44 $\pm$ 0.28   &    -          \\
\hline
Total background               & 1.3 $\pm$ 0.3     & 0.65 $\pm$ 0.15 \\
\hline
\hline
$t\bar{t}$ expectation, M$_{top}$ = 180 GeV/c$^{2}$  & 2.4    & 1.2           \\
\hline 
Data           & 5 $e\mu$   1$ \mu\mu$ & 2 $e\mu$   1 $\mu\mu$ \\
\hline
\hline
\end{tabular}
\end{center}
\caption{Background sources, $t\bar{t}$ expectations, and event yields in 
the CDF (Roser, 1995)
and D0 (Narain, 1995) dilepton analyses.  One additional CDF candidate event,
consistent with the $Z \rightarrow \mu\mu\gamma$ hypothesis, 
is not included in this table, see the 
discussion in the text.  The integrated luminosity for the CDF data is 
67 pb$^{-1}$.  For the D0 data, the integrated luminosity is 55.7 pb$^{-1}$
for $ee$, 44.2 pb$^{-1}$ for $\mu\mu$, and 47.9 pb$^{-1}$ for $e\mu$.
The $t\bar{t}$ expectations are obtained using the calculation of
$\sigma(t\bar{t})$ from Laenen, Smith, and van Neerven, 1994.}
\label{dilsummary}
\end{table}

CDF finds 7 dilepton candidates in 67 pb$^{-1}$ with $1.3\pm0.3$ expected
background events. 
Of these 7 candidate events, five are
$e\mu$+ jets events and two are $\mu\mu$+ jets events. 
One of the two $\mu\mu$ events looks very much like a radiative Z decay,
$Z \rightarrow  \mu\mu\gamma$, since
the invariant mass of the
$\mu\mu\gamma$ is 86 $\pm$ 7 GeV/c$^{2}$. 
Although the background from radiative $Z$ decays was
estimated to be less than .04 events, CDF conservatively a-posteriori removes
this event from the top candidate sample and is left with 6 events (5
$e\mu$ and one $\mu\mu$).
Acceptances for top
are such that
60\% of $t\bar{t}$ are expected in $e\mu$ and 40\% in $ee/\mu\mu$.
Figs.~\ref{mm} and ~\ref{em}
show the CDF data in the
$\Delta\phi-\met$ plane.  As can be seen from these figures, the 
$e\mu$ channel is much cleaner due to the absence of 
the Drell-Yan contribution.

D0 finds 3 candidates in approximately 
50 pb$^{-1}$ with $.65\pm0.15$ expected background
events. 
Of these three events, two are $e\mu$+ jets and one is $\mu\mu$+jets.
This is consistent with the ratio of expected number of events, 
$e\mu : ee : \mu\mu = .34:.25:.11$ for a 200 GeV top quark.


\begin{figure}[htb]
\hspace*{0.7in}
\epsffile{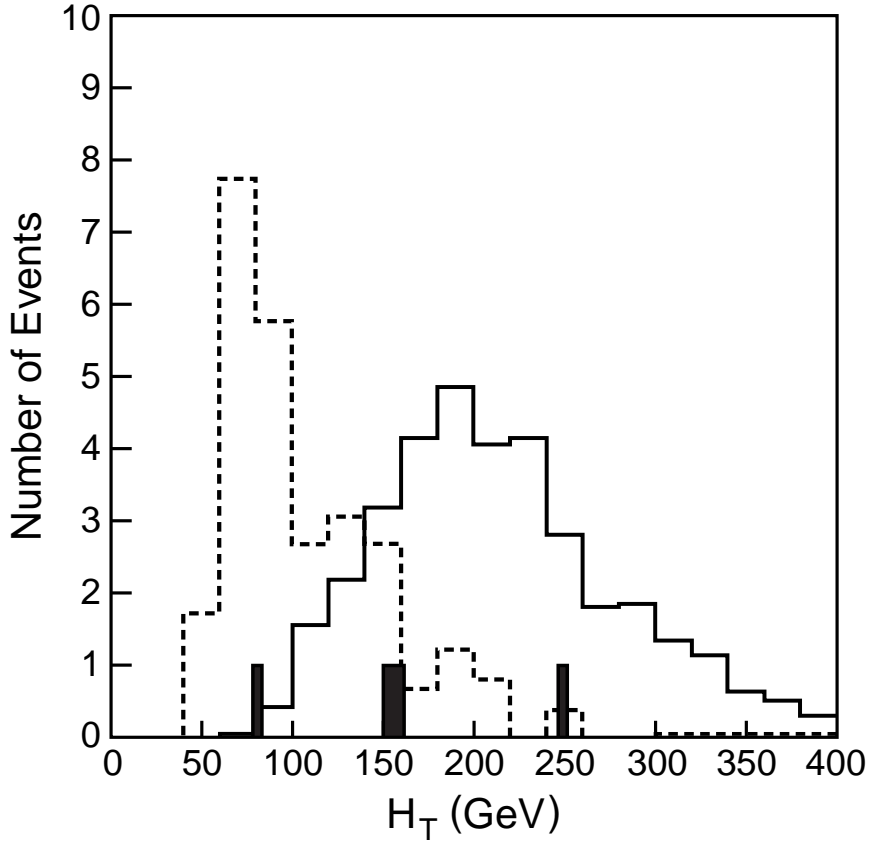}
\caption{\protect\baselineskip 12pt Distributions of 
H$_{T}$ for dilepton data for D0 (Grannis, 1995) 
The integrated luminosity is $\approx$ 50 pb$^{-1}$. The solid line is the
expected H$_{T}$ distribution for top events in the dilepton channel.
The dashed line is the expected backgrounds and the solid histogram is the data.
See text for details.}  
\label{cdfht}
\end{figure}

Thus, both experiments see an excess of dilepton events over the 
total background prediction.  
The size of this excess is loosely
consistent with expectations for $t\bar{t}$.  It is interesting
to notice that there are some significant differences in the
CDF and D0 background sources, see Table~\ref{dilsummary}.
The highest background source in the CDF analysis is due to
Drell-Yan events.  This background is negligible in D0, probably due to 
the D0 H$_{T}$ requirements, the stricter \MET~ requirement 
in the $ee$ channel, and
the hermeticity of the D0 detector.  Also, $Z \rightarrow ee$ and
$\mu\mu$ are totally eliminated by the CDF invariant mass requirement, while
they make up a substantial fraction of the total background in D0.

As partial evidence that
the excess of events is not due to a background fluctuation or to 
underestimation of the backgrounds, we present in Fig.~\ref{cdfht} the
H$_{T}$ distribution of dilepton events from D0.  
As was discussed earlier, in
D0 H$_{T}$ is defined as the scalar sum of the transverse energy
of all of the jets in the event $+$ the E$_{T}$ of the highest
E$_{T}$ electron (for the $ee$ and $e\mu$ channels).  
As we
argued throughout this Section, and as shown in Fig.~\ref{ht10},
H$_{T}$ is a powerful discriminant between signal and background.  The 
D0 candidate events are distributed in H$_{T}$ in a manner
more consistent with $t\bar{t}$ than background.   

In addition, in the
six CDF dilepton candidate events, there are five jets (in three events)
that are tagged as $b$-jets by the algorithms which will be described in
Section~\ref{ljtag}.  This is to be compared with the expectation
of approximately 0.5 $b$-tagged jets if all six events were background, and
approximately 3.6 $b$-tagged jets if all six events were $t\bar{t}$.
This suggests that the excess of dilepton $+$ jets events over the
background prediction is correlated with the presence of $b$-quarks
in the event, as expected if the excess were due to $t\bar{t}$.

\clearpage

\subsection{Lepton $+$ jets $+$ $b$-tag}
\label{ljtag}

As was discussed in Section~\ref{tagging}, one of the most effective
ways of isolating a top signal in the lepton $+$ jets mode
($t\bar{t} \rightarrow q\bar{q}~ l\nu~ b \bar{b}$) is to tag
the $b$-quarks in $t\bar{t}$ events.  This can be achieved by
searching for leptons from the semileptonic decay of $b$-quarks
(lepton tagging), or by exploiting the long lifetime of $b$-hadrons
(vertex tagging).  Lepton tagging has been used by both CDF and
D0 to extract a top signal; vertex tagging, which requires momentum
analysis, as well as precise vertex detection capabilities, has
been performed by CDF only.  In Section~\ref{sel} we will review
the CDF and D0 selections of lepton $+$ jets data, before the
$b$-tag requirement.  The results of the $b$-tag analyses on these
data are then discussed in Sections~\ref{slt} and~\ref{svx}
for lepton and vertex tagging respectively.  
A brief summary of the tagging searches, as well as
results of 
cross-checks performed
on samples of $Z +$ jets, are presented in Section~\ref{zcheck}.

\subsubsection{Selection of lepton $+$ jets data before $b$-tagging}
\label{sel}

The CDF and D0 selections of the isolated
e or $\mu$ $+$ jets data for the
$b$-tagging analyses are summarized in Table~\ref{ljsel}.
The event selection employed by the two collaborations are similar.
In principle, one would expect to detect four jets in a 
$t\bar{t} \rightarrow$ lepton $+$ jets event;  however, as
discussed in Section~\ref{quarkdetect}, the detected number of jets
can be smaller (see for example Fig.~\ref{njets_mc}).
This is because jets from top decay sometime
have E$_{T}$ below threshold, or two or more of these jets can be
close enough to each other that they are reconstructed as a single jet.
Therefore, in order to maintain high efficiency, the minimum number
of jets required by both CDF and D0 is three rather than four.
The one significant difference in the selection requirements
between the two experiments is that D0 imposes a further requirement
on the minimum scalar sum of the transverse energy of all jets in
the event (H$_{T}$, see Table~\ref{ljsel}).  This requirement
is expected to improve the rejection against the dominant W $+$ jets
background, especially for high top quark masses
(see Fig.~\ref{htpart}).   No such requirement is imposed on the
CDF data in order to maintain high $t\bar{t}$ detection
efficiency for M$_{top}$ as
low as 100 GeV/c$^{2}$, and because of the superior background
rejection capabilities of the CDF vertex tagging analysis.

\begin{table}
\begin{center}
\begin{tabular}{ccc} \hline \hline
   &  CDF  & D0 \\ \hline
Lepton E$_{T}$ or P$_{T}$ &  $\geq$ 20 GeV  & 
$\geq$ 20 Gev (e); $\geq$ 15 GeV ($\mu$) \\
Lepton rapidity coverage  & $|\eta| \leq 1^{ (a)}$ & $|\eta| \leq 2$ (e);
$|\eta| \leq 1.7$ ($\mu$)$^{ (b)}$ \\ 
\MET & $\geq$ 20 GeV & $\geq$ 20 GeV \\
Number of jets & $\geq$ 3 & $\geq$ 3  \\
Jet E$_{T}$ & $\geq$ 15 GeV (uncorrected) & $\geq$ 20 GeV (corrected) \\
Jet rapidity coverage & $|\eta| \leq 2$ & $|\eta| \leq 2$ \\
Jet cone clustering radius & 0.4 & 0.5 \\
H$_{T}$ & - & $\geq$ 140 GeV \\
Integrated luminosity & 67 pb$^{-1}$ & 48 pb$^{-1}$ (e); 44 pb$^{-1}$ ($\mu$) \\
No. of events in data sample & 203 & 66 \\
No. of expected $t\bar{t}$ events, M$_{top}$ = 140 GeV/c$^{2}$ &
$\approx$ 95 & $\approx$ 22 \\ 
No. of expected $t\bar{t}$ events, M$_{top}$ = 200 GeV/c$^{2}$ &
$\approx$ 13 & $\approx$ 6 \\ 
\end{tabular}
\end{center}
\caption{\protect \baselineskip 12pt
CDF and D0 lepton $+$ jets requirements for the $b$-tag analysis.
The variable H$_{T}$ in D0 is defined as the scalar sum of the E$_{T}$ of
all the jets.  The expected number of top events are derived
from the theoretical estimate of the $t\bar{t}$ cross-section
(Laenen, Smith, and van Neerven, 1994) and the calculated
acceptances. Notes : $^{(a)}$ the CDF muon chambers only cover
$\approx$ 2/3 of the solid angle for $0.6 < |\eta| < 1$; 
$^{(b)}$ Muons in D0 are restricted to $|\eta| < 1$ for the last
$\approx$ 70\% of data, due to ageing of the forward muon chambers.}
\label{ljsel}
\end{table}

The selections summarized in Table~\ref{ljsel} yield data samples 
consisting mostly of $W +$ jets events, as well as a contamination
from QCD events, and hopefully also a $t\bar{t}$ component.
The QCD contamination is due to events with fake leptons,
as well as semileptonic decays of $b$-quarks in 
$p\bar{p} \rightarrow b\bar{b}$ events.   In almost all of the
channels, these events are estimated to contribute approximately 10\%
to the event sample; the exception is the D0 $\mu +$ jets
channel, where this background is a factor of 2-2.5
larger, due to the poor resolution of the D0 muon momentum measurement.
These data samples also include small contributions from 
$Z$ and diboson events.  

The numbers of expected $t\bar{t}$ events in
the data samples (see Table~\ref{ljsel})
depend on M$_{top}$, since both the 
$t\bar{t}$ production cross-section and acceptance depend on M$_{top}$.
Given the theoretical expectations for the $t\bar{t}$ production
cross-section, the expected signal-to-background in these data samples
before applying $b$-tagging ({\em pre-tag} samples)
varies between approximately 1/16 and 1/3, depending on top mass, and
selection details.

The dominant physics
background in the lepton $+$ jet $+$ $b$-tag channel is due to the
associated production of a $W$ boson and a pair of heavy quarks
($WQ\bar{Q}$, $Q = b$ or $c$, see Section~\ref{tagging},
Fig.~\ref{gspfig}).  As discussed in Section~\ref{tagging}, the
fraction of $W + \geq 3$ jets events containing a $Q\bar{Q}$ pair
is expected to be of order 3\% for $Wb\bar{b}$ and 
5\% for $Wc\bar{c}$.  Therefore, given the 
signal-to-background levels in the lepton $+ \geq$ 3 jets
samples before demanding a $b$-tag, 
a $t\bar{t}$ signal is expected to stand out 
after application of a $b$-tag requirement, provided that
instrumental backgrounds can be kept under control.

\begin{figure}[htb]
\epsfxsize=4.0in
\epsfysize=2.0in
\vskip 0.5cm
\gepsfcentered[20 200 600 600]{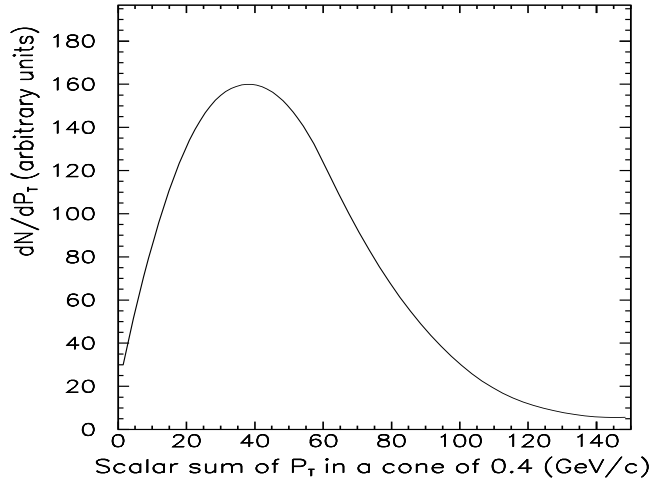}
\vskip 0.5cm
\caption{\protect \baselineskip 12pt
The expected scalar sum of the transverse momenta of all particles
in the $t\bar{t}$ final state within a cone of radius 
0.4 of the lepton from $b \rightarrow c l \nu$ or
$b \rightarrow c \rightarrow s l \nu$.  From the ISAJET Monte Carlo,
for M$_{top}$ = 175 GeV/c$^{2}$.  The momenta of the lepton
and the neutrino are not included in the sum.}
\label{slt_iso}
\end{figure}

\subsubsection{Lepton tagging in CDF and D0}
\label{slt}

Leptons from $b$-quarks in $t\bar{t}$ arise from direct 
($b \rightarrow c l \nu$) or cascade 
($b \rightarrow c \rightarrow s l \nu$) decays.
Given the $b$- and $c$-quark semileptonic branching ratios,
and including the contribution from semileptonic decays
of $c$-quarks from $W \rightarrow c\bar{s}$, there is
on average approximately one lepton (e or $\mu$) from $b$
or $c$ decay in each 
$t\bar{t} \rightarrow $ lepton $+$ jets event in addition
to the lepton from $W$ decay.
These additional leptons tend to have low transverse momentum, see
Fig.~\ref{sltpt}, and to be non-isolated because of the
presence of nearby hadrons from the $b$-quark fragmentation and
the $b$-hadron decay, (see Fig.~\ref{slt_iso}).
Detection of these leptons is therefore
more difficult than detection of the isolated high P$_{T}$
leptons from $W$ decays.

There are two main differences in the lepton-tagging algorithms developed by
the two collaborations.  The first difference is that the D0 
analysis only tags muons, whereas CDF tags both muons and electrons.
This is because detection of non-isolated electrons in D0 is not
as effective as in CDF,
mostly due to the absence of a magnet for momentum
measurement.  
In CDF electrons from $b$-decays
can be identified with sufficient
background rejection.  However, the efficiency for detecting these
electrons is lower than the efficiency for detecting muons, and
only of order 30\% of the lepton tagging efficiency in CDF 
is due to electron tags.
The second difference between the two experiments resides in
the minimum P$_{T}$ requirement for these tagging leptons.
In D0, this requirement is set at 4 GeV/c, which corresponds to
the minimum P$_{T}$ for a muon to penetrate the 
D0 muon detectors; in CDF this minimum P$_{T}$ requirement is
set at 2 GeV/c.  The lower P$_{T}$ requirement improves the
efficiency for detecting muons from cascade decays, especially
for low top quark masses, (see Fig.~\ref{sltpt}), however
it also results in higher background levels.

Additional requirements are imposed in the D0 analysis to
reject events with tagging muons collinear or back-to-back
with the direction of the \MET~ vector.  These requirements are
designed to reject QCD events.
The rapidity coverages for tagging leptons in the two experiments
are the same as those for the high P$_{T}$ leptons from $W$ decays,
see Table~\ref{ljsel}.  For M$_{top} > 140$ GeV/c$^{2}$,
the $t\bar{t}$ lepton-tagging efficiency is 20\% for both experiments.
The greater muon coverage of the D0 detector makes up for
the higher P$_{T}$ threshold and the absence of electron-tags.

\begin{figure}[htb]
\epsfysize=4.0in
\gepsfcentered[27 250 300 550]{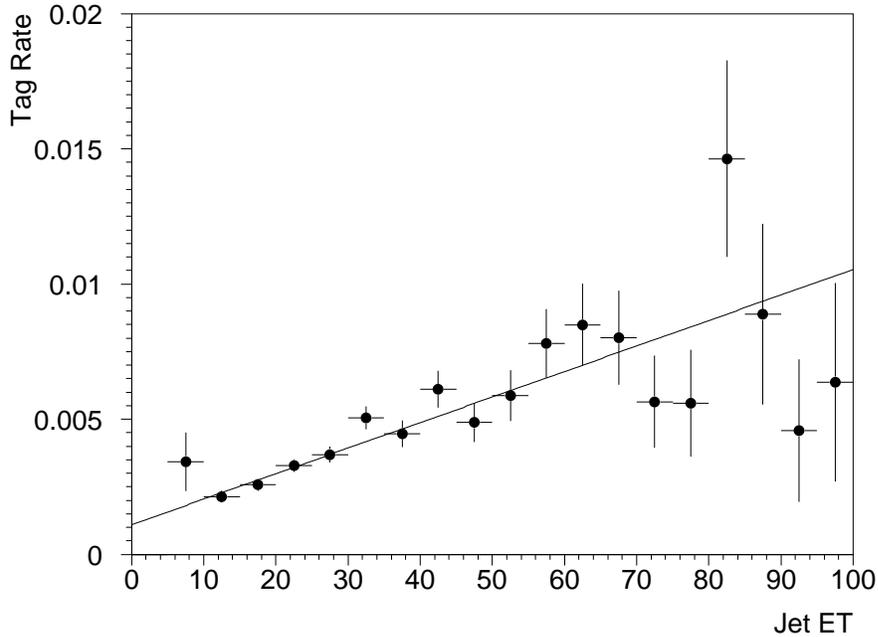}
%
%
\vskip -1.5cm
\caption{\protect \baselineskip 12pt
Tagging rate for jets as a function of E$_{T}$ in the D0
detector as measured from a sample of fake electron $+$ jets events.
From Snyder, 1995a.}
\label{d0_fake}
\end{figure}

Backgrounds to $b$-tagging were discussed extensively in 
Section~\ref{tagging}.  The $WQ\bar{Q}$ 
backgrounds in both CDF and D0 are calculated by assuming 
that the heavy flavor content of jets in W $+$ jets events is
the same as that of generic jets.  This corresponds to the
Method I background estimate of Section~\ref{tagging}, and
is expected to yield an over-estimate of the background.
In D0, a tagging rate per jet as a function of jet E$_{T}$
is defined 
as the ratio of the number of tagged jets divided by the 
total number of jets as a function of E$_{T}$,
(see Fig.~\ref{d0_fake}).
The background to the lepton tagged signal
is then calculated by convoluting this tagging
rate with the E$_{T}$ spectrum of jets in the 
$W + \geq$ 3 jets sample.   As discussed in Section~\ref{tagging},
this procedure yields simultaneously an estimate of the
instrumental and $WQ\bar{Q}$ backgrounds.

\begin{figure}[htb]
\epsfxsize=4.0in
\epsfysize=2.0in
\vskip 1cm
\gepsfcentered[20 200 600 600]{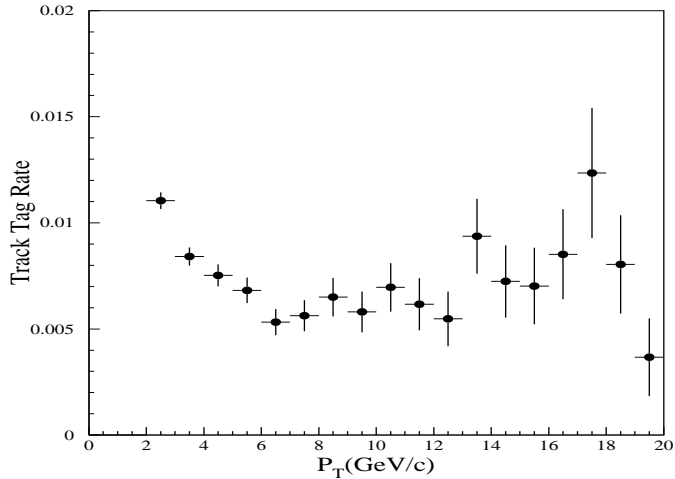}
\vskip 1cm
\caption{\protect \baselineskip 12pt
Muon track tagging rate as a function of P$_{T}$ in the CDF
detector as measured from a sample of jet events.
From F. Abe {\em et al.}, 1994a.}
\label{prd_fig39}
\end{figure}

A similar procedure is used in CDF, with the difference
that track-tag rates instead of jet-tag rates are used.
The track-tag rate is defined as the ratio of the number of
tracks in jet events that are tagged as leptons divided by
the total number of tracks.
To calculate the background,
electron and muon track-tag rates as a function of P$_{T}$ are 
convoluted with the P$_{T}$ spectrum of tracks in the 
$W + \geq$ 3  jets
sample.
The CDF muon track-tag rate is displayed in Fig.~\ref{prd_fig39};
the analogous
electron track-tag rate is considerably smaller, see
F. Abe {\em et al.}, 1994a.

An additional physical background to the tagged $W +$ jets search
is due to production of a $W$ and a single $c$-quark, $W +$ charm,
(see Fig.~\ref{wc}).  In the D0 analysis it is assumed that this background 
is automatically included when 
the jet-tag rate is convoluted with the jet E$_{T}$ spectrum.
On the other hand, 
this background is calculated explicitly by CDF and is added
in separately.

A background contribution
from $Z \rightarrow \mu \mu$ events is also present in D0.
The two muons can result in misinterpreting
such an event as $W \rightarrow \mu\nu$ event with a $\mu$ tag. 
The poor muon momentum
resolution does not allow for a clean removal of these 
events via an invariant mass cut, as is done in CDF.
Just as in the D0 dilepton analysis (see Section~\ref{dildisc}),
these events are identified using a global event
$\chi^{2}$-test for consistency with the $Z$ hypothesis.  
Since the $Z$ removal procedure is not 100\% efficient,
the D0
background estimate in the lepton $+$ jets $+$ $b$-tag search
also includes a contribution from residual $Z \rightarrow \mu \mu$ events
in the sample.
Additional small backgrounds are due to all sources of dileptons discussed
in Section~\ref{dildisc} (e.g. $Z \rightarrow \tau \tau$, dibosons,
$b\bar{b}$).  These backgrounds are expected to be small.
They are included in the CDF background estimate, and taken as
negligible by D0.

\begin{figure}[htb]
%
\epsfxsize=4.0in
\vskip 1cm
\hskip 3cm
\epsffile{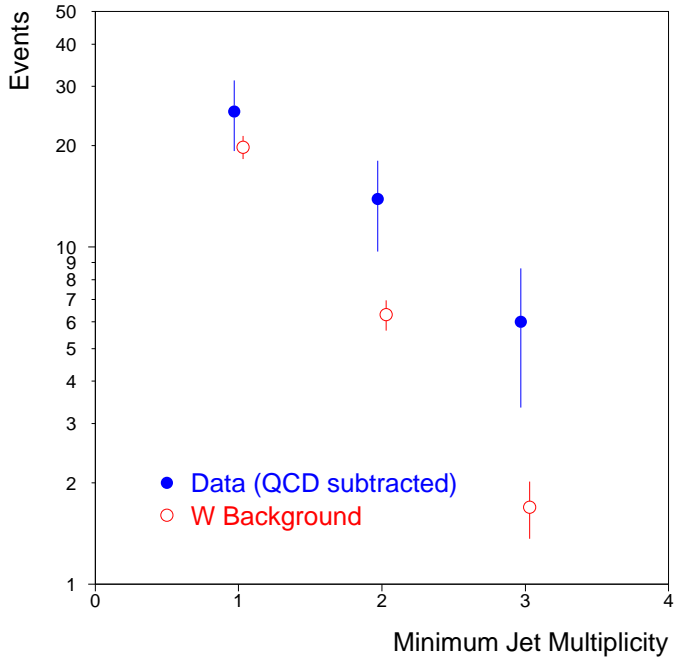}
\caption{\protect \baselineskip 12pt
Comparison between the observed number of muon tags and 
the background expectation in $W +$ jets event as a function 
of jet multiplicity for D0 data.  The H$_{T}$ requirement
(see Table~\protect\ref{ljsel})
has been removed for these data.  Note that the horizontal
axis is in terms of {\em minimum jet multiplicity}, e.g.
the 2-jet bin includes all events with 2 or more jets.
From Snyder, 1995a.}
\label{btag_d0}
\end{figure}

\begin{figure}[hbt]
\epsfxsize=4.0in
\epsfysize=2.0in
\gepsfcentered[20 200 600 600]{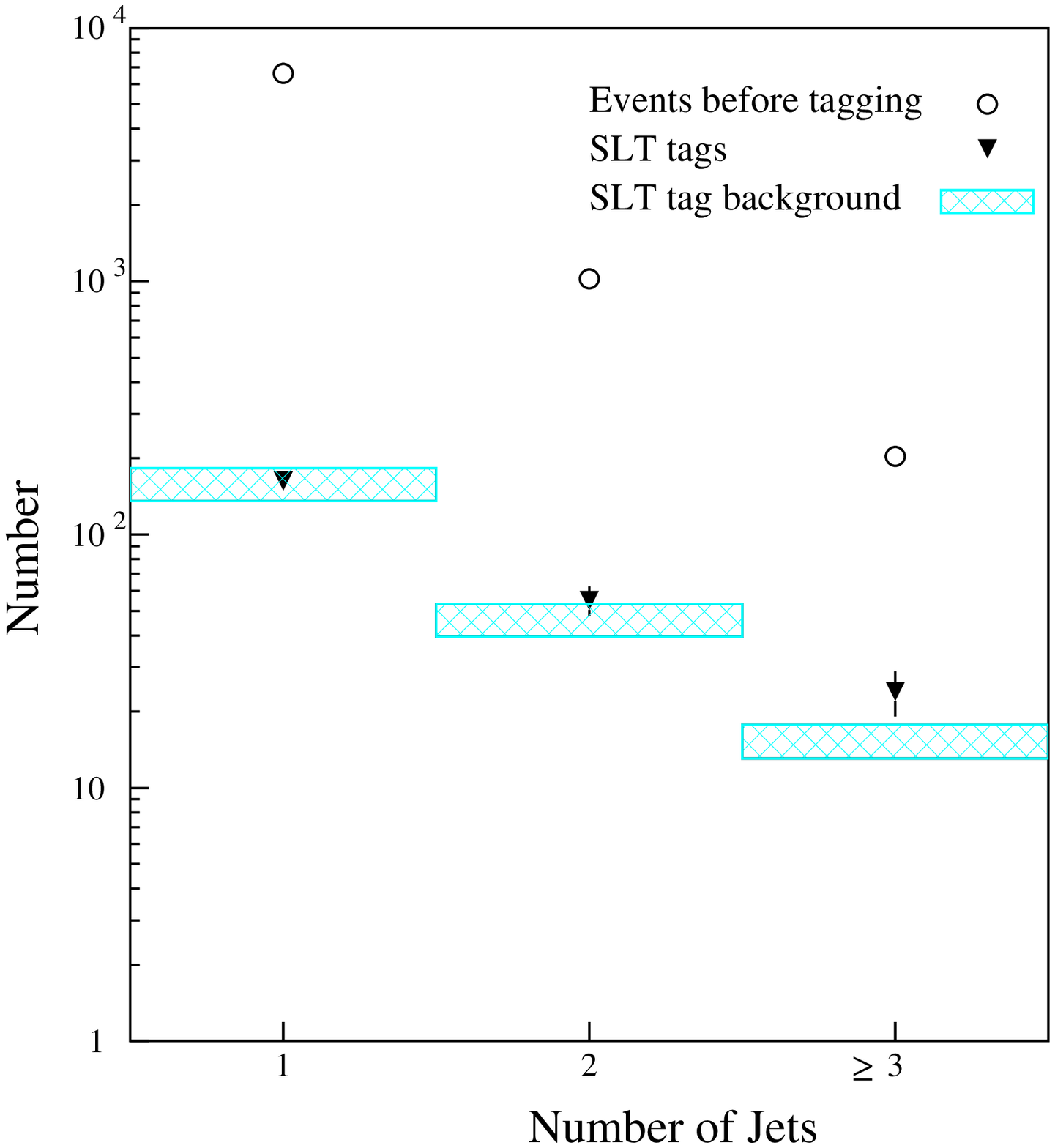}
\vskip 1cm
\caption{\protect \baselineskip 12pt
Comparison between the observed number of lepton tags and 
the background expectation in $W +$ jets event as a function 
of jet multiplicity for CDF data (Kestenbaum, 1996).}
\label{slt_njet}
\end{figure}

Both CDF and D0 see an excess of tags in the $W + \geq$ 3 jets samples 
over their respective background estimates.  The D0
collaboration finds 6 events on an expected background of 
1.2 $\pm$ 0.2;  the CDF collaboration finds 23 tags in 22 events,
with an expected background of 15.4 $\pm$ 2.0 tags.
The most powerful check of the background calculation procedure
is to repeat the exercise for $W$ events with only one or two
jets, where the $t\bar{t}$ content of the data sample is expected to
be very small.
This is summarized in Fig.~\ref{btag_d0} for
D0 and Fig.~\ref{slt_njet} for CDF.  The background
calculations reproduce the expected tagging rates in $W$ events
with low jet multiplicity. 

In this Section and in Section~\ref{tagging},
we have argued that the $WQ\bar{Q}$ background is overestimated
in both the CDF and D0 analyses, yet the background predictions
in the $W +$ 1 and $W +$ 2 jets samples are in good agreement with the
data.  The reason for this is that the dominant background
in these analyses is due to fake leptons, not $WQ\bar{Q}$.  For example,
if the CDF backgrounds were to be calculated using the best theoretical
input for $WQ\bar{Q}$, the background estimate would have been reduced by
only about 10\%, independently of jet multiplicity
(F. Abe {\em et al.}, 1994a).   
This is well within the 20\% systematic uncertainty
assigned to the background calculation.

The sizes of the excesses of events seen by the two collaborations 
are loosely consistent with expectations for $t\bar{t}$.
The D0 collaboration sees an excess of 4.8 events; the number
of expected tagged $t\bar{t}$ events varies between 4.4 and 
1.2 for M$_{top}$ between 140 and 200 GeV/c$^{2}$.
The CDF excess is 7.6 tags, to be compared with a $t\bar{t}$
expectations of between
19 tags (M$_{top}$ = 140 GeV/c$^{2}$) and
2.6 tags (M$_{top}$ = 200 GeV/c$^{2}$).

Despite the fact that the lepton tagging efficiencies are comparable
in the two experiments,
the signal-to-background is better for the D0 analysis.
This is mostly because backgrounds to the detection of low P$_{T}$
muons are lower in D0 than in CDF.  This fact is best illustrated
by comparing Fig.~\ref{d0_fake} with Fig.~\ref{prd_fig39}.
The muon tagging rate in D0 is a fraction of 1\% per jet;
in CDF it is a fraction of 1\% per track.  There are three main reasons
for the lower muon background in D0 : (i)
muons in D0 have to traverse more steel
than in CDF, so that background from hadronic 
punch-through is lower in D0 than CDF; (ii) the
D0 detector is more compact, resulting in a lower probability
for decays in flight of pions and kaons;
(iii) the momentum of the
muon is measured after the decay-in-flight in D0, whereas in CDF,
some average of the momentum of the parent pion and daughter 
muon is often measured in the
drift chamber.
The background rate
per event, dominated by decays in flight and punch-through, 
is 1.8\% in D0 and 7.6\% in CDF. 

In addition to these instrumental effects, the different
choice of requirements between CDF and D0 contribute to
differences in signal-to-background.  The 
D0 pre-tag event sample with the
H$_{T}$ requirement is expected to contain a higher fraction
of $t\bar{t}$ events, for sufficiently high top mass.
On the other hand, there is no requirement on H$_{T}$ in
CDF, in order to maintain 
good efficiency 
for low top masses (M$_{top} <$ 120 GeV/c$^{2}$).  
For example, for M$_{top}$ = 200 GeV/c$^{2}$, the 
$t\bar{t}$ contents of the D0 and CDF pre-tag samples are expected to be
approximately 9\% and 6\% respectively, see 
Table~\ref{ljsel}.

Furthermore, the minimum P$_{T}$ requirement
for lepton tags is lower in the CDF analysis than in the
D0 analysis (2 GeV/c vs. 4 GeV/c).  Again, the CDF requirement
was chosen to maintain efficiency for low top mass, where
leptons from cascade decays have very low transverse momenta,
(see Fig.~\ref{sltpt}).
If the CDF requirement were to be raised to 4 GeV/c, the background
would be reduced by a factor of $\approx$ 1.8, while the tagging 
efficiency for M$_{top} > 130$ GeV/c$^{2}$
would be lowered by approximately 20\% only (F. Abe {\em et al.}, 1994a).
An excess of events in the CDF lepton tag analysis is also present 
when the minimum P$_{T}$ requirement is raised to 4 GeV/c.
In this case, 15 tags in 14 events are observed, with a background of 
8.7 $\pm$ 1.8 tags (Kestenbaum, 1996).

\clearpage

\subsubsection{Displaced-vertex tagging in CDF}
\label{svx}
The most powerful method employed by the CDF collaboration to extract a 
top signal in the lepton $+$ jet channel is to search for secondary
vertices from $b$-quark decay (vertex tagging).  This is made possible
with the precise tracking information obtained from
a silicon vertex detector.  

CDF is the first, and
so far the only, experiment to operate such a detector in a hadron
collider.  The first silicon vertex detector (SVX) at CDF was installed
in 1992 prior to the beginning of Run Ia.  The SVX (Amidei {\em et al.},
1994) is a four-layer cylindrical
detector which is 51 cm long.  The four layers are at distances
of 3.0, 4.2, 5.7, and 7.9 cm from the beamline.  Axial microstrips
with 60 $\mu$m pitch on the three innermost layers and 55 
$\mu$m pitch on the outermost layer provide precision track reconstruction 
in the plane transverse to the beam.  The single hit resolution is 
13 $\mu$m, and the impact parameter resolution, for high momentum
tracks is 17 $\mu$m.  The SVX detector suffered significant radiation
damage, and
was replaced in 1993, during the
accelerator shutdown period between Runs Ia and Ib, by a very similar
detector (SVX$'$, Azzi, 1994) equipped with radiation-hard electronics.

The vertex tagging algorithm used by CDF is based on reconstruction
of displaced vertices using information from both the central
tracking chamber and the SVX (or SVX$'$).  
At the Tevatron, $p\bar{p}$ interactions are spread along 
the beamline with standard deviation $\sigma \approx$ 30 cm, so that
the geometrical acceptance of the 
vertex detector is about 60\% for 
$p\bar{p}$ interactions.  This geometrical effect turns out to be
the largest source of efficiency loss in tagging $t\bar{t}$ events.

The position of the primary vertex is needed before 
searching for a possible secondary vertex.  The transverse spreads
of the colliding
$p$ and $\bar{p}$ beams result in a luminous region
in the transverse plane
which is Gaussian in shape with $\sigma \approx$ 36 $\mu$m.
The position and size of this region varies somewhat from store
to store, and is monitored with an accuracy of order 10 $\mu$m.
On an event-by-event basis, knowledge about the
position of the primary vertex 
is improved by performing a fit using information
from tracks consistent with
originating from the primary vertex.  The accuracy
of the event-by-event
determination of the position of the primary vertex in the transverse
plane depends on the number of tracks available to the fit, and 
varies between 6 and 36 $\mu$m.  

\begin{figure}[htb]
\epsfxsize=4.0in
\epsfysize=2.0in
\vskip 1cm
\gepsfcentered[20 200 600 600]{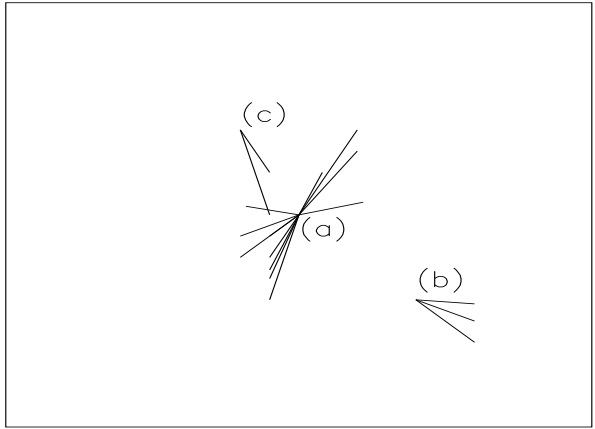}
\vskip 1cm
\caption{An idealized sketch of vertex tagging in 
the transverse plane.  The tracks
from the primary vertex (a) are used to improve on the
accuracy of the determination of the position of the primary vertex.
Vertex (b) is a secondary vertex which is consistent with the
hypothesis that it originates from the decay of a long-lived particle 
produced at the primary.  This is a {\em positive tag}.
On the other hand, vertex (c), although detached from the
primary, is not consistent with the decay of a long-lived particle
from the primary.  Vertex (c) is an example of a {\em negative tag},
which is due to track mismeasurements.
In a $t\bar{t} \rightarrow $ lepton $+$ jets event, the tracks
from vertex (a) in general originate from the underlying event or 
from the hadronization
of the $q$ or $\bar{q}$ from the decay of the $W$ in
$t \rightarrow Wb, W \rightarrow q\bar{q}$.}
\label{vsketch}
\end{figure}

The vertex tagging algorithm operates on combinations of at least
two tracks with impact parameter at least three standard deviations
different from zero.
Constrained vertex fits are performed on these
track combinations in an attempt to find one or more sets
of tracks which are consistent with originating from a secondary
vertex.  Selection criteria are applied to reject tracks from
decays of $\Lambda$ and K$^{0}_{s}$.
Results of the vertex fit include the distance in the
transverse plane of the secondary
vertex from the primary vertex (L$_{xy}$) and
its uncertainty ($\sigma_{lxy}$).  For a good secondary vertex
which results in a $b$-tag,
L$_{xy}$/$\sigma_{lxy}$ is required to be $> 3$.  
The typical accuracy on the determination
of the position of a secondary vertex is $\sigma_{lxy} \approx$
130 $\mu$m.  This is much smaller than the distance travelled by
a $b$-hadron in a top event (typically a few mm, see
Fig.~\ref{dl}), thereby allowing for efficient identification of
secondary vertices from $b$-hadron decay.
There are two possible kind of vertex tags,
{\em positive tags} and {\em negative tags}, (see Fig.~\ref{vsketch}).
Only positive tags are consistent with originating
from the decay of a long-lived particle produced at the primary
vertex.
Negative tags, however, provide useful information on the performance 
of the $b$-tagging algorithm, as we will discuss below.

\begin{figure}[htb]
\epsfxsize=4.0in
\vskip 1cm
\gepsfcentered[20 200 600 600]{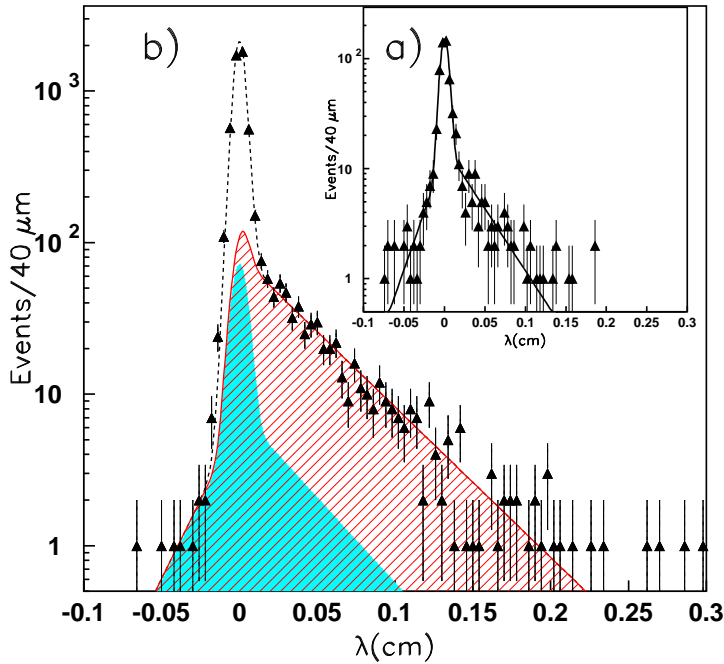}
\vskip 1cm
\caption{\protect \baselineskip 12pt
CDF measurement of the $b$-hadron lifetime
from $B \rightarrow J/\Psi + X$ decays.
(a) The distribution in $\lambda$, the proper decay length,
of data in the $J/\Psi$ sideband regions.  The solid line shows the result
of the fit. (b) The distribution in $\lambda$, of data in
the $J/\Psi$ region.  The curves are the result of the fit.  The
lightly shaded area is for $B \rightarrow J/\Psi + X$ events.
The darkly shaded area is for the non-$J/\Psi$ background.
The large $J/\Psi$ component at $\lambda \approx 0$ is due
to prompt $J/\Psi$ production and to $\chi \rightarrow J/\Psi$
decays.  From F. Abe {\em et al.}, 1993d.}
\label{psilife}
\end{figure}

The capabilities of the CDF silicon vertex detector for 
detection of secondary $b$-vertices are best illustrated in
Fig.~\ref{psilife}, which summarizes the CDF measurement of
the lifetime of $b$-hadrons from the decay
$B \rightarrow J/\Psi + X; J/\Psi \rightarrow \mu\mu$.
The proper decay length ($\lambda$)
of the $b$-hadron is reconstructed
from the position of the $\mu\mu$ vertex as

$$\lambda = L_{xy}~~ \frac{M_{J/\Psi}}{P_{T}^{J/\Psi} F(P_{T}^{J/\Psi})}$$
where $M_{J/\Psi}$ is the $J/\Psi$ mass, $P_{T}^{J/\Psi}$ is the
$J/\Psi$ transverse momentum, and $F$ is a Monte-Carlo determined
correction factor which accounts for the undetected particles in the
$B \rightarrow J/\Psi + X$ decay.  
The result of the lifetime fit,
$\tau_{B} = 1.46 \pm 0.06 \pm 0.06$ ps, is one of the world's
most precise measurements of this quantity, and is consistent with
the results from LEP.

The CDF vertex tag efficiency in top events is 
$\epsilon_{tag}$ = (42 $\pm$ 5)\%.
This efficiency is
defined as the probability of finding at least one (positive) displaced
vertex within one of the jets in a $t\bar{t}$ event with $\geq 3$
jets.  The vertex tag efficiency 
is a factor of two larger than the analogous
efficiency of the CDF and D0 lepton tag
algorithms discussed in the previous Section.

The value of $\epsilon_{tag}$
is determined from a Monte Carlo simulation of the 
response of the detector to
$t\bar{t}$ events.  In order to verify the reliability of the
detector simulation, a number of studies of $b$-tagging are
performed in samples of $b \rightarrow l$ events.  The results of these
studies are compared with expectations from a Monte Carlo 
simulation of this process, and the level of agreement found between
the data and Monte Carlo is used to set the systematic uncertainty on 
$\epsilon_{tag}$.  The largest source of inefficiency is
due to the long luminous region of the Tevatron ($\sigma = 30$ cm). 
Since the vertex detector is only 51 cm long, 
approximately 40\% of the 
interactions occur outside its geometrical coverage.

Note that the tagging efficiency reported in the first CDF publications
on vertex tagging in top events
(F. Abe {\em et al.}, 1994a) was considerably smaller,
$\epsilon_{tag}$ =  (22 $\pm$ 6)\%.  This value of the tagging
efficiency was underestimated by 15\%, due
to a mistake in the Monte Carlo simulation.  In addition, the tagging
algorithm used in the earlier analysis has been substantially
improved, and the performance of the SVX$'$ in terms of efficiency and
signal-to-background is somewhat better than that of SVX.

In the lepton tag analyses described in the previous Section, the 
instrumental and $WQ\bar{Q}$ backgrounds were estimated simultaneously
under the assumption that the heavy flavor content of generic jets
is the same as the heavy flavor content of jets in $W$ events
(Method I, see Section~\ref{tagging}).  As has been argued before,
this method overestimated the size of the $WQ\bar{Q}$ background.
On the other hand, the Method II
background calculation, which was also discussed at length
in Section~\ref{tagging}, relies on a theoretical
estimation of $WQ\bar{Q}$.  To the extent 
that the instrumental background in the lepton tag analysis
is larger than the $WQ\bar{Q}$ background, numerically the 
Method I and Method II backgrounds are not very different.
For the vertex tag analysis, however, this is not the case.  Therefore,
the CDF collaboration has chosen to calculate the background
in this channel using Method II.

\begin{figure}[htb]
\epsfxsize=4.0in
\vskip 1cm
\gepsfcentered[18 158 522 644]{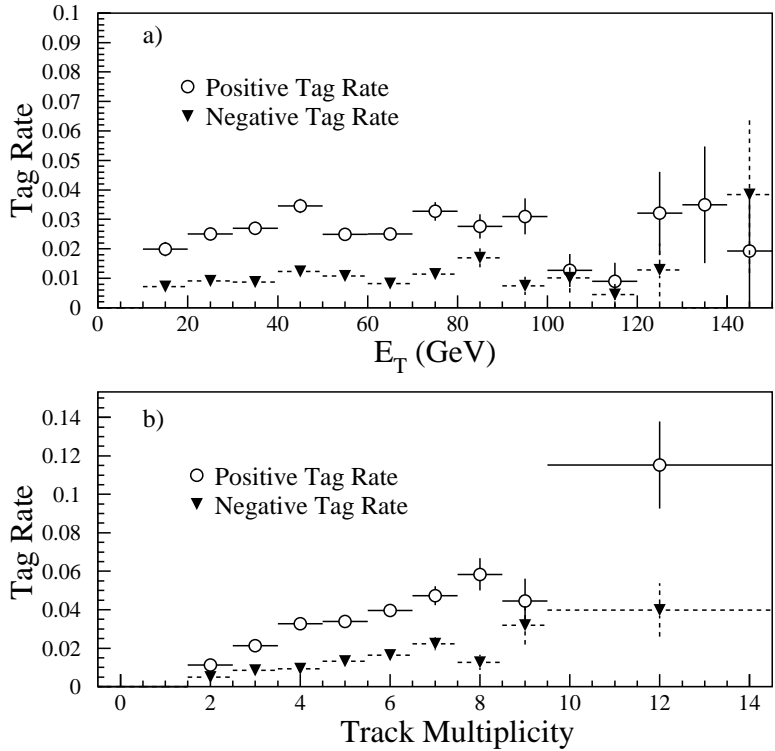}
%
%
\caption{\protect \baselineskip 12pt
The tagging rate, defined to be the number of tagged jets divided 
by the number of jets having two or more tracks with
P$_{T} > 2$ GeV/c
reconstructed in 
the SVX, as a function of (a) the jet E$_{T}$, and (b) the
number of tracks associated with the jet.  Shown are the rates
for positive tags (circles) and negative tags (triangles)
as measured in a sample of $p\bar{p} \rightarrow $ jets events.
The positive tag rate is higher due to the fact that these
jets contain a small fraction of heavy quarks ($b$ and $c$).
The positive tag rate is used to calculate the sum
of backgrounds due to mistags and $WQ\bar{Q}$ in the Method I
background calculation.  The negative track rate is used to
calculate the mistag background separately in the Method
II version of the background calculation.
From the CDF collaboration, F. Abe {\em et al.}, 1994a.
Note that these tag rates are not for the final version
of the vertex tagging algorithm used by CDF.}
\label{tagrate}
\end{figure}

The instrumental background, i.e. the background due to false tags
in light quarks or gluon jets, is estimated from the negative tagging 
rate as measured in a sample of generic jets.  Since negative tags
are almost exclusively due to tracking mismeasurements, and since
these mismeasurements are equally likely to produce a positive
or a negative tag, the negative tagging rate in generic jet
is a measure of the mistag probability.  This mistag probability
is parametrized as a function of jet E$_{T}$, jet pseudo-rapidity,
and track multiplicity, (see Fig~\ref{tagrate}).  Based on this
probability, the number of expected mistagged jets in the 
$W +$ jets samples is calculated by summing the mistag probabilities
for all the jets in the sample.

\begin{figure}[htb]
\epsfxsize=4.0in
\vskip 1cm
\gepsfcentered[20 200 600 600]{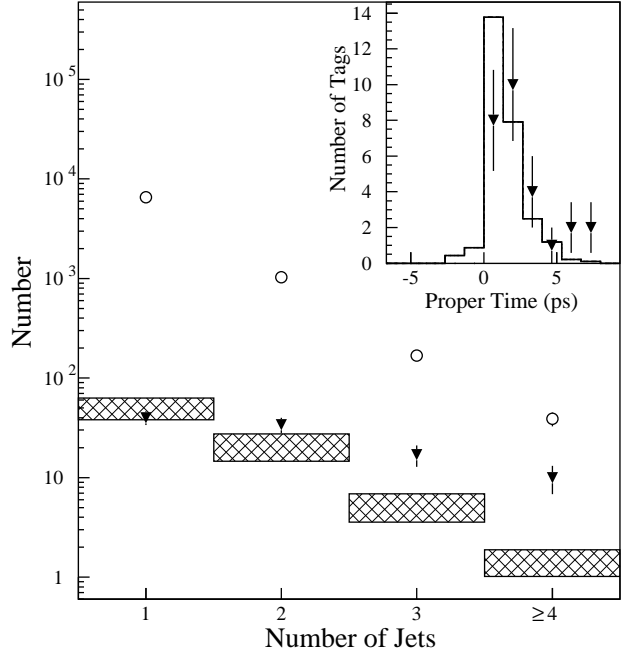}
%
%
\vskip 1cm
\caption{\protect \baselineskip 12pt
Number of events before vertex tagging (circles),
number of tags observed (triangles), and expected number of
background tags (hatched) as a function of jet multiplicity.
The inset shows the secondary vertex proper time distribution for
the 27 tagged jets in the $W + \geq$ 3 jets data (triangles)
compared to the expectation for $b$-quark jets from $t\bar{t}$
decays.  From the CDF collaboration, F. Abe {\em et al.}, 1995a.}
\label{svxfig}
\end{figure}

The Method II $WQ\bar{Q}$ 
backgrounds is calculated using 
theoretical expectations for the rates of $WQ\bar{Q}$.
The fraction of $W$ events containing a $Q\bar{Q}$ pair
is taken from theory, and multiplied by the number
of observed $W$ events to estimate the number of
$WQ\bar{Q}$ in the pre-tag samples.  The product of
this number with the expected $WQ\bar{Q}$ tagging efficiency
yields the expected tagged $WQ\bar{Q}$ background.
Note that the
theoretical input is the fraction of $W$ events that include
a $Q\bar{Q}$ pair rather than the much more uncertain
absolute rate prediction,
see the discussion in Section~\ref{tagging}.
The $W + c$ background (see Fig.~\ref{wc})
is calculated in a similar manner, i.e. by multiplying the number
of $W +$ jets events by the fraction of $W$ events which are expected
to contain a single $c$-quark, and by the tagging efficiency
for these events.  Other backgrounds, such as
$Z \rightarrow \tau \tau$, dibosons, and $b\bar{b}$ are 
computed mostly from Monte Carlo.  
The background expectations and event yields are summarized in
Table~\ref{svxbg}, and displayed in Fig.~\ref{svxfig}

\begin{table}
\begin{center}
\begin{tabular}{ccccc} \hline \hline
Source & $W +$ 1 jet & $W +$ 2 jets & $W +$ 3 jets 
& $W + \geq$ 4 jets \\ \hline \hline
$WQ\bar{Q}$ & 13.8 $\pm$ 11.1 & 7.8 $\pm$ 6.2 & 2.0 $\pm$ 1.6 & 0.5 $\pm$ 0.4 \\
Mistags    & 14.8 $\pm$ 3.0  & 5.3 $\pm$ 1.1 & 1.4 $\pm$ 0.3 & 0.5 $\pm$ 0.1 \\
$W + c$    & 15.3 $\pm$ 4.6  & 4.2 $\pm$ 1.3 & 0.9 $\pm$ 0.4 & 0.2 $\pm$ 0.1 \\
$Z \rightarrow \tau \tau$, Dibosons &
0.8 $\pm$ 0.3 & 0.8 $\pm$ 0.3 & 0.2 $\pm$ 0.1 & 0.04 $\pm$ 0.2 \\
$b\bar{b}$ & 5.7 $\pm$ 1.4   & 3.0 $\pm$ 0.8 & 0.8 $\pm$ 0.2 
& 0.18 $\pm$ 0.04 \\ \hline \hline
Total       & 50 $\pm$ 12 & 21.1 $\pm$ 6.5 & 5.2 $\pm$ 1.7 & 1.45 $\pm$ 0.43 \\
Tagged jets & 40          & 34             & 17            & 10 \\
\end{tabular}
\end{center}
\caption{\protect \baselineskip 12pt
Summary of Method II backgrounds and tags in the CDF vertex tag analysis
(Carithers, 1995).  The 27 tags in the $\geq$ 3 jets sample occurr in 
21 events.  Assuming Standard Model top production, one would
expect between approximately 40 (M$_{top} = 140$ GeV/c$^{2}$)
and 5 (M$_{top} = 200$ GeV/c$^{2}$)
tagged $t\bar{t}$ events 
in the $\geq$ 3 jets sample.}
\label{svxbg}
\end{table}

In the $W + \geq$ 3 jets sample, CDF finds 27 tagged jets in 21
events, with a background expectation of 6.7 $\pm$ 2.1 tags.
The power of the vertex tag algorithm is such that
only $1.9 \pm 0.4$ of these tags can be attributed to
mistags, see Table~\ref{svxbg}.  The six events with two tagged 
jets can be compared with four expected for the top $+$ background
hypothesis, and $\leq$ 1 for background alone.  
Furthermore, six of the vertex tagged events also include a lepton tag.
This is in much better agreement with expectations for
top $+$ background (about four events) than with background
alone (about one event).
As further evidence of a $b$-contribution to this
sample, the proper time
distribution for tagged jets is also found to be consistent with expectations
from $b$-jets, see the inset of Fig.~\ref{svxfig}.
As we will show in Section~\ref{xsecmeas}, the size
of the excess is consistent with expectations from $t\bar{t}$ for
a top mass in the neighborhood of 160 GeV/c$^{2}$.

The Method II background calculation in the $W +$ 1 jet 
sample is in good agreement
with the data, providing a very important check of the
reliability of the background estimation.  We note that
the Method I background estimate in the $W +$ 1 jet sample yields
80 $\pm$ 10 tags, in clear disagreement with the observed 40 tags.
In the $W +$ 2 jets sample, the Method II background calculation is 
lower than the number of tags observed in the data.
However, approximately 5 tags from $t\bar{t}$ are 
expected in this sample based on the excess of tags seen in the 
higher jet multiplicity samples.  
After accounting for these events, the
background calculation in the $W +$ 2 jet sample 
is then found to be in satisfactory agreement with the data.

%
%

\subsubsection{Summary and 
cross checks of the tagging background calculation on $Z +$ jets}
\label{zcheck}

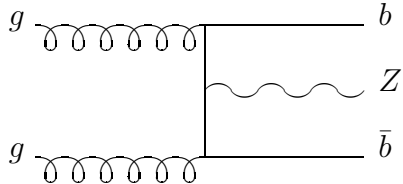
\begin{figure}
\vskip 2cm
\begin{picture}(32000,5000)(0,-3000)

\drawline\gluon[\E\REG](10000,6000)[6]
\drawline\fermion[\E\REG](\pbackx,\pbacky)[6000]
\drawline\fermion[\S\REG](\pfrontx,\pfronty)[5000]
\drawline\photon[\E\REG](\pmidx,\pmidy)[6]
\drawline\fermion[\E\REG](\fermionbackx,\fermionbacky)[6000]
\drawline\gluon[\W\FLIPPED](\pfrontx,\pfronty)[6]
\put(9000,6000){$g$}
\put(23000,6000){$b$}
\put(9000,1000){$g$}
\put(23000,1000){$\bar{b}$}
\put(23000,3500){$Z$}

\end{picture}
\caption{Feynman diagram for $gg \rightarrow Z b\bar{b}$.}
\label{zbb}
\end{figure}

Samples of $Z +$ jets in principle provide an ideal testing ground
for the calculation of tagging backgrounds in the $W +$ jets sample.
The properties of jets in $W$ and $Z$ events are very similar, due
to the similarities in the $W$ and $Z$ production mechanism.
One exception is that whereas in $W$ events $Q\bar{Q}$ pairs 
can be produced only through gluon splitting, see Fig.~\ref{gspfig},
in $Z$ events
additional mechanisms are also expected to contribute (e.g.
$gg \rightarrow Z b\bar{b}$, see Fig.~\ref{zbb}).  
A study of expected
tagging rates in $W$ or $Z$ $+$ 4 jets, including 
a model that approximates
experimental efficiencies and backgrounds, indicates that
the probabilities of tagging a $W$ or a $Z$ event 
are expected to be the same within 10-15\% (Barger 
{\em et al.}, 1994).

Unfortunately, the combined cross-section times branching ratio
for $p\bar{p} \rightarrow Z \rightarrow ll$ 
is an order of magnitude lower than
the cross-section for 
$p\bar{p} \rightarrow W \rightarrow l\nu$, giving only limited
statistics in the $Z$ channel.  Nevertheless, it is still instructive
to compare the yield of tags in the $Z$ sample with the results of
the background calculations.  This is especially true in light
of the fact that two out of the three $Z + \geq$ 3 jets events
collected by CDF during run Ia contained one vertex-tagged jet
(F. Abe {\em et al.}, 1994a).

\begin{table}
\begin{center}
\begin{tabular}{cccc} \hline \hline
  & $Z +$ 1 jet & $Z +$ 2 jets & $Z + \geq$ 3 jets \\ \hline
CDF lepton and vertex tag background expectation & 17.5 & 4.2 & 1.5 \\
CDF lepton and vertex tag data                   & 15   & 3   & 2  \\ \hline
DO muon tag background expectation  & 0.97 $\pm$ 0.08 & 0.35 $\pm$ 0.05 & 
0.09 $\pm$ 0.03 \\
D0 muon tag data                    & 0 & 0 & 0 \\
\end{tabular}
\end{center}
\caption{\protect \baselineskip 12pt
Comparison between the number of observed tags in $Z +$ jets data
and the background expectations.  The CDF results are from Gerdes, 1995,
and correspond to an integrated luminosity of 67 pb$^{-1}$.
The D0 results are from Abachi {\em et al.}, 1995d, and are based on the 
Run Ia data set only, with an integrated luminosity of 13.5 pb$^{-1}$.}
\label{ztable}
\end{table}

The results of the tagging algorithms on $Z +$ jets events are displayed 
in Table~\ref{ztable}.  With the higher luminosity data sample, the
number of tags in the $Z +$ jets data is fully consistent with
the background expectations, within the limited statistics.  There is
no evidence for an anomalously high tagging rate in $Z$ events.

To summarize, the CDF and D0 background calculations have been proved to
be reliable and, within the statistical and systematic
uncertainties, they have been shown to be able to account for the
rate of tagged jets in both $Z +$ jets and $W +$ 1 and 2 jets.
There is significant evidence for the presence
of an excess of $b$-jets in the 
$W + \geq$ 3 jets sample.  The largest excess is seen in the CDF
vertex tag analysis, which has the highest efficiency and the
best signal-to-background.  The statistical significances of 
these excesses, which are consistent in size with the expected
$t\bar{t}$ contribution,
will be discussed in Section~\ref{signif}.

It is very natural to attribute these excesses to 
a $t\bar{t}$ component in the data, although they could also
be due to some source of $W +$ heavy flavor production beyond
standard QCD processes.  As we will show in Section~\ref{ljkin},
additional evidence for the existence of the top quark can be
obtained by kinematic studies of lepton $+$ jets events. 
Furthermore, studies of invariant masses in lepton $+$ jets events
show evidence for both
$t \rightarrow Wb; W \rightarrow l\nu$ and
$t \rightarrow Wb; W \rightarrow q\bar{q}$, see Section~\ref{mass}.

\clearpage

\subsection{Lepton + jets}
\label{ljkin}
Both CDF and D0 have performed analyses based on event shapes or 
kinematic variables rather than 
$b$-quark identification in order to increase their $t\bar{t}$ acceptance. 
They have both used some form of a 
powerful discriminator between $W +$ jets background and top signal, which
is the scalar sum of the transverse energies of the jets and in some cases 
leptons. 
D0 has performed an analysis on a data set
complementary to the lepton + jet + $b$-tag data set. 
That is, it starts with the 
sample of lepton + jet events in which a $b$-tag is not found. The results
from this counting experiment are used in calculating the significance of
the top signal.
CDF on the other hand, performs two separate kinematic analyses 
on the complete lepton+jet data 
sample which includes $b$-tagged events. These analyses are however
not used in 
calculating the significance of the top signal both because their results
are 
correlated with
the $b$-tag result,  
and because kinematic analyses depend to a greater extent on 
Monte Carlo generation details and theoretical assumptions.

In comparing the data with the theoretical expectations for $W +$ jets,
the CDF and D0 analyses use the VECBOS Monte Carlo.  As discussed in 
Section~\ref{bg}, VECBOS is a leading order QCD parton Monte Carlo.
Models of the underlying event and of jet fragmentations have been added
to VECBOS by both collaborations to allow for comparisons
with experimental data. 

\subsubsection{D0 lepton + jet kinematic analysis}

\begin{table}
\begin{tabular}{c|c|c|c|c|c|c|c}
& \multicolumn{2}{c|}{Leptons} & \multicolumn{2}{c|}{Jets} & & & \\
\cline{2-5}
Channel & E$_{T}$(e) & P$_{T}$($\mu$) & $N_{\rm jet}$ & E$_{T}$ 
& $\rlap{\kern0.25em/}E_T$ & $H_T$ & $\cal A$ \\
\hline
\hline
$e + {\rm jets}$ & $\geq$ 20 & &  $\geq$ 4 & $\geq$ 15 & $\geq$ 25 & 
$\geq$ 200 & $\geq$ 0.05 \\
$\mu + {\rm jets}$ & & $\geq$ 15 & $\geq$ 4 & $\geq$ 15 & $\geq$ 20 & 
$\geq$ 200 & $\geq$ 0.05 \\
\hline
\end{tabular}
\caption{D0 kinematic requirements for the standard event selection
(energies in GeV, momenta in GeV/c). From Abachi {\em et al.}, 1995b.}
\label{d0lepjetreq}
\end{table}

\begin{figure}[htb]
\vspace*{0.5in}
\mbox{\hspace*{0.1in}\epsfysize=3.6in\epsffile{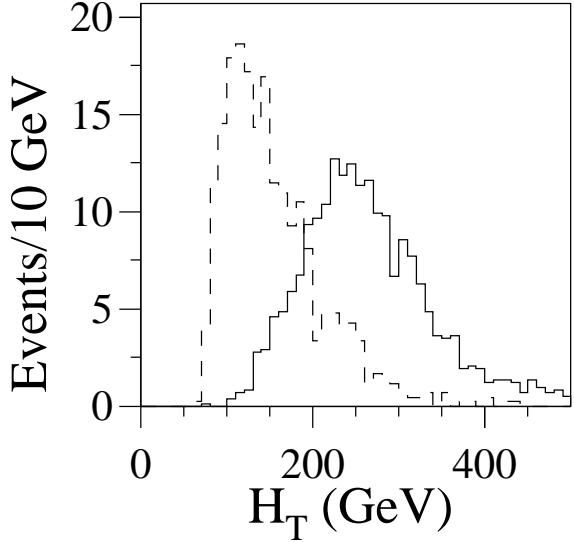}}
\vspace*{-0.6in}
\vskip -2cm
\caption{\protect Expected H$_{T}$ distributions in D0 
for top Monte Carlo events with M$_{top}$ = 200 GeV/c$^{2}$ in the lepton + jets +
no $b$-tag sample
(solid) and                                        
the principal backgrounds to this channel (dashed).  From Abachi {\em et al.}, 1995b.}
\label{d0ht}
\end{figure}

The requirements for this analysis are given in Table~\ref{d0lepjetreq}.
This analysis requires that there be 
at least four jets in the event, in contrast to the $b$-tag analysis
which includes events with three jets.
The two main kinematic requirements are 
that the scalar sum of the transverse energies
of the jets, H$_{T}$, be $>$ 200 GeV and that the event be aplanar such that,
$\cal A$ $> 0.05$. Aplanarity, $\cal A$, was defined in 
Section~\ref{kinematics} as ${\cal A}  \equiv 3/2 \lambda_{1}$,
where $\lambda_{1}$ is the smallest eigenvalue of
the matrix $M_{ab} = \sum P_{a}P_{b} / \sum P^{2}$, where
$P_{i}$ are the cartesian components of momentum of the parton,
$P$ is the magnitude of the three-momentum, and the sum is over
all partons.  In the D0 analysis, $\cal A$ is calculated from the jets
in the event.  
Note that the definition of H$_{T}$ in the D0 lepton $+$ jets analysis is 
different than the definition of H$_{T}$ in the dilepton analysis,
where the E$_{T}$ of the highest transverse energy electron was also
included, see Section~\ref{dildisc}.
Fig.~\ref{d0ht} shows the distribution H$_{T}$ for 
top quarks produced by Monte Carlo and expectations for principal backgrounds.
There is expected to be a clear separation between signal and background. 

\begin{figure}[htb]
\vspace*{0.5in}
\mbox{\hspace*{0.1in}\epsfysize=3.6in\epsffile{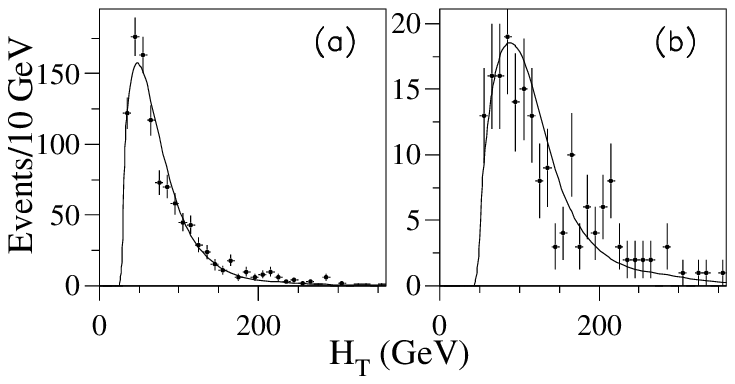}}
\vspace*{-0.6in}
\caption{\protect D0 H$_{T}$ distributions for 
(a) e + $\geq 2$ jets, and (b) e + $\geq 3$ jets
The curves are expectations from the VECBOS Monte Carlo normalized to the data.
From Snyder, 1995a, for an integrated luminosity of 48 pb$^{-1}$.}
\label{htchecks}
\end{figure}

\begin{table}
\begin{center}
\begin{tabular}{ccc}
\hline
\hline
         &       &Events        \\
\hline
e + jets &  Expected background& $1.22 \pm 0.42$\\
         &  Expected for M$_{top}$ = 140 GeV/c$^{2}$ & $4.05\pm 0.94$\\
         &  Expected for M$_{top}$ = 200 GeV/c$^{2}$ & $1.8 \pm 0.31$\\
         &  D0 data            &  5\\
\hline
$\mu$ + jets & Expected backgrounds & $0.71\pm 0.28$ \\
         &  Expected for M$_{top}$ = 140 GeV/c$^{2}$ & $ 2.47\pm 0.68$\\
         &  Expected for M$_{top}$ = 200 GeV/c$^{2}$ & $ 0.95\pm 0.24$\\
         &  D0 data            &  3\\
\hline
\end{tabular}
\end{center}
\caption{\protect D0 Lepton + jets results.  From Abachi {\em et al.}, 1995b.
The integrated luminosities are 48 pb$^{-1}$ for e $+$ jets and 44 pb$^{-1}$
for $\mu +$~jets.  The $t\bar{t}$ expectations are normalized to the
$\sigma(t\bar{t})$ calculation of Laenen, Smith, and van Neerven, 1994.}
\label{d0lepjetres}
\end{table}

\begin{figure}[htb]
\gepsfcentered[20 200 600 600]{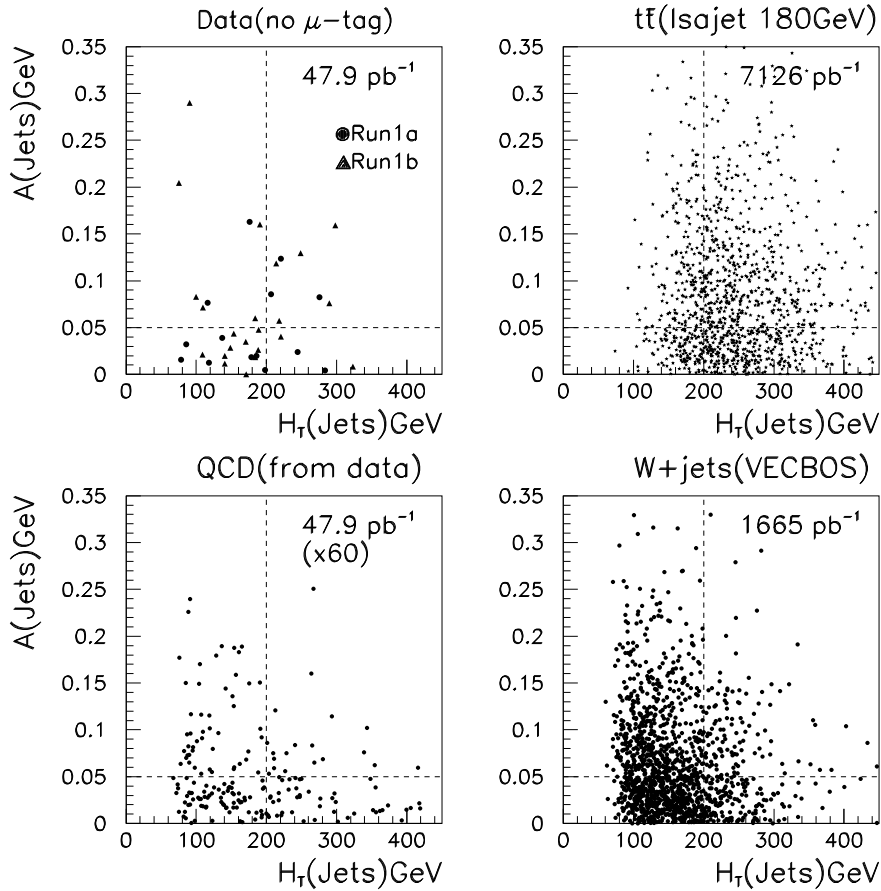}
\vskip 2cm
\caption{\protect \baselineskip 12pt
Aplanarity ($\cal A$) versus H$_{T}$ for single-lepton events 
for data, $t\bar{t}$ Monte Carlo, multijet background from data
(with an effective luminosity = 60 x data luminosity),
and background from $W + $ jets VECBOS Monte Carlo.
From the D0 collaboration, Abachi {\em et al.}, 1995f.}
\label{d0_kinematics}
\end{figure}

\begin{figure}[htb]
\epsfxsize=4.0in
\gepsfcentered[20 200 600 600]{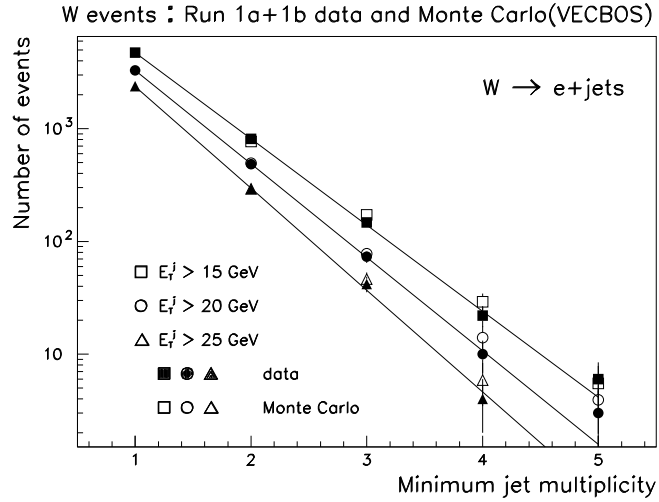}
\caption{\protect \baselineskip 12pt
Inclusive jet multiplicity spectrum for $W \rightarrow e\nu$ + jets
events for several jet energy thresholds.  Data are shown by the 
solid symbols.  Monte Carlo predictions are shown by the open symbols.
From the D0 collaboration, Abachi {\em et al.}, 1995f.  The integrated
luminosity is 48 pb$^{-1}$.}
\label{d0_njet}
\end{figure}

\begin{figure}[htb]
\hskip -1cm
\centerline{\psfig{figure=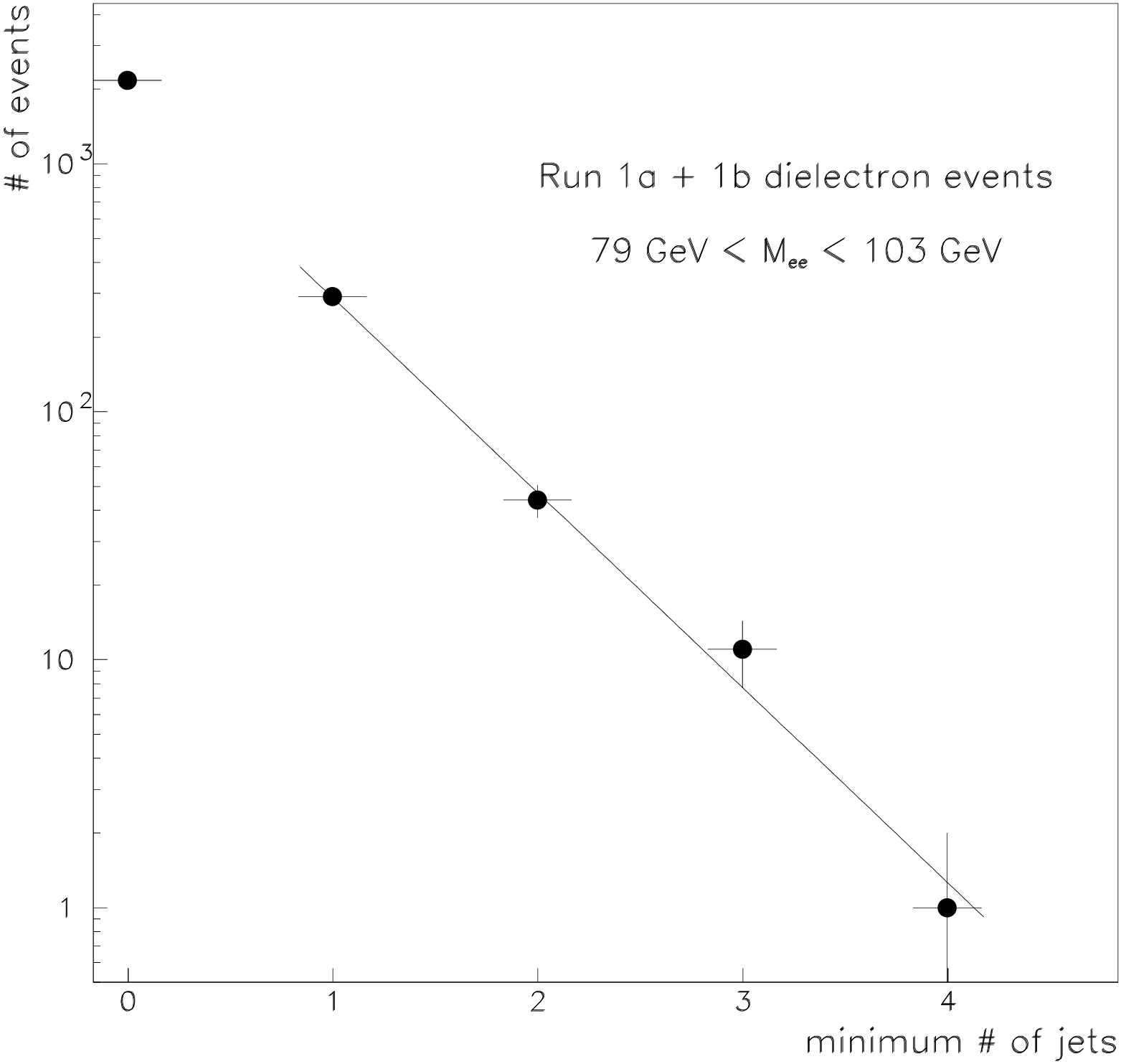,height=3.0in,width=4.0in}}
\caption{\protect \baselineskip 12pt
Inclusive jet multiplicity spectrum for $Z \rightarrow ee +$ jets
events. 
From the D0 collaboration, Snyder, 1995a.  The integrated luminosity is 
56 pb$^{-1}$.}
\label{d0_zmult}
\end{figure}

This analysis must rely on the correctness of both the Monte Carlo 
kinematics and the energy scale of the calorimeters. To check the combined 
effect, comparisons are performed between data and Monte Carlo
distributions of H$_{T}$ 
for two samples dominated by background,
$W + \geq 2$ jets and $W + \geq 3$ jets. 
Fig.~\ref{htchecks} shows
the data and VECBOS Monte Carlo prediction for these two samples. These samples
may contain some top events,
but are dominated by $W +$ jets background if the theoretical expectations
for the $t\bar{t}$ production cross-section are correct. 
The agreement between the data and the Monte Carlo prediction
is quite good. 

The acceptance, including branching ratio (24/81, see Table~\ref{tdec}),
for the D0 $t\bar{t}$~ kinematic search ranges from
0.50\% for M$_{top}$ = 140 GeV/c$^{2}$ to 1.70\% for 
M$_{top}$ = 200 GeV/c$^{2}$.
The expected background, the expected signal and the data
are shown in Table~\ref{d0lepjetres}, and Fig.~\ref{d0_kinematics}.

The two main sources of background in this analysis are
QCD events in which one jet fakes the lepton signature and, especially,
$W + \geq 4$ jet events.  The QCD background is calculated by studying a sample
of events containing electromagnetic clusters
that fail the lepton selection requirements.
The estimation of the
$W +$ jets background is clearly a crucial issue.  This background
is estimated in two ways (Abachi {\em et al.}, 1995d; Grannis, 1995).  
In the first method, the number of $W + \geq 4$
jets events in the data is estimated by extrapolating from the
number of $W + \geq$ 2 and 3 jets events assuming a scaling law, i.e.
assuming that the ratio N$_{n}$/N$_{n-1}$ is independent of $n$, 
where N$_{n}$ is the number of QCD $W + \geq n$ jets events, 
(see Fig.~\ref{d0_njet} and Fig.~\ref{d0_zmult}).
Then, the number of $W + \geq 4$ jets events expected to satisfy
the aplanarity and H$_{T}$ requirement is obtained from N$_{4}$
and the shape of the VECBOS Monte Carlo distribution for 
$W + \geq$ 4 jets in the $\cal A$ vs. H$_{T}$ plane, (see Fig.~\ref{d0_kinematics}).
An alternative estimate is obtained by fitting the number of
events in the four regions of the $\cal A$ vs. H$_{T}$ plane
indicated in Fig.~\ref{d0_kinematics}
to contributions from $t\bar{t}$, $W +\geq $ 4 jets and QCD events,
where the shapes of the $t\bar{t}$ and $W + \geq$ 4 jets components
are taken from Monte Carlo.  The two methods are in reasonable
agreement.  The total expected background is $1.9 \pm 0.5$ events
for the scaling method, and $2.6^{+0.5}_{-0.8}\pm 0.5$ events for the fit
in the $\cal A$ vs. H$_{T}$ plane (Grannis, 1995).

\clearpage


\subsubsection{CDF lepton + jets kinematic analysis}
\label{CDFKIN}
Two kinematic analyses of lepton $+$ jets data have been published
by the CDF collaboration.  The first analysis 
({\em H analysis}, F. Abe {\em et al.},
1995c) is similar to the D0 analysis.  The second analysis 
({\em cos$\theta^{*}$ analysis}, F. Abe {\em et al.}, 1995b)
relies on the fact that jets in 
$t\bar{t}$ events are more central than jets in $W +$ jets.

The variable H in CDF is 
defined as the scalar sum of the lepton transverse momentum, the 
\met~ (i.e. the P$_{T}$ of the $\nu$) and the E$_{T}$'s of all jets 
with E$_{T} \geq 8$ Gev and pseudo-rapidity $|\eta| \leq 2.4$.
Note that while the 
event selection uses uncorrected jet transverse energies, 
H is calculated using corrected energies.
The difference between the variables H used by CDF and
H$_{T}$ used by D0 is the inclusion of the measurements of \met~
and lepton momentum in H. 

The H analysis is performed by comparing the observed H distribution in the data 
with prediction from the $W +$ jets and $t\bar{t}$ Monte Carlos.  A deviation
from the $W +$ jets expectation would then signal the presence of $t\bar{t}$
in the data.  Note that this approach is totally independent of theoretical
expectations for the $W +$ jets cross-section 
since it relies only 
on the predicted shape of the H distribution.  In order to check the
reliability of the theoretical prediction, the data H distribution is 
compared with theoretical predictions in four different samples, 
two control samples and two signal samples, see
Table~\ref{cdfh}. 

\begin{table}
\begin{center}
\begin{tabular}{cccccc}
\hline
\hline
Sample& Threshold& N$_{jets}$& E$_{T}$ (GeV) & $|\eta_{jet}|$ & Events\\
\hline
\hline
Control& low  & =3        &$\geq 8$ & $\leq 2.4$& 814\\
   &    &   Veto jet 4& $\geq 8$& $\leq 2.4$&   \\
\hline
Control& high & =3        &$\geq 15$&$\leq 2.0$ &104 \\
             &   & Veto jet 4&$\geq 8$ &$\leq 2.4$ &   \\
\hline
Signal & low  & $\geq 4$  &$\geq 8$ &$\leq 2.4$ &267 \\
\hline
Signal & high & $\geq 3$  &$\geq 15$ &$\leq 2.0$ &99  \\
       &      & $\geq 1$  &$\geq 8$ &$\leq 2.4$ &    \\
\hline
\hline
\end{tabular}
\end{center}
\caption{CDF definition of the two control and two signal 
samples in the H analysis. 
The third, fourth and fifth columns list
the criteria placed on the jets in each event.  The integrated luminosity is
67 pb$^{-1}$.}
\label{cdfh}
\end{table}


\begin{figure}
\gepsfcentered[20 200 600 400]{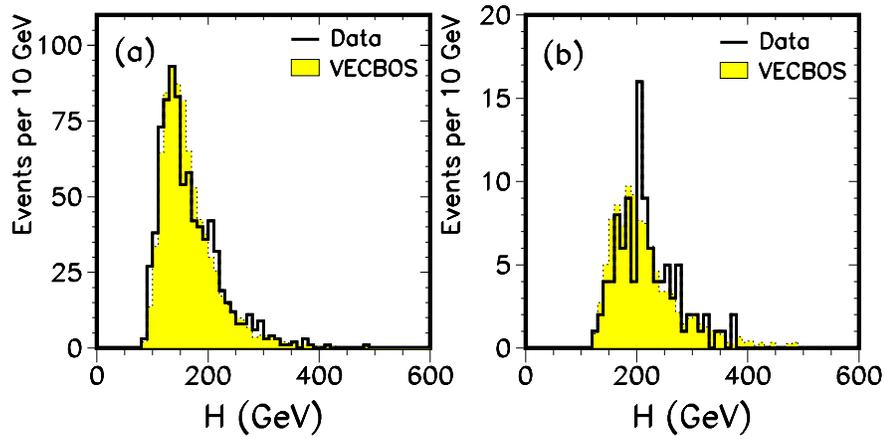}
\vskip 1cm
\caption{Comparison of the H distribution for the control
data samples and the VECBOS Monte Carlo prediction.
(a) $W + 3$ jets events passing the low E$_{T}$ threshold requirements;
(b) $W + 3$ jets events passing the high E$_{T}$ threshold requirements.
The VECBOS prediction, including a 1\% $t\bar{t}$ contribution
for (a) and 10\% $t\bar{t}$ contribution for (b), has been
normalized to the data. From the CDF collaboration, 
F. Abe {\em et al.}, 1995c.    The integrated luminosity is
67 pb$^{-1}$.}
\label{cdfh1}
\end{figure}

\begin{figure}
\gepsfcentered[20 200 600 400]{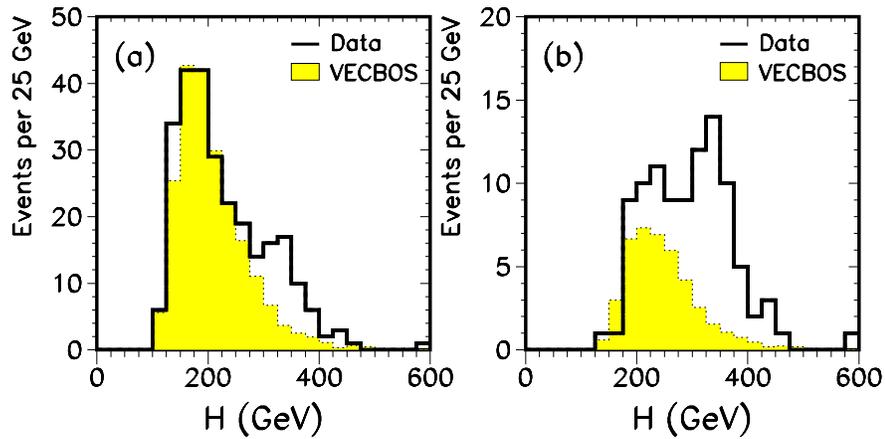}
\vskip 1cm
\caption{Comparison of the H distribution for the signal
data samples and the VECBOS Monte Carlo prediction.
(a) $W + \geq 4$ jets events passing the low E$_{T}$ threshold requirements;
(b) $W + \geq 4$ jets events passing the high E$_{T}$ threshold requirements.
The VECBOS prediction is normalized to a fit to the sum of $t\bar{t}$
and $W + 4$ jets Monte Carlo predictions. From the CDF collaboration, 
F. Abe {\em et al.}, 1995c.   The integrated luminosity is
67 pb$^{-1}$.}
\label{cdfh2}
\end{figure}

\begin{figure}
\epsfxsize=4.0in
\gepsfcentered[20 200 600 600]{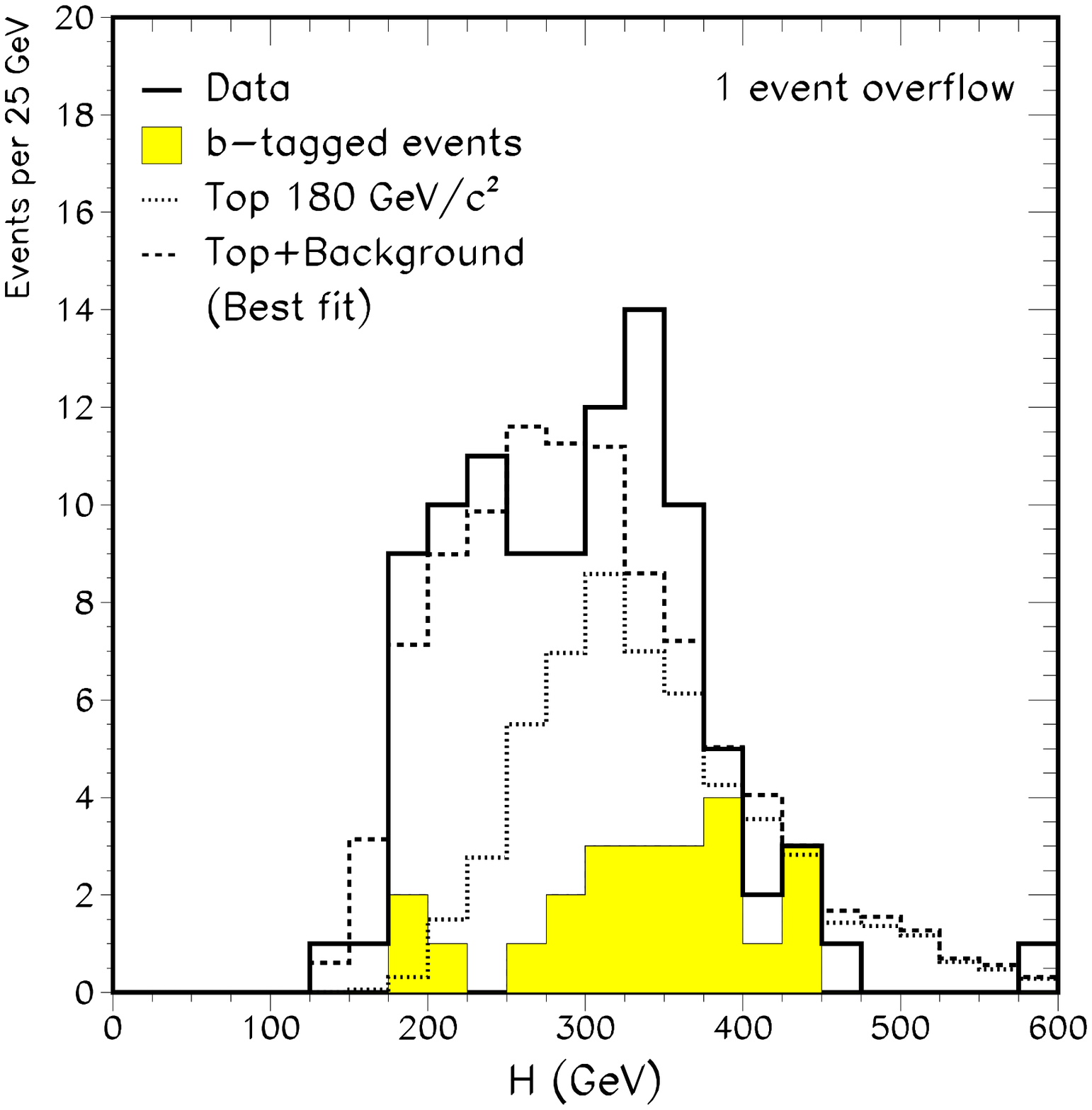}
\vskip 1.5cm
\caption{The binned likelihood fit of the high E$_{T}$ threshold signal sample
(solid line) to a linear combination of the VECBOS $W + 4$ jets and HERWIG
$t\bar{t}$ Monte Carlo prediction.  The H distribution of $b$-tagged
events is shown in the shaded histogram. From the CDF collaboration, 
F. Abe {\em et al.}, 1995c.   The integrated luminosity is
67 pb$^{-1}$.}
\label{cdfh3}
\end{figure}

The two control samples reject events with a fourth jet.  These 
samples are designed to be dominated by background, and are a 
good testing ground for the VECBOS $W +$ jets calculation. 
There are less than 10 
$t\bar{t}$ expected in each sample.
The first control sample has a low requirement on the transverse
energy of the jets, to further increase the proportion of $W +$ jet background 
relative to top events. The two signal samples require the presence
of a fourth jet.  The first signal sample is obtained with a
low transverse energy 
threshold on the jets (8 GeV); in the second signal samples, which is 
expected to be enriched in $t\bar{t}$ events, at least
three of the four jets must have 
transverse energies
greater than 15 GeV. 

Data-Monte Carlo comparisons for the control samples and
the signal samples are displayed in Fig.~\ref{cdfh1} and Fig.~\ref{cdfh2}.
The shape of the H distribution in the control samples agrees with 
the $W +$ jets VECBOS
Monte Carlo, 
while the two signal samples are not consistent with $W +$ jets production
alone.
The differences between the high threshold
signal sample and the VECBOS Monte Carlo
is too large to be explained entirely by
a systematic error in the experimental energy scale.

The shape of the H distribution for $t\bar{t}$ events
is expected to be a function of M$_{top}$.
Fitting the data to a linear combination of $W +$ 4 jets and $t\bar{t}$
Monte Carlo prediction yields M$_{top}$ = 
$180\pm 12$(stat.)$^{+19}_{-15}$(syst.) GeV/c$^{2}$.  This value
for M$_{top}$ is in good agreement with both the expected value from the 
Standard Model and with the directly measured value, see
Section~\ref{mass}.  Furthermore, the $t\bar{t}$ cross-section
extracted from this measurement is also in good agreement with
the cross-section extracted from studies of dilepton and $b$-tagged events,
see Section~\ref{xsecmeas}.

As a further evidence for $t\bar{t}$, the events containing $b$-tags
are highlighted in Fig.~\ref{cdfh3}.  These events are concentrated
at high values of H, as would be expected for $t\bar{t}$. 
Note that in background events there is a small
dependence on the b-tag probability as a function of jet E$_{T}$, and hence as
a function of H, which biases the tagged events to high H.
However, if one assumes that there are no $t\bar{t}$ events in the sample,
this bias is not sufficient to account for the concentration
of tagged events at high values of H.



\begin{table}
\begin{tabular}{c|c|c|c|c|c|c|c}
& \multicolumn{2}{c|}{Leptons} & \multicolumn{3}{c|}{Jets} & & \\
\cline{2-6}
Channel & E$_{T}$(e) & P$_{T}$($\mu$) & $N_{\rm jet}$ & E$_{T}$
& $|$cos$\theta^{*}|$ & \MET & $l\nu$ transverse mass \\
\hline
Signal $e + {\rm jets}$ & $\geq$ 20 & &  $\geq$ 3 & $\geq$ 20 
& $\leq$ 0.7 & $\geq$ 25 & $\geq$ 40 \\
Signal $\mu + {\rm jets}$ &  &$\geq$ 20 &  $\geq$ 3 & $\geq$ 20
& $\leq$ 0.7 & $\geq$ 25 & $\geq$ 40 \\
Control $e + {\rm jets}$ & $\geq$ 20 & &  $\geq$ 3 & $\geq$ 20 
& $\geq$ 0.7 & $\geq$ 25 & $\geq$ 40 \\
Control $\mu + {\rm jets}$ &  & $\geq$ 20 &  $\geq$ 3 & $\geq$ 20 
& $\geq$ 0.7 & $\geq$ 25 & $\geq$ 40 \\
\hline
\end{tabular}
\caption{CDF kinematic requirements for the cos$\theta^{*}$
kinematic analysis.  Cos$\theta^{*}$ is the cosine of the
polar angle of the jet
in the center of mass frame of the lepton $+$ neutrino $+$ jets system.
The cos$\theta^{*}$ requirement applies to the three highest E$_{T}$ 
jets only.  Unlike all other CDF analyses, in this analysis 
jet energies are corrected at the event selection stage.
Energies are in GeV, momenta are in GeV/c. 
From F. Abe {\em et al.}, 1995b.
The integrated luminosity is
67 pb$^{-1}$.}
\label{cthsel}
\end{table}

The second CDF kinematic analysis of lepton $+$ jets 
data, (cos$\theta^{*}$ analysis, F. Abe {\em et al.}, 1995b), 
is based on the fact that jets in 
top events are expected to be more central than in $W +$ jets events
(Cobal, Grassman, and Leone, 1994).  This analysis is also
based on control
and signal samples, see Table~\ref{cthsel}.  The only difference
in the two samples is in the requirement on the cosine of the
polar angle ($\theta^{*}$) of the jets
in the center of mass frame of the two incoming partons. Events
in the signal sample have three high E$_{T}$ central jets, events
in the control sample have at least one jet that is emitted in the
forward or backward region.  Monte Carlo studies indicate that
top events should be approximately equally split between
signal and control samples, whereas of order 75\% of $W +$ jets
events should be in the control sample.  Thus, the signal sample should
be enriched in top quarks with respect to the control sample.

\begin{figure}
\epsfxsize=4.0in
\gepsfcentered[20 200 600 400]{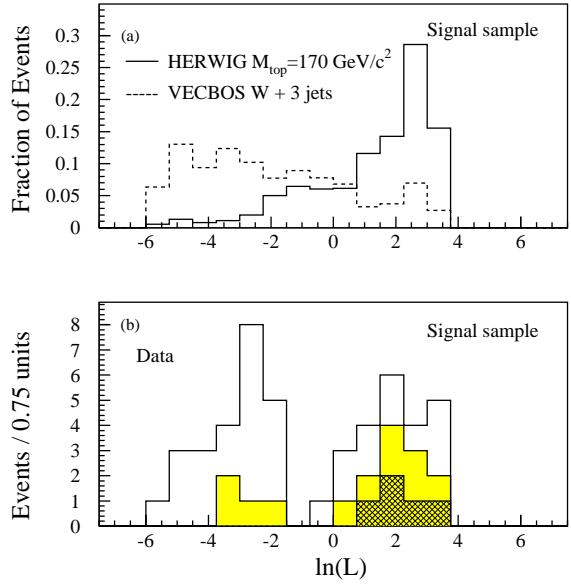}
\vskip 3cm
\caption{The distribution in natural log-likelihood for events in the
signal sample.  (a) Monte Carlo expectations for $W +$ jets (VECBOS)
and $t\bar{t}$ (HERWIG); (b) Data.  The lightly shaded histogram shows the 
$b$-tagged events.  The darkly shaded histogram shows events with 
two $b$-tags. From the CDF collaboration,
F. Abe {\em et al.}, 1995b.   The integrated luminosity is
67 pb$^{-1}$.}
\label{grass1}
\end{figure}

\begin{figure}
\epsfxsize=4.0in
\gepsfcentered[20 200 600 400]{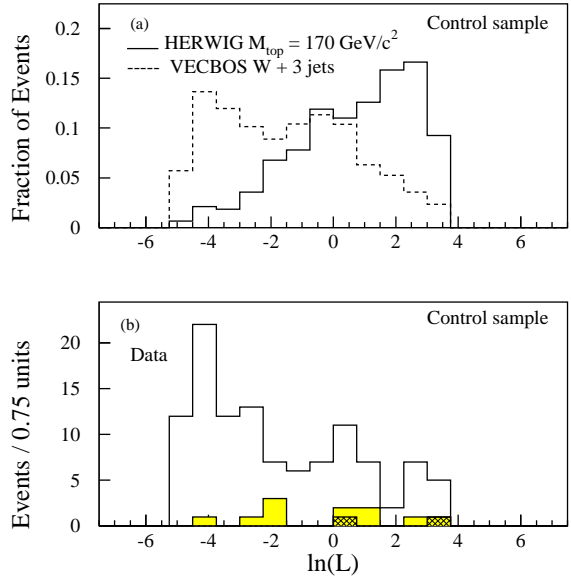}
\vskip 3cm
\caption{The distribution in natural log-likelihood for events in the
control sample.  (a) Monte Carlo expectations for $W +$ jets (VECBOS)
and $t\bar{t}$ (HERWIG); (b) Data.  The lightly shaded histogram shows the
$b$-tagged events.  The darkly shaded histogram shows events with
two $b$-tags. From the CDF collaboration,
F. Abe {\em et al.}, 1995b.    The integrated luminosity is
67 pb$^{-1}$.}
\label{grass2}
\end{figure}

An event-by-event Monte Carlo-based relative likelihood for the $t\bar{t}$  
versus the $W +$ jets hypothesis is constructed for all events based on the
E$_{T}$ of the second and third highest E$_{T}$
jets in the event.  Distributions
of log-likelihood for data and Monte Carlo, for both signal and control
sample, are displayed in Fig.~\ref{grass1} and Fig.~\ref{grass2}.
There is a clear excess of events at high log-likelihood in the signal
sample.  The probability that the shape of the data log-likelihood
distribution can be explained by $W +$ jets alone is $<$ 0.26\%,
assuming the VECBOS model of $W +$ jets production.  Furthermore, the 
$b$-tagged events are concentrated in the high log-likelihood region,
as one would expect for $t\bar{t}$.  

\subsubsection{Summary of lepton $+$ jets kinematic analyses}
\label{ljetsumm}
The CDF and D0 collaborations have performed extensive studies of the 
kinematic properties of lepton $+$ jet events.  The $W +$ jets data 
are inconsistent with the expectations from leading order QCD
as implemented in the VECBOS Monte Carlo program, for high jet multiplicities
and high jet transverse energies.  On the other hand, there seems to 
be satisfactory agreement between data and theory in regions of 
low jet multiplicity and/or low jet transverse energies.  The conclusion
that can be drawn is that either the available QCD calculation is incorrect or
incomplete, or that there is a source of events beyond standard QCD
production of $W +$ jets.  The most natural interpretation of the
data is to ascribe this discrepancy to a $t\bar{t}$ component
in the data.  The correlation between high jet transverse energies and
$b$-tags reported by the CDF collaboration strengthens this conclusion.

As was mentioned in Section~\ref{kinematics}, samples of $Z +$ jets 
events are ideal to test the predictive power of the vector-boson
$+$ jets QCD calculation.  A number of such tests have been performed,
and they seem to yield results consistent with the QCD calculation.
However, the statistics in the $Z$ samples
are not sufficient to decisively validate the QCD calculation in the 
kinematic region most relevant for the top search.

\subsection{Significance of the top signal}
\label{signif}

As discussed in Sections~\ref{dildisc},~\ref{ljtag}, and~\ref{ljkin},
the CDF and D0 top searches in a number of different channels
find excesses of top-like events over the background predictions.
The significance of the excess for a given
channel is defined as the probability
that a background fluctuation would yield a number of events equal to
or larger than the number of observed events.  This
probability is calculated by convoluting the Poisson
probability for the fluctuation of the mean expected number
of background events with its uncertainty, which is assumed to
be Gaussian.  

When combining results from more than one channel, D0 and CDF
use different procedures.  In the D0 case, 
events from all channels as well as backgrounds from all channels are added, 
and the combined significance is defined
as the probability that the sum of all the backgrounds fluctuates
to give a total number of events greater than or equal the 
number of observed events.  On the other hand, 
the CDF combined significance
is defined as the probability that the product of the significances
for the different channels be less than or equal to the measured
value for this product.  This is necessary since the
CDF vertex-tag channel has so much better efficiency and signal
to noise than the dilepton and (especially) the lepton-tag channels.
In fact, if the D0 prescription were to be applied to compute
the total significance of the CDF excesses in the three
channels, the combined excess would appear less significant
than the excess seen in the vertex-tag channel alone, despite
the fact that excesses are seen in the other two channels as well.
When combining different channels, correlations are accounted for.
For example,
both the CDF lepton and vertex tag searches are sensitive
to the size of the W$Q\bar{Q}$ background, resulting
in a correlation between significances in the two channels.
These effects are taken into account using Monte Carlo simulations.

The significances for the CDF and D0 results
are displayed in
Table~\ref{significance}.  These significances, which as discussed above
are just probabilities, are
also expressed in number of standard deviations (No. of $\sigma$).
The number of $\sigma$ corresponds to the
point in a Gaussian probability function of mean zero and 
unit standard deviation where the integral of the probability function
between that point and infinity is equal to the significance.

\begin{table}
\begin{center}
\begin{tabular}{cccccc}
\hline \hline
Experiment & Channel & Events observed & Expected background
& Significance & No. of $\sigma$ \\
\hline
D0  & Muon $b$-tag       & 6 & 1.2$\pm$0.2  & 2 x 10$^{-3}$ & 2.9$\sigma$ \\
D0  & Lepton $+$ jets& 8 & 1.9$\pm$0.5  & 2 x 10$^{-3}$ & 2.9$\sigma$ \\
D0  & Dileptons      & 3 & 0.7$\pm$0.2  & 3 x 10$^{-2}$ & 1.9$\sigma$ \\
D0  & All combined   &17 & 3.8$\pm$0.6  & 2 x 10$^{-6}$ & 4.6$\sigma$ \\
\hline
CDF & Vertex $b$-tag    &21 (27) &  6.7$\pm$2.1  & 2 x 10$^{-5}$ & 4.2$\sigma$ \\
CDF & Lepton $b$-tag    &22 (23) & 15.4$\pm$2.0  & 6 x 10$^{-2}$ & 1.5$\sigma$ \\
CDF & Dileptons     & 6      &  1.3$\pm$0.3  & 3 x 10$^{-3}$ & 2.7$\sigma$ \\
CDF & All combined  &43 (56) & 23.4$\pm$2.9  & 1 x 10$^{-6}$ & 4.8$\sigma$ \\
\hline \hline
\end{tabular}
\end{center}
\caption{\protect \baselineskip 12pt
CDF (F. Abe {\em et al.}, 1995a) and 
D0 (Abachi {\em et al.}, 
1995b) 
event yields, background expectations,
and significances
of the observed excesses of top-like events in different channels.
In the CDF $b$-tagging channels,
we show both the number of events and, in brackets,
the number of tags.  
There are 6 events which are tagged
with both a vertex $b$-tag and a lepton $b$-tag in CDF.
The two {\em all combined} entries for CDF data refer to 43 
events and 56 {\em objects}, where an object is defined as either a 
dilepton event or a tag in a lepton $+$ jets event.
The CDF $b$-tag 
backgrounds are expressed in number of tags; the CDF all combined
background is expressed in number of objects.
Tags in CDF dilepton events are not included anywhere in this table.}
\label{significance}
\end{table}

The results from both experiments are such that the probability
that the data can be explained as background fluctuations
is exceedingly small.  We note that,
in the calculation of the
combined significance of their top observation,
the CDF collaboration very
conservatively does not include the excess of top-like events
seen in their purely kinematic analyses of lepton $+$ jets
data (see Section~\ref{CDFKIN}).  Furthermore,
as will be discussed in Section~\ref{mass}, both CDF and D0 
find additional evidence for a $t\bar{t}$ contribution to their
lepton $+$ jet data samples in the top mass reconstruction
analyses.  To summarize, the data from the counting experiments at the Tevatron
provide overwhelmingly convincing evidence for the existence of
new physics. 
We shall show in the following section that the invariant mass and
kinematics of this new physics are consistent with a top quark
of M$_{top} \approx$ 175 GeV/c$^{2}$.

\subsection{Measurement of the $p\bar{p} \rightarrow t\bar{t}$
cross-section}
\label{xsecmeas}

The $t\bar{t}$ production cross-section is calculated from the luminosity, the
background-subtracted event yields, the acceptances, and the $t\bar{t}$
decay branching fractions for the various channels.
The calculated cross-section depends on the top mass, since
the acceptances are in general also a function of the top mass.

The measurement is complicated by the fact that in some cases 
the expected number of background events
depends on the cross-section itself.
For example, in the CDF $b$-tag analysis
the background due to $WQ\overline{Q}$, which is quoted in
computing the significance of the $t\bar{t}$ observation,
is calculated under the assumption that the pre-tag $W + \geq$ 3 jets
sample does not contain any top events.
Having established the existence of the top quark, this background
estimate needs to be revised, since the untagged sample does indeed
include a $t\bar{t}$ component.   This is accomplished by first
calculating the cross-section using the overestimated background
contribution, then recalculating the background based on 
the value of the cross-section, and iterating until the 
result becomes stable.

\begin{table}
\begin{center}
\begin{tabular}{cccc}
\hline \hline
Experiment & Top mass & Channel & Measured cross-section \\
\hline
D0  & 180 GeV/c$^{2}$ & Muon tag        & 6.8 $\pm$ 3.2 pb \\
D0  & 180 GeV/c$^{2}$ & Lepton $+$ jets & 3.9 $\pm$ 1.9 pb \\
D0  & 180 GeV/c$^{2}$ & Dileptons       & 4.6 $\pm$ 3.1 pb \\
D0  & 180 GeV/c$^{2}$ & All combined    & 4.7 $\pm$ 1.6 pb \\
\hline
D0  & 170 GeV/c$^{2}$ & All combined    & 5.2 $\pm$ 1.8 pb \\
\hline
\hline
CDF & 176 GeV/c$^{2}$ & Vertex tag   & 6.8$^{+2.3}_{-1.8}$ pb \\
CDF & 176 GeV/c$^{2}$ & Lepton tag   & 8.0$^{+4.4}_{-3.6}$ pb \\
CDF & 176 GeV/c$^{2}$ & Dileptons    & 9.3$^{+4.4}_{-3.4}$ pb \\
CDF & 176 GeV/c$^{2}$ & All combined & 7.5$^{+1.9}_{-1.6}$ pb \\
\hline \hline
\end{tabular}
\end{center}
\caption{CDF (Tartarelli, 1996) and D0 (Narain, 1996; Klima, 1996)
measurements of 
$\sigma(p\bar{p} \rightarrow t\bar{t})$ at a center-of-mass energy
of 1.8 TeV.  The integrated luminosities for these measurements
are 110 pb$^{-1}$ for CDF and 100 pb$^{-1}$ for D0. Note that the 
most recent D0 top mass value is M$_{top} = 170 \pm 18$ GeV/c$^{2}$,
see Section IX.A.4. Cross section values for this value of M$_{top}$
for the individual channels have not been released by the D0 
collaboration.  Here we list these cross-sections at 
M$_{top} = 180$ GeV/c$^{2}$ to demonstrate the consistency between
the measurements in the different channels.}
\label{xsectable}
\end{table}

\begin{figure}[htb]
\epsfxsize 4.0in
\epsfysize 2.0in
\vskip 1cm
\gepsfcentered[20 200 600 600]{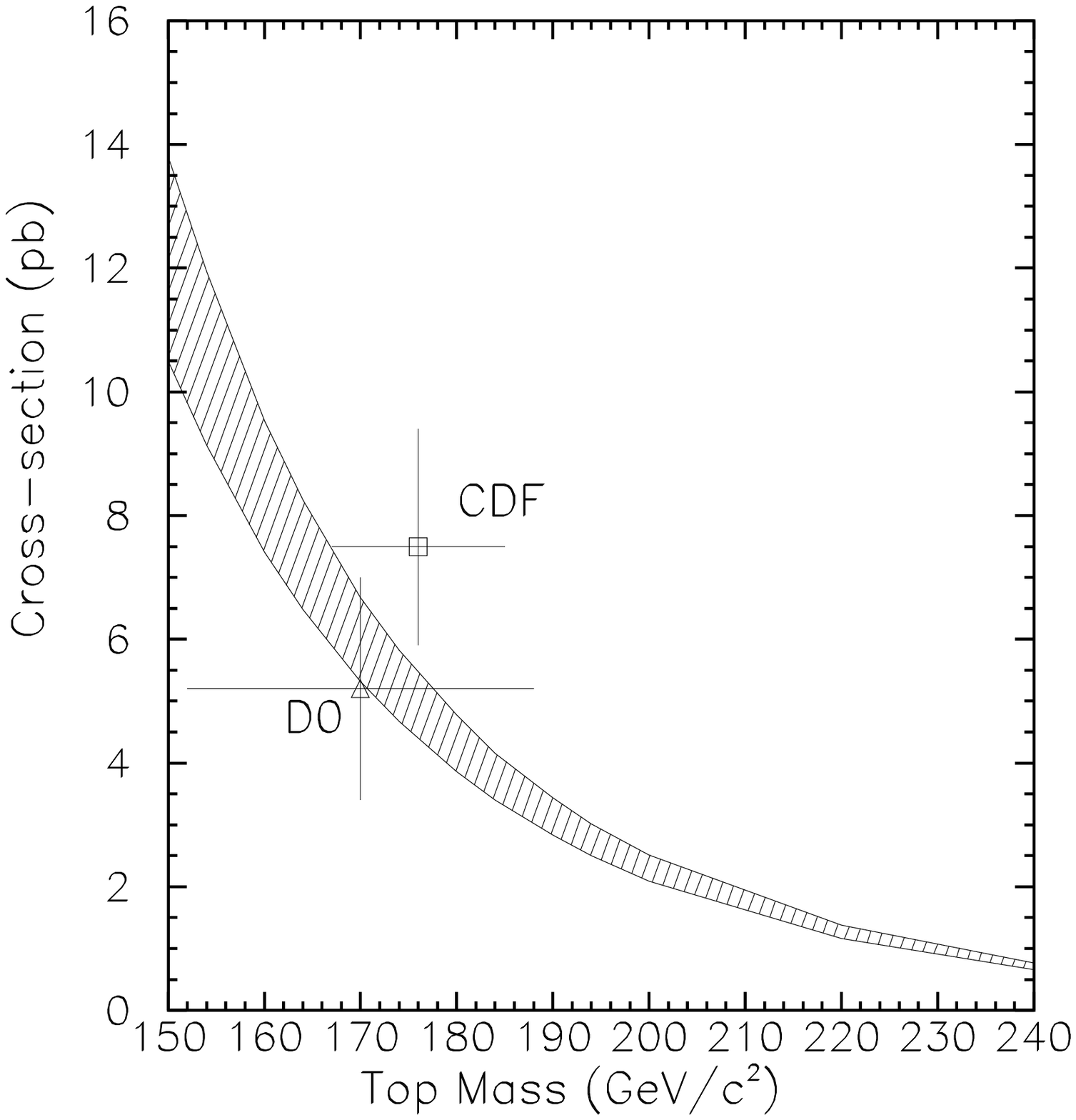}
\vskip 1cm
\caption{\protect\baselineskip 12pt
Combined CDF and D0 
top cross-section and mass measurements. (The mass measurements will
be summarized in Section~\protect\ref{mass}).
Also shown, as a band, is the theoretical expectation for
$\sigma(p\bar{p} \rightarrow t\bar{t})$ as a function of M$_{top}$
with its
uncertainty
(Laenen, Smith, and van Neerven, 1994).}                      
\label{xsec_d0_cdf}
\end{figure}

The CDF and D0 $t\bar{t}$ production cross-sections, are listed
in Table~\ref{xsectable}, and displayed in Fig.~\ref{xsec_d0_cdf}.
As discussed at the beginning of this Section, the cross-section
calculation depends on the assumed value of M$_{top}$.  
Here we
report $\sigma(p\bar{p} \rightarrow t\bar{t})$ computed at the
values of M$_{top}$ measured by the two collaborations,
see Section~\ref{mass}.  The CDF and D0 cross-sections reported here
are obtained from data sets 
of 110 pb$^{-1}$ and 100 pb$^{-1}$
respectively. 

The top mass dependence of the cross-section
measurement is $\delta\sigma/\delta$M$_{top} \approx $ 
$-0.05$ pb/GeV/c$^{2}$ (D0) and $-0.04$ pb/GeV/c$^{2}$ (CDF).
For both experiments the dominant uncertainty on
$\sigma(p\bar{p} \rightarrow t\bar{t})$ is due to 
the limited statistics of the data samples.
Also, for both experiments the separate cross-section measurements in the 
different channels are nicely consistent with each other.
The CDF measurement is lower than, although still
consistent with, the cross-section value of 13.9$^{+6.1}_{-4.8}$ pb
obtained from the analysis of the first 19 pb$^{-1}$
of CDF data (F. Abe {\em et al.}, 1994a).  

Both the CDF and D0 $t\bar{t}$  cross-section measurements are
in agreement with the theoretical prediction.
To summarize, both collaborations have not only reported excesses of top-like
events in several different channels, but the sizes of these
excesses are also consistent with each other.  From these
measurements, they have extracted values of
$\sigma(p\bar{p} \rightarrow t\bar{t})$ which are consistent, within
the quoted uncertainties, with Standard Model expectations.

\section{Measurement of the top quark mass}
\label{mass}
The CDF and D0 top quark searches described in Section~\ref{discintro}
yield an excess of top-like events over the background expectations.
In this Section we will present measurements of the top mass
performed on a sub-set of these events. 
As was discussed in Section~\ref{ind}, the mass of the 
top quark is one of the free parameters of the
Standard Model, and its value enters in the calculation of
radiative corrections to a large number of electroweak
observables.  Comparison of the measurement of M$_{top}$
with indirect determinations from electroweak
measurements at LEP and SLC, 
as well as measurements of the $W$ mass at the Tevatron and
LEP200, and analysis of neutral current neutrino data,
allow for rather stringent tests of the Standard
Model.

There are a number of possible ways to extract the top mass
from the CDF and D0 data.   An indirect method is to simply 
compare the cross-section for the observed $t\bar{t}$ 
signal with its theoretical expectation, since the latter
depends very strongly
on the top mass, (see Fig.~\ref{xsec_d0_cdf}).  However,
it is more informative to {\em measure} both the top
mass and the $t\bar{t}$ cross-section, and then compare with the
theoretical prediction for $\sigma(t\bar{t})$.  
This approach results in a
test of the QCD calculation of top production, and is
sensitive to possible non-Standard Model production mechanisms
for top quarks, see the discussion in Section~\ref{future}.

Both CDF (F. Abe {\em et al.}, 1994a and 1995a)
and D0 (Abachi {\em et al.}, 1995b)
have reported direct measurements of the
top mass from lepton $+$ jets data and these will be
reviewed in Section~\ref{massdir}.  Other less direct methods
to extract the value of M$_{top}$ from both the lepton $+$
jets data  and the dilepton data will be summarized in 
Section~\ref{massindir}.  Finally, in Section~\ref{wevid},
we will present results of attempts to reconstruct the 
$W$ mass from $W \rightarrow q\bar{q}$ in lepton $+$
jets events.

\subsection{Direct measurements of the top quark mass from lepton $+$
jets events}
\label{massdir}

The method used by both CDF and D0 to measure the top quark mass is
the standard method used in particle physics
to measure particle masses, namely one measures 
the momenta of all the decay products, assigns masses to them, 
and then reconstructs the invariant mass of the original particle.  
In the case of top quarks, the decay products include neutrinos and
hadron jets. 
The first example of
the reconstruction of resonances using
jets was UA2's observation of a bump consistent with 
the $W$ and $Z$ in the jet-jet
invariant mass distribution (Alitti {\em et al.}, 1991a).
In this case there
was an an enormous background from QCD production of jets not involving a
vector boson.  The field of jet-spectroscopy
is still in its infancy, although
much progress has been made in the past year. 

\begin{figure}[htb]
\epsfxsize=4.0in
\gepsfcentered[200 200 600 600]{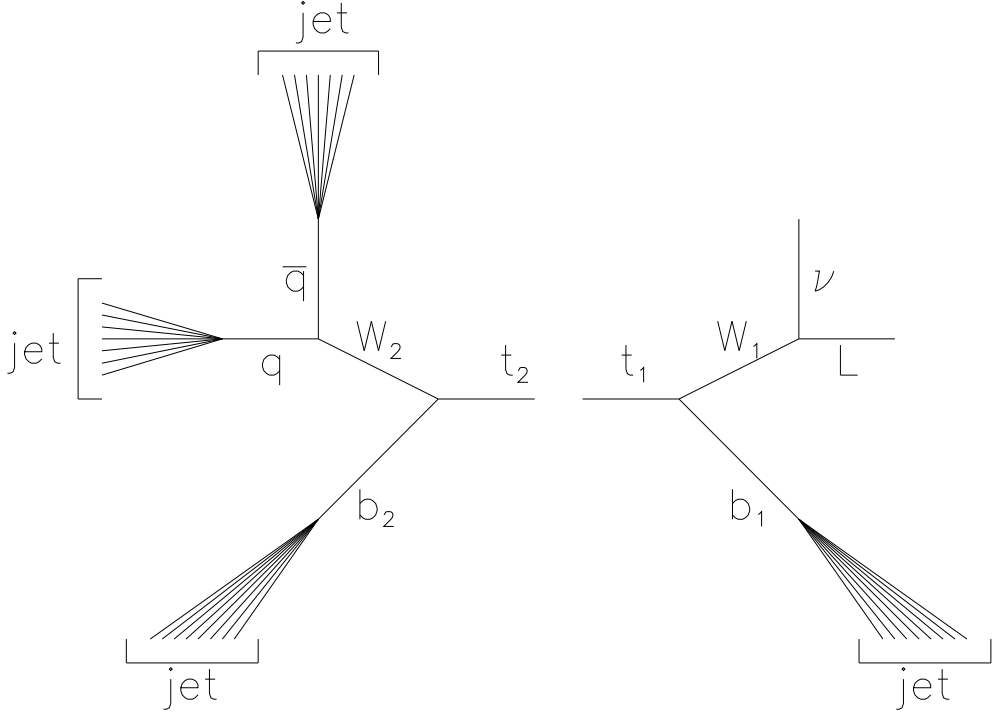}
\caption{Sketch of the $t\bar{t}$ decay chain in the lepton $+$
jets channel.}
\label{tpict}
\end{figure}

\subsubsection{Constrained fits, combinatorics, and top mass
resolution}
\label{fits}

In the $t\bar{t} \rightarrow$ lepton $+$ jets mode, one of the 
two top quarks ($t_{1}$) decays semi-leptonicaly and the other
one ($t_{2}$) decays hadronicaly, (see Fig.~\ref{tpict}) :

\begin{eqnarray*}
t_{1} \rightarrow W_{1} b_{1}~~~~~;~~~~~W_{1} \rightarrow l \nu 
\end{eqnarray*}
\begin{eqnarray*}
t_{2} \rightarrow W_{2} b_{2}~~~~~;~~~~~W_{2} \rightarrow q\bar{q}
\end{eqnarray*}
The mass measurement is performed on the sample of events
with a lepton, \MET~and four jets.  The four jets are identified
with the four quarks in the final state ($b_{1},b_{2},q$ and $\bar{q}$).
Without further information there is no way of knowing which
jet originates from which quark. All possible combinations 
must be considered.
A constrained fit to the $t\bar{t}$ hypothesis
is then performed for each jet-quark assignment in
a given event, assuming energy-momentum conservation at each vertex in the
$t\bar{t}$ decay chain.  The fit uses the following constraints :

\begin{itemize}
\item The invariant mass of the jets assigned
to the $q$ and $\bar{q}$ is constrained to the $W$ mass.

\item The \MET~gives the transverse momentum of the neutrino.
The longitudinal momentum of the neutrino, P$_{L\nu}$, 
is obtained by requiring the mass of the lepton and the neutrino to
equal the $W$ mass.  This condition results in a quadratic equation
for P$_{L\nu}$, which has in general two distinct solutions.

\item The invariant masses of the decay products of $t_{1}$
and $t_{2}$ must be equal, i.e. the invariant mass of the three
jets assigned to $b_{2}$, $q$, and $\bar{q}$ (M$_{2}$) must be equal to
the invariant mass (M$_{1}$)
of the lepton, neutrino and fourth jet 
($b_{1}$, see Fig.~\ref{tpict}).  The top
mass for the jet-quark assignment under consideration
is then M$_{top}$ = M$_{1}$ = M$_{2}$.

\end{itemize}
All of the components of momentum for the final state
particles are measured, except P$_{L\nu}$.  With
one unmeasured quantity and three constraints, the fit is
a two-constraint (2C) fit.
The constrained
fit also yields a $\chi^{2}$ for each combination which is a 
measure of the goodness-of-fit to the $t\bar{t}$ hypothesis.
The fitted value of the top quark mass for a given combination is given
by M$_{top}$ = M$_{1}$ = M$_{2}$ at the point where $\chi^{2}$
is minimized.

There are $4!$ = 24 possible ways of assigning the four jets to the
four final-state quarks.  Since there are two solutions
for P$_{L\nu}$, this would result in 24 x 2 = 48 configurations.
However, the interchange of jet assignments
between the $q$ and $\bar{q}$ from the $W$ has no effect
(see Fig.~\ref{tpict}), so that the number of truly distinct
configurations is 48/2 = 24.
If one or more of the jets is $b$-tagged, the number of configurations
can be reduced by allowing only configurations where 
$b$-tagged jets are associated with the $b$-quarks in the
event.  With one $b$-tagged jet, the number of combinations is twelve;
with two $b$-tagged jets, this number is reduced to four.

\begin{figure}[htb]
\vskip 1cm
\epsfxsize=4.0in
\gepsfcentered[20 200 600 600]{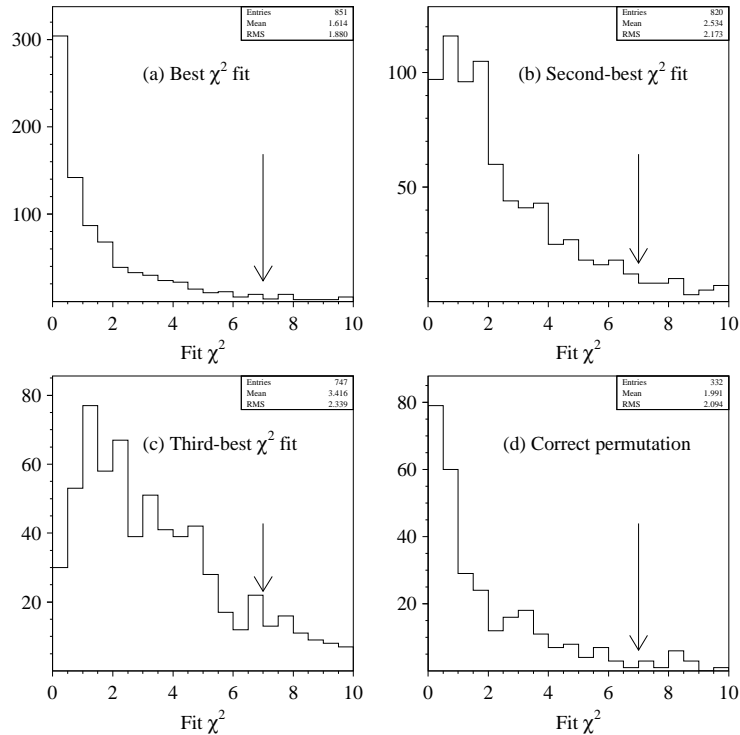}
\vskip 2cm
\caption{\protect \baselineskip 12pt
Fit $\chi^{2}$ distributions for (a) best, (b) second-best,
(c) third-best, and (d) the correct jet permutation.
For e$+$jets events, from the ISAJET Monte Carlo, 
M$_{top}$=180 GeV/c$^{2}$.  The arrow shows the cut value. Note that the plot
of correct permutations has less than half as many entries because 
only the cases in which a correct permutation could be found unambiguously
from the Monte Carlo are included.
From the D0 collaboration, Snyder, 1995a.}
\label{D0chi}
\end{figure}

The large number of possible jet-quark assignments, 
the poor jet energy resolution (see Section~\ref{quarkdetect} and
Fig.~\ref{jetres}),
and the effects
of initial and final state gluon radiation,
greatly complicate the top quark mass measurement.
Monte Carlo studies indicate that there is often at least one
combination with incorrect
quark-jet assignments which yields a better fit to the $t\bar{t}$
hypothesis than the combination with the correct assignment,
(see Fig.~\ref{D0chi}).  
Gluon radiation presents a problem because it
can give rise to additional jets in the event.
Both CDF and D0 consider only the four highest E$_{T}$ jets in the
event, since inclusion of a fifth jet would increase
the number of possible combinations by a factor of five.
However, if one of these four jets is from gluon radiation,
the constrained fit will be operating on
the wrong objects.  Because of the poor energy resolution,
the goodness-of-fit variable is not very effective at eliminating
this kind of event from the data sample.
An additional effect of gluon radiation is that the lepton $+$ 4
jets sample also includes $t\bar{t}$ events
from the {\em wrong} decay mode, for instance
events of the type
$t\bar{t} \rightarrow l \nu b~\tau \nu \bar{b}$
can pass a lepton $+$ 4 jets selection
if both $b$-jets are found,
the tau decays hadronically and is reconstructed as a jet,
and an additional gluon jet is present.

The size of these effects depends somewhat on the
details of the event selection.  In the CDF
analysis, Monte Carlo studies indicate that
approximately 7\% of $t\bar{t}$ lepton $+$ 4 jets events
are from the wrong decay mode, and of order 50\% of the
events have at least one of the four highest 
P$_{T}$ jets from gluon radiation. In the 
remaining events, the combination with the
lowest $\chi^{2}$ corresponds to the correct parton-jet
assignments only about one half of the times.

Constrained fits to incorrect
parton-jet assignments or to $t\bar{t}$ events from
the wrong decay mode,
in general yield incorrect values
of M$_{top}$.   
The CDF and D0 groups have chosen to deal
with the problem of combinatorics in slightly different ways.  In the CDF
analysis, only the lowest $\chi^{2}$ combination in a
given event is considered.  In the D0 analysis,
the top quark mass for a given event is taken as the 
$\chi^{2}$-weighted average of all combinations (up to three)
that have acceptable values of $\chi^{2}$.
The advantage of the D0 approach is that the M$_{top}$
value for a given event is more stable under small changes in the
measurements or the fitting procedure.  These changes
can cause the fit to converge to a different jet-parton 
configuration, and result in a value of M$_{top}$ which 
can be considerably different from the original.

\begin{figure}[htb]
\vskip 1cm
\epsfxsize=4.0in
\epsfysize=2.0in
\gepsfcentered[20 200 600 600]{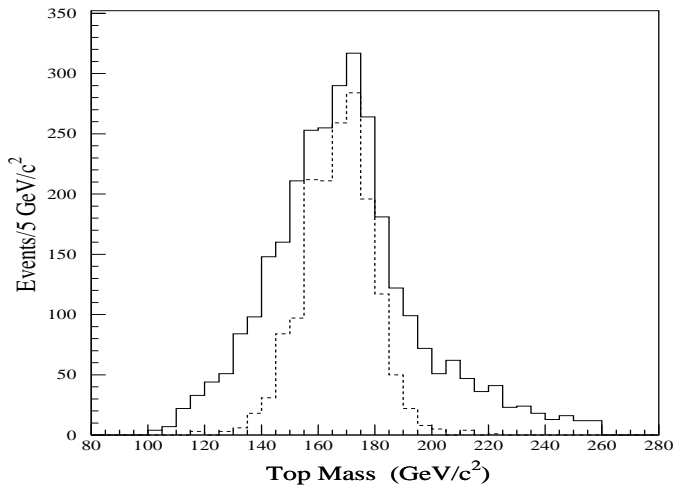}
\vskip 1cm
\caption{\protect \baselineskip 12pt
Reconstructed top mass distribution for Monte Carlo events
generated with the HERWIG Monte Carlo program, and simulated
with the CDF detector simulation.  The input value of the top
mass in the Monte Carlo is
M$_{top} = 170$ GeV/c$^2$.  The solid line corresponds to
the result of the constrained fit when requiring that one of the
$b$-jets is a $b$ in the fit. The dashed histogram
refers to the fit with the correct assignment for each of the jets.
The width of the top quark, which for this top
mass is $\Gamma$(t) $\approx$ 1 GeV/c$^{2}$
see Fig.~\protect\ref{topw}, is negligible compared to the experimental
resolution.
From the CDF collaboration, F. Abe {\em et al.}, 1994a.}
\label{CDF60}
\end{figure}

Wrong combinations result in 
a significant broadening of the expected mass resolution, see
Fig.~\ref{CDF60}.  In CDF, the mass resolution for the
correct jet assignment is expected to
be 12 GeV/c$^{2}$.  As a result of gluon radiation
and wrong parton-jet assignments, the mass resolution
is a factor of two worse, with significant
non-Gaussian tails. Very similar results have also been reported 
by the D0 collaboration.

Despite the effects of gluon radiation and
the high probability to choose a
wrong combination, the peak in the mass distribution
of Monte Carlo events is not shifted significantly,
(see Fig.~\ref{CDF60}).
This is partly because
of order one-half of the wrong combinations involve
interchange of one of the quarks from $W$ decay ($q$
or $\bar{q}$) with the $b$ from the hadronic top
decay ($b_{1}$, see Fig.~\ref{tpict}).  
For this class of events, the reconstructed
top mass distribution is broader but still peaks at the correct
value.

It may be possible to reduce the effect of wrong combinations
by including more information in the event fitting
procedure.  Examples of additional pieces of information that
could be included are
the expected angular distributions derived from the
V-A structure of the  top quark decay or 
the rapidity distributions of top quarks predicted by 
the $t\bar{t}$ cross-section calculation
(Kondo, 1991;
Goldstein, Sliwa, and Dalitz, 1993).


Another effect of the large number of combinations and the
poor jet energy resolution is that the $\chi^{2}$ goodness-of-fit
variable does not provide significant background rejection.
(If it did, this variable would have been used to separate 
the top signal from the background).
As discussed in Sections~\ref{bg},~\ref{ljtag},
and~\ref{ljkin}, the main background to the 
lepton $+$ jets top signal is due to $W +$ jets production.
In a $W +$ 4 jets event, it is almost always possible
to find one jet-quark assignment with $\chi^{2}$ 
low enough to be consistent with the $t\bar{t}$ hypothesis.
Therefore, in order to perform the top mass measurement,
both the background contamination of the event sample,
and the mass distribution of the background events,
need to be understood.

\subsubsection{CDF and D0 top mass measurements}
\label{likelihood}
The CDF and D0 top mass measurements are based on the constrained
fit procedure described in the previous Section.  
What is measured in these detectors is the energy and direction of 
the jets. In order to measure the top mass,
the jet energies must be corrected to infer the 
original momentum of the partons. The correction procedure
takes into account effects of
non-linearities in the hadron energy response of the calorimeters, 
underlying event contributions to the energy of the calorimeter cluster, 
and
gluon radiation outside the clustering cone.  Special corrections
are applied to lepton-tagged $b$-jets, since additional
energy is carried away by the neutrino emitted in the
semileptonic $b$-decay (typically a few GeV).

A value of M$_{top}$ is calculated for each event using the
constrained-fit procedure described above. Events are rejected 
if the fit $\chi^{2}$ is inconsistent with the 
$t\bar{t} \rightarrow$ lepton + jets hypothesis.  The top quark mass
is extracted by performing a likelihood fit to the M$_{top}$
distribution for the remaining events.  This M$_{top}$
distribution is fit
to the
sum of background and $t\bar{t}$ components, with the value of the top
quark mass allowed to vary.
The M$_{top}$ distribution for the
dominant background, ($W +$ 4 jets events),
is obtained by performing the same constrained
fit procedure on a sample of
events obtained from
the VECBOS Monte Carlo and the detector simulation.
The shape of the expected M$_{top}$ distribution, as a 
function of the mass of the top quark, is also taken from
Monte Carlo event generators (ISAJET, HERWIG, or PYTHIA) and the
detector simulation.  In the fit, 
the size of the background contribution can be constrained, within
errors, to its calculated value, or can be left free to float.  In the
latter case, the size of the background contribution
returned by the fit serves as a consistency check of the procedure.

\begin{figure}[htb]
\centerline{\psfig{figure=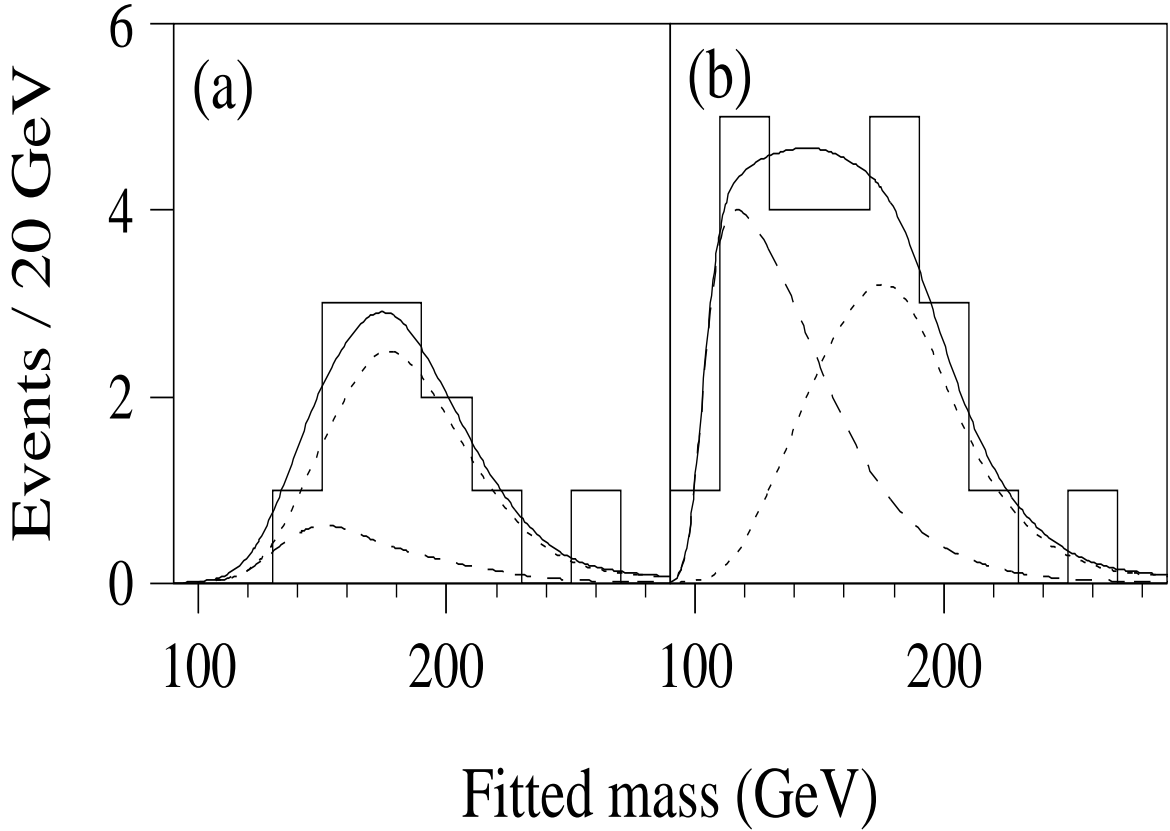,height=5.0in,width=7.0in}}
\caption{\protect \baselineskip 12pt
Mass distribution for D0 top candidate events (histogram)
compared with the expected mass distribution for 199 GeV/c$^{2}$
top quark events (dotted curve), background (dashed curve),
and the sum of top $+$ background (solid curve) for (a) standard
and (b) loose event selection.  $b$-tagging information is used when
available. From Abachi {\em et al.}, 1995b.  The integrated luminosity
is $\approx$ 45 pb$^{-1}$.}
\label{D0mass}
\end{figure}

The results of the likelihood fit from the D0 collaboration
are shown in Fig.~\ref{D0mass}(a) and~\ref{D0mass}(b) for two
different event selections, {\em tight} and {\em loose}.  In both cases
the four jets are required to have E$_{T} > 15$ GeV. 
The tight selection includes requirements on H$_{T}$ and aplanarity.
The tight sample includes the
eight lepton $+$ four jets events from the analysis described in
Section~\ref{ljkin}, as well as the 6 $b$-tagged
events from Section~\ref{ljtag}.  Only 11 out of these
14 events yield an acceptable fit to the $t\bar{t}$ hypothesis.
For the loose selection, the H$_{T}$ requirement
is removed and the aplanarity requirement is loosened.
The H$_{T}$ requirement selects events with high E$_{T}$ jets, and
introduces a bias that favors events with high reconstructed
top mass.  By removing this requirement, however, the background
contribution is enhanced.  The number of events increases from 
14 to 27; 24 of these have at least 
one combination with good $\chi^{2}$ (Snyder, 1995a).
As discussed in Section~\ref{detect}
and illustrated in Fig.~\ref{tj1}, in $t\bar{t} \rightarrow$
lepton $+$ jets events there is a significant
probability for two quarks in the final state to merge into a single jet.
To minimize this effect, the
jet-cone clustering radius ($\Delta$R, see Section~\ref{detect}) 
in the D0 mass reconstruction
is changed from the
0.5 used in the selection of the top signal to 0.3.
Likelihood fits to the two distributions result in top mass values
of M$_{top} = 199^{+31}_{-25}$ GeV/c$^{2}$ and
M$_{top} = 199^{+19}_{-21}$ GeV/c$^{2}$ for the tight and loose
selections respectively (statistical errors only).
The M$_{top}$ data distributions
are not well described by the background hypothesis
alone.  This provides
further kinematic evidence for the existence of the top quark.

\begin{figure}[htb]
\epsfxsize=3.0in
\vskip 1cm
\gepsfcentered[20 200 600 600]{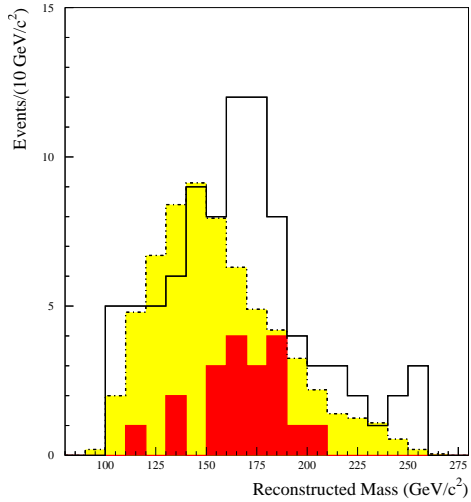}
\vskip 1cm
\caption{\protect \baselineskip 12pt
Mass distribution for the 88 CDF lepton $+ \geq$ 4 jets events
with a good $\chi^{2}$ fit to the $t\bar{t}$ hypothesis, before
the $b$-tag requirement (solid line). The darkly
shaded histogram is for the 19 events with a $b$-tag.
The expected $W +$ 4 jets
contribution to the pre-tag sample
is shown in the lightly shaded histogram.
From F. Abe {\em et al.}, 1995a. The integrated luminosity
is 67 pb$^{-1}$.}
\label{CDFPRL2}
\end{figure}

\begin{figure}[htb]
\epsfxsize=3.0in
\vskip 1cm
\gepsfcentered[20 200 600 600]{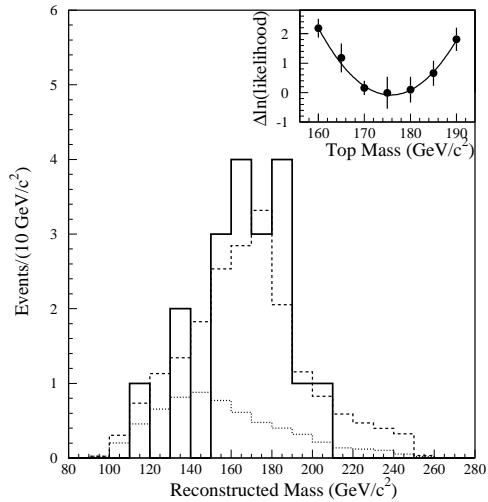}
\vskip 1cm
\caption{\protect \baselineskip 12pt
Mass distribution for the 19 CDF $b$-tagged
lepton $+ \geq$ 4 jets events with a good
$\chi^{2}$ fit to the $t\bar{t}$ hypothesis
(solid).  Also shown are the background shape (dotted), and
the sum of background and $t\bar{t}$ Monte Carlo
(dashed) for M$_{top}$ = 175 GeV/c$^{2}$.
The inset shows the likelihood fit used to determine the
top mass. From F. Abe {\em et al.}, 1995a. The integrated luminosity
is 67 pb$^{-1}$.}
\label{CDFPRL3}
\end{figure}

The CDF top observation in the lepton $+$ jets channel was based
on a sample of $b$-tagged events with at least three jets
of uncorrected E$_{T} >$ 15 GeV, clustered with a cone of
radius $\Delta$R = 0.4.  As it is clear from Fig.~\ref{tpict},
the constrained fit can only be applied to events with at least
four jets.  Therefore, the CDF
mass measurement
is performed on the sub-sample of events with a fourth jet.
To maintain high efficiency, the E$_{T}$ threshold on the fourth
jet is lowered from 15 to 8 GeV (uncorrected).   We stress that 
uncorrected jet energies are
used at the event-selection stage only;  for fitting purposes,
all jet energies are corrected, see the discussion in 
Section~\ref{systematics}.  
In Fig.~\ref{CDFPRL2} we show
the M$_{top}$ distribution of the pre-tag CDF lepton $+$
4 jets sample.  Based on the CDF $t\bar{t}$ cross-section measurement,
this sample is expected to be a mixture of approximately
30\% $t\bar{t}$ and 70\% $W +$ jets.  The probability for the shape 
of the data 
M$_{top}$ distribution to be consistent with background only is 
approximately 2\%.  Hence, just as in the D0 case, this distribution
provides additional evidence for the top quark.
By demanding a $b$-tag, the bulk of events
at low M$_{top}$ is removed, leaving a cluster of events 
between M$_{top}$ = 150 and 210 GeV/c$^{2}$, see 
Fig.~\ref{CDFPRL3}.  The result of the likelihood
fit to these events is 
M$_{top} = 176 \pm 8$ GeV/c$^{2}$ (statistical error only).

When including systematic effects, the mass values reported
by the two collaborations are 
M$_{top} = 199^{+19}_{-21}$(stat.)$\pm 22$(syst.) GeV/c$^{2}$ (D0) and
M$_{top} = 176 \pm 8$(stat.)$\pm 10$(syst.) GeV/c$^{2}$ (CDF).   The results
are consistent with each other.
In the following Section we will discuss the important issue of
systematic uncertainties.

\clearpage

\subsubsection{Jet energy corrections and systematics on the M$_{top}$ 
measurement}
\label{systematics}
Although the data samples of top candidates
are rather small, the size of the
statistical error on the top quark mass measurement is
already comparable to the size of the systematic uncertainties.
It is expected that in the next decade much larger 
$t\bar{t}$ samples will be available, see Section~\ref{future}.  
The systematic uncertainties will then be the limiting factor in 
the precision of the top mass measurement.  

The dominant uncertainty in the top mass measurement
is due to the uncertainty in the jet energy scale,
i.e. to the transfer function ({\em correction factor})
that relates the measured jet
energies to the energies of the original quarks from top decay.
This uncertainty has two components, (i) an instrumental
uncertainty related to the response of the calorimeter
to hadrons, and (ii) an uncertainty in the understanding
of the fragmentation and gluon radiation processes.
In order to discuss these systematic uncertainties, we
begin by describing the many steps in the
jet energy correction procedure (F. Abe {\em et al.}, 1992d
and 1993c; Abachi {\em et al.}, 1995d).

The jet energy response is in general not uniform
across the detector because, for example, of cracks at the
boundaries between calorimeter modules.
The first task of the correction procedure is to equalize response
across the calorimeter.
The size of the effect is measured {\em in-situ}
by di-jet or photon balancing techniques.  Di-jet
balancing is performed on a sample of 
$p\bar{p} \rightarrow$ 2 jets events, with one jet
restricted to a well-understood region of the calorimeter.
Since the transverse energies of the two jets are expected to
be equal, any inbalance as a function of the position
of the second jet is a measure of the position-dependent
non-uniformity of the calorimeter response.  A similar
study can be performed using $p\bar{p} \rightarrow \gamma +$ jet
events.  In these events the accurate measurement of the
E$_{T}$ of the photon can be compared to the measurement of
the jet E$_{T}$.
These effects can be measured, and hence corrected
for, with high precision, because of the very
large number of di-jet and photon-jet
events that can be used for this study.

\begin{figure}[htb]
\epsfysize=4.0in
\gepsfcentered[0 20 522 550]{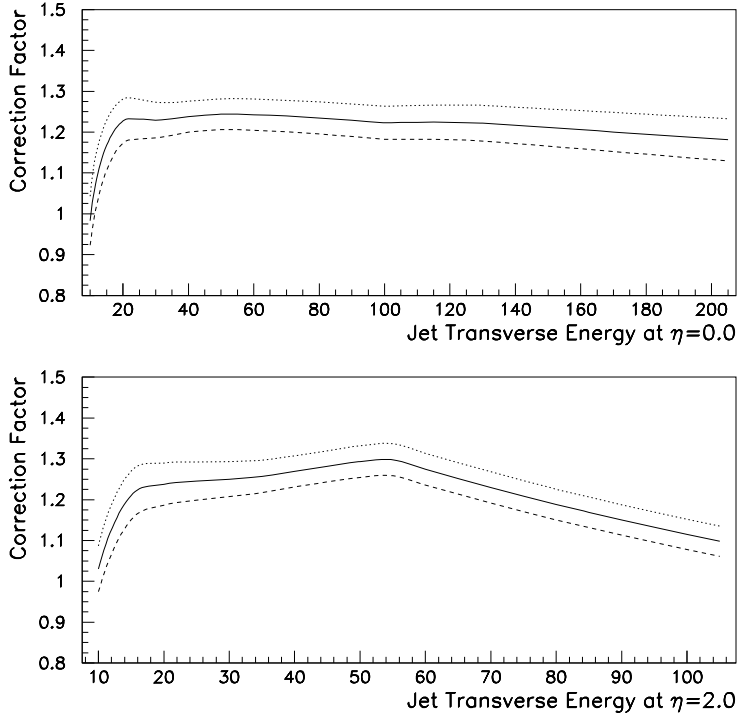}
\caption{\protect \baselineskip 12pt
D0 energy scale correction for jets as a function of jet
transverse energy in the central and forward regions.
Results are for jets reconstructed using a cone-size
$\Delta$R$=$ 0.5.  The dashed curves represent the error bands.
From Abachi {\em et al.}, 1995d.}
\label{D0_JCORR}
\end{figure}

\begin{figure}[htb]
\epsfxsize=4.0in
\gepsfcentered[27 158 522 644]{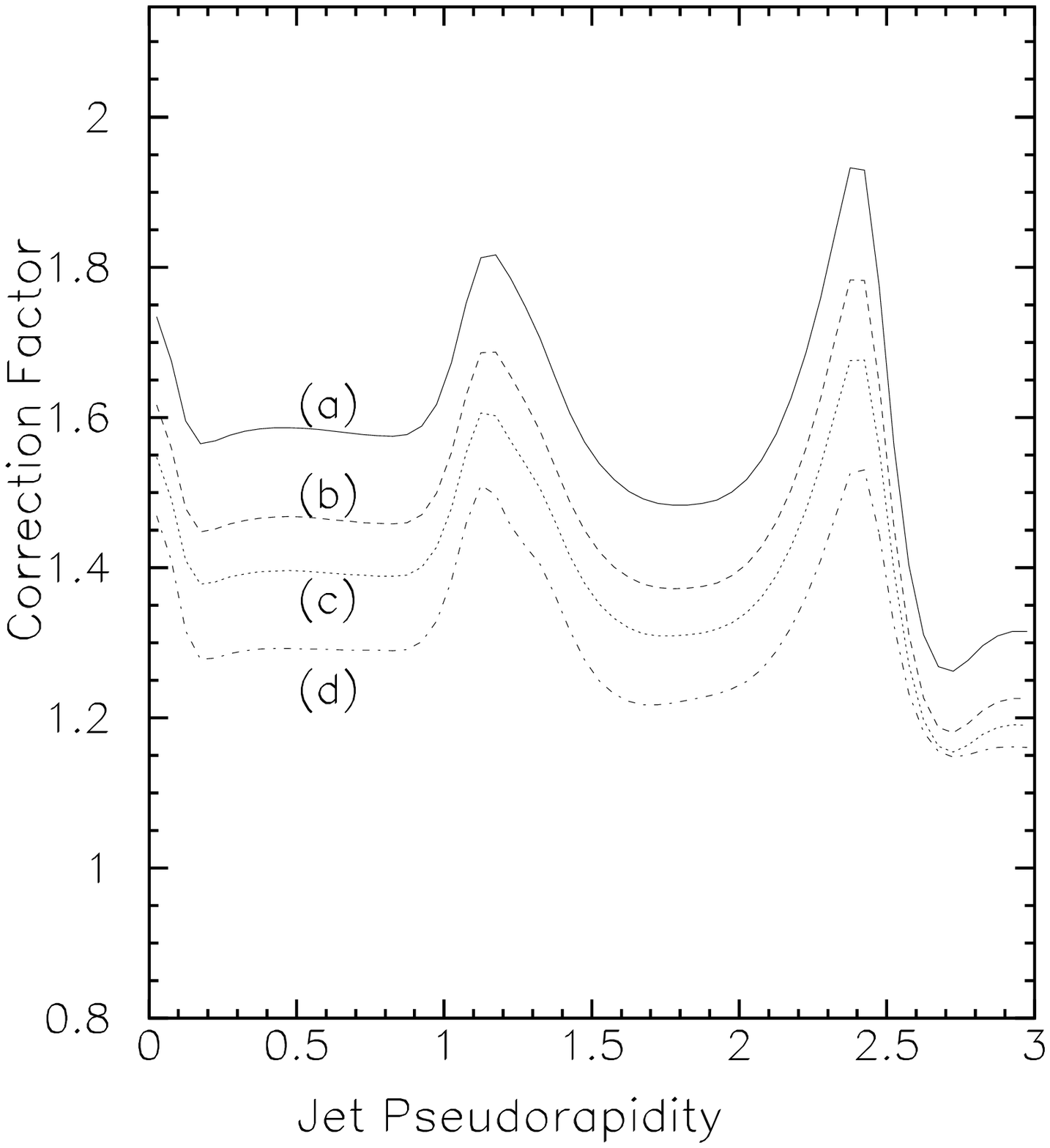}
\vskip 0.5cm
\caption{\protect \baselineskip 12pt
CDF energy scale correction for jets as a function of jet pseudorapidity.
Results are for jets reconstructed using a cone-size
$\Delta$R $=$ 0.4.
(a) observed E$_{T}$ = 15 GeV;
(b) observed E$_{T}$ = 30 GeV;
(c) observed E$_{T}$ = 50 GeV;
(d) observed E$_{T}$ = 100 GeV.
The cracks between calorimeter modules (at $\eta \approx 0., 1.1, 2.4$)
are apparent.}
\label{qd}
\end{figure}

The jet cluster will in general include energy deposited from
particles unrelated to the parent parton, for example particles from
the underlying event.  The jet energy is therefore
corrected by subtracting
off the average underlying event deposition as measured
in minimum bias events.  For jets clustered with a
cone radius $\Delta$R $=$ 0.4, this correction amounts to
approximately 600 MeV for the CDF detector.
Note that in $t\bar{t}$
events the amount of extra energy can be higher, because
of cross-talk between the many jets in the final state.
An additional, 
Monte Carlo based, correction for this effect has been developed by the
CDF collaboration.  A further correction needs to
be applied in D0 due to noise from radioactivity of the
uranium plates used in the calorimeter.

The next ingredient in the jet energy correction procedure
involves understanding the absolute energy response of the
calorimeter.  The response to individual hadrons is measured
in test-beams.  In CDF the hadron response of the calorimeter
as a function of momentum is also obtained from
samples of isolated particles from $p\bar{p}$ collisions.
The detector simulation is adjusted to incorporate
information from these measurements.  Finally, the jet energy
scale is measured by simulating the calorimeter
response to jets generated using a QCD-based model of jet
fragmentation.

The absolute energy scale
can also be derived from photon balancing,
since the energy of the photon is precisely measured
in the electromagnetic calorimeter. 
The electromagnetic calorimeter is precisely calibrated
by studying the distribution of E/P for electrons (CDF), or
from reconstruction of the $Z \rightarrow ee$ resonance (D0).
As we will discuss shortly, the photon 
balancing technique simultaneously tests instrumental
as well as gluon radiation effects.

Finally, in order to go back to the original parton energy,
an additional correction has to be applied for 
energy radiated outside the clustering cone,
or carried away by low momentum hadrons swept away by the
magnetic field (CDF).
This correction is based on a Monte Carlo model
of the QCD process. 
The overall jet energy correction factors
are displayed in Fig.~\ref{D0_JCORR} and
Fig.~\ref{qd} for D0 and CDF respectively. 
The observed jet energies are to be multiplied by these
correction factors to obtain the corrected energy.

\begin{figure}[htb]
\epsfxsize=4.0in
\vskip 1cm
\gepsfcentered[27 158 522 644]{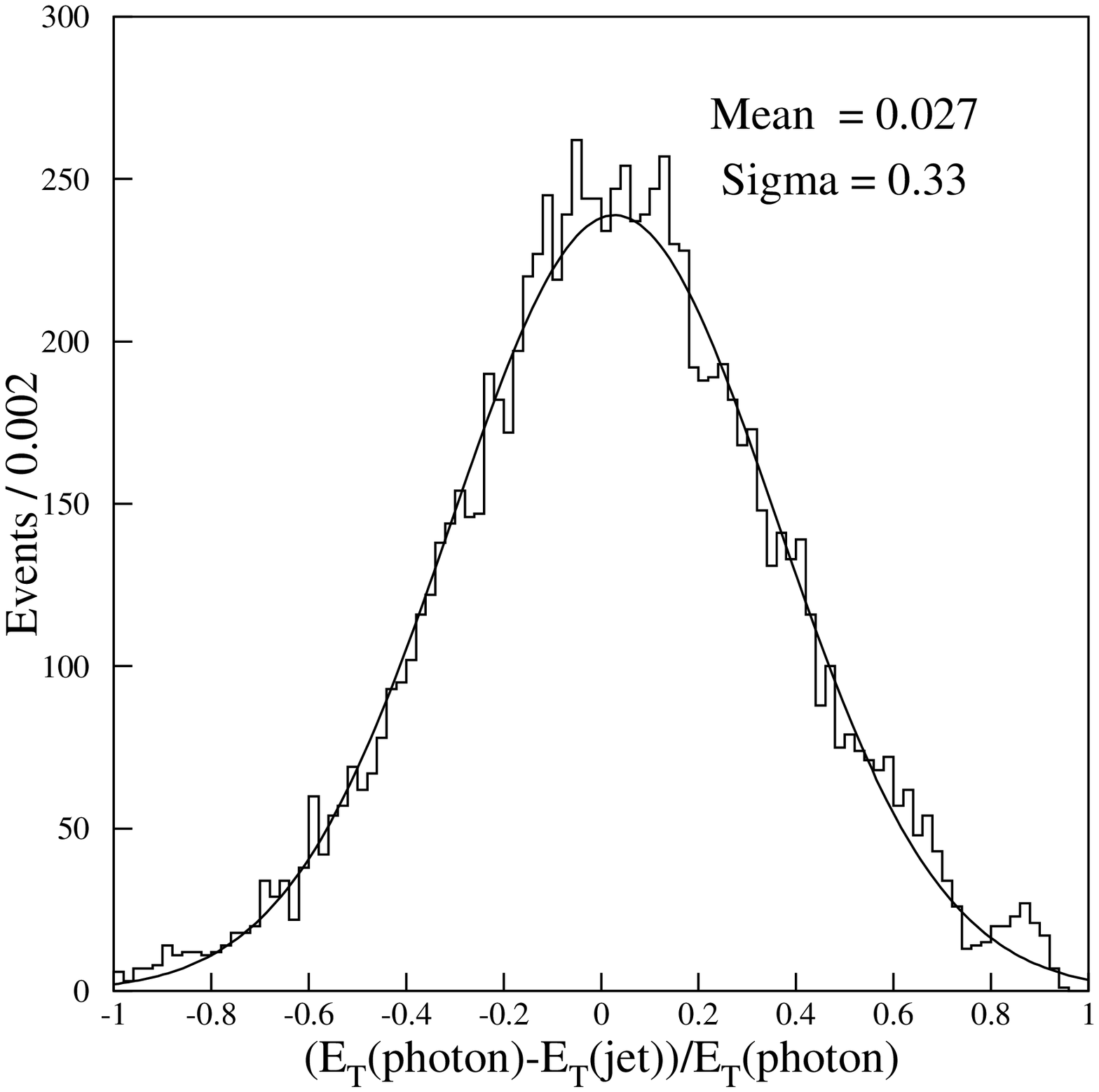}
\vskip 1cm
\caption{\protect \baselineskip 12pt
$\Delta =$ (E$_{T}$ (photon) - 
E$_{T}$ (recoiling-jet))/E$_{T}$ (photon). The jet
transverse energy is corrected.
Jets are clustered in a cone $\Delta$R $=$ 0.4.
From the CDF collaboration, F. Abe {\em et al.}, 1994a.}
\label{cdffig4}
\end{figure}

\begin{figure}[htb]
\epsfxsize=4.0in
\vskip-3cm
\gepsfcentered[27 158 522 644]{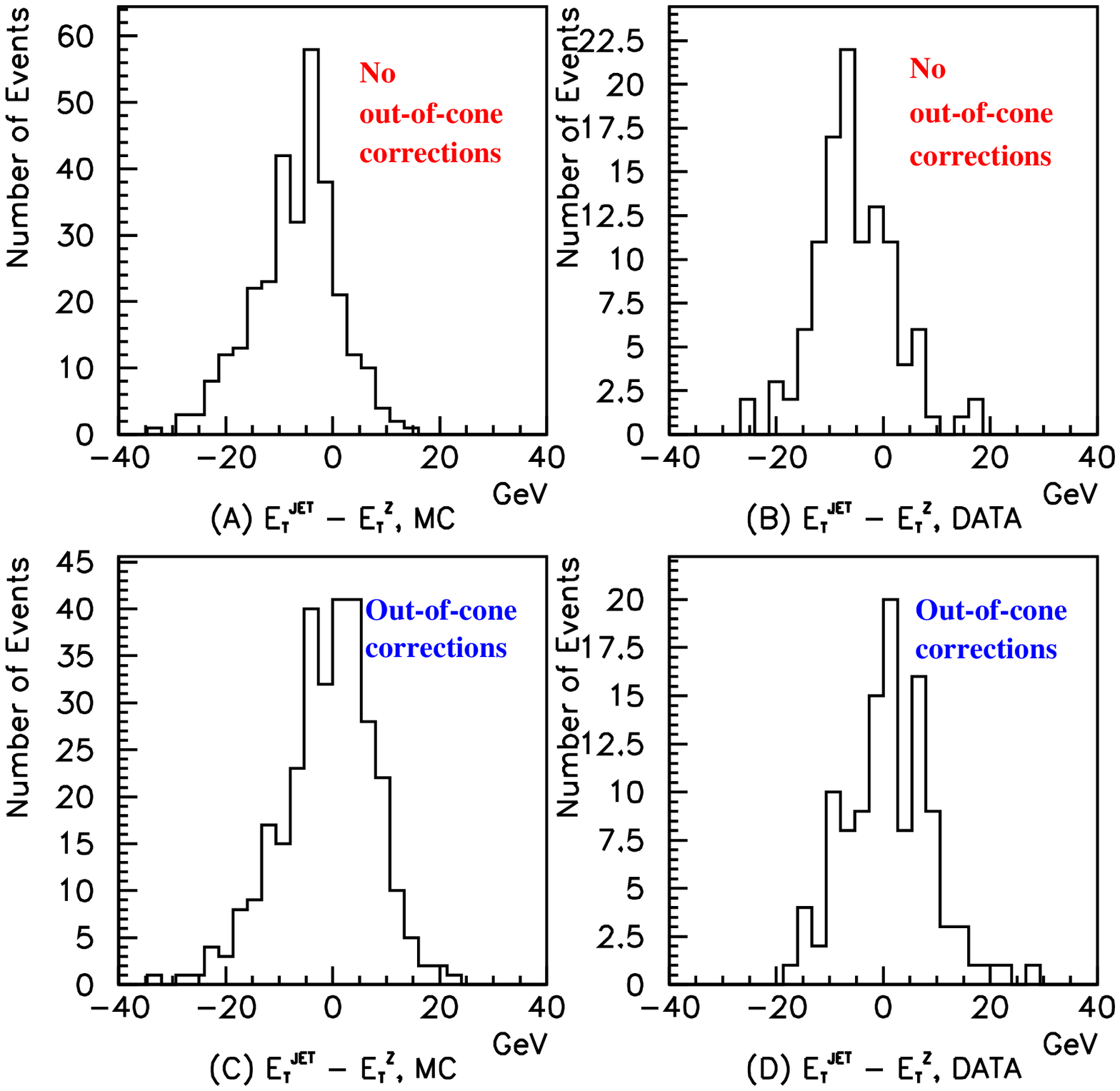}
\caption{\protect \baselineskip 12pt
Difference between the transverse energy of the electron
pair and the jet in ($Z \rightarrow ee$)~$+$~1 jet events. 
For Monte Carlo and data, with and without out-of-cone corrections,
but including all other corrections.  The minimum jet E$_{T}$ is 10 GeV.
Jets are clustered in a cone $\Delta$R $=$0.3.
From the D0 collaboration, Snyder, 1995a.}
\label{D0ZJET}
\end{figure}

Photon-jet balancing provides a powerful probe of
the behavior of the overall jet energy correction function,
(see Fig.~\ref{cdffig4}).  Unfortunately this test is not
free from its own systematic uncertainties.  Photon samples
collected at the collider are contaminated at about the 50\%
level by two-jet events, with one of the jets fragmenting to 
a high momentum
$\pi^{0}$ or $\eta$, which is misidentified as a single photon.
In these cases the transverse energy of the photon candidate
is not expected to equal the transverse energy of the recoiling
jet, because of the presence of additional fragmentation hadrons
in the jet that fakes the photon signature.
Furthermore the balancing results can be affected by
undetected low-energy initial-state gluon radiation.  It is
estimated that these effects introduce an uncertainty
of order 5\% on the determination of the absolute energy scale.
Similar studies are performed
using $Z +$~jet events, (see Fig.~\ref{D0ZJET}).  
The background effects which systematically limit the usefulness of
photon-jet balancing are not present in this case; however, the 
available statistics are considerably lower.

The correction procedure described above applies to gluon
and light-quark jets.  Monte Carlo studies show no significant
difference for $b$-jets, except in the case of 
semileptonic decays where an additional correction for the undetected
neutrino needs to be applied.

The uncertainty in the jet energy scale for 
both the CDF and D0 mass analyses is estimated to be 10\%.
This value is somewhat larger than one would infer from
photon-jet and Z-jet balancing studies, (see Fig.~\ref{cdffig4}
and Fig.~\ref{D0ZJET}).
There are however additional questions concerning the
applicability of this correction procedure to the
hadronic environment in top events which contribute
to the 10\% uncertainty estimate.
The 10\% jet energy scale uncertainty translates to
a top mass uncertainty of 8 GeV/c$^{2}$ and 21 GeV/c$^{2}$
for CDF and D0 respectively.  The difference
between the two experiments in 
the size of the reported M$_{top}$ uncertainties is not fully
understood at this time.

There are additional smaller systematic uncertainties on the top mass 
measurement.  A related uncertainty comes from the Monte Carlo modeling
of gluon radiation in $t\bar{t}$ events.
As was discussed in the previous Section,
gluon radiation broadens the reconstructed M$_{top}$ distribution.
Since the top mass is extracted from a fit to the M$_{top}$ data based on
the expected $t\bar{t}$ M$_{top}$ distribution, different
assumptions on its shape can result in shifts of the measured top
mass.  For example, differences between the ISAJET and HERWIG
models result in shifts of 4 and 1 GeV/c$^{2}$ for D0 and
CDF respectively.  
Further uncertainties at the 1-2 GeV/c$^{2}$
level are present in the CDF measurement
from uncertainties in the background shape, which
is taken mostly from the VECBOS Monte Carlo, as well as the details
of the likelihood fitting technique.

The CDF and D0 M$_{top}$ measurements are the first 
examples of the application of jet spectroscopy techniques
to the determination of the mass of an elementary particle.
The understanding and control of the systematic 
uncertainties due to the jet energy measurements are
expected to improve in the future.  This will be
crucial to allow for more precise measurements of the top
mass.  We will discuss a number of possible approaches in
Section~\ref{future}.

\subsubsection{Updated CDF and D0 top mass measurements}
\label{massupdate}
As this review article was being completed, both the CDF (Tartarelli, 1996)
and D0 (Narain, 1996) top mass results have been updated.
In both experiments, the statistical errors have decreased
due to the larger data samples and, more importantly, the systematic
uncertainties have been reduced as a result of more detailed
studies of the jet energy scales in the two experiments.
Since numerically the D0 result is somewhat different than
the 1995 result discussed in Section~\ref{likelihood}, 
we briefly include these updated results here.

\begin{table}[htb]
\begin{center}
\begin{tabular}{cc}
\hline \hline
Source & M$_{top}$ uncertainty \\
\hline
Jet energy scale (4\% $\pm$ 1 GeV) & 7 GeV/c$^{2}$ \\
Different Monte Carlo $t\bar{t}$ generators & 6 GeV/c$^{2}$ \\
Fitting                            & 3 GeV/c$^{2}$ \\
Background uncertainty             & 2 GeV/c$^{2}$ \\
\hline 
Total                              & 10 GeV/c$^{2}$ \\
\hline \hline
\end{tabular}
\end{center}
\caption{Systematic uncertainties in the updated D0 mass measurement.
From Narain, 1996.}
\label{d0newsyst}
\end{table}

\begin{table}[hbt]
\begin{center}
\begin{tabular}{cc}
\hline \hline
Source & M$_{top}$ uncertainty \\
\hline
Jet energy scale (detector effects) & 3.1 GeV/c$^{2}$ \\
Soft gluon effects                  & 1.9 GeV/c$^{2}$ \\
Hard gluon effects                  & 3.6 GeV/c$^{2}$ \\
Different Monte Carlo $t\bar{t}$ generators  & 0.9 GeV/c$^{2}$ \\
$b$-tagging bias                    & 2.3 GeV/c$^{2}$ \\
Background spectrum                 & 1.6 GeV/c$^{2}$ \\
Fit configuration                   & 2.5 GeV/c$^{2}$ \\
Likelihood method                   & 2.0 GeV/c$^{2}$ \\
Monte Carlo statistics              & 2.3 GeV/c$^{2}$ \\
\hline
Total                               & 7.1 GeV/c$^{2}$ \\
\hline \hline
\end{tabular}
\end{center}
\caption{Systematic uncertainties in the updated CDF mass measurement.
From Tartarelli, 1996.}
\label{cdfnewsyst}
\end{table}

The new D0 measurement is based on an integrated luminosity
of 100 pb$^{-1}$.  Besides the improvement in statistics,
there are four differences between the old and new D0 mass measurements:
(i) the selection of the sample on which the mass fit is performed has
been changed, (ii) an error in the out-of-cone jet energy corrections
has been fixed, (iii) the HERWIG Monte Carlo is used instead of ISAJET,
and (iv) the jet energy scale has been shifted downwards by 5\%.

The new sample is derived from the
loose lepton $+$ four jets sample, removing the aplanarity cut. 
The sample is then split into two pieces, those events with and without
a low P$_{T}$ muon $b$-tag. 
The untagged events must satisfy two additional requirements:
(i) the transverse energy of the leptonic $W$ must be greater than 60 GeV, 
and (ii) the pseudorapidity of the $W$ must be in the interval
$\pm$ 2.
The longitudinal momentum of the 
neutrino from the $W$ is chosen from two possibilities as the one having 
the lowest momentum.  Furthermore, the events are required to pass a 
kinematic liklihood test.  This was done in order to reduce background for the 
loose sample, while reducing the 
mass bias due to the previous H$_{T}$ requirement. No further requirements
are made on the sample with a $b$-tag, except for additional
improvements to the algorithm for removing
$Z \rightarrow \mu\mu$, and for the application of a tighter 
low P$_{T}$ muon selection.

\begin{figure}[htb]
\epsfxsize=4.in
\gepsfcentered[0 0 532 532]{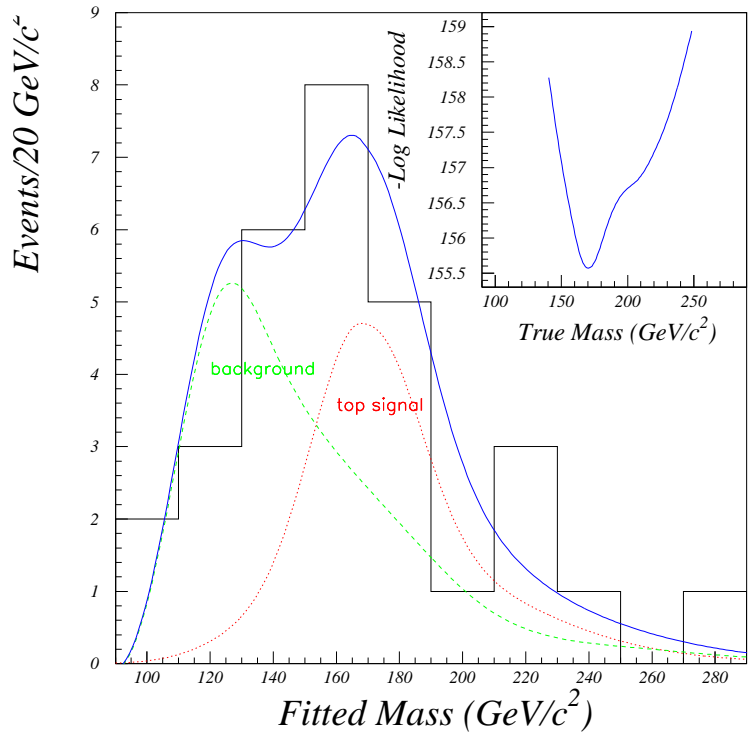}
\caption{\protect \baselineskip 12pt
Mass distribution for the 30 D0 lepton $+ \geq$ 4 jets events with a good
$\chi^{2}$ fit to the $t\bar{t}$ hypothesis, from the 100 pb$^{-1}$ data sample.
The solid line is a fit to top plus expected background; the dashed line is
the expected background from VECBOS (W $+$ jets) 
and QCD multi-jet background;
the dotted line is the fitted top contribution. The inset shows the liklihood
distribution. From Narain, 1996.}
\label{d02}
\end{figure}

This analysis results in 30 events which satisfy the fit $\chi^{2}$,
five of which have a $b$-tagged jet.   The background in this sample is
estimated to be
17.4 $\pm$ 2.2 events; the fit is shown in Fig.~\ref{d02}, and
gives 
M$_{top} = 170 \pm 15$(stat.) $\pm 10$(syst.) GeV/c$^{2}$.  
This is to be compared with the earlier result
M$_{top} = 199^{+19}_{-21}$(stat.)$ \pm 22$(syst.) GeV/c$^{2}$.
The systematic
uncertainties are summarized in Table~\ref{d0newsyst}.

The updated CDF measurement, based on an integrated luminosity of 
110 pb$^{-1}$, is performed in the same manner as described
in Section~\ref{likelihood}.  The result is 
M$_{top} = 176 \pm 6$(stat.)$ \pm 7$(syst.) GeV/c$^{2}$; the 
systematic uncertainties are summarized in Table~\ref{cdfnewsyst}.

The combined top mass measurement from the two Tevatron collaborations
is M$_{top} = 175 \pm 8$ GeV/c$^{2}$.  Here we have added the errors
in quadrature and neglected correlations in the systematic uncertainties.
These correlations are due, for example, to the modelling of 
$t\bar{t}$ production, and to the common assumptions made by the
two collaborations in the determination of the jet energy scale.

\clearpage

\subsection{Top mass from dilepton events and kinematic distributions}
\label{massindir}
The top mass can also be reconstructed, in a less direct way,
from dilepton events (Dalitz and Goldstein, 1992; Kondo, 1988 and 1991).
In a $t\bar{t}$ dilepton event, both top quarks decay semi-leptonicaly :

\begin{eqnarray*}
t_{1} \rightarrow W_{1} b_{1}~~~~~;~~~~~W_{1} \rightarrow l_{1} \nu_{1} 
\end{eqnarray*}
\begin{eqnarray*}
t_{2} \rightarrow W_{2} b_{2}~~~~~;~~~~~W_{2} \rightarrow l_{2} \nu_{2}
\end{eqnarray*}
Because of the presence of two neutrinos, a direct event-by-event 
reconstruction of the top mass based only on the measurements of the 
momenta of the leptons and the jets is not possible.  
The system is underconstrained, as can be seen from a simple accounting
of the degrees of freedom.  To fully describe the event, one needs
the momenta of all the quarks and leptons in the final state.
The momenta of the charged leptons are measured, and those of the
quarks are inferred from the jet energies.
Six parameters are needed to fully
describe the two neutrinos; only two measurements (from the two
components of \MET), and three constraints are available : 
Mass($l_{1} \nu_{1}$) = M$_{W}$;~~Mass($l_{2} \nu_{2}$) = M$_{W}$;~~
Mass($l_{1} \nu_{1} b_{1}$) = Mass($l_{2} \nu_{2} b_{2}$).  This leaves
6$-$5 = 1 parameter undetermined.
In order to measure the top mass, additional information need
to be included.  Possibilities include the
Standard Model V-A expectations
for the angular distributions in the top decay chain, or
theoretical expectations
for the kinematic properties
of the produced $t$ and $\bar{t}$.

The D0 collaboration (Snyder, 1995b) recently reported
a preliminary measurement of the top mass from dilepton events.
The D0 dilepton mass measurement uses five dilepton events.
It is based on a data sample with a higher integrated
luminosity than the D0 dilepton top search
described in Section~\ref{dildisc} which yielded three candidate events.

\begin{figure}[htb]
\hskip -1cm
\vskip  5cm
\epsfxsize=1.25in
\gepsfcentered[50 0 100 50]{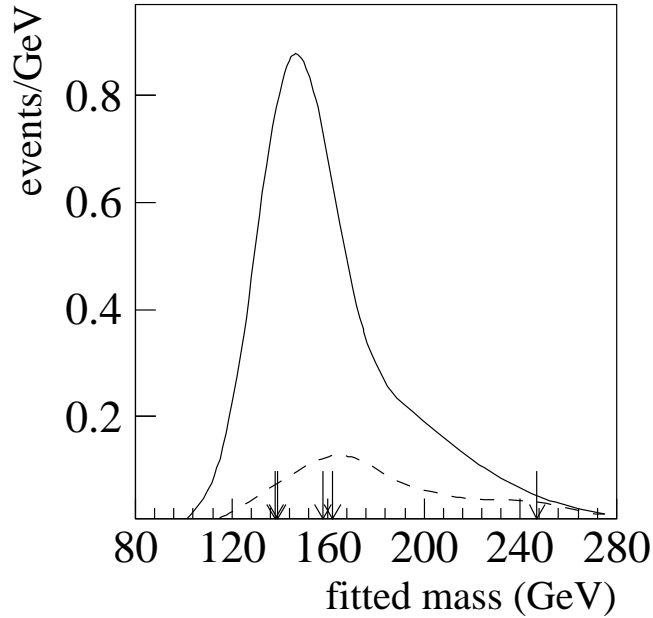}
\caption{Dilepton mass measurement from the D0 experiment.
The arrows show the M$_{peak}$ values for the  five D0 dilepton
candidate events. 
The solid curve is the expected signal distribution
for M$_{top}$ = 145 GeV/c$^{2}$, the dashed curve is the expected background
distribution.  See text for details.  From Snyder, 1995b.}
\label{d0dilmass}
\end{figure}

The top quark mass reconstruction for a single event goes as follows.
By {\bf assuming} a value of M$_{top}$, and for a given jet-$b$-quark
assignment,
the momenta of $\nu_{1}$ and 
$\nu_{2}$ are determined up to a possible four-fold ambiguity
from the two quadratic constraints 
M$^{2}$($l_{1}\nu_{1}$) = M$_{W}^{2}$
and M$^{2}$($l_{2}\nu_{2}$) = M$_{W}^{2}$.
For each configuration (two jet-$b$-quark assignments,
and possible four-fold neutrino ambiguity), the event is
fully reconstructed.
A probability (Prob$_{lep}$)
is assigned to each configuration
based on the energy of the leptons
in the rest frame of the top quarks.
This probability is calculated
for the assumed top mass, and is 
based on the expected structure of the decay.
The momentum fractions $x_{1}$ and $x_{2}$ of the incoming
quarks $q_{1}$ and $\bar{q}_{2}$ in the reaction
$q_{1} + \bar{q}_{2} \rightarrow t\bar{t}$ are also reconstructed from
the invariant mass (M$_{t\bar{t}}$) and momentum (P$_{t\bar{t}}$)
of the $t\bar{t}$ system.  Modulo the effects of gluon radiation,
$x_{1}$ and $x_{2}$ can be obtained from
M$_{t\bar{t}} = \sqrt{x_{1}x_{2}s}$ and 
P$_{t\bar{t}} = 0.5 \sqrt{s} (x_{1} - x_{2})$,
where $\sqrt{s}=$ 1.8 TeV is the center-of-mass energy
of the $p\bar{p}$ collision.
An additional probability (Prob$_{x}$)
is assigned to the configuration,
still as a function
of the assumed top mass, based on the parametrization
of the parton distribution functions at momentum
fractions $x = x_{1}$ and $x = x_{2}$, (see Fig.~\ref{pdf}).
Because
the parton distribution functions are decreasing
functions of $x$, Prob$_{x}$
includes a correction to remove
biases towards low values of M$_{top}$.  
The total probability for the configuration
is given by the product Prob$_{lep}$Prob$_{x}$.
The total probability for the event is
the sum of the probabilities for all possible configurations.
By varying the value of the assumed top mass,
this procedure yields a probability distribution for each 
event as a function of top mass.  
To include resolution effects,
the procedure is repeated many times after smearing the 
measurements of the lepton and jet momenta according to their
expected resolutions.  The total probability distribution
for a single event is
then defined as the sum of the probability distributions for
the smeared events. 
An event top mass value is
defined as the value of M$_{top}$ at which the                
probability distribution is maximum (M$_{peak}$).  Finally, the top
mass is extracted from likelihood fits of the data
M$_{peak}$ distribution to the superposition of the expected
$t\bar{t}$ and background contributions, (see Fig~\ref{d0dilmass}).

The preliminary D0 result for this procedure is 
M$_{top} = 145 \pm 25$(stat.)$\pm 20$(syst.) GeV/c$^{2}$.
The systematic uncertainties include energy scale effects, Monte Carlo
modeling, and uncertainties in the background contributions to the
fit.  Within the large errors, this value is consistent
with both the CDF 
(M$_{top} = 176 \pm 9$ GeV/c$^{2}$)
and D0
(M$_{top} = 170 \pm 18$ GeV/c$^{2}$)
measurements in the lepton $+$ jets
channel described in Section~\ref{massupdate}.

\begin{figure}[htb]
\epsfxsize=4.0in
\vskip 1cm
\gepsfcentered[20 200 600 600]{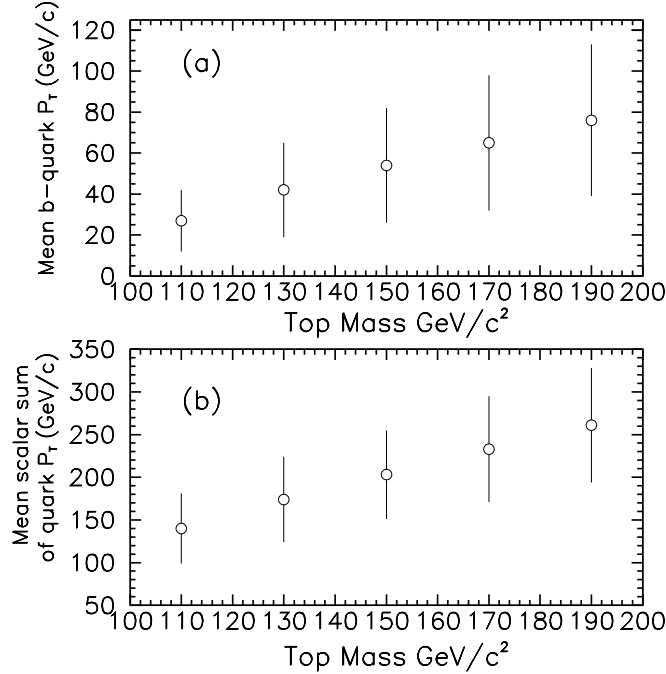}
\vskip 1cm
\caption{\protect \baselineskip 12pt
Predictions from the 
ISAJET $p\bar{p} \rightarrow t\bar{t}$ Monte Carlo at $\surd{s}$ = 1.8 TeV,
as a function of the top mass,
for (a) the mean transverse momentum of $b$-quarks, and
(b) the mean of the scalar sum of 
the transverse momenta of the four quarks in
lepton $+$ jets events.  The vertical error bars represent
the expected RMS for these quantities.}
\label{ptbmass}
\end{figure}

A number of kinematic quantities in both dilepton
and lepton $+$ jet events have also been shown to
be sensitive to the top mass (see for example Baer {\em et al.}, 1990).
It is then in principle possible to perform a measurement of 
the top mass by comparing kinematic distributions to expectations 
for $t\bar{t}$.
Some possibilities are the event transverse mass in dilepton
events, the mean of the lepton-$b$ mass, the transverse
momentum of $b$-jets, or the total transverse energy,
(see Fig.~\ref{ptbmass}).

These more indirect measurements are less statistically
powerful than the direct measurement of the top mass described
in Section~\ref{massdir}, since the mass peak used in the direct
measurement
provides optimal discrimination
between different M$_{top}$ hypotheses.  However, indirect
measurements are sensitive to different systematic 
effects.
For instance, the combinatorics problem which plagues the direct
top mass measurement, would not play a role in most of
these techniques.
A measurement of the top
mass based on the $b$-quark transverse momentum spectrum
or the $b$-lepton invariant mass, is
almost entirely independent of initial state radiation in 
$t\bar{t}$ events.  

The CDF collaboration has also reported measurements of the top mass
based on kinematic distributions
in both dilepton events (Tartarelli, 1996) and lepton $+$ jets events 
(F. Abe {\em et al.}, 1995c).
The dilepton mass determination is based on a fit
to the jet E$_{T}$ spectrum
in these events and yields
M$_{top} = 159^{+24}_{-22}$(stat.)$\pm 17$(syst.) GeV/c$^{2}$.
The lepton $+$ jet measurement is based on a fit to the 
the total transverse energy distribution
(see Section~\ref{ljkin}), and gives 
M$_{top} = 180 \pm 12$(stat.)$^{+19}_{-15}$(syst.) GeV/c$^{2}$.
Both these results are in agreement
with the results derived by direct mass
reconstruction techniques. 

The direct top mass measurement in $b$-tagged
lepton $+$ jets events
is, and is likely to remain, the method of choice for determining
the top mass.  Nevertheless, alternative methods such as
the under-constrained dilepton mass reconstruction or 
methods based on kinematic
distributions, serve as useful consistency checks.

\clearpage

\subsection{Reconstruction of the $W$ mass from hadronic decays in
lepton $+$ jets events}
\label{wevid}

In a $t\bar{t} \rightarrow$ lepton $+$ jet event, one of the two
$W$ bosons from the top decay chain is expected to
decay hadronically ($W \rightarrow q\bar{q}$, see Fig.~\ref{tpict}).
Both the CDF and D0 collaborations have reported evidence for
this decay mode by reconstructing a peak in the invariant mass
distribution of two jets in the event.  Hadronic $W$ reconstruction
is interesting for two reasons.  First of all, the presence of such a peak
in the data is further strong evidence that the excess of events
over the background prediction is indeed due to $t\bar{t}$ production.
Furthermore, the hadronic $W$
peaks provide the most ideal calibration for the jet energy scale of
the two experiments.
Note that it is extremely difficult
to get the scale information from inclusive $W$ decays to two jets,
because of the very high QCD di-jet background which would both
mask the signal and saturate the data acquisition bandwidth.

\begin{figure}[htb]
\epsfxsize=4.0in
\vskip 3cm
\gepsfcentered[20 200 600 600]{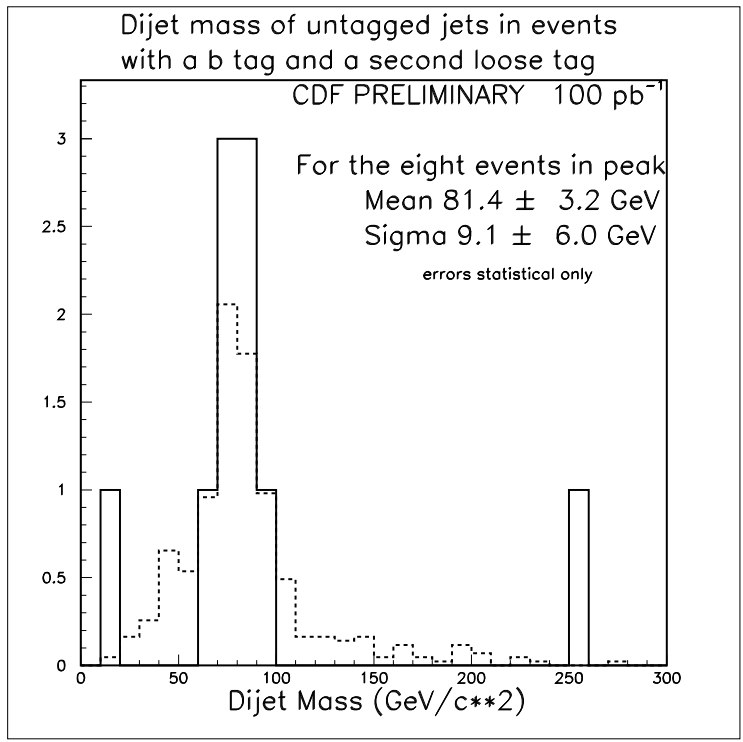}
\vskip 1cm
\caption{\protect \baselineskip 12pt
The di-jet mass of untagged jets in data events with a double $b$-tag
(histogram), compared to the expected signal from the top Monte Carlo
(dotted).
The integrated luminosity for this data set is 100 pb$^{-1}$.
From the CDF collaboration, Yao, 1995.}
\label{weiming1}
\end{figure}

The hadronic $W$ has been reconstructed by the CDF 
collaboration 
(Yao, 1995).
The 
method uses $W + \geq$ 4 jet events with
two $b$-tagged jets.  One of the jets is tagged using either of
the standard CDF lepton or vertex tagging algorithms.
In order to increase the statistics
of the sample,
the second jet is tagged using a looser vertex tag algorithm.
Such a double-tag requirement considerably reduces
many sources of background, and the $t\bar{t}$ purity
of the sample is very high.
Just as in the constrained fit procedure described
in Section~\ref{massdir}, only the four highest E$_{T}$ jets are
considered.  With this restriction, in a doubly $b$-tagged 
$t\bar{t}$ event there is no ambiguity in assigning
jets to the hadronic $W$-decay.   The invariant mass
distribution of the two untagged jets in this sample is
shown in Fig.~\ref{weiming1}.  In eight of the ten events,
the invariant masses of the two untagged jets cluster 
tightly around M$_{jj} \approx$ 80 GeV/c$^{2}$.  This provides
overwhelming evidence for the presence of hadronic decays of
$W$ bosons in the lepton $+$ jets $+$ $b$-tag sample. The two outlying 
events are
most likely due to cases where one of the untagged jets considered to be
from the $W$ is really a gluon jet from initial or final state radiation.



\begin{figure}[htb]
\epsfxsize=4.0in
\gepsfcentered[27 158 522 644]{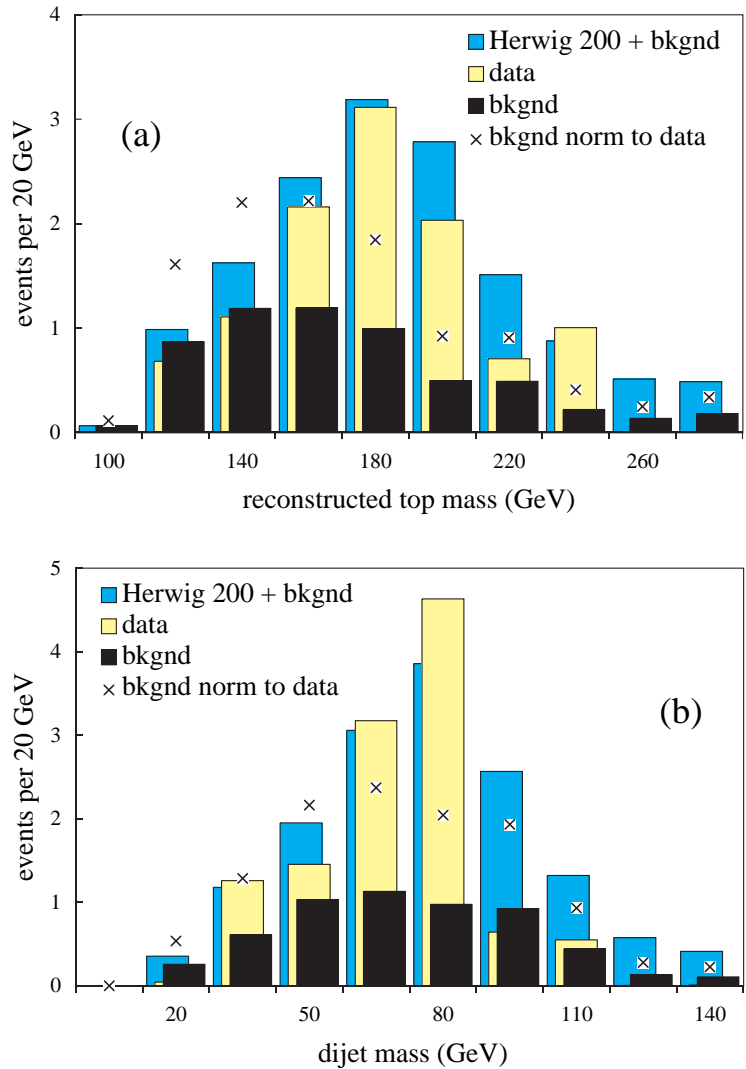}
\vskip 4cm
\caption{\protect \baselineskip 12pt
Distributions of (a) reconstructed top quark mass, and (b) reconstructed 
di-jet-mass.  The reconstructed top-mass is plotted only for
di-jet-mass $>$ 58 GeV/c$^{2}$.  The reconstructed di-jet mass
is plotted only for top mass $>$ 150 GeV/c$^{2}$.
Distributions are shown for data, sum of background and HERWIG
$t\bar{t}$ Monte Carlo, background alone, and background
normalized to match the area of the data.  From the D0
collaboration, Strovink, 1995.}
\label{strovink}
\end{figure}

The D0 hadronic $W$ mass reconstruction procedure (Strovink, 1995)  
in lepton $+$ jets events is much more complicated,
since the $b$-tag in D0 is not as efficient as in CDF.
The analysis uses lepton $+ \geq$ 4 jets data, and considers only
the four jets with the highest E$_{T}$.  Just as in the constrained
fit procedure (see Section~\ref{massdir}), the neutrino longitudinal
momentum (P$_{L\nu}$)
is obtained by constraining the lepton-neutrino invariant
mass to the $W$-mass, and only the solution with the smallest
$|$P$_{L\nu}|$ is considered.  No constraint is placed on the di-jet mass,
so that there are four different ways of partitioning the event into
($l\nu j_{1}$) and ($j_{2}j_{3}j_{4}$), 
unless one of the jets is $b$-tagged, in which
case the number of combinations is two.  A two-dimensional scatter-plot
of di-jet mass versus top mass is filled for each combination
with weight proportional to exp($-\chi^{2}/2$), where 
$\chi^{2} \equiv$ ln$^{2}$(mass($l\nu j_{1}$)/mass($j_{2}j_{3}j_{4}$)).  
The normalization
is chosen in such a way that the weights for all combinations in a 
given event sum to unity.  The top mass is defined as the 
average of mass($l\nu j_{1}$) and mass($j_{2}j_{3}j_{4}$) for
electron $+$ jets events.  For muon $+$ jets events, the
top mass is calculated by taking a weighted average of the two, with
the values of hadronic and leptonic
masses weighted in the ratio $60:40$.
The di-jet mass
is defined as follows.  If one of the jets in ($j_{2}j_{3}j_{4}$)
is tagged,
the di-jet mass is the invariant mass of the two other jets.
Otherwise, the most energetic jet in the top rest
frame is used as the $b$.  But if no jets in ($j_{2}j_{3}j_{4}$)
are
tagged, and their energy in the top rest frame are such that
(E$_{2}$ $-$ E$_{3}$) $<$ (E$_{3}$ $-$ E$_{4}$), then the permutation
is plotted twice, with equal weights, with di-jet mass chosen as
mass($j_{3}j_{4}$) or mass($j_{2}j_{4}$).  
Studies of $t\bar{t}$ Monte Carlo events, indicate that
this procedure results in a peak in the di-jet-mass vs.
top-mass scatter plot at di-jet-mass $=$ M$_{W}$ and 
top-mass $=$ M$_{top}$.  On the other hand, background events are
expected to peak at lower values.  Projections
from the di-jet-mass vs. top mass scatter plot are shown 
in Fig.~\ref{strovink}.  The top mass and di-jet mass
distributions are peaked around
180 and 80 GeV/c$^{2}$ respectively.   The probability
for these distributions to be consistent with the background-only
hypothesis is 1.3\%.

In conclusion, both collaborations have shown evidence for hadronic 
$W$ decays in their respective top event samples.  
With the aid of vertex tagging, the CDF $W \rightarrow q\bar{q}$ peak
is very clean and straightforward to understand.  
Future higher statistics samples of 
$W \rightarrow q\bar{q}$ in top events
will provide a very important calibration to the top mass measurement,
see Section~\ref{future}.

\clearpage

\section{Future prospects}
\label{future}

One of the goals of particle physics in the next decade is
to perform detailed experimental studies of the properties of the
top quark.   Since the mass of the top quark is of the same order 
as the scale for electroweak symmetry breaking, it is possible
that new physics effects will manifests themselves in the top
sector.  While we do not know what these new effects might be,
it is clear that top quark physics represents an opportunity
to uncover physics beyond the Standard Model.

In this Section we will concentrate on prospects for top physics
at the Tevatron.  We will begin by briefly summarizing plans for
the upgrade of the accelerator and the detectors
in Section~\ref{upgrade}.
Detailed studies of the possibilities for 
the Tevatron top physics program
are well under way 
(Amidei and Brock, 1996).  Prospects for improving the
accuracy of the top mass measurement will be discussed in
Section~\ref{massfut}; the potential for study of the
$Wtb$ vertex will be addressed in Section~\ref{tfwidth};
finally, additional tests of the Standard Model, as well as
searches for new physics involving the top quark will
be discussed in Section~\ref{new}.

\subsection{Accelerator and detector upgrades}
\label{upgrade}

The CDF (CDF collaboration, 1995) and 
D0 (Tuts, 1996) detectors at the Tevatron will
be undergoing major upgrades in the next few years.  The
most significant improvements for top physics will be the 
installation of a magnet for charged particle momentum determination  in D0 and 
new 3D silicon vertex detectors in both detectors.
Since vertex-tagging of $b$-quarks in top events has been shown
to be such a powerful tool, the capabilities
of the upgraded D0 detector for top physics will be considerably enhanced.
The new CDF silicon vertex
detector will also have a significant impact in the top
physics program at CDF.  With three dimensional information,
the $b$-tagging efficiency will be improved, and the instrumental 
background level will be reduced; furthermore,
the new vertex detector
will be long enough to cover the whole luminous region of 
the Tevatron, increasing the acceptance for vertex tags
(recall that the geometrical coverage
of the present CDF vertex detector is only about 60\%, see 
Section~\ref{svx}).  

At the same time, the Fermilab accelerator 
complex will be upgraded with the construction of a new high intensity
120 GeV proton accelerator ({\em The Main Injector}).
The Main Injector will replace the Main Ring, whose aperture
currently limits the luminosity,
as the 
injector to both the Tevatron and the $\bar{p}$ source. 
In addition,
a new 8 GeV permanent magnet ring ({\em Recycler}) has been proposed
to achieve a more efficient accumulation of antiprotons
by recycling (hence the name) the $\bar{p}$ from the previous
store, store the anti-protons when the accumulator is full,
and protect them from power glitches in the accelerator
(Foster, 1995; Jackson, 1995).
The number of bunches will also be increased from 6 to 36, allowing higher 
luminosity without increasing the number of interactions per crossing. This
increase shortens the bunch crossing interval from the current value
of 3.5 $\mu$sec
to 396 ns.  This requires 
an upgraded trigger, data
acquisition and 
front-end electronics systems,
which must be pipelined to handle the increased rate.
After completion of these upgrades in 1998-99, the Tevatron luminosity
will be increased by one order of magnitude to 2 x 10$^{32}$ 
cm$^{-2}$s$^{-1}$,  with further
luminosity improvements likely to occur in the following years.
The center-of-mass energy of the Tevatron is also expected
to increase from 1.8 TeV to 2 TeV, resulting in a 
increase of approximately 30\% in the $t\bar{t}$ cross-section. 

This series of improvements in the Fermilab collider program will allow
for much more detailed studies of the top quark than those
that are possible with the low statistics data samples that 
are available now.  The projected sizes of the $t\bar{t}$ 
samples in each experiment for an integrated luminosity of
1 fb$^{-1}$ will be of order 100 events in the dilepton
channel and of order 500 events in the lepton $+$ 4 jets channel
with one $b$-tag, with half of these events having both
$b$-jets tagged (Amidei and Brock, 1996).

In addition, towards the middle of the
next decade, the Large Hadron Collider (LHC) at CERN will
become operational.  The LHC is a very high luminosity
($ > 10^{34}$ cm$^{-2}$ s$^{-1}$) $pp$ machine with $\sqrt{s}$ = 14 TeV.
At this energy, $\sigma(pp \rightarrow t\bar{t}) \approx$ 700 pb, a factor of
100 higher than $\sigma(p\bar{p} \rightarrow t\bar{t})$
at the Tevatron.  With the high luminosity and
the high cross-section, the LHC can be considered to be a top-factory.
In the even more distant future, top physics will also be pursued at 
a very high energy electron collider (e.g. 
{\em The Next Linear Collider}, NLC), assuming that such a 
machine will be built.


\subsection{Improving the top mass measurement}
\label{massfut}
As was discussed in Section~\ref{ind}, the top quark mass is a fundamental
parameter of the Standard Model.  Its value enters in the calculation
of radiative corrections to a large number of electroweak observables.
It is therefore very important to measure the top quark mass as 
accurately as possible to allow for precise tests of the Standard Model.

The best value of M$_{top}$ as obtained from fits to the LEP and SLC
measurements, as well as the measurements of the $W$ mass from 
$p\bar{p}$ experiments and deep inelastic neutrino scattering,
is M$_{top} = 179 \pm 9^{+17}_{-10}$ GeV/c$^{2}$ 
where the second uncertainty comes from varying the Higgs mass
between 60 and 1000 GeV/c$^{2}$ (see Section~\ref{ind}).
This is in good agreement with the values reported by the
CDF (M$_{top} = 176 \pm 9$ GeV/c$^{2}$)
and D0 collaborations (M$_{top} = 170 \pm 18$ GeV/c$^{2}$),
see Section~\ref{mass}.

The accuracies of the neutrino measurements and
of the LEP and SLC measurements at the $Z$ are not
expected to dramatically improve in the coming years.
On the other hand, the accuracy in the determination of the
$W$ mass will improve by almost one order of magnitude, from
the Tevatron experiments as well as from $e^{+}e^{-} \rightarrow 
W^{+}W^{-}$ at LEP200.   Within the Standard Model, radiative 
corrections to the $W$ propagator 
(see Fig. 11)
result in definite predictions for the $W$ mass as a function
of the top and Higgs mass (see Fig.~\ref{topvsw}).  
It is therefore very interesting to measure both the $W$ mass and
the top mass as precisely as possible.  

How accurately can the top mass be measured?  Experience from
CDF and D0 indicates that the method of choice for measuring the top mass
is to perform constrained fits on the $b$-tagged lepton
$+$ jets data sample, see Section~\ref{massdir}.  We can extrapolate
the statistical accuracy of a future top mass measurement from the 
present CDF measurement.  This measurement has a statistical 
uncertainty of 6 GeV/c$^{2}$ for an integrated luminosity
of 110 pb$^{-1}$.  After the first Main Injector run of the
Tevatron, we can expect the integrated luminosity to be
of order 1 fb$^{-1}$ per experiment, with a 30\% increase in top cross-section
from running at the higher center of mass energy of 2 TeV, and
a 40\% increase in geometrical
acceptance from the new CDF vertex detector.
Since the statistical uncertainty varies inversely as the square root
of the number of events, we can expect a statistical 
uncertainty of order 1.5 GeV/c$^{2}$ from CDF.  A similar
uncertainty can be expected from the upgraded D0 detector.  
Systematic effects, which at present are at the level of 7 GeV/c$^{2}$,
will be the limiting factor in the precision
of the top mass determination.

The systematic uncertainties in the top mass measurements have
been described in Section~\ref{systematics}.  The largest uncertainty
is due to the understanding of the jet energy scale, as well as the
related issue of additional jets from gluon radiation.
At this time, it is not entirely clear what the ultimate precision
will be.  The dominant component of the energy scale
uncertainty is related to the reliability of the extrapolation from 
jet energies to parton energies, and it is the understanding of the QCD
process, rather than the understanding of the detector, which limits
the measurement.  Higher statistics tests of the understanding of 
the energy scale will be performed in
$\gamma$ $+$ jet and $Z$ $+$ jet events (see Section~\ref{systematics},
Figs.~\ref{cdffig4}, and~\ref{D0ZJET}).  The Monte Carlo
modeling of gluon radiation will be more precisely
checked and/or tuned 
by examining the energy flow within a jet.
There will however remain systematic uncertainties 
related to the transfer of this calibration from the control samples
to the hadronic environment in $t\bar{t}$ events.  The 
ultimate size of these
uncertainties is at the moment not well understood.

The additional statistics that will become available will be
important to reduce the systematic uncertainty.  With enough
statistics, the number of lepton $+$ jets events with two $b$-tagged
jets will be sizable, and these events will provide a very important
tool for understanding systematic uncertainties in the 
top mass measurement.
In these events, there are no ambiguities in assigning
the two jets to the hadronic $W$ decay.  The invariant mass
of these two jets, which ideally should reconstruct to the $W$ mass,
provides an in-situ calibration of 
$W \rightarrow$ jet-jet invariant mass
reconstruction.  First results from
this kind of study are very promising, (see Fig.~\ref{weiming1}).
Besides the energy scale issue, studies of doubly-$b$-tagged
events will provide useful handles on other systematic effects
that limit the precision of the top mass measurement.
Since mis-tag backgrounds in this data sample are very small,
events in the tails of the jet-jet
invariant mass distribution will be $t\bar{t}$
events with one jet from gluon radiation and W$Q\bar{Q}$
events.  Therefore doubly-$b$-tagged
events will be useful in directly measuring
these components of the data set.  This will improve the
understanding of the top resolution function, which 
also depends on the number of jets from gluon
radiation in the sample.  The doubly-$b$-tagged sample
will also allow for a test
of the modeling of the $WQ\bar{Q}$ background component, which affects
the top mass measurement since the top mass is extracted from
a likelihood fit of the data to the sum of $t\bar{t}$ and $W$
background.
The number of jet-parton
combinations for doubly-$b$-tagged events is only four, as opposed
to twelve combinations that must be considered in the
present CDF analysis which only demands one $b$-tag,
see Section~\ref{fits}.
Since statistics are
not expected to be the limiting factor, it is possible that 
a measurement of the top mass using these
events will be ultimately more accurate than a measurement
based on
events with one $b$-tag.  Alternatively,
these events can be used to study the effect of wrong combinations.
Using
these events, it will also become possible to 
measure the probability for
the constrained fit to
converge to a combination with a $b$-tagged jet assigned
to a light quark jet.

Further improvements in the 
understanding of the systematic uncertainties will
be possible due to the high statistics $t\bar{t}$ samples that will be 
available.  It will become possible to better quantify the effects of 
gluon radiation by measuring the relative rates of
$t\bar{t} \rightarrow$ dilepton $+$ 2 vs. 3 jets, and
$t\bar{t} \rightarrow l +$ 4 vs. 5 jets.  The modeling of the
$W +$ jets background, which at present is entirely based 
on the VECBOS Monte Carlo, will be tested using
the large sample of $Z +$ jets.

\begin{figure}[htb]
\epsfxsize=4.0in
\gepsfcentered[27 158 522 644]{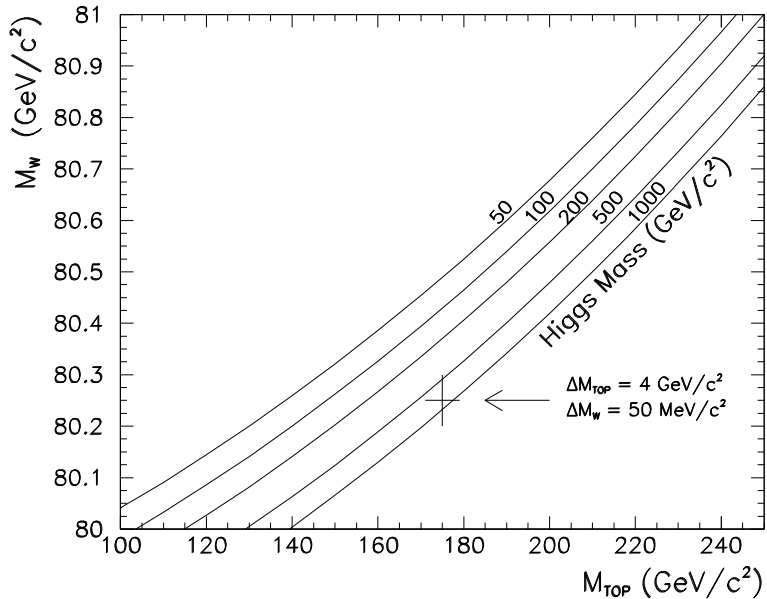}
\caption{\protect \baselineskip 12pt
The expected correlation between the masses of the top, the $W$
and the Higgs.  We also show, at an arbitrary point, 
results of measurements of the $W$ mass to 50 MeV/c$^{2}$ and the
top mass to 4 GeV/c$^2$.}
\label{MWMTOPFUTURE}
\end{figure}

In summary, while the ultimate precision of the top quark mass measurement 
is not fully known at this time, it is likely that an accuracy of
order of a few GeV/c$^{2}$ will be achievable.  
In conjunction
with a $W$ mass measurement with a precision of tens of MeV/c$^{2}$,
this measurement will be sensitive to physics beyond the Standard
Model, and will provide useful information on the value of the
Higgs mass.  In Fig.~\ref{MWMTOPFUTURE}
we show what the M$_{top}$ vs. M$_{W}$ measurements might look
like by the year 2000.  
If the experimental
point in the M$_{W}$ vs. M$_{top}$ plane 
was to fall outside the region allowed by the Standard Model,
this kind of measurement would provide indirect evidence for 
physics beyond the Standard Model. 

\subsection{Probing the $Wtb$ vertex}
\label{tfwidth}
The structure of the $Wtb$ vertex can be probed by studying
top decay and/or production of single top quarks.
Up to this point we have mostly discussed $t\bar{t}$ pair production.
However, as mentioned in Section~\ref{prod}, top quarks in
$p\bar{p}$ collisions can also be produced singly, in the
Drell-Yan process $q\bar{q} \rightarrow W^{*} \rightarrow t\bar{b}$,
(see Fig. 17), 
and in the $W$-gluon fusion process
$qg \rightarrow t\bar{b}q'$, or
$qb \rightarrow tq'$, 
(see Fig. 18).
The expected cross-sections for Drell-Yan and $W$-gluon fusion 
single top production at the Tevatron are displayed in Fig.~\ref{xsec}.
These cross-sections are smaller than the strong 
$p\bar{p} \rightarrow t\bar{t}$ production cross-section, but they
are sizable enough that single top production is expected to
be observable at the Tevatron.
Production of $tW$ is also allowed, but its cross-section is 
much smaller.

The potential for studying single top production 
has attracted a lot of attention for
the future top physics program at the Tevatron.
The cross-section is proportional to the top quark 
width, $\Gamma$($t \rightarrow Wb$), and within the
context of the Standard Model, proportional to the square
of the CKM-matrix element $|V_{tb}|$.   Note that the
top lifetime is too short to be directly measured,
(see Fig.~\ref{topw}); furthermore, the top width is very hard
if not impossible to measure
from the reconstruction of the Breit-Wigner, since the 
experimental resolution is one order of magnitude
worse than the width itself (see Fig.~\ref{CDF60}).
Hence, only indirect measurements of $\Gamma$($t \rightarrow Wb$)
and $|V_{tb}|$ can be performed.

Assuming three
generations and unitarity
of the CKM matrix, the value of $|V_{tb}|$ is expected 
to be very near unity, in the range 0.9988$-$0.9995 (Montanet
{\em et al.}, 1994).  It is clearly very interesting to 
test this result.  An additional potentially interesting 
measurement using the single top sample
would be a comparison of the rates for
$p\bar{p} \rightarrow tX$ and $p\bar{p} \rightarrow \bar{t}X$, which
can be used to search for CP violating effects in the top sector.

The possibility of extracting a single top-quark
signal has been examined by many authors 
(Yuan, 1990; Cortese and Petronzio, 1991;
Jikia and Slabospitsky, 1992; Ellis and Parke, 1992;
Carlson and Yuan, 1993; Stelzer and Willenbrock, 1995;
Amidei and Brock, 1996).  
Because of the high multi-jet QCD background, only events
with $t \rightarrow Wb$ followed by
$W \rightarrow l\nu$ are useful.
The signature then is a lepton, missing energy, and two $b$-jets
for Drell-Yan $q\bar{q} \rightarrow W^{*} \rightarrow t\bar{b}$ production,
and one or two $b$-jets $+$ one light quark jet for $W$-gluon
fusion, ($qg \rightarrow t\bar{b}q'$ or
$qb \rightarrow tq'$, see Fig. 18).
Just as in the $t\bar{t}$ search, the main background is
from $W +$ jets production (see Section~\ref{bg}), and $b$-tagging
must be used to reduce this background to a manageable level.
Rejection of the tagged $WQ\bar{Q}$ background can be achieved by requiring
the mass of the lepton, neutrino, and $b$-jet to reconstruct to the
known top mass.  Pair production of $t\bar{t}$ also constitutes a 
significant background to observation of single top production.  
Since the jet multiplicity in $t\bar{t}$ events 
is higher, the optimal sample in which to
isolate the single top signal seems to be that of events
with one lepton $+$ \MET $+$ two and only two jets.

\begin{figure}[htb]
\epsfxsize=4.0in
\hskip 3cm
\epsffile{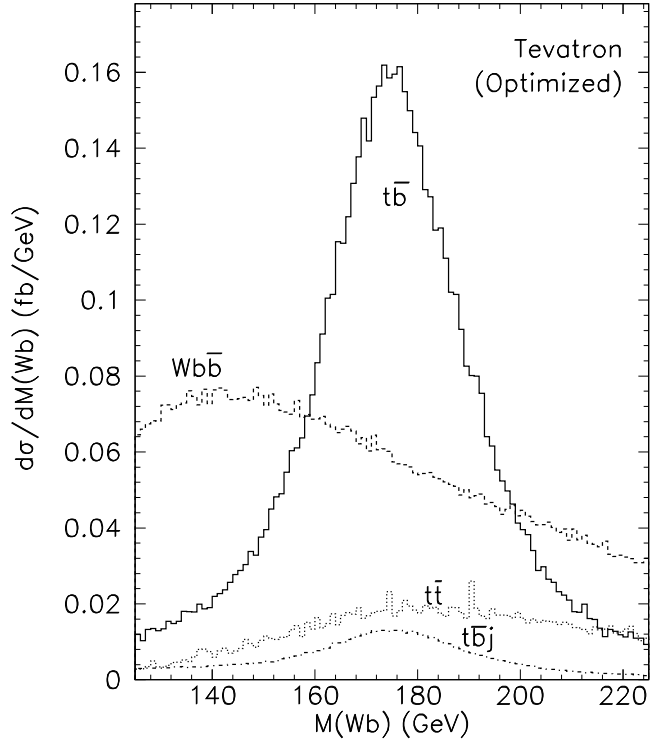}
\vskip -1cm
\caption{\protect \baselineskip 12pt
Expected observed cross-section for 
$p\bar{p} \rightarrow W^{*} \rightarrow t\bar{b}$ ,
$t \rightarrow Wb$, $W \rightarrow l\nu$
as a function
of the invariant mass of the $Wb$.  Also shown are expectations for
the most important backgrounds ($t\bar{b}j$ denotes $W$-gluon fusion).
This analysis is based on the $W +$ 2 jets sample.
From Stelzer and Willenbrock, 1995.}
\label{willenbrock}
\end{figure}

A study of the expected signal and background for single top 
production at the Tevatron
shows that a signal-to-background of order
1 to 2 can be achieved.  The number of signal events 
for $|V_{tb}| \approx$ 1
would be of order
120 per fb$^{-1}$, resulting in a statistical uncertainty
in the cross-section measurement of order 17\% in
a 1 fb$^{-1}$ data set (Amidei and Brock, 1996).
An additional study, optimized for the detection of the Drell-Yan 
$q\bar{q} \rightarrow W^{*} \rightarrow t\bar{b}$ process, 
suggests that with the requirement of a double $b$-tag the
signal can be isolated, (see Fig.~\ref{willenbrock}
Stelzer and Willenbrock, 1995).  Indications from this study are that,
assuming
that $|V_{tb}|$ is indeed close to unity, a 3 fb$^{-1}$ exposure
would yield a 20\% measurement of the cross-section, and therefore
a 10\% measurement of $|V_{tb}|$.  If on the other hand no signal
is seen in 3 fb$^{-1}$, then $|V_{tb}| <$ 0.60 at the 95\% confidence
level.  The authors of this study
also suggest that the Drell-Yan process may be more useful
in extracting $|V_{tb}|$ than the $W$-gluon fusion process.
This is because expectations for the cross-section for 
$p\bar{p} \rightarrow W^{*} \rightarrow t\bar{b}$ are well
understood, and can be also normalized to the observed rate of
$p\bar{p} \rightarrow W^{*} \rightarrow l\nu$.  In contrast,
the calculation of $W$-gluon fusion suffers from uncertainties
in the higher order corrections as well as in the input gluon density.

The CKM matrix element $|V_{tb}|$ can also be measured from
the ratio of branching ratios

$$\frac{Br(t \rightarrow Wb)}{Br(t \rightarrow Wq)}  =  
\frac{|V_{tb}|^{2}}{|V_{tb}|^{2}+|V_{ts}|^{2}+|V_{td}|^{2}}$$
$$~~$$
The $b$-tag can differentiate between $t \rightarrow Wb$
and $t \rightarrow Wd$ or $Ws$.  Therefore, 
this ratio can be measured 
by comparing the number of single- and double-$b$-tagged 
$t\bar{t} \rightarrow $ lepton $+$ jets events and by measuring
the tagging rate in $t\bar{t} \rightarrow $ dilepton events.
A preliminary analysis by the CDF collaboration finds (Yao, 1995):

$$\frac{Br(t \rightarrow Wb)}{Br(t \rightarrow Wq)}  =  
0.94 \pm 0.27 \pm 0.13$$
$$~~$$
and sets a (not-yet-very-interesting)
limit $|V_{tb}| >$ 0.022 at the 95\% confidence level.
The statistical sensitivity for this branching ratio
measurement is projected to be 3\% for a 
1 fb$^{-1}$ exposure (Amidei and Brock, 1996).

Top decays also provide a unique opportunity to test the
the structure of the charged weak current at the
$t \rightarrow Wb$ vertex.  Because the top quark is so heavy,
it is possible that new physics may manifest itself at the
$Wtb$ vertex.  The most general form of the $Wtb$ interaction
is (Kane, Ladinsky, and Yuan, 1992) :

$$L~~ =~~ \frac{g}{\sqrt{2}}~~ \biggl[ W^{-}_{\mu}\bar{b}\gamma^{\mu}
(f_{1}^{L}P_{-} + f_{1}^{R}P_{+})t~~
-~~\frac{1}{M_{W}} \partial_{\nu}W^{-}_{\mu}\bar{b}\sigma^{\mu\nu}
(f_{2}^{L}P_{-} + f_{2}^{R}P_{+})t \biggr]~~~+$$
$$~~~~~~~~~ \frac{g}{\sqrt{2}}~~ \biggl[  W^{+}_{\mu}\bar{t}\gamma^{\mu}
(f_{1}^{L*}P_{-} + f_{1}^{R*}P_{+})b~~
-~~\frac{1}{M_{W}} \partial_{\nu}W^{+}_{\mu}\bar{t}\sigma^{\mu\nu}
(f_{2}^{R*}P_{-} + f_{2}^{L*}P_{+})b \biggr]$$
$$~~$$
where $P_{\pm} = \frac{1}{2}(1 \pm \gamma_{5})$ and
$i\sigma^{\mu\nu} = -\frac{1}{2}[\gamma^{\mu},\gamma^{\nu}]$.

The quantities $f_{1}^{L}$ and $f_{1}^{R}$ parametrize the strength
of the left-handed and right-handed weak charged current.  The
$f_{2}$'s can be interpreted as giving
rise to an anomalous weak magnetic moment.
In the Standard Model at tree level $f_{1}^{L} = 1$ and 
$f_{1}^{R} = f_{2}^{L} = f_{2}^{R} = 0$.  There is obviously 
no direct experimental
information on these form factors, although consistency with the
measured branching ratio for $b \rightarrow s\gamma$ constrains 
$f_{1}^{R}$ to be at most a few percent (Fujikawa and Yamada, 1994).

The polarization of the $W$ in the $t \rightarrow Wb$ decay probes
the values of the form factors.  Denoting the
left-handed, right-handed, and longitudinal polarization states of the
$W$ by $\lambda_{-}$, $\lambda_{+}$, and $\lambda_{0}$ respectively, the
expected relative polarizations of the $W$ boson after averaging
over the top and bottom polarization states, are 
(Kane, Ladinsky, and Yuan, 1992) :

$$\lambda_{-}~~ =~~ |f^{L}_{1} + \beta f^{R}_{2}|^{2}$$
$$\lambda_{+}~~ =~~ |f^{R}_{1} + \beta f^{L}_{2}|^{2}$$
$$\lambda_{0}~~ =~~ \frac{1}{2}|f^{R}_{2} + \beta f^{L}_{1}|^{2}~~ +~~~
\frac{1}{2}|f^{L}_{2} + \beta f^{R}_{1}|^{2}$$
$$~~$$
with $\beta = \frac{M_{top}}{M_{W}}$.  Therefore, the Standard Model
predicts $\lambda_{+} = 0$, and the fraction of
longitudinally polarized $W$-bosons in top decays is
$\frac{1}{2} \beta^{2}$.  These polarizations 
can be measured from the angle of emission of the lepton
in $W$-decays.  Studies indicate that the statistical accuracies
in the measurements of $\lambda_{0}$ and $\lambda_{+}$ in an
exposure of 1 fb$^{-1}$ at the Tevatron should be 5\% and 
2\% respectively (Amidei and Brock, 1996).
These form factors can also be probed by measuring the 
single top production cross-section (Carlson, Malkawi, and Yuan, 1994;
Malkawi and Yuan, 1994).

The longitudinal polarization state of the $W$ is directly 
connected with the breaking of electroweak symmetry, since
it arises from the Goldstone boson degree of freedom.
This kind of study therefore
provides a rather unique and particularly interesting
test of the Standard Model.

\subsection{Further tests of the Standard Model and searches for new
physics in the top quark sector}
\label{new}
The $p\bar{p} \rightarrow t\bar{t}$ cross-section can be calculated 
in QCD.  Its measurement tests the predictive power of QCD 
and is sensitive to new physics.  Considerable theoretical interest
in the subject was triggered by the initial measurement
of the $t\bar{t}$ cross-section by the CDF collaboration
(F. Abe {\em et al.}, 1994a).
The measured value was higher
than expected, although still consistent with the QCD
calculation, within the large experimental uncertainties.
The more recent higher statistics measurement by both
CDF and D0 are in better agreement with the
calculation
(see Fig.~\ref{xsec_d0_cdf}).

The precision of the $t\bar{t}$ cross-section measurement
depends on the accuracy of the luminosity normalization (3.5\%), the
background estimate, and the acceptance calculation.  
The uncertainty in the acceptance calculation is mostly
due to the uncertainty in the $b$-tag efficiency and
the uncertainties in the modeling of $t\bar{t}$ production (e.g.
the effects of gluon radiation).  
The high statistics $t\bar{t}$ data samples that will be collected
at the Tevatron will provide several handles to reduce this
uncertainty.  It is not clear what the ultimate systematic
uncertainty on the $t\bar{t}$ cross-section 
will be; our guess is that a precision
of 10\%, comparable to the uncertainty in the QCD calculation
of $t\bar{t}$ production, should be achievable.

\begin{figure}[htb]
\hskip 1.5cm
\epsffile{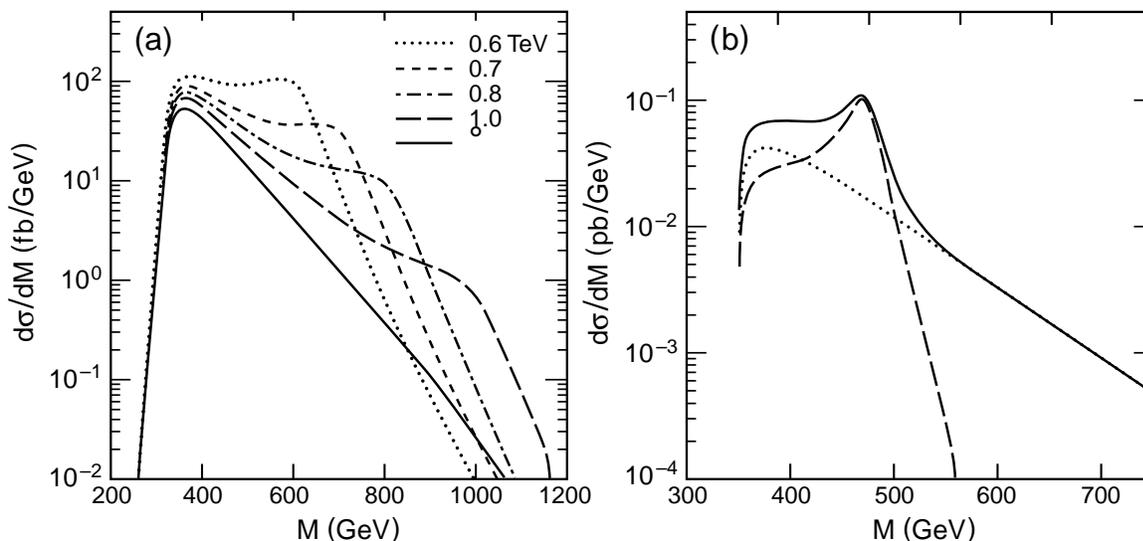}
\caption{\protect \baselineskip 12pt
Expected invariant mass of the $t\bar{t}$ pair : (a) 
including the contribution
of a color octet
vector meson in the top color model (from Hill and Parke, 1994). 
The different curves
show the expectation as the vector meson mass varies from 600 GeV/c$^{2}$
to infinity.
(b) including the
effect of a 475 GeV/c$^{2}$ color-octet technipion in multiscale
walking technicolor (from Lane, 1995). The dotted line is the QCD prediction;
the dashed line is the technipion contribution; and the solid line is 
the sum of the two.} 
\label{parke}
\end{figure}

Large enhancements to the $t\bar{t}$ cross-section and, more dramatically,
resonances in the $t\bar{t}$ invariant mass spectrum are expected
in a number of models, (see Fig.~\ref{parke}).  
These could be due to
color octet vector mesons (Hill and Parke, 1994) in models where
the electroweak symmetry breaking is realized via top condensation
(Hill, 1991; Martin, 1992a and 1992b), or to technipions (Eichten
and Lane, 1994) in multiscale models of walking technicolor
(Lane and Eichten, 1989; Lane and Ramana, 1991).

\begin{figure}[htb]
\epsfxsize=4.0in
\gepsfcentered[20 200 600 600]{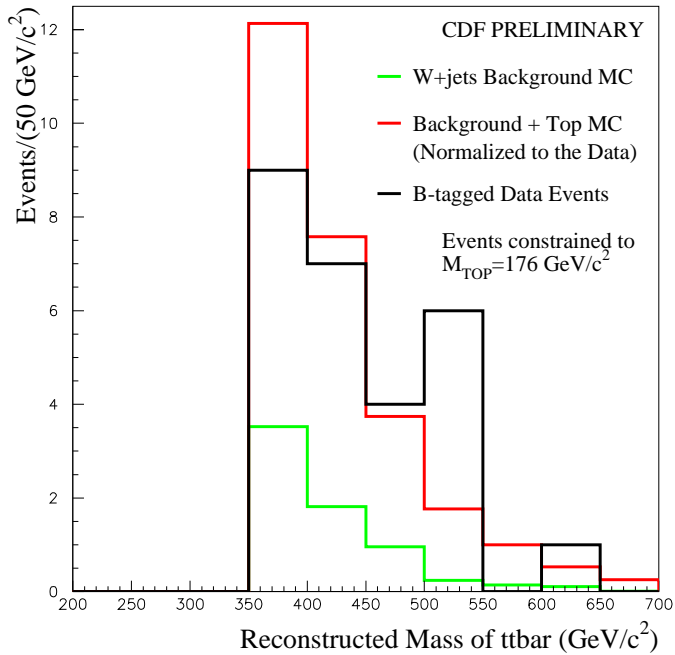}
\vskip 1cm
\caption{\protect \baselineskip 12pt
Reconstructed invariant mass of the $t\bar{t}$ pair compared to
the background and $t\bar{t}$ $+$ background expectations. From 
the CDF collaboration, Yao, 1995.  The integrated luminosity is
$\approx$ 100 pb$^{-1}$.}
\label{kirsten}
\end{figure}

Measurements of the $t\bar{t}$ invariant mass distributions
are possible in the lepton $+$ jets channel.  Preliminary
CDF (Yao, 1995) and D0 (Narain, 1996) analyses are
consistent with QCD expectations, (see Fig.~\ref{kirsten}).
With the high statistics that will become available, these 
models will be critically tested in the future.

\begin{figure}[htb]
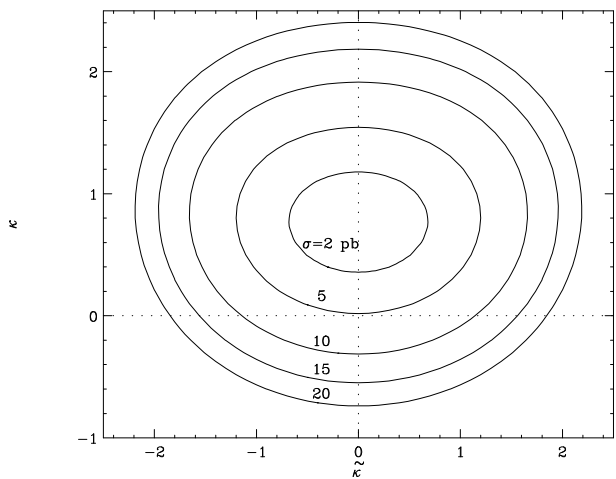

\epsfxsize=4.0in
\hskip 3cm
\gepsfcentered[0 200 612 600]{cheung.ps}
\caption{\protect \baselineskip 12pt
Contours of $t\bar{t}$ cross-section (M$_{top}$ = 176 GeV/c$^{2}$)
as a function the chromoelectric ($\tilde{\kappa}$) and
chromomagnetic ($\kappa$) dipole moments of the top quark.}
\label{cheung}
\end{figure}

Chromoelectric and chromomagnetic dipole moments of
the top quark affect the structure of the $t\bar{t}$-gluon vertex,
and hence affect the $t\bar{t}$ production cross-section and $t\bar{t}$
transverse momentum (Atwood, Kagan, and
Rizzo, 1994;  Rizzo, 1994; Cheung, 1995; Haberl, Nachtmann, and Wilch, 1995).
A top-quark chromomagnetic 
dipole moment can occur in composite and technicolor models, with
magnitude of order M$_{top}^{2}/\Lambda^{2}$, where $\Lambda$
is the characteristic scale for new physics.
As can be
seen from Fig.~\ref{cheung}, the present measurement of the $t\bar{t}$
cross-section is already accurate enough to probe the scale 
$\Lambda \approx$ 200 GeV.  A chromoelectric dipole moment would
be CP-violating, and could arise from large couplings between
top quarks and Higgs bosons in the multi-Higgs doublet model
(Atwood, Aeppli, and Soni, 1992;  Brandenburg and Ma, 1993;
Cheung, 1995; Haberl, Nachtman, and Wilch, 1995).

CP violation effects in $t\bar{t}$ production
would manifest themselves in different polarizations for
the $t$ and the $\bar{t}$ (Kane, Ladinsky, and Yuan, 1992;
Schmidt and Peskin, 1992, Kao, Ladinsky, and Yuan, 1994).
Because the top quark lifetime is short, the top decays before
hadronization and polarization information is preserved
in the $t \rightarrow Wb$ decay.  Assuming Standard Model
$Wtb$ couplings, the polarization of the top quark is then
analyzed by the polarization of the $W$ from the top decay.  Examples of
CP violating observables that can be studied 
are differences in rates between $t_{L}\bar{t}_{L}$
and $t_{R}\bar{t}_{R}$, or $t_{L}\bar{t}_{R}$
and $t_{R}\bar{t}_{L}$, where the subscripts $L$ and $R$ denote
left- and right-handed polarizations respectively.  

In the single top production process, top quarks are almost 100\%
longitudinally polarized, since they are produced through the 
weak interaction (Carlson and Yuan, 1993).
Searches for CP violation in the top quark decay can then also be 
carried out in the $t \rightarrow W^{+}b \rightarrow l^{+}\nu b$
decay by studying the quantity 
{$\bf \sigma$} $\cdot$ ({\bf P}$_{b}$ $\times$ {\bf P}$_{l}$), where
{\bf $\sigma$} is the top polarization vector 
(Kane, Ladinsky, and Yuan, 1992; Grzadowski and Gunion, 1992).

Precise measurements of polarization effects in the top
sector will however be rather difficult.  These measurements require very
high statistics and good control of the systematics.
Most likely, only very large asymmetries will be accessible
experimentally. 

\begin{figure}[htb]
\epsfxsize=4.0in
\gepsfcentered[20 200 600 600]{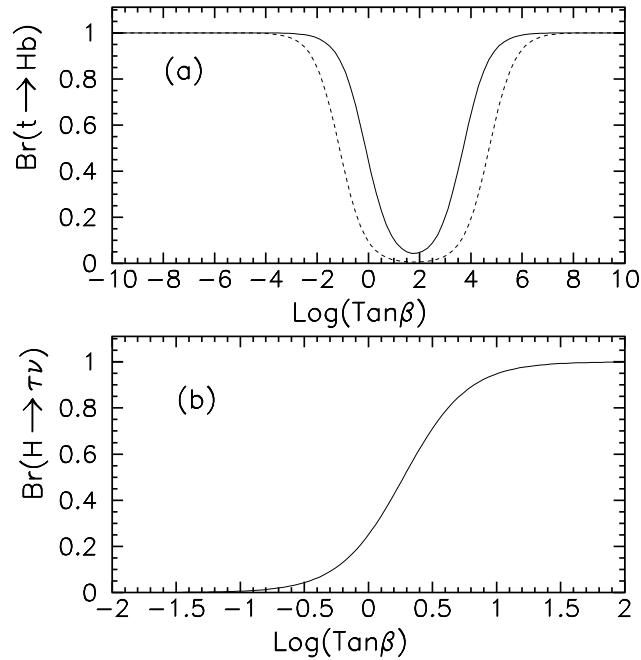}
\vskip 1cm
\caption{\protect \baselineskip 12pt
(a) Expected branching ratios of $t \rightarrow H^{+}b$ in the two Higgs
model for M$_{top} =$ 180 GeV/c$^{2}$
as a function of log(tan$\beta$). Solid line: M$_{H}$ = 70 GeV/c$^{2}$;
dashed line: M$_{H}$ = 150 GeV/c$^{2}$.
(b) Expected branching ratio for $H^{+} \rightarrow \tau\nu$ 
as a function of log(tan$\beta$).}
\label{higgs_rmp}
\end{figure}

Physics beyond the Standard Model can also give rise to exotic decays
of the top quark. 
The measurement of the ratio of branching ratios

$$R ~~~=~~~\frac{Br(t\bar{t} \rightarrow l + jets)}
{Br(t\bar{t} \rightarrow ll + jets)}$$
$$~~~~$$
is quite generally sensitive to decays different from $t \rightarrow WX$,
provided that the new top quark decay mode includes jets.  More
model-dependent searches for new top quark decays can also be carried
out.

One example of a non-Standard Model decay of the top quark is
the decay $t \rightarrow H^{+}b$, which can occur in models
with two Higgs doublets, see Sec~\ref{NSMdecay}.
The H$^{+}$ would then decay into the heaviest fermion pairs,
$c\bar{s}$ or $\tau\nu$.  The branching ratios depend on the ratio
of vacuum expectation values for the two doublets, tan$\beta$,
(see Fig.~\ref{higgs_rmp}).  If tan$\beta$ is large, the signature for
this decay mode would be an excess of 
lepton $+$ hadronic $\tau + b$-tag events; if tan$\beta$ is small,
one could search for the $H^{+} \rightarrow c\bar{s}$ peak
in the invariant mass distribution of lepton $+$ jets events.
Other possible exotic decays of the top quark which can be searched
for include flavor changing neutral current decays such as
$t \rightarrow Zc$ (Han, Peccei, and Zhuang, 1995) and 
$t \rightarrow \gamma c$, and
decays into a supersymmetric top quark ({\em stop}) and a neutralino
(Mrenna and Yuan, 1995).  Some preliminary limits on these FCNC
have been presented by CDF (LeCompte, 1995).

\section{Conclusion}

\begin{figure}[htb]
\vskip 1cm
\epsfxsize=4.0in
\epsfysize=2.0in
\gepsfcentered[20 200 600 600]{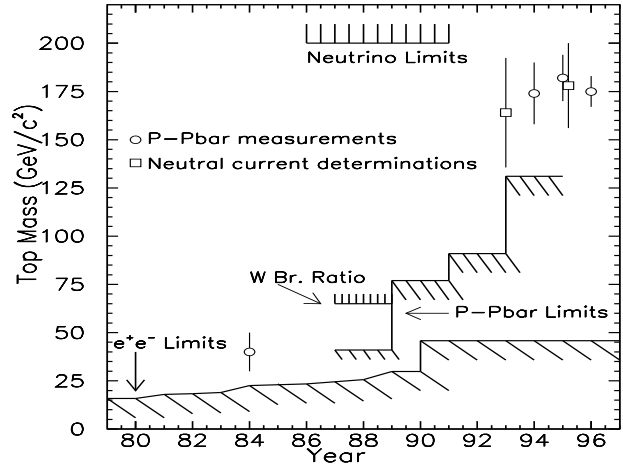}
\vskip 1cm
\caption{\protect \baselineskip 12pt
A brief history of the search for the top quark.  Here we show the 
limits from searches in $e^{+}e^{-}$ (Section VI.A) and
$p\bar{p}$ (Section VI.B).  The square data points are predictions
from neutral current studies, mostly at LEP and SLC (Section III).
We also show the limit from neutrino experiments that were obtained 
prior to the commissioning of LEP and SLC (Amaldi {\em et al.}, 1987).
The circular data points show the recent measurements
of the top mass (Section IX.A), as well as the UA1 1984 result
M$_{top} = 40 \pm 10$ GeV/c$^{2}$, which turned out to be wrong 
(Section VI.B).  Just for fun, we also show the {\bf upper} limit on 
the top mass from early measurements of the $W$ leptonic branching ratio at
the S$p\bar{p}$S (Section VI.C).  More recent measurements of this quantity at
both the Tevatron and the S$p\bar{p}$S are consistent with a high top quark
mass.}
\label{ykkim}
\end{figure}

The evidence for the existence of the top quark from the CDF
and D0 collaborations at the Tevatron
is persuasive.  The mass of the top quark
is M$_{top}$ = 175 $\pm$ 8 GeV/c$^{2}$, in agreement
with expectations from precision electroweak measurements.
This mass is a factor of forty higher than the mass of the
second heaviest fundamental fermion, and is of the same order
of magnitude as the scale for electroweak symmetry breaking.
The high mass of the top quark is a somewhat of a surprise;
in Fig.~\ref{ykkim} we show the evolution of the top mass 
limits and measurements
since the discovery of the companion $b$-quark.
It is interesting to notice that theoretical arguments based
on local supersymmetry from the early eighties, when the experimental
lower limit on the top quark mass was only
approximately 20 GeV/c$^{2}$, favored a rather high top mass, see 
Section~\ref{role}.
It is however still far from clear whether the high value of the
top mass is an accident, or a consequence of physics
at a higher mass scale.

The properties of the top quark will be studied much more precisely
at the upgraded Tevatron starting in 1999.  There is a possibility 
that effects beyond the Standard Model will manifest themselves in
the top sector.  If that is the case, the CDF and D0 collaborations
are well positioned to observe them.  In the more distant future,
the LHC $pp$ collider at CERN, with the higher energy and luminosity, 
will serve as a $t\bar{t}$ factory.

\acknowledgments

First of all, we wish to thank our CDF colleagues with whom
we shared the excitement of the hunt for the top quark.
A number of people helped us to write this review.
N. Hadley, B. Klima, M. Narain, 
S. Protopopescu, and S. Snyder
explained many aspects
of the D0 top analysis.  E. Laenen kindly provided
us with a modern
calculation of the $t\bar{t}$ cross-section at
S$p\bar{p}$S energies.  A. Blondel helped us understand the implications
of the very precise neutral current measurements.
We also benefited from discussions
with D. Amidei, P. Giromini, 
A. Heinson, S. Holmes, D. Kestenbaum, Y.K.Kim, J. Konigsberg,
J. Kroll, M. Kruse, M. Mangano, M. Mannelli, 
and J. Polchinski.  We also wish to thank D. Ceder
and D. McLaren for assistance in the preparation of the manuscript.

\end{document}